\documentclass{tADP2e}
\usepackage{epstopdf}
\usepackage{subfigure}
\usepackage{hyperref}
\usepackage{color,xcolor}
\usepackage{multirow}
\usepackage{slashbox}
\usepackage{rotating}
\theoremstyle{plain}

\theoremstyle{definition}

\theoremstyle{remark}

\begin{document}
\jvol{64} \jnum{519-626} \jyear{2015} \jmonth{December}
\articletype{REVIEW ARTICLE}
\title{Multiferroic Materials and Magnetoelectric Physics:\\
Symmetry, Entanglement, Excitation, and Topology}

\author{\name{Shuai Dong,\textsuperscript{1}$^{\ast}$\thanks{$^\ast$Email: sdong@seu.edu.cn} Jun-Ming Liu,\textsuperscript{2}$^{\dag}$\thanks{$^\dag$Email: liujm@nju.edu.cn} Sang-Wook Cheong,\textsuperscript{3}
and Zhifeng Ren\textsuperscript{4}} \affil{\textsuperscript{1}Department of Physics \& Jiangsu Key Laboratory of Advanced Metallic Materials, Southeast University, Nanjing 211189, China\\\textsuperscript{2}Laboratory of Solid State Microstructures \& Collaborative Innovation Center of Advanced Microstructures, Nanjing University, Nanjing 210093, China\\\textsuperscript{3}Rutgers Center for Emergent Materials \& Department of Physics and Astronomy, Rutgers University, New Jersey 08854, USA\\ \textsuperscript{4}Department of Physics \& TcSUH, University of Houston, Houston, Texas 77204, USA} \received{October 2015}}

\maketitle

\begin{abstract}
Multiferroics are those materials with more than one ferroic order, and magnetoelectricity refers to the mutual coupling between magnetism (spins and/or magnetic field) and electricity (electric dipoles and/or electric field). In spite of the long research history in the whole 20th century, the discipline of multiferroicity has never been so highly active as that in the first decade of the 21st century, and it has become one of the hottest disciplines of condensed matter physics and materials sciences. A series of milestones and steady progress in the past decade have enabled our understanding of multiferroic physics substantially comprehensive and profound, which is further pushing forward the research frontier of this exciting area. The availability of more multiferroic materials and improved magnetoelectric performance are approaching to make the applications within reach. While seminal review articles covering the major progress before 2010 are available, an updated review addressing the new achievements since that time becomes imperative. In this review, following a concise outline of the basic knowledge of multiferroicity and magnetoelectricity, we summarize the important research activities on multiferroics, especially magnetoelectricity and related physics in the last six years. We consider not only single phase multiferroics but also multiferroic heterostructures. We address the physical mechanisms regarding magnetoelectric coupling so that the backbone of this divergent discipline can be highlighted. A series of issues on lattice symmetry, magnetic ordering, ferroelectricity generation, electromagnon excitations, multiferroic domain structure and domain wall dynamics, and interfacial coupling in multiferroic heterostructures, will be re-visited in an updated framework of physics. In addition, several emergent phenomena and related physics, including magnetic skyrmions and generic topological structures associated with magnetoelectricity will be discussed. The review is ended with a set of prospectives and forward-looking conclusions, which may inevitably reflect the authors' biased opinions but are certainly critical.
\end{abstract}

\begin{classcode}
75.85.+t Magnetoelectric effects, multiferroics, 77.55.Nv Multiferroic/magnetoelectric films, 77.80.-e	 Ferroelectricity and antiferroelectricity, 75.70.Tj	Spin-orbit effects, 75.80.+q Magnetomechanical effects, magnetostriction
\end{classcode}


{\abstractfont\centerline{\bfseries Index of this review}\vspace{12pt}
\hspace*{-12pt}{1.}   Introduction - history, concepts, \& representative materials\\
\hspace*{7pt} {1.1.}  Why multiferroics are interesting\\
\hspace*{7pt} {1.2.}  Trajectory of magnetoelectricity \& multiferroicity\\
\hspace*{7pt} {1.3.}  Two representative materials\\
\hspace*{24pt}  {1.3.1.} BiFeO$_3$\\
\hspace*{24pt}  {1.3.2.} TbMnO$_3$\\
\hspace*{7pt} {1.4.}  Classifying multiferroics\\
\hspace*{7pt} {1.5.}  Motivation for an updated review\\
\hspace*{24pt}  {1.5.1.} Existing reviews\\
\hspace*{24pt}  {1.5.2.} Why do we need the current one?\\
{2.}    Phenomenological approaches\\
\hspace*{7pt} {2.1.}  Time-reversal \& space-inversion symmetries\\
\hspace*{7pt} {2.2.}  Magnetoelectric couplings based on order parameters\\
\hspace*{24pt}  {2.2.1.} Helicity\\
\hspace*{24pt}  {2.2.2.} Parity\\
\hspace*{24pt}  {2.2.3.} Other forms\\
\hspace*{7pt} {2.3.}  A unified model on magnetism-induced polarization\\
{3.}    Magnetoelectric coupling in single phases: from physics to materials\\
\hspace*{7pt} {3.1.}  Magnetoelectric coupling I: based on spin-orbit coupling\\
\hspace*{24pt}  {3.1.1.} Dzyaloshinskii-Moriya interaction\\
\hspace*{24pt}  {3.1.2.} Spin-dependent metal-ligand hybridization\\
\hspace*{7pt} {3.2.}  Magnetoelectric coupling II: based on spin-lattice coupling\\
\hspace*{24pt}  {3.2.1.} One dimensional Ising spin chain\\
\hspace*{24pt}  {3.2.2.} Two dimensional E-type antiferromagnets\\
\hspace*{7pt} {3.3.}  Magnetoelectric properties of some selected type-I multiferroics\\
\hspace*{24pt}  {3.3.1.} Hexagonal $R$MnO$_3$ \& $R$FeO$_3$\\
\hspace*{24pt}  {3.3.2.} $A_3M_2$O$_7$\\
\hspace*{24pt}  {3.3.3.} Fluorides\\
\hspace*{24pt}  {3.3.4.} non-$d^0$ perovskites\\
\hspace*{24pt}  {3.3.5.} Other typical type-I materials\\
\hspace*{7pt}  {3.4.}  Magnetoelectric properties of some selected type-II multiferroics\\
\hspace*{24pt}  {3.4.1.} $R$Mn$_2$O$_5$\\
\hspace*{24pt}  {3.4.2.} Orthoferrites \& orthochromites\\
\hspace*{24pt}  {3.4.3.} Quadruple perovskites\\
\hspace*{24pt}  {3.4.4.} Cu-based oxides\\
\hspace*{24pt}  {3.4.5.} Complex hexaferrites\\
\hspace*{24pt}  {3.4.6.} Organic molecules \& polymers\\
\hspace*{24pt}  {3.4.7.} Other typical type-II multiferroics\\
\hspace*{7pt}  {3.5.}  Magnetoelectric coupling III: multiple contributions\\
\hspace*{7pt}  {3.6.}  Thermodynamic formulations \& phase diagrams\\
\hspace*{7pt}  {3.7.}  Magnetoelectric excitations \& dynamics\\
\hspace*{24pt}  {3.7.1.} Electromagnons: spin-lattice coupling \textit{vs} spin-orbit coupling\\
\hspace*{24pt}  {3.7.2.} Dynamics of multiferroic domains in o-$R$MnO$_3$\\
{4.}    Magnetoelectric coupling in thin films and heterostructures: from science to devices\\
\hspace*{7pt}  {4.1.}  Strain-mediated magnetoelectricity in thin films\\
\hspace*{24pt}  {4.1.1.} Tetragonal BiFeO$_3$\\
\hspace*{24pt}  {4.1.2.} Strain driven ferromagnetic \& ferroelectric EuTiO$_3$\\
\hspace*{24pt}  {4.1.3.} Piezostrain tuning of magnetism\\
\hspace*{7pt}  {4.2.}  Carrier-mediated interfacial magnetoelectricity\\
\hspace*{24pt}  {4.2.1.} Spin-dependent screening, bonding, \& oxidization\\
\hspace*{24pt}  {4.2.2.} Electrical tuning of magnetocrystalline anisotropy\\
\hspace*{24pt}  {4.2.3.} Ferroelectric control of magnetic phases\\
\hspace*{24pt}  {4.2.4.} Ferroelectric-magnetic tunneling junctions\\
\hspace*{7pt}  {4.3.}  Electrically controllable exchange bias\\
\hspace*{7pt}  {4.4.}  Hybrid improper ferroelectricity in superlattices\\
\hspace*{7pt}  {4.5.}  Magnetoelectricity at domain walls\\
{5.}    Magnetoelectric topology: monopole \& vortex\\
\hspace*{7pt}  {5.1.}  Magnetic monopole\\
\hspace*{24pt}  {5.1.1.} Magnetoelectric response of monopole\\
\hspace*{24pt}  {5.1.2.} Monopole excitation in spin ices\\
\hspace*{7pt}  {5.2.}  Skyrmion\\
\hspace*{7pt}  {5.3.}  Ferroelectric domain vortex\\
\hspace*{7pt}  {5.4.}  Magnetoelectricity of topological surface state\\
{6.}    Summary \& perspective\\
\hspace*{7pt}  {6.1.}  Relook on multiferroicity \& magnetoelectricity\\
\hspace*{7pt}  {6.2.}  Novel multiferroics on the way\\
\hspace*{24pt}  {6.2.1.} BaFe$_2$Se$_3$: crossover with superconductor family?\\
\hspace*{24pt}  {6.2.2.} High temperature ferrimagnetic ferroelectrics by design?\\
\hspace*{24pt}  {6.2.3.} Ferromagnetic \& ferroelectric titanates\\
\hspace*{7pt}  {6.3.}  New microscopic mechanisms needed\\
\hspace*{7pt}  {6.4.}  New territories to explore\\
\hspace*{7pt}  {6.5.}  Any applications\\
}

\begin{flushright}
We do not do the possible\\
While we do do some impossible\\
When an impossible goes to the possible\\
We are then asked what the possible deserves for\\
(Written by Jun-Ming Liu)
\end{flushright}

\section{Introduction - history, concepts, \& representative materials}
\subsection{Why multiferroics are interesting}
\begin{figure}
\centering
\includegraphics[width=\textwidth]{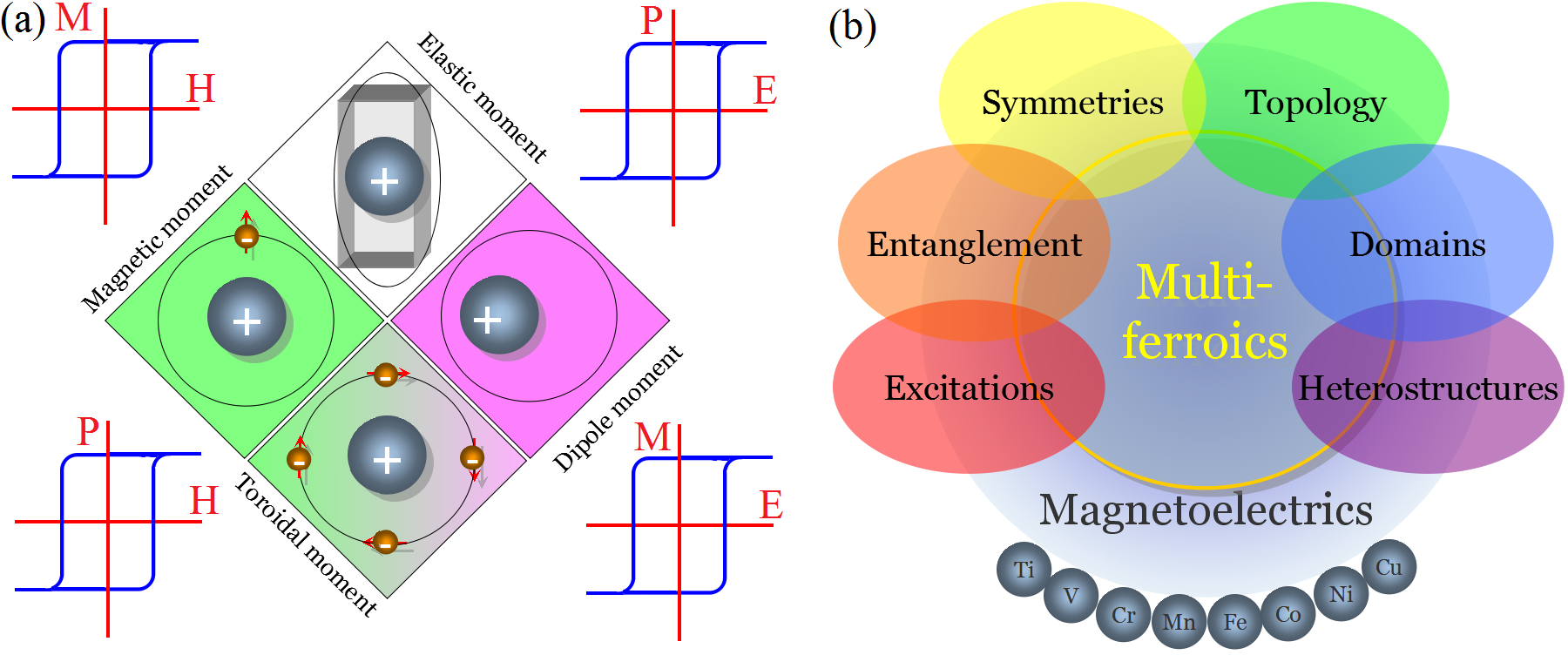}
\caption{(Colour online) Multiferroicity and magnetoelectricity. (a) A schematic of four primary ferroic moments and their microscopic origins. In general, a magnetic moment, which breaks the time-reversal symmetry, originates from unpaired electrons and thus corresponding partially occupied orbitals (usually $d$ or $f$ orbitals). In contrast, an electric dipole moment, which breaks the space-inversion symmetry, is a relative displacement of positive and negative charges. Such a displacement usually occurs due to the preference for a coordinate (covalent) bond between an ion (cation) with empty $d$ orbitals and another ion (anion) with fully occupied $p$ orbitals. An elastic moment, which breaks neither the time-reversal symmetry nor the space-inversion symmetry, comes from lattice distortions. A toroidal moment, which breaks both the time-reversal and space-inversion symmetries, can be visualized as a toroidal arrangement of magnetic moments. However, this intuitive interpretation seems to non-support the definition of toroidal moment as a primary ferroic order. An alternative scheme has been recently proposed, where a magnetic monopole is defined as the fourth ferroic moment (See Sec. 5.1.1. for more details). In the past decade, attention of the multiferroic/magnetoelectric community has been paid to the entanglement between magnetic and electric dipole moments. The four hysteresis loops, each as the most important characteristic of one ferroic-order, can be classified into two categories. The upper two loops are canonical: magnetization ($\textbf{M}$) switched by magnetic field ($\textbf{H}$) and polarization ($\textbf{P}$) switched by electric field ($\textbf{E}$). The lower two loops denote the magnetoelectric cross-controls, which are the desired targets for magnetoelectric research. (b) A grand garden of multiferroics and magnetoelectrics, where several important physical issues, sub-disciplines, and crossovers with other disciplines to be presented in this review, are cycled. The small spheres on the bottom denote the most-commonly involved elements (ions) in magnetoelectric materials, which cover the whole $3d$ family excepts the head and end.}
\label{multiferroic}
\end{figure}

As defined by Schmid in 1994, multiferroics refer to a big class of materials which exhibit simultaneously more than one primary ferroic order parameter in a single phase \cite{Schmid:Fe}. We illustrate the main features of four primary ferroic orders (moments) and their microscopic origins in Fig.~\ref{multiferroic}(a). The scope of this definition has been continuously extended to cover a much broader spectrum of relevant materials. As a successive hot topic in correlated electronic materials after high-$T_{\rm C}$ superconducting cuprates and colossal magnetoresistive manganites, multiferroics have attracted enormous attention in the past decade, due to their fascinating physics and applicable magnetoelectric functionalities. A variety of promising technological applications include energy transformation, signal generation and processing, information storage, and so on. Nevertheless, the main driving force for multiferroic research in the past decade has been triggered more by scientific curiosity rather than by application anticipation and investment.

So far most well studied multiferroic materials belong to complex transition metal compounds, and this fact enables multiferroics to be the third part in parallel to high-$T_{\rm C}$ superconducting cuprates and colossal magnetoresistive manganites in this big family. The difference lies in the fact that cuprates and manganites are limited to Cu-based and Mn-based oxides with perovskite or layered perovskite (Ruddlesden-Popper series) structures, but multiferroics cover a much broader spectrum of transition metal compounds including most $3d$ electronic transition metal systems (from Ti to Cu) with various structures and beyond oxides. Thus, on one hand, multiferroicity evidences substantial complexity without a doubt, challenging our conventional understanding of electrons in crystals since a huge package of materials with different structures and compositions are involved and plethoric novel physical phenomena are manifested. On the other hand, the physics of multiferroicity is fascinatingly unparalleled since these diverse materials share common physical rules, some of which are sketched in Fig.~\ref{multiferroic}(b).

Basically, magnetism and electricity are the two fundamental concepts in physics, and they co-setup the unified framework of electromagnetism via the four Maxwell equations. However, in crystals, magnetic moment and electric dipole are usually mutually exclusive, namely only one or neither of them is available in one single material. Similar physical phenomena in crystals include the mutual exclusion between magnetism and superconductivity. However, the discovery of high-$T_{\rm C}$ superconductivity in doped cuprates and iron-based materials breaks through this exclusion condition, in which the superconductivity always twists with magnetism. Similarly, the discovery of multiferroicity breaks the principle of exclusion between magnetic moment and electric dipole. Extensive research activities in the past decade have been enabling a comprehensive theoretical framework of multiferroicity to be gradually established, which can explain many existing experimental observations and lead to new predictions and experimental findings. In this sense, the past decade of multiferroicity can be considered as a successful example of exploring the physics of correlated electronic materials.

\subsection{Trajectory of magnetoelectricity \& multiferroicity}
The earliest study on magnetoelectric effect in crystals can be traced back to the 19th century. In 1894, based on the lattice symmetry argument, Pierre Curie predicted the possibility of an intrinsic magnetoelectric effect in some crystals. While the terminology ``magnetoelectric effect" was defined by Debye in 1926, the speculation remained inactive until 1960 when the first real magnetoelectric material Cr$_2$O$_3$ was discovered \cite{Astrov:Jetp}, following Dzyaloshinskii's generic prediction one year earlier \cite{Dzyaloshinskii:Jetp}.

In the whole 20th century, research on magnetoelectric physics and materials was quite slow. On one hand, available magnetoelectric materials were rare and inactive, and the observed magnetoelectric performance was poor. On the other hand, magnetoelectric theories were basically phenomenological, lacking ingredients of modern electronic theory based on quantum mechanics. Modern electronic theory of ferroelectricity was not established until 1990s \cite{Smith:Prb,Resta:Rmp}. Extensive research on correlated electronic materials including high-$T_{\rm C}$ superconducting cuprates from late 1980s \cite{Dagotto:Rmp94} and colossal magnetoresistive manganites in 1990s  \cite{Tokura:Bok} laid a good foundation for the advanced multiferroic research. In 1994, in a proceeding paper of MEIPIC2 (MagnetoElectric Interaction Phenomena In Crystals) Workshop \cite{Schmid:Fe}, Schmid coined a new terminology: multiferroics, which denotes the coexistence of multiple ferroic orders in a single phase material. In 2000, in a seminal paper entitled ``Why are there so few magnetic ferroelectrics?", Hill (Spaldin) clarified the intrinsic challenges for accommodating magnetism and ferroelectricity in single phase perovskite oxides \cite{Hill:Jpcb}, shedding a pessimistic prediction of this quiet discipline.

The long incubation period was finally over because of the two unexpected breakthroughs both occurred in 2003. The first one was the discovery of room-temperature large ferroelectric polarization in coexistence with notable magnetization in BiFeO$_3$ thin films \cite{Wang:Sci}, which stimulated numerous subsequent investigations on BiFeO$_3$ bulks, films, and heterostructures. Although the reported strong magnetization was later found to be non-intrinsic \cite{Eerenstein:Sci,Wang:Rep}, it was the first time to obtain a single phase magnetoelectric compound offering excellent multiferroic performance with potential room temperature applications. The second material is the orthorhombic TbMnO$_3$, which has only a weak polarization ($\sim0.1\%$ of that of BiFeO$_3$) developed only in low temperature (below $28$ K) \cite{Kimura:Nat}. In spite of its poor performance in terms of ferroelectric polarization and magnetism, TbMnO$_3$ represents a milestone material which offers intrinsically a strong magnetoelectric coupling, giving a $100\%$ polarization flip driven by a magnetic field of several Tesla. We will review these two unique materials later.

In the next year, another two interesting multiferroic materials were discovered: orthorhombic TbMn$_2$O$_5$ \cite{Hur:Nat} and hexagonal HoMnO$_3$ \cite{Lottermoser:Nat}. Similar to TbMnO$_3$, TbMn$_2$O$_5$ also demonstrated a strong magnetoelectric coupling, giving a switchable polarization upon magnetic field, although its ferroelectricity is poor too (low ferroelectric Curie temperature of $38$ K and small polarization of $\sim0.04$ $\mu$C/cm$^2$). In contrast, the magnetoelectric behavior of hexagonal HoMnO$_3$ is highly appreciated due to its high ferroelectric Curie temperature (up to $875$ K) although its antiferromagnetic N\'eel temperature is relatively low ($75$ K). The magnetoelectric manifestation was evidenced with electro-control of antiferromagnetic domains. In fact, as early as 2002, the coupling between antiferromagnetic domains and ferroelectric domains was observed in hexagonal YMnO$_3$ \cite{Fiebig:Nat}, a cousin of hexagonal HoMnO$_3$,  using the second harmonic generation technique.

Since then, the activities to this exciting discipline have been enormous, featured by reports of a huge number of novel materials and magnetoelectric phenomena, revolutionary understanding of the microscopic mechanisms, and gradual establishment of a new framework of multiferroicity, thus making multiferroicity to be understandable and predictable in a microscopic quantum level, much different from the phenomenological scenario.

\begin{figure}
\centering
\includegraphics[width=\textwidth]{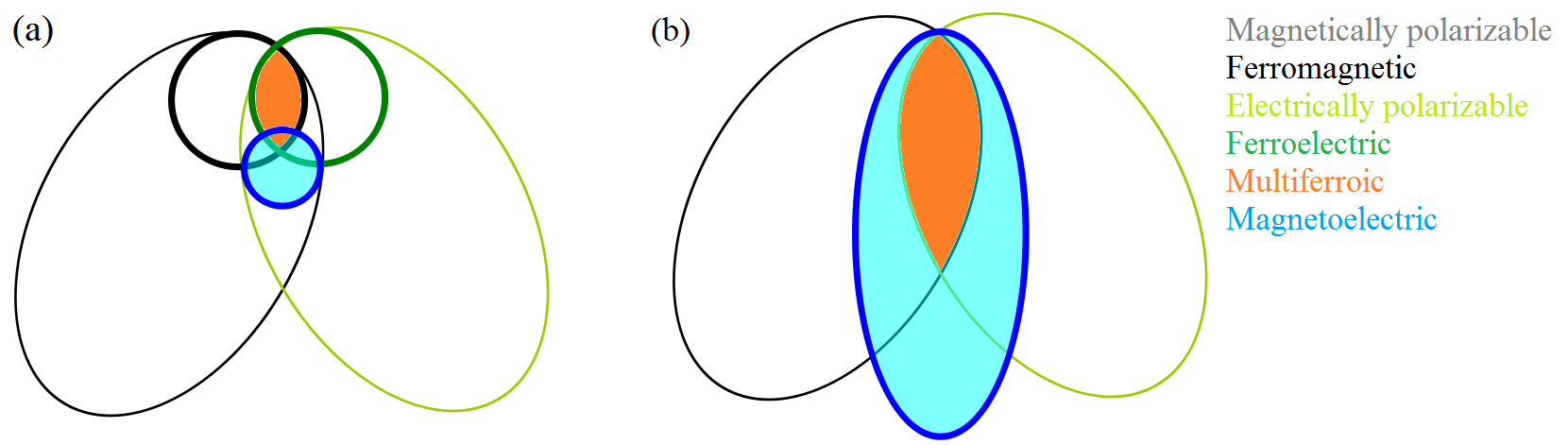}
\caption{(Colour online) A schematic landscape of multiferroic and magnetoelectric materials. (a) The version proposed by Eerenstein, Mathur, and Scott in 2006 \cite{Eerenstein:Nat}, where only the ferromagnetic + ferroelectric materials are called multiferroics. (b) An updated version (according to the authors' biased viewpoint), where the multiferroic zoo contains more members, not limited to the ferromagnetic + ferroelectric materials. For example, if a material is antiferromagnetic and ferroelectric, it can be certainly considered as a multiferroic. Magnetoelectric materials, which show magnetoelectric response/coupling, cover an even broader spectrum beyond ferroic materials.}
\label{MME}
\end{figure}

In concomitant with the research progress, the essence and extension of multiferroics and magnetoelectrics (also multiferroicity and magnetoelectricity) have been in gradual development. In the beginning, terminology ``multiferroic" in the narrow sense defines a material exhibiting both ferromagnetism and ferroelectricity, as schematically shown in Fig.~\ref{MME}(a). Nowadays, ``multiferroic" as a well-accepted concept can cover a material with one antiferroic-order plus another ferrioic-order, and ``multiferroicity" refers to the coexistence of these ferroicities, as illustrated in Fig.~\ref{MME}(b) where the multiferroic/multiferroicity territory has been substantially expanded. In fact, most type-II multiferroics to be addressed in this article are antiferromagnetic and ferroelectric. Another terminology ``magnetoelectric" (including magnetoelectric coupling, magnetoelectric response, and magnetoelectricity etc), once popular in the multiferroic community, has been much more acknowledged nowadays, and its inclusion becomes even broader. First, magnetoelectricity can be achieved at an interface between two non-multiferroic materials constituting a heterostructure. Second, some non-multiferroic single phase compounds may exhibit magnetoelectric response. In fact, the first discovered magnetoelectric material, Cr$_2$O$_3$, is not a multiferroic. Third, magnetoelectricity may emerge, e.g. as a consequence of the surface state of a topological insulator which usually has no magnetism or electric dipoles. Therefore, ``magnetoelectric" as a terminology covers a much broader spectrum than ``multiferroic" does, as schematically illustrated in Fig.~\ref{MME}(b). In this review, the two terminologies will be extensively and crosswise used, following the above definitions.

\subsection{Two representative materials}
As aforementioned, BiFeO$_3$ and TbMnO$_3$ were the first two multiferroics leading to revival of the current excitement of magnetoelectricity in the 21st century. The two perovskites, belonging to different types of multiferroics, have been extensively studied not only for their magnetoelectric performance but also more for their representativeness leading to new physics of multiferroicity. A number of experimental findings and relevant physical models stemming from them have been proven to be of broad generality applicable to other multiferroics. A parallel revisiting of the two materials and relevant findings would be helpful for a generic roadway on how physics proceeds from a particular to a general scenario of multiferroicity.

\subsubsection{BiFeO$_3$}
\begin{figure}
\centering
\includegraphics[width=0.85\textwidth]{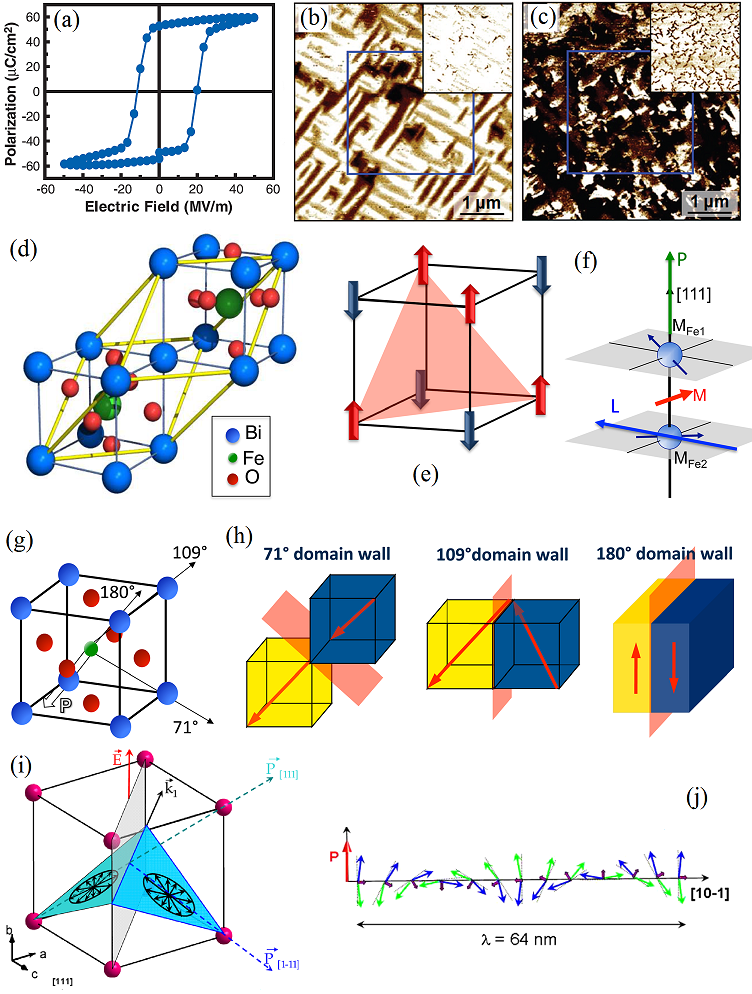}
\caption{(Colour online) Multiferroic properties of BiFeO$_3$. (a) A typical room-temperature ferroelectric hysteresis measured along the pseudo-cubic [001]-axis. The full polarization, pointing along the pseudo-cubic [111]-axis, should be $\sqrt{3}$ times larger. (a) From \href{http://dx.doi.org/10.1126/science.1080615}{J. Wang \textit{et al.}, Science, 299, pp. 1719-1722, 2003} \cite{Wang:Sci}. Reprinted with permission from the American Association for the Advancement of Science. (b-c) Room-temperature piezo-response force microscopy images of ferroelectric domains. The images are from the in-plane signals and the inserts are from the out-of-plane signals. The domains in (b) are stripe-like with the $71^\circ$ domain walls, while the domains in (c) are mostly mosaic-like with considerable amount of the $109^\circ$ and $180^\circ$ domain walls. (b-c) Reprinted with permission from \href{http://dx.doi.org/10.1021/nl801391m}{L. W. Martin \textit{et al.}, Nano Letters, 8, pp. 2050-2055, 2008} \cite{Martin:Nl}. Copyright \copyright (2008) by the American Chemical Society. (d) A schematic of rhombohedral structural frame (yellow) and pseudocubic frame (blue). (e) The G-type antiferromagnetic pattern with arrows for the Fe magnetic moments. The red shadow marks the pseudo-cubic (111) plane. (f) The geometrical (perpendicular) relationship among three vectors: polarization ($\textbf{P}$) along the pseudo-cubic [111] axis, antiferromagnetic order parameter ($\textbf{L}$), and canted magnetization ($\textbf{M}$). (g) A schematic of the equivalent axes of polarizations and their inclined angles. (h) A schematic of the three-types of domain walls. (d-h) Reprinted with permission from \href{http://dx.doi.org/10.1063/1.4870957}{J. T. Heron \textit{et al.}, Applied Physics Reviews, 1, p. 021303, 2014} \cite{Heron:Apr}. Copyright \copyright (2014) by the American Institute of Physics. (i) A schematic of the spin rotation and cycloidal vector ($\textbf{k}_1$). Here two polarization domains with a domain wall (light gray plane) are illustrated. (j) A schematic of the antiferromagnetic circular cycloid  along the pseudo-cubic [10-1] axis, whose period is about $64$ nm. (i-j) Reprinted with permission from \href{http://dx.doi.org/10.1103/PhysRevLett.100.227602}{J. T. Lebeugle \textit{et al.}, Physical Review Letters, 100, p. 227602, 2008} \cite{Lebeugle:Prl}. Copyright \copyright (2008) by the American Physical Society.}
\label{BiFeO3}
\end{figure}

The room-temperature crystal structure of bulk BiFeO$_3$ is a rhombohedral perovskite, in which the oxygen octahedra rotate alternatively along the pseudo-cubic [111]-axis (i.e. the $a^-a^-a^-$ type distortion in the Glazer notation \cite{Glazer:Acb}), as shown in Fig.~\ref{BiFeO3}(d). Although BiFeO$_3$ as a magnetoelectric material with high critical temperatures ($T_{\rm C}\sim1103$ K for ferroelectricity and $T_{\rm N}\sim643$ K for antiferromagnetism, both of which are above room temperature) was reported long time ago, its ferroelectric polarization was quite small (i.e. $\sim6$ $\mu$C/cm$^2$ for single crystal) and its magnetization was also very weak at that time \cite{Teague:Ssc}. However, these facts were completely changed in 2003 by Ramesh group in Maryland at that time, and Wang \textit{et al.} observed a residual polarization as large as $P_{\rm r}\sim55$ $\mu$C/cm$^2$ along the [001] direction (as shown in Fig.~\ref{BiFeO3}(a-c)) in the high-quality epitaxial BiFeO$_3$ thin films deposited on SrTiO$_3$ substrate \cite{Wang:Sci}. Meanwhile, a large magnetization was also claimed, which can reach $\sim0.5-1.0$ $\mu_{\rm B}$/Fe. The large spontaneous polarization, strong magnetization, and high critical temperatures, are almost all the required properties for practical applications, stimulating the research on every aspect of BiFeO$_3$ as a multiferroic in both bulk and film/heterostructure forms in the subsequent years till today. Its large ferroelectric polarization has been confirmed to be intrinsic \cite{Neaton:Prb}, in spite of the questions on the nature of magnetization \cite{Eerenstein:Sci,Wang:Rep}.

Different from thin film BiFeO$_3$, bulk BiFeO$_3$ (in both ceramic and single crystal forms) has been for a long time much less interesting due to the notorious leakage caused by nonstoichiometry of species and high density of defects. Substantial efforts have been devoted to synthesis trials which can produce stoichiometric and defect-free samples, including the rapid liquid phase sintering \cite{Wang:Apl}. This issue has not been resolved until the successful synthesis of high quality single crystals \cite{Choi:Sci}, and relevant measurements did confirm the large spontaneous polarization: $\sim60$ $\mu$C/cm$^2$ along the [001] direction and $\sim90-100$ $\mu$C/cm$^2$ along the [111] axis or its equivalent axes (see Fig.~\ref{BiFeO3}(g-h)), which made BiFeO$_3$ competitive and a superstar multiferroic.

It is nowadays known that the large polarization of BiFeO$_3$ mainly comes from the Bi$^{3+}$ ion, but additional source associated with the Fe$^{3+}$ spin order is still under debate. The $6s^2$ lone pair of Bi$^{3+}$, just like the $d^0$ orbitals, has a strong tendency to form the coordinate bond along one direction. This $6s^2$ lone pair mechanism for ferroelectricity is not new and similar case can be found in well known normal ferroelectric PbTiO$_3$. As for magnetism of BiFeO$_3$, it is now clear that the Fe$^{3+}$ is in the high spin state, giving a local moment of $\sim5$ $\mu_{\rm B}$/Fe. The spins form a G-type antiferromagnetic order (as shown in Fig.~\ref{BiFeO3}(e)) below $T_{\rm N}\sim643$ K, namely all the nearest-neighbor moments are antiparallel. However, an ideal G-type antiferromagnetism does not show any nonzero net magnetization, a weakness for applications. Interestingly, careful investigations revealed a spiral modulation superposed onto the G-type antiferromagnetic order as shown in Fig.~\ref{BiFeO3}(f). This modulation lasts a spatial period as long as $62-64$ nm (Fig.~\ref{BiFeO3}(i-j)) \cite{Sosnowska:Jpc,Przenioslo:Jpsj,Lebeugle:Prl} due to the accumulation of Fe$^{3+}$ spin canting \cite{Ederer:Prb}. This spiral modulation can be suppressed by spatial confinements, as found in thin films and nanostructures \cite{Gao:Am}, giving rise to a weak magnetization as a result of the Dzyaloshinskii-Moriya (DM) interaction. It is implied that a large magnetization may not be expected in spite of a large number of reports on the ferromagnetism in BiFeO$_3$. The G-type antiferromagnetic alignment plus the spiral spin canting of BiFeO$_3$ allows opportunities for interfacial magnetic coupling in multiferroic heterostructures, where BiFeO$_3$ plays several roles as ferroelectric substrate, antiferromagnetic pinning layer for exchange bias, interfacial quantum modulation donor, and so on. These are the reasons for BiFeO$_3$ to be the best multiferroic material so far.

In fact, in a long period even after 2003, BiFeO$_3$ was the only multiferroic showing both the magnetism and ferroelectricity above room temperature. Together with its large ferroelectric polarization, BiFeO$_3$ is no doubt the first superstar in the multiferroics family. Besides, BiFeO$_3$ has been studied for its photovoltaic effect \cite{Choi:Sci}, photocatalytic effect \cite{Gao:Am}, and fascinating role in domain wall nanoelectronics \cite{Catalan:Rmp}, all of which are associated with its prominent ferroelectricity. More physical mechanisms involved in BiFeO$_3$ and its heterostructures will be introduced in the following sections. Readers can also refer to several excellent topical reviews, e.g. Ref.~\cite{Catalan:Am} for bulk,  Ref.~\cite{Heron:Apr} for heterostructures, Refs.~\cite{Sando:Jpcm,Yang:Armr} for thin films and devices.

\subsubsection{TbMnO$_3$}
\begin{figure}
\centering
\includegraphics[width=0.95\textwidth]{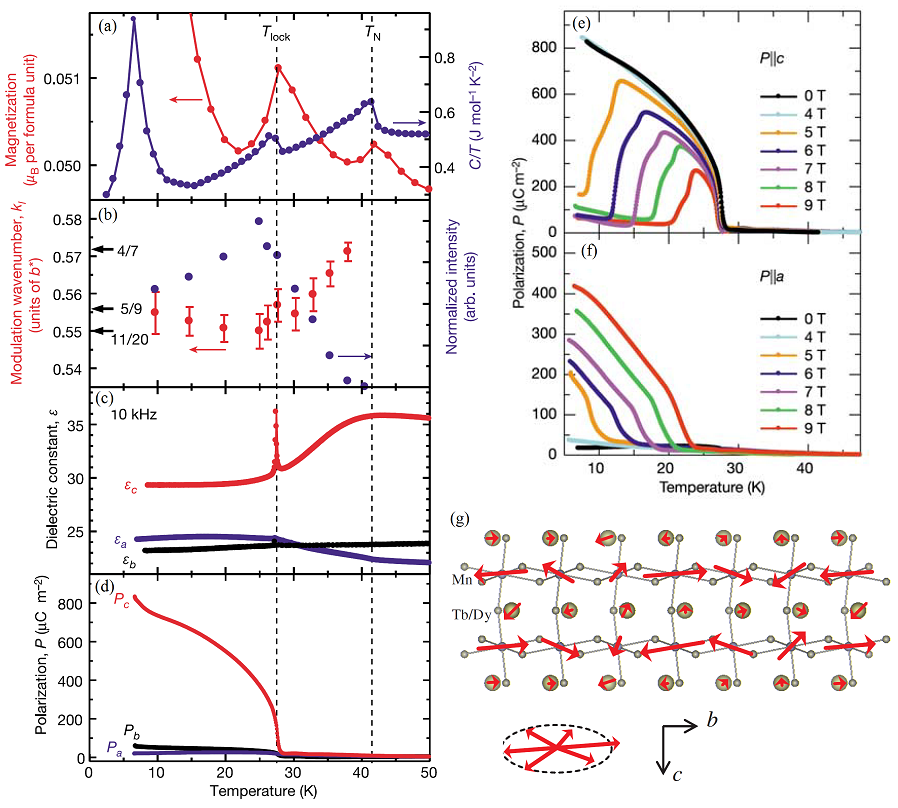}
\caption{(Colour online) (a-f) Temperature dependent of multiferroic and other physical properties of TbMnO$_3$ single crystal. (a) Magnetization (left) and specific heat/temperature (right). Two transitions (at $T_{\rm N}$ and $T_{\rm lock}$) can be identified. (b) Incommensurate modulation of magnetic moments. With decreasing temperature, the magnetic signal appears since $T_{\rm N}$ and the wave number is (almost) fixed below $T_{\rm lock}$. (c) The dielectric constants along different crystalline axes. Only the one along the $c$-axis shows a sharp peak at $T_{\rm lock}$. (d) The pyroelectric polarization emerges below $T_{\rm C}$, with an additional tiny anomaly at $\sim 7$ K corresponding to the independent ordering of Tb$^{3+}$ magnetic moments. (e-f) The modulation of polarizations along the $c$-axis (e) and $a$-axis (f) upon increasing magnetic field applied along the $b$ axis. (a-f) Reprinted by permission from Macmillian Publishers Ltd: \href{http://dx.doi.org/10.1038/nature02018}{T. Kimura \textit{et al.}, Nature, 426, pp. 55-58, 2003} \cite{Kimura:Nat}. Copyright \copyright (2003). (g) A schematic of the Mn$^{3+}$ moment orders below $T_{\rm lock}$. (g) Reprinted with permission from \href{http://dx.doi.org/10.1103/PhysRevLett.96.097202}{T. Arima \textit{et al.}, Physical Review Letters, 96, p. 097202, 2006} \cite{Arima:Prl}. Copyright \copyright (2006) by the American Physical Society.}
\label{TbMnO3}
\end{figure}

As the other unique material discovered in 2003, TbMnO$_3$, another perovskite oxide with orthorhombic structure, has also attracted sufficient attention. An impulsive analog to compare TbMnO$_3$ and BiFeO$_3$ is that they are the two sides of a perfect multiferroic. In physics, TbMnO$_3$ may be superior to BiFeO$_3$. Instead, for promising applications, BiFeO$_3$ is much more powerful than TbMnO$_3$. TbMnO$_3$ only exhibits a weak ferroelectricity \cite{Kimura:Nat}: 1) low critical (Curie) temperature $\sim28$ K; 2) small polarization ($\sim0.06-0.08$ $\mu$C/cm$^2$), one thousand times smaller than that of BiFeO$_3$, as summarized in Fig.~\ref{TbMnO3}(a-d). In addition, the temperatures for magnetic ordering in TbMnO$_3$ are low, as shown in Fig.~\ref{TbMnO3}(a-b). The strong activities on TbMnO$_3$ is mainly due to its unbeatable significance of physics, and most of the fundamental findings with the so-called type-II multiferroics (definition to be given below), conceptually different from BiFeO$_3$ which belongs to the type-I multiferroics, were originated from TbMnO$_3$.

As revealed in Kimura \textit{et al.}'s experiment \cite{Kimura:Nat}, TbMnO$_3$ becomes antiferromagnetically ordered below temperature $T_{\rm N}\sim40$ K (Fig.~\ref{TbMnO3}(a-b)). This antiferromagnetism is quite complex, with a sinusoidal-type of modulation of the $b$-axis components of Mn$^{3+}$ magnetic moments. The modulation period is incommensurate to the lattice constant and decreases with decreasing temperature \cite{Arima:Prl}. At temperature $T_{\rm lock-in}=28$ K, this incommensurate modulation is locked-in and the sinusoidal-type of modulation turns to be a cycloid spiral in the $b-c$ plane (Fig.~\ref{TbMnO3}(g)), as revealed by neutron studies \cite{Kenzelmann:Prl}. Further reducing temperature leads to the independent Tb$^{3+}$ spin ordering below temperature $T_{\rm Tb}\sim7-8$ K, which is even more complex and has not been well understood so far. The most interesting property of TbMnO$_3$ is the emergence of a ferroelectric polarization right below $T_{\rm lock-in}$, coinciding with the cycloid spiral ordering (Fig.~\ref{TbMnO3}(c-d)). In the other words, the ferroelectricity shares the identical transition temperature with the magnetic ordering, implying an intrinsic entanglement between the spiral-type antiferromagnetism and ferroelectricity in TbMnO$_3$. A magnetic field up to several Tesla can tune the spiral plane from the $b-c$ plane to the $a-b$ plane, together with a simultaneous switching of the polarization from the $c$-axis to the $a$-axis as demonstrated in Fig.~\ref{TbMnO3}(e-f), further confirming the strong coupling between magnetism and ferroelectricity. This effect is absent in BiFeO$_3$-like multiferroics. Similar multiferroic behaviors were also found in isostructural DyMnO$_3$ and Eu$_{1-x}$Y$_x$MnO$_3$ \cite{Kimura:Prb05,Ishiwata:Prb}. The polarization of DyMnO$_3$, somehow larger ($\sim0.2$ $\mu$C/cm$^2$) than that of TbMnO$_3$, remains far smaller than the value of BiFeO$_3$. The $T_{\rm lock-in}$ of DyMnO$_3$, $\sim18$ K,  is even lower than that of TbMnO$_3$. For Eu$_{0.75}$Y$_{0.25}$MnO$_3$, the phase competition between the A-type antiferromagnetic phase with weak canted ferromagnetism and the spiral spin phase allows possibilities to cross-control of magnetization and polarization by electric and magnetic fields respectively \cite{Choi:Prl10}, paving a route towards colossal magnetoelectricity \cite{Argyriou:Phy}.

The details of physical mechanisms involved in TbMnO$_3$ and related systems will be discussed in the following sections. Readers can also consult to the excellent review by Kimura \cite{Kimura:Armr}. Here, we have intentionally discussed BiFeO$_3$ and TbMnO$_3$ as two representatives to introduce how the current research on multiferroics happened, noting again that BiFeO$_3$ and TbMnO$_3$ are indeed quite different from each other in terms of the origin of ferroelectricity and manifestation of magnetoelectric coupling. These discussions portray the physical motivations for a classification of multiferroics, foreshadowing the attempts in classifying multiferroics in the subsequent years by various researchers.

\subsection{Classifying multiferroics}
In 2007, Cheong and Mostovoy classified ferroelectrics into two families: 1) proper ferroelectrics; 2) improper ferroelectrics \cite{Cheong:Nm}. In proper ferroelectrics, the coordinate bonding between $3d^0$ metal ion (e.g. Ti$^{4+}$) and anion (e.g. O$^{2-}$) is believed to be the necessary ingredient for inducing ferroelectric polarization. However, in this case, the $d^0$ property usually implies the absence of magnetism, excluding a traditional ferroelectric to be a multiferroic. Another source for the proper ferroelectricity is the $6s^2$ lone pair, as found in BiFeO$_3$, PbTiO$_3$, and Pb(Fe$_{1/2}$Nb$_{1/2}$)O$_3$ \cite{Yang:Prb04}, where Bi$^{3+}$ or Pb$^{2+}$ contributes to the ferroelectric polarization but magnetism comes from the B-site transition metal like Fe$^{3+}$, which ensures the coexistence of multiple ferroic orders in a single material. For the improper ferroelectrics, the origin of ferroelectricity is no longer so traditional, which can be structural transition driven (the so-called geometric ferroelectricity), charge-ordering driven (the so-called electronic ferroelectricity), or magnetic ordering driven (the so-called magnetic ferroelectricity). Indeed, a number of improper ferroelectrics do have magnetism, while exceptional examples include non-magnetic geometric ferroelectrics such as (Ca,Sr)$_3$Ti$_2$O$_7$ discovered recently \cite{Oh:Nm}.

In 2009, Khomskii revisited the issue of multiferroic classification, and he proposed a classification of multiferroics into two categories \cite{Khomskii:Phy}. The type-II multiferroics refer to those magnetic ferroelectrics, in which ferroelectricity is induced by some specific magnetic orders, as evidenced in TbMnO$_3$. All the other multiferroics are called the type-I multiferroics, in which ferroelectricity does not have magnetic origin, with BiFeO$_3$ as an example although it does contain magnetic ion. Khomskii highlighted the physical significance of the type-II multiferroics, and acclaimed the discovery of the type-II multiferroics to be ``the biggest excitement nowadays". Both the type-I and type-II families can be further sub-classified according to the delicate difference in the origin of ferroelectricity.

Besides these two widely accepted classifications, other choices were once discussed. For example, also in 2009, Picozzi and Ederer proposed a partition of electronic magnetic ferroelectrics, a sub-class of multiferroics, into two types (spin-ordering driven and charge-ordering driven), according to the underlying driving forces of ferroelectric polarization \cite{Picozzi:Jpcm}. There exist some overlaps between the two types, and the spin-ordering driven family can be further partitioned into smaller classes. In principle, these classifications have no essential distinction and the origin of ferroelectricity is the core criterion.

\subsection{Motivations for the present review}
\subsubsection{Existing reviews}
In the past decade, a series of review articles have been published in concomitance with the significant progress of multiferroic materials and magnetoelectric physics. In 2005, Fiebig published his seminal review entitled ``Revival of the magnetoelectric effect" \cite{Fiebig:Jpd}, which presents an integrated introduction to the history and the latest progress till that time. As a leading review, it boosted the multiferroic research community which has been continuously expanding. At that time, our understanding of BiFeO$_3$ and TbMnO$_3$ etc was yet in the early stage, and magnetoelectric composites occupied the main characters of that review. The magnetoelectric physics was mainly discussed in the framework of phenomenological theory. In the same year, Prellier, Singh, and Murugavel published their review paper entitled ``The single-phase multiferroic oxides: from bulk to thin film" \cite{Prellier:Jpcm}, in which the available experimental results on BiMnO$_3$, BiFeO$_3$, $R$MnO$_3$, and $R$Mn$_2$O$_5$ at that time were outlined in details.

More attention to single phase multiferroics was given in the short review article full of wisdom by Eerenstein, Mathur, and Scott in 2006: ``Multiferroic and magnetoelectric materials" \cite{Eerenstein:Nat}. Both the single phase and composite materials were reviewed, focusing on the symmetry issue of multiferroic order parameters. Subsequently, in 2007, Cheong (one of the current authors) and Mostovoy published their well-known review article entitled ``Multiferroics: a magnetic twist for ferroelectricity" \cite{Cheong:Nm}. In this review, specific attention was paid onto magnetic ferroelectrics, complying with the substantial progress both theoretically and experimentally in 2004-2006. The keyword, as coined ``magnetic twist", stimulated researchers in a vivid and imaginative way. Meanwhile, Ramesh and Spaldin published another short but impressive review entitled ``Multiferroics: progress and prospects in thin films" \cite{Ramesh:Nm}, which was instead devoted to various aspects of multiferroic thin films (mainly BiFeO$_3$ thin films) besides magnetoelectric coupling based novel device physics and relevant issues. A comprehensive survey on all aspects of multiferroic researches was not available until 2009 when Wang and two of the current authors (J.M.L and Z.F.R) published their  review article entitled ``Multiferroicity: the coupling between magnetic and polarization orders" \cite{Wang:Ap}. In 2011, Velev, Jaswal, and Tsymbal published a review entitled ``Multi-ferroic and magnetoelectric materials and interfaces", covering both bulks and heterostructures \cite{Velev:Ptrsa}.

Besides the above mentioned several review papers, quite a few of excellent topical reviews focusing on particular branches of multiferrocity are available. For example, Kimura published two reviews: one on multiferroic spiral magnets \cite{Kimura:Armr} and one on magnetoelectric hexaferrites \cite{Kimura:Arcp}.  Tokura and his collaborators summarized the milestone progresses on multiferroics with spiral spin orders \cite{Tokura:Am,Tokura:Rpp14} and on electrical control of magnetism (which even goes beyond multiferroics) \cite{Matsukura:Nn}. Johnson and Radaelli wrote a review on diffraction studies of multiferroics \cite{Johnson:Armr}. Dong and Liu presented a short review on type-II multiferroic manganites \cite{Dong:Mplb}.

For magnetoelectric heterostructures as a specific branch of multiferroic discipline, several excellent reviews deserve for mention, including Vaz's article on electrical control of magnetism in multiferroic heterostructures \cite{Vaz:Jpcm}, Ramesh and collaborators' papers on multiferroic oxide thin films and heterostructures \cite{Yu:Ptrsa,Heron:Apr}, and Martin's papers on synthesis of multiferroic thin films \cite{Martin:Mse,Martin:Cossms}. A recent survey article on similar topics by Lu \textit{et al.} \cite{Lu:Apr} and a mini-review by Huang and Dong on ferroelectric control of magnetism and transport in oxide heterostructures \cite{Huang:Mplb} are also noted. Besides, there exist several interesting prospects which can give beginners a simple introduction of multiferroicity \cite{Spaldin:Sci,Tokura:Sci06,Spaldin:Pto}.

\subsubsection{Why do we need the current one?}
Given a number of already available review and survey articles, why do we need the current one? The answer is in fact simple. After more than ten years of booming development, this promising discipline has gradually entered into a new era, featured by several aspects. First, advanced investigations in the last six years since 2010 found more exciting materials and demonstrated many novel phenomena, laying a solid fundamental framework of multiferroic physics which was far more than what was done before 2010. Second, a number of individual and isolated findings before 2010 have been coherently linked from one and another by clarifying those uncertainties and discarding those improper observations. A logistically self-consistent description of multiferroic physics and materials science becomes possible, which even meets the boundaries with other condensed matter branches, e.g. superconductors and topological matters. Third, in parallel to these developments, a series of issues and challenges have been emerging, calling for new understanding of ``old" subfields in recent years, and some of them can be essential. These motivations enable a view of the whole discipline from a relatively high level. Along this line, an updated review of the major advancements since 2010 is definitely imperative.

The present review will be devoted to a physically sound platform on which most of those experimental and theoretical findings accomplished since 2003 will be laid into a common framework: multiferroicity, while major attention is paid to the contributions since 2010. Several satellite discoveries associated with magnetoelectric effect in various classes of materials, in particular some topological structures such as skyrmion, vortex, and specific topological domain structures etc, will be covered. Nevertheless, it should be noted that due to topical choice limitation, some branches of magnetoelectricity such as magnetoelectric composites dominated with interfacial strain mediated coupling and related magnetoelectric devices, is not considered. Readers who interest in these topics can find relevant details in several reviews \cite{Nan:Jap,Ma:Am,Wang:Mt,Hu:Mrs}. Besides, ferroelastic and ferrotoroidic materials are not covered either, and readers may consult to Salje's review on ferroelastic materials \cite{Salje:Armr}, and Spaldin \textit{et al.}'s brief review on toroidal moments \cite{Spaldin:Jpcm}, as well as the first paper on ferrotoroidic materials \cite{Aken:Nat}. We don't touch much of those magnetic materials with improper ferroelectricity driven by charge-ordering, e.g. LuFe$_2$O$_4$,\cite{Ikeda:Nat}, noting van der Brink and Khomskii's brief review on this issue \cite{Brink:Jpcm}. These choices are made more or less compelled to the authors' bias and knowledge, and the authors are responsible for the choices.

It should be mentioned that some overlapping of the present review with the previous articles is inevitable although duplicate description is avoided as far as possible. Several issues which have been covered by the previous reviews may be re-visited due to an alternative or updated understanding afterwards. The two examples are electromagnon excitations and new physics in hexagonal manganites. For accessing this review, some basic knowledge of quantum mechanics and elementary condensed matter physics are required. More than five hundred of references are cited here, covering the most relevant literature on multiferroicity and related topics. However, considering that there have been thousands of technical reports and papers (certainly even more) on this topic published in the past decade, it is likely that some important papers may have been unintentionally missed. We apologize in advance for these oversights, and encourage colleagues whose works are not mentioned to contact us, so that their contributions can be properly cited in our future publications if any.

\section{Phenomenological approach: symmetry \& coupling}
We start from phenomenological approaches incorporated with symmetry analysis to magnetoelectricity. This section is thus devoted to the Landau phenomenological theory. It goes beyond individual materials and scattered microscopic mechanisms, rendering the general applicability. Nevertheless, it should be mentioned that the Landau theory on magnetoelectricity has been continuously updated by incorporating more energy terms associated with newly found microscopic mechanisms, and thus there have emerged several different versions of this theory. One of the representatives is the comprehensive theory developed by Harris \cite{Harris:Prb}, which reasonably handles every aspect for a few typical materials \cite{Harris:Prb11,Harris:Prb12}. For a consideration of readability and generality, we will follow the Mostovoy's version \cite{Mostovoy:Prl} with proper extensions.

\subsection{Time-reversal symmetry \& space-inversion symmetry}
Symmetries are the most fundamental physical ingredients or mathematical characters, corresponding to particular invariant quantities under some transformations. The primary ferroic orders, e.g. ferroelectric order and ferromagnetic order, can be characterized according to their symmetries. The two basic symmetries associated with magnetism and ferroelectricity are time-reversal symmetry and space-inversion symmetry. The electric (charge) dipole $\textbf{P}$, defined as $\sum_iQ_i\textbf{r}_i$ where $Q_i$ denotes charge at position $i$, breaks the space-inversion symmetry due to the involved space vector $\textbf{r}$, but keeps the time-reversal symmetry. In contrast, the magnetic moment $\textbf{M}$, or namely spin $\textbf{S}$, breaks the time-reversal symmetry but keeps the space-inversion symmetry, because a magnetic moment can be expressed as $\sim\frac{dQ\textbf{r}}{dt}\times \textbf{r}'$ ($t$: time), in analogy to a current loop. Therefore, the charge dipole $\textbf{P}$ and magnetic moment $\textbf{M}$ associate different symmetries and thus are naturally independent.

Besides the ferroelectric order and ferromagnetic order, there exist two more primary ferroic orders \cite{Aken:Nat}: ferroelastic order which breaks neither the time-reversal symmetry nor the space-inversion symmetry \cite{Salje:Armr}; and ferrotoridal order $\textbf{T}\sim\textbf{M}\times\textbf{r}$ which breaks both the time-reversal symmetry and the space-inversion symmetry \cite{Spaldin:Jpcm}. One can find some more extended ferroic orders, e.g. antiferromagnetic order parameter, defined as $\textbf{L}=\textbf{M}_1-\textbf{M}_2$ where the suffixes $1$ and $2$ denote the sublattices. Here the time- and space-symmetry consensus constitutes the basis for Landau theory on magnetoelectricity.

\subsection{Magnetoelectric coupling based on order parameters}
The Landau theory is an elegant approach to formulate a general expression of continuous (i.e. second-order) phase transitions. The free energy near the phase transition points can be formulated as the Taylor expression of some order parameters. For example, for a ferromagnet/ferroelectic without external magnetic/electric fields, this formulation can be written as:
\begin{eqnarray}
\nonumber F_{m}&=&F_{m0}+\alpha_m\textbf{M}^2+\beta_m\textbf{M}^4+......,\\	
F_{p}&=&F_{p0}+\alpha_p\textbf{P}^2+\beta_p\textbf{P}^4+......,
\label{m2p2}
\end{eqnarray}
where $F_{m0}$ ($F_{p0}$) is the ``original" free energy without magnetism (polarization), $\alpha$ and $\beta$ are the Landau coefficients. The absence of odd power orders of $\textbf{M}$ and $\textbf{P}$ is required by the invariance of energy as a scalar quantity under both time-reversal and space-inversion operations. For example, a term like $\textbf{M}^3$ breaks the time-reversal symmetry, and thus $F_m$ will change if the time sequence is reversed, which is physically forbidden. If an external magnetic field $\textbf{H}$ (electric field $\textbf{E}$) is applied, terms like $\textbf{M}\cdot\textbf{H}$ ($\textbf{P}\cdot\textbf{E}$) does not violate the symmetry and thus is allowed in the free energy. The symmetry restriction makes the ``lowest" power order of magnetoelectric coupling term nothing other than $\textbf{P}^2\textbf{M}^2$. However, such a coupling term takes already the fourth power order, higher than the primary ferroic order term $\textbf{P}^2$ or $\textbf{M}^2$, implying that the magnetoelectric effect is indirect (e.g. via strains in magnetoelectric composites) and weak.

One of the great progresses in the past decade has gone beyond this conventional magnetoelectric coupling term. For example, although a single magnetic moment only breaks the time-reversal symmetry, a spatial collection of many magnetic moments, which forms specific magnetic orders $\textbf{M}(\textbf{r})$, can break the space-inversion symmetry and thus couple with ferroelectric order. Two representative magnetic orders will be discussed here in details, and others which have been addressed in literature will only be briefly mentioned. Basically, these specific magnetic orders can be classified into two categories: helicity and parity, although other possibilities can't be excluded.

\subsubsection{Helicity}
\begin{figure}
\centering
\includegraphics[width=\textwidth]{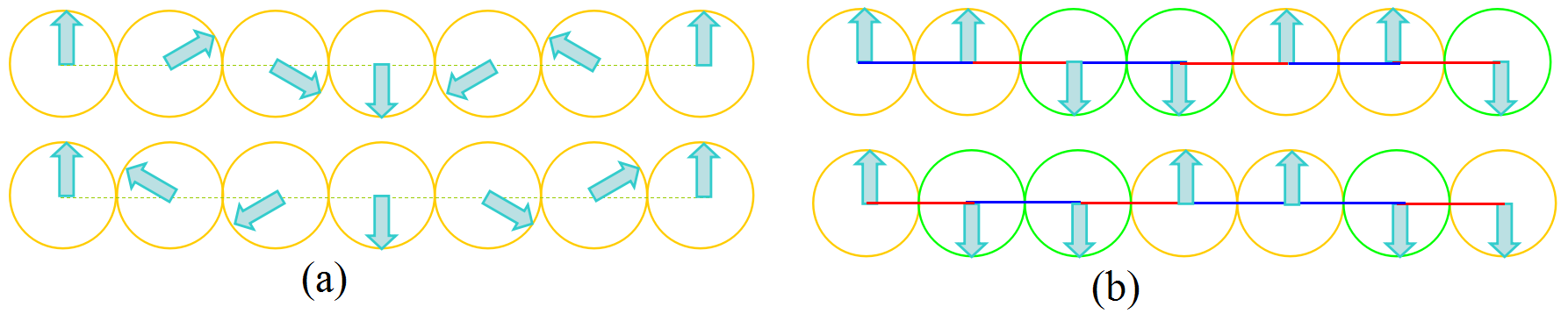}
\caption{(Colour online) A schematic of two specific magnetic orders, and both of them may generate ferroelectricity. The arrows denote magnetic moments. (a) A spin chain with spiral order: upper for clockwise helicity and lower for anticlockwise helicity. These helicities may break the space inversion symmetry. (b) An E-type antiferromagnetic order. If the one-unit-cell translational symmetry is broken, a parity can be defined to distinguish the upper and lower patterns.}
\label{spin}
\end{figure}

Helicity is a widely-existing phenomenon in nature. For example, morning glories in pattern always obey the right-hand rule, namely winding in the anti-clockwise direction. This right-hand rule applies to most winding plants with only a few exceptions. The winding helicity also exists in DNA structures, hurricanes, and solar systems. In a quantum world, if spins form a spiral order with a clockwise or anti-clockwise helicity, as schematically shown in Fig.~\ref{spin}(a), the magnetic ordering may break the space-inversion symmetry. A space inversion operation ($\textbf{r}\rightarrow-\textbf{r}$) may on the other hand alter the helicity between the clockwise and anti-clockwise rotations.

In 2006, Mostovoy proposed a phenomenological theory on the coupling between a noncollinear spiral spin order and ferroelectric polarization \cite{Mostovoy:Prl}. The free energy for a magnetoelectric coupling reads as:
\begin{equation}
F_{em}=\gamma\textbf{P}\cdot[\textbf{M}(\nabla\cdot\textbf{M})-(\textbf{M}\cdot\nabla)\textbf{M})+...],
\end{equation}
where $\gamma$ is the coefficient. Since the space derivative operation $\nabla$ breaks the space-inversion symmetry, neither space-inversion symmetry nor time-reversal symmetry will be violated in the free energy. This magnetoelectric coupling takes the cubic power order of ferroic order parameters, lower than the fourth order ($\textbf{P}^2\textbf{M}^2$). Therefore, the upper-limit of coefficient $\gamma$ may be large. To provide an evidence to this prediction, one considers the electro-static energy in the quadratic term of polarization $\textbf{P}^2/2\epsilon$ ($\epsilon$ is the dielectric constant) to be included in the total free energy. It is straightforward to obtain the induced polarization by minimizing the total energy, yielding:
\begin{equation}
\textbf{P}=\epsilon\gamma[\textbf{M}(\nabla\cdot\textbf{M})-(\textbf{M}\cdot\nabla)\textbf{M}].
\label{pmm}
\end{equation}

According to Eq.~\ref{pmm}, if all magnetic moments are uniformly aligned, namely for an ideal ferromagnetic order, the derivation of \textbf{M} is zero, giving zero polarization. If these moments are frustrated to be spatially modulated, Eq.~\ref{pmm} may give rise to a nonzero \textbf{P}. For example, one considers a noncollinear spiral/conical magnetic order which is expressed as:
\begin{equation}
\textbf{M}(\textbf{r})=M_x\cos(\textbf{Q}\cdot\textbf{r})\textbf{x}+M_y\sin(\textbf{Q}\cdot\textbf{r})\textbf{y}+M_z\textbf{z},
\label{spiral}
\end{equation}
where $\textbf{Q}$ is the propagation vector in reciprocal space and $\textbf{r}$ is the position vector in real space; $M_x$, $M_y$, and $M_z$ are the components of magnetic moment along the $\textbf{x}$, $\textbf{y}$, and $\textbf{z}$ axes, respectively. If $M_z=0$, a coplanar spiral order is expressed, otherwise a three-dimensional conical order with a net magnetic moment along the $\textbf{z}$ axis is obtained. An averaged polarization immediately obtained from Eq.~\ref{pmm} is:
\begin{equation}
<\textbf{P}>=\epsilon\gamma M_xM_y(\textbf{z}\times\textbf{Q}).
\label{pconical}
\end{equation}

This equation clearly indicates that a nonzero polarization perpendicular to the propagation vector $\textbf{Q}$ and the spiral axis $\textbf{z}$ can be generated in such a noncollinear spiral/conical spin order. In this sense, a cycloid spiral, with $\textbf{z}\bot\textbf{Q}$, allows a nonzero $<\textbf{P}>$ but a screw spiral, with $\textbf{z}||\textbf{Q}$, is incapable of doing so \footnote{A few exceptions will be discussed later.}.

This phenomenological scenario of cycloid spiral driven ferroelectricity explains reasonably experimental observations on TbMnO$_3$. As highlighted in Sec. 1.3.2, the Mn$^{3+}$ moments develop a cycloid spiral lying in the $b-c$ plane with a wave vector along the $b$-axis ($\textbf{Q}||b$), as shown in Fig.~\ref{TbMnO3}(e). According to Eq.~\ref{pconical}, the induced $\textbf{P}$ must be aligned along the $c$-axis, consistent with the experimental observations. The magnetic field driven polarization reversal can be understood as a consequence of the spiral plane rotation from the $b-c$ plane to the $a-b$ plane. For the $a-b$ plane spiral, $\textbf{Q}$ remains along the $b$-axis and thus the induced $\textbf{P}$ is aligned along the $a$-axis. In fact, TbMnO$_3$ is a representative case in which almost all the experimental observations can be well explained by this simple phenomenological model, including the electric/magnetic control of spin helicity \cite{Yamasaki:Prl,Murakawa:Prl}.

It is noted that this phenomenological model is basically material-independent but a quantitative agreement with experiments remains elusive, partially because of absence of relevant thermodynamic parameters. Surely, besides TbMnO$_3$, several other frustrated magnets with the spiral spin order also exhibit weak ferroelectric polarization, as summarized in Kimura's review \cite{Kimura:Armr}. The main features do fit qualitatively the model predictions. The success of this phenomenological model allows a sketch of an elegant and unified physics scenario possible, although these materials are actually very complex with distinctions from case to case.

An extension of this model can be made to a conical order which is certainly more interested than the coplanar spiral order. Although component $M_z$ does not explicitly appear in Eq.~\ref{pconical}, it provides a net magnetization along the $z$-axis, which is advantageous for multiferroic tuning by magnetic field \cite{Yamasaki:Prl06,Ishiwata:Sci}. The Zeeman energy $H_zM_z$ can tune the conical's spiral plane and the magnitudes of $M_x$ and $M_y$. There exist plenty of magnets, e.g. hexaferrites, which have non-coplanar conical order and thus induced polarization. This issue will be discussed later in Sec. 3.4.5 and readers may consult to Kimura's another review for details \cite{Kimura:Arcp}.

\subsubsection{Parity}
Besides the noncollinear spin orders, some special collinear spin orders can also break the space-inversion symmetry. Here, the key figure with these collinear orders is the parity instead of the helicity. Similar to helicity, parity is also an important concept in the symmetry consensus. A parity transformation denotes a point reflection, which changes vector $\textbf{r}$ to $-\textbf{r}$. In this sense, a polarization (electric dipole), is asymmetric under the parity operation. What can a spin order do here? We discuss a one-dimensional Ising spin chain as the simplest example. As shown in Fig.~\ref{spin}(b), all the spins form an ...$\uparrow\uparrow\downarrow\downarrow$... type order. For any site in this chain, its left and right neighbors are non-equivalent. In the other words, the parity symmetry is broken for a point reflection operation at any site. A parity-relevant order parameter ($\Psi_i$) for each site $i$ is defined as ($\textbf{M}_{i-1}-\textbf{M}_{i+1}$). Obviously, a time-reversal invariant can be constructed as:
\begin{equation}
\Omega=\Psi_1\cdot\Psi_4=\Psi_2\cdot\Psi_3=(\textbf{M}_1-\textbf{M}_3)\cdot(\textbf{M}_2-\textbf{M}_4).
\label{parity}
\end{equation}

One sees that a shift of spin pattern by one span, giving $\uparrow\downarrow\downarrow\uparrow$ or $\downarrow\uparrow\uparrow\downarrow$, reverses only one of $\Psi_2$ and $\Psi_3$ but not both. Clearly, the sign of $\Omega$ must be reversed by this shift operation but remains invariant upon the time-reversal operation, which is denoted as a parity reversal, a kind of symmetry breaking. Consequently, a space-inversion and time-reversal invariant free energy term can be constructed as:
\begin{equation}
F_{em}=[\alpha(\textbf{r})\cdot\textbf{P}]\Omega,
\label{energy-parity}
\end{equation}
where $\alpha$ is a coefficient containing odd power orders of space vector $\textbf{r}$, determined by the lattice symmetry. For example, $\alpha$ can be written as $\frac{\partial J}{\partial x}$ where $J$ is the exchange interaction between the nearest-neighbor sites and $x$ is the bond length. Eq.~\ref{energy-parity} is a third power order function of ferroic order parameters. A minimization of the total energy (including $F_{em}$ and polarization energy $\textbf{P}^2/2\epsilon$) gives rise to an induced polarization:
\begin{equation}
\textbf{P}=-\epsilon\alpha(\textbf{r})\Omega.
\label{p-parity}
\end{equation}
It is seen that an E-type antiferromagnetic order with the $\uparrow\uparrow\downarrow\downarrow$ type spin alignment fits to this mechanism, as observed by experiments on e.g. o-HoMnO$_3$ \cite{Sergienko:Prl} and Ca$_3$CoMnO$_6$ \cite{Choi:Prl} which will be discussed in details later.

As shown in Eq.~\ref{p-parity}, the direction of $\textbf{P}$ is governed by the lattice-relevant coefficient $\alpha(\textbf{r})$ instead of the time-reversal invariant $\Omega$ which is dependent of magnetic order. This implies that the direction of ferroelectric polarization is robust against magnetic field, which is distinctly different from the case of spiral spin order driven polarization. As a result, any material with a magnetic order in this category may exhibit ferroelectricity but the magnetoelectric effect in response to magnetic stimuli would be weak if any. Sure enough, the magnitude of polarization can still be tuned by magnetic field. For example, for Ca$_3$CoMnO$_6$, a magnetic field of several Tesla does drive a transition of the original antiferromagnetic $\uparrow\uparrow\downarrow\downarrow$ configuration to a ferrimagnetic $\uparrow\uparrow\uparrow\downarrow$, or even to a ferromagnetic $\uparrow\uparrow\uparrow\uparrow$, in accompanying with suppressed polarization \cite{Jo:Prb}.

\subsubsection{Other possibilities}
In addition to the two types of magnetoelectric responses dependent of the helicity and parity of magnetic orders, other possible combinations of magnetic and electric-dipole order parameters in order to pursue the magnetoelectricity can be proposed. Two examples are presented here:
\begin{equation}
F_{em}=[\textbf{P}\cdot\beta(\textbf{r})][\textbf{M}\cdot\gamma(\textbf{r})]^2,
\label{other1}
\end{equation}
and
\begin{equation}
F_{em}=[\nabla\cdot\textbf{P}][\textbf{M}\cdot\gamma(\textbf{r})]^2,
\label{other2}
\end{equation}			
where $\beta$ and $\gamma$ are the functions of space vector $\textbf{r}$ respectively. Function $\beta$ must be of the odd power of $\textbf{r}$, while no special requirement for function $\gamma$ is needed. $\gamma$ can be even a plain function $\gamma=1$. Some examples fitting the two coupling schemes will be discussed later.

In fact, not only a coordinate vector can be incorporated into the free energy, a time arrow ($t$) may be also included into the energy for a dynamic magnetoelectric process. For example, a magnetoelectric coupling term yielding a time-dependent ferroelectric polarization can be written as:
\begin{equation}
F_{em}=\alpha\frac{d\textbf{P}}{dt}\cdot\nabla\textbf{M},
\label{other3}
\end{equation}								
which is in the second power order, lower than all the aforementioned third power order. One of the consequences is electromagnon excitation, i.e. AC electric field excited spin wave. The charge dipole $\textbf{P}$ in synchronization with an AC electric field gives rise to a nonzero $d\textbf{P}/dt$, which may couple with a space modulation of magnetic moments, i.e. spin wave. More detailed analysis of electromagnon in TbMnO$_3$ and other multiferroics can be found in Sec. 3.7.1.

In short, any combination of $\textbf{P}$ and $\textbf{M}$ may act as a source of magnetoelectric coupling as long as it is invariant upon a space-reversal plus time-inversion operation. Many more magnetoelectric phenomena may be expected by properly manipulating the magnetic and ferroelectric orders in real materials or artificial structures so that the symmetry consensus can be satisfied.

\subsection{A unified model on magnetism-induced polarization}
The noncollinear spiral order and collinear $\uparrow\uparrow\downarrow\downarrow$ order associated with a helicity and a parity respectively evidence the great progress in our understanding of multiferroicity. Nevertheless, experiments did reveal some exceptions in which the observed magnetoelectric phenomena can't be reasonably explained by the two simple phenomenological models. According to Eq.~\ref{pmm}, a screw-type spiral spin order cannot generate a macroscopic polarization. However, a magnetic field or substitution induced ferroelectric polarization was indeed observed in CuFeO$_2$ which does have an in-plane screw type spiral order \cite{Kimura:Prb06}. Similar situation occurs in Ba$_3$NiNb$_2$O$_9$ (and a number of isostructures) with an in-plane triangle $120^\circ$ noncollinear spin order, in which a weak ferroelectric polarization was measured \cite{Hwang:Prl}.

To accommodate these exceptions, a more general phenomenological framework was proposed recently by Xiang \textit{et al.} so that all these phenomena on magnetism driven ferroelectricity can be properly interpreted. It seems, these exceptions together with the conventional facts can be well accommodated \cite{Xiang:Prl11,Xiang:Prb13}. To proceed, one considers a magnetic ferroelectric whose total polarization is approximately partitioned into two parts:
\begin{equation}
\textbf{P}=\textbf{P}_{\rm ion}(\textbf{U})+\textbf{P}_{e}(\textbf{S},\textbf{U}=0),
\label{xiang1}
\end{equation}
where the first part $\textbf{P}_{\rm ion}$ is from ionic displacements, and the second part $\textbf{P}_{e}$ is pure electronic contribution. $\textbf{S}$=($\textbf{S}_1$, $\textbf{S}_2$, ... , $\textbf{S}_n$) denotes the collection of spins. $\textbf{U}$=($\textbf{u}_1$, $\textbf{u}_2$, ..., $\textbf{u}_n$) denotes the ions' displacements from the centrosymmetric positions. Without losing a generality, $\textbf{P}_{\rm ion}(\textbf{U})$ can be simply obtained by summating all the local dipoles, following the treatment for traditional displacive-type ferroelectrics:
\begin{equation}
\textbf{P}_{\rm ion}(\textbf{U})=\sum_i\textbf{u}_iZ_i,
\label{xiang2}
\end{equation}
where $Z_i$ is the Born effective charge. Dealing with term $\textbf{P}_{e}$ is a bit more complex, and a generalized form of $\textbf{P}_{e}$ can be written as:
\begin{equation}
\textbf{P}_{e}=\sum_{i,\alpha\beta}\textbf{P}_{i,\alpha\beta}S_i^\alpha S_i^\beta+\sum_{<ij>}\textbf{P}_{ij}^{es}\textbf{S}_i\cdot\textbf{S}_j+\sum_{<ij>}\textbf{M}_{ij}\cdot(\textbf{S}_i\times\textbf{S}_j),
\label{xiang3}
\end{equation}
Here the first term is the single-site component containing the quadratic term of spin components at the same site; the second is the scalar product between neighboring spin-pairs, related to the conventional spin exchange; and the last component is the so-called spin-current term containing the cross product of neighboring spin-pairs, where $\textbf{M}$ is a $3\times3$ matrix:
\begin{equation}
\textbf{M}_{ij}=\left(
\begin{array}{ccc}
(P_{ij}^{yz})_x &  (P_{ij}^{zx})_x & (P_{ij}^{xy})_x\\
(P_{ij}^{yz})_y &  (P_{ij}^{zx})_y & (P_{ij}^{xy})_y \\
(P_{ij}^{yz})_z &  (P_{ij}^{zx})_z & (P_{ij}^{xy})_z
\end{array}
\right),
\label{xiang4}
\end{equation}
These coefficients, e.g. $\textbf{P}_i$, $\textbf{P}^{es}$, and $\textbf{M}$, must obey the restriction of space-inversion and time-reversal symmetries as well as the lattice symmetry if applied. The helicity (chirality) and parity related magnetoelectric coupling mechanisms can be self-consistently covered by this unified model.

To the end of this section, one is convinced with the fact that the Landau phenomenological theory based on the symmetry consensus constitutes a generalized platform on which the coupling between magnetism and ferroelectricity can be presented in various mathematically graceful forms, somehow irrelevant with concrete materials and chemical structures. Surely, a modern theory on multiferroicity, however, must have its quantum fundamentals, with which the microscopic mechanisms for various magnetoelectric phenomena can be understood and then manipulated via advanced physical approaches such as band structure engineering, domain engineering, and microstructural variations. This mission will be reasonably illustrated in the subsequent two sections on magnetoelectric phenomena and underlying microscopic mechanisms in compounds and artificial microstructures among others.

\section{Magnetoelectric coupling in single phases: from physics to materials}
In this section, we are mainly concerned with the quantum-level ingredients of multiferroic physics for magnetoelectric coupling in single phase multiferroics. For each case, we start from the microscopic origin and materials, followed by a comprehensive discussion on the observed phenomena and magnetoelectric properties.

In fact, for each phenomenological energy term mentioned in Sec. 2.2., there must be a set of corresponding microscopic mechanisms. An understanding of these microscopic mechanisms is also the key progress in the past decade, which can be traced back to the phenomenological expressions. For example, in Mostovoy's theory regarding the magnetic helicity driven polarization (Sec. 2.2.1.), the ``absolute" directions of magnetic moments are needed, implying that the spin-orbit coupling is essential in the corresponding microscopic mechanism. In contrast, for the parity-related mechanism (Sec. 2.2.2.), only the inner production between spins is involved while the ``absolute" directions of these two spins do not affect the polarization. In this case, the underlying microscopic mechanism seems to be based on normal exchanges instead of the spin-orbit coupling. These differences should be accounted for in order to categorize the microscopic mechanisms.

\subsection{Magnetoelectric coupling I: based on the spin-orbit coupling}
\subsubsection{Dzyaloshinskii-Moriya interaction}
In 1958, Dzyaloshinskii proposed a thermodynamic theory to explain the weak ferromagnetism observed in Cr$_2$O$_3$ \cite{Dzyaloshinsky:Jpcs}. This was the first time to postulate an asymmetrical exchange interaction which was subsequently elaborated as a consequence of the spin-orbit coupling by Moriya using quantum perturbation theory in 1960 \cite{Moriya:Prl,Moriya:Pr}, and further re-interpreted by Shekhtman, Entin-Woodman, and Aharony in 1992 \cite{Shekhtman:Prl}. It is now coined as the Dzyaloshinskii-Moriya (DM) interaction. The microscopic origin of this exchange is the relativistic correction to the exchanges in presence of the spin-orbit coupling. The Dzyaloshinskii-Moriya interaction plays a crucial role not only in the physics of multiferroics (not limited to the so-called type-II ones), but also in many sub-fields of magnetism. For instance, it is believed to be one of the origins for skyrmion quasi-particle generation.

\begin{figure}
\centering
\includegraphics[width=\textwidth]{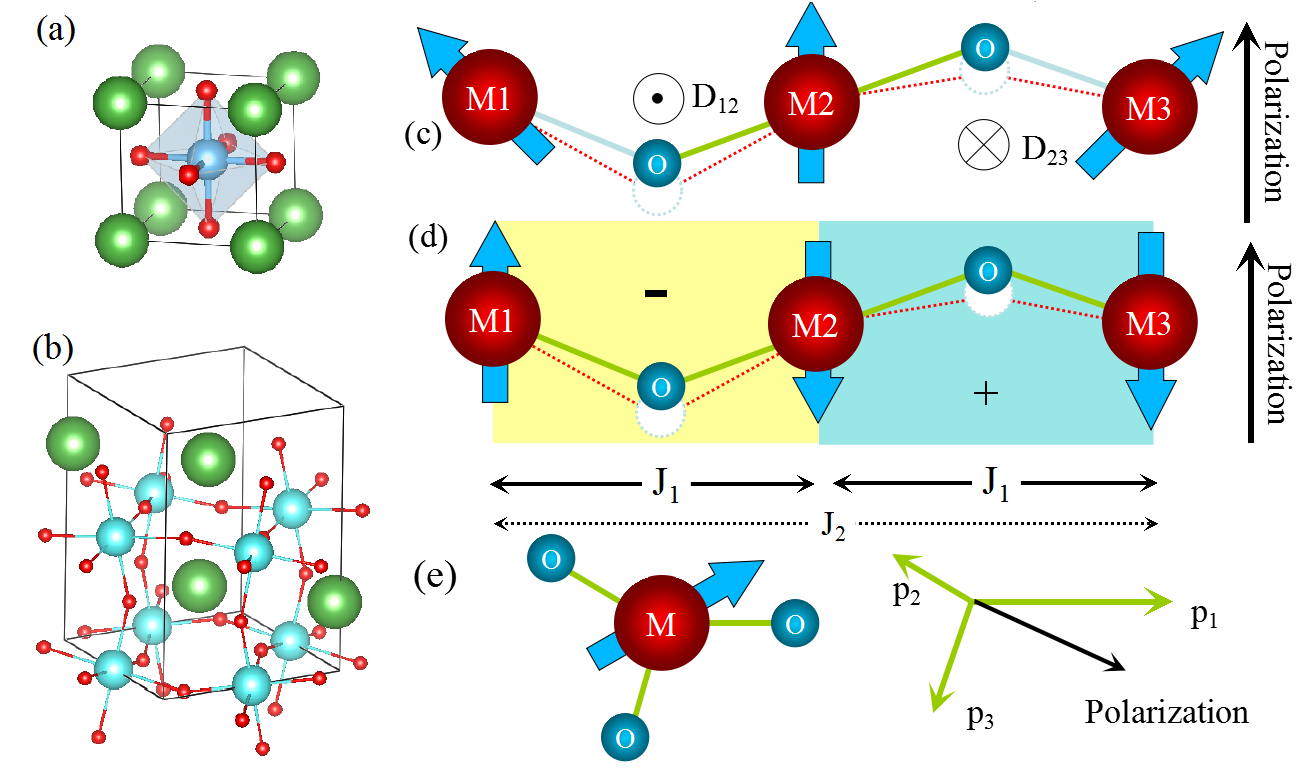}
\caption{(Colour online) (a-b) Crystal structure of an $AB$O$_3$ perovskite. Green: $A$; Cyan: $B$. (a) an ideal cubic perovskite. All the nearest-neighbor $B$-O-$B$ bonds are straight. (b) an orthorhombic perovskite lattice with the GdFeO$_3$-type distortion. All the nearest-neighbor $B$-O-$B$ bonds are bent. (c) A breaking of the rotation symmetry regarding the $M_i$-$M_j$ axis ($M$ denotes metal ion) due to the bond-bending. The asymmetric Dzyaloshinskii-Moriya interaction is allowed, with the $\textbf{D}_{ij}$ vector perpendicular to the $M_i$-$M_j$ axis and oxygen displacement vector, e.g. pointing in/out the paper plane, as sketched. A noncollinear spin pattern with a fixed helicity will uniformly modulate the $\textbf{D}_{ij}$ vector to lower the energy, generating aligned ferroelectric dipoles. (d) A schematic of the exchange striction induced ferroelectric polarization. The ionic displacements caused by the $\uparrow\downarrow$ and $\downarrow\downarrow$ spin pairs, are not compensated, giving rise to a net dipole moment. The exchange frustration, namely antiferromagnetic $J_2$, favors the spin orders as shown in (c) or (d). (e) A metal ion is surrounded in an anion cage,breaking the inversion symmetry (left), and the metal-ligand hybridization can give rise to the three spin-dependent dipoles for the three bonds, allowing a net polarization (right).}
\label{microscopic}
\end{figure}

We take perovskite oxides as examples to illustrate how the Dzyaloshinskii-Moriya interaction allows a coupling between magnetism and ferroelectricity. For a perovskite $AB$O$_3$ with ideal cubic structure, the $B$-O-$B$ bonds are straight, giving a $180^\circ$ bond angle, as illustrated in Fig.~\ref{microscopic}(a). Each of these bonds has a rotation symmetry with respect to one $B$-$B$ axis. However, for most cases, the size mismatch between $A$ and $B$ ions usually makes the oxygen octahedra to tilt and rotate, resulting in the distortion shown in Fig.~\ref{microscopic}(b), which can be characterized using the Glazer notation \cite{Woodward:AcbI,Woodward:AcbII,Glazer:Acb}. Therefore, each oxygen ion sandwiched between two neighboring $B$ ions may move away from the middle point, giving a bent $B$-O-$B$ bond and breaking the $B$-$B$ axis rotation symmetry. This bent $B$-O-$B$ bond will induce the Dzyaloshinskii-Moriya interaction as a relativistic correction to the superexchange between magnetic $B$ ions, schematically shown in Fig.~\ref{microscopic}(c). The Hamiltonian can be expressed as:
\begin{equation}
H_{\rm DM}=\textbf{D}_{ij}\cdot(\textbf{S}_i\times\textbf{S}_j),
\label{dm}
\end{equation}
where $\textbf{D}_{ij}$ is the coefficient of the Dzyaloshinskii-Moriya interaction between spins $\textbf{S}_i$ and $\textbf{S}_j$. For a perovskite structure with bent $B$-O-$B$ bonds, vector $\textbf{D}_{ij}$ must be perpendicular to the $B$-O-$B$ plane and determined by the symmetry restrictions. In the first order approximation, the magnitude of $\textbf{D}_{ij}$ is proportional to the displacement of oxygen ion ($\textbf{d}_{\rm O}$) away from the ``original" middle point \cite{Sergienko:Prb}:
\begin{equation}
\textbf{D}_{ij}=\zeta\textbf{e}_{ij}\times\textbf{d}_{\rm o},
\label{d}
\end{equation}
where $\zeta$ is a coefficient; $\textbf{e}_{ij}$ is the unit vector pointing from site $i$ to site $j$. The quantum-level derivation of the Dzyaloshinskii-Moriya interaction can be found in Moriya's original paper \cite{Moriya:Pr}. In the cubic limit with the highest symmetry, $\textbf{D}_{ij}=0$. For a distorted structure with lower symmetry, the cooperative rotation or tilting of oxygen octahedra leads to a reversal of the direction of $\textbf{D}_{ij}$ between the nearest-neighbor $B$-O-$B$ bonds if all the O-$B$-O (not $B$-O-$B$) bond angles are $180^\circ$ (i.e. rigid octahedra).

A prediction of the noncollinear spin alignment and the easy spin plane due to the Dzyaloshinskii-Moriya interaction is straightforward from Hamiltonian Eq.~\ref{dm}. Without losing a generality, we consider a one-dimensional $B$-O-$B$-O-$B$-O spin chain and the corresponding Hamiltonian reads:
\begin{equation}
H=\sum_{ij}[J_{ij}\textbf{S}_i\cdot\textbf{S}_j+\textbf{D}_{ij}\cdot(\textbf{S}_i\times\textbf{S}_j)],
\label{jd}
\end{equation}
where $J$ is the standard exchange. If $\textbf{D}_{i,i+1}$ and $J_{ij}$ are respectively identical at each bond, the ground state must be a spiral spin order. The nearest-neighbor spin angle is $\arctan(|D|/J)$ if $J$ is negative (or $\pi-\arctan(|D|/J)$ if $J$ is positive).

The inverse effect of the Dzyaloshinskii-Moriya interaction for a spiral spin order is to generate a uniform bias of $\textbf{D}$ (consequently the corresponding uniform bias of oxygen shift). The underlying mechanism can be described by Hamiltonian:
\begin{equation}
H=(\zeta\textbf{e}_{ij}\times\textbf{d}_{\rm o})\cdot(\textbf{S}_i\times\textbf{S}_j)+\frac{\kappa}{2}\textbf{d}_{\rm o}^2,
\label{dmd}
\end{equation}
where the second term is the elastic energy with $\kappa$ the stiffness. A minimization of the total energy results in a displacement $\textbf{d}_{\rm o}=-\frac{\zeta}{\kappa}\textbf{e}_{ij}\times(\textbf{S}_i\times\textbf{S}_j)$. Since $\textbf{S}_i\times\textbf{S}_j$  in a spiral magnet with a fixed helicity uniformly points to one direction, all the induced ionic biases (displacements) are along the same direction, as sketched in Fig.~\ref{microscopic}(c), giving a macroscopic ferroelectric polarization.

This scenario was first proposed by Sergienko and Dagotto in 2006 (the Sergienko-Dagotto model) to illustrate the origin of ferroelectricity in TbMnO$_3$ \cite{Sergienko:Prb}. In fact, TbMnO$_3$ does exhibit a cycloid $b-c$ plane spiral order with the spiral wave vector along the $b$-axis below $28$ K. The induced polarization $\sim\textbf{e}_{ij}\times(\textbf{S}_i\times\textbf{S}_j)$ aligns along the $c$-axis, consistent with the experimental observations. This scenario can also explain the polarization flip from the $b$-axis to the $a$-axis in accordance with the cycloid plane flop from the $b-c$ plane to the $a-b$ plane, driven by a magnetic field.

Besides the above discussed ionic displacements, the spin-orbit coupling (not necessarily the Dzyaloshinskii-Moriya interaction) may distort the electron cloud surrounding an ionic core for a noncollinear spin pair, leading to a pure electronic charge dipole. This issue was discussed by Katsura, Nagaosa, and Balatsky (the so-called KNB theory) using the quantum perturbation theory to the Hubbard model with spin-orbit coupling. It was proved that the induced charge dipole is proportional to \cite{Katsura:Prl}:
\begin{equation}
\textbf{P}\sim\textbf{e}_{ij}\times(\textbf{S}_i\times\textbf{S}_j).
\label{knb}
\end{equation}
which is the main prediction of this KNB theory on multiferroicity. No ionic displacement contribution is considered here and this theory can be viewed as a counterpart of the Sergienko-Dagotto model. More details of this theory can be found in the following interpretations \cite{Jia:Prb,Jia:Prb07}.

Although the KNB theory is complicated, the final outcome is elegant, in agreement with the Sergienko-Dagotto prediction based on the Dzyaloshinskii-Moriya interaction. In this sense, the two theories are syngeneic although the routes are different: the Sergienko-Dagotto model is based on the ionic shift while the KNB theory is based on the electronic bias. Both the sources contribute to the total ferroelectric polarization, while they may add or cancel out depending on whether the two polarizations are parallel or antiparallel. Nevertheless, density functional calculations suggest that the contribution from the ionic displacements is dominant in TbMnO$_3$ \cite{Malashevich:Prl,Xiang:Prl,Malashevich:Prb}, as confirmed experimentally \cite{Walker:Sci}. The mechanism based on the Dzyaloshinskii-Moriya interaction also matches the phenomenological consideration based on the helicity argument, and explains reasonably the magnetism driven ferroelectricity in many multiferroics with spiral type spin order.

\subsubsection{Spin-dependent metal-ligand hybridization}
Different from the Dzyaloshinskii-Moriya interaction scenario (the Sergienko-Dagotto model), Arima once proposed an alternative mechanism based on the spin-orbit coupling \cite{Arima:Jpsj}. For each transition metal ion, the local polarization can be expressed as:
\begin{equation}
\textbf{P}=\sum_i(\textbf{e}_{i}\cdot\textbf{S}_i)^2\textbf{e}_i,
\label{singleion}
\end{equation}
where the summation is over all the bonds and $\textbf{e}_i$ denotes the bond direction, as shown in Fig.~\ref{microscopic}(e). A distinct difference here is that only a single magnetic site is considered for a local electric dipole, while the Dzyaloshinskii-Moriya interaction involves two magnetic sites. The underlying mechanism is the spin-orbit coupling induced perturbation to the hybridization between metal's $d$ orbitals and anion's $p$ orbitals. This $p-d$ hybridization may be responsible for the polarization measured in Ba$_2$CoGe$_2$O$_7$ \cite{Murakawa:Prl10}, while more examples to support this mechanism would be appreciated. In fact, a similar $p-d$ hybridization mechanism was claimed for the measured polarization in CuFeO$_2$, but a careful analysis of the crystal symmetry rules out this mechanism, since the summation over all the bonds gives no net polarization due to a cancellation of the dipoles.

It is noted that a generalized quantum microscopic mechanism associated with the spin-orbit coupling remains yet to access. So far proposed mechanisms are more or less materials-dependent. For instance, although the unified model given in Sec. 2.3 does give a phenomenological basis for polarization generation in CuFeO$_2$ \cite{Xiang:Prb13}, the microscopic mechanism is not yet well understood. Furthermore, magnetoelectric effects based on the spin-orbit coupling are favored because the spin orientation is flexible to external magnetic stimuli, making a magneto-control of ferroelectricity and magnetic helicity easy. However, it is clear that the spin-orbit coupling of $3d$ orbitals is usually weak and the as-generated ferroelectric polarizations in this category are small, e.g. $\sim0.08$ $\mu$C/cm$^2$ for TbMnO$_3$, a disadvantage of materials in this category.

\subsection{Magnetoelectric coupling II: based on spin-lattice coupling}
In parallel to the spin-orbit coupling, the spin-lattice coupling among some others is concerned too. For a transition metal with active $3d$ electrons, the spin-orbit coupling is intrinsically weak, and coefficient $\zeta$ in Eq.~\ref{dmd} should be a small quantity giving only a tiny electric dipole for a noncollinear spin order. To overcome this drawback, Sergienko, \c{S}en, and Dagotto proposed an alternative microscopic mechanism, namely the symmetric exchange striction (Fig.~\ref{microscopic}(d)) induced ferroelectric polarization by investigating orthorhombic HoMnO$_3$ \cite{Sergienko:Prl}. Later, this symmetric exchange striction was applied to explain the magnetism induced ferroelectricity in Ca$_3$CoMnO$_6$ \cite{Choi:Prl}. This exchange striction effect is essentially based on the spin-lattice coupling.

\subsubsection{One dimensional Ising spin chain}
\begin{figure}
\centering
\includegraphics[width=\textwidth]{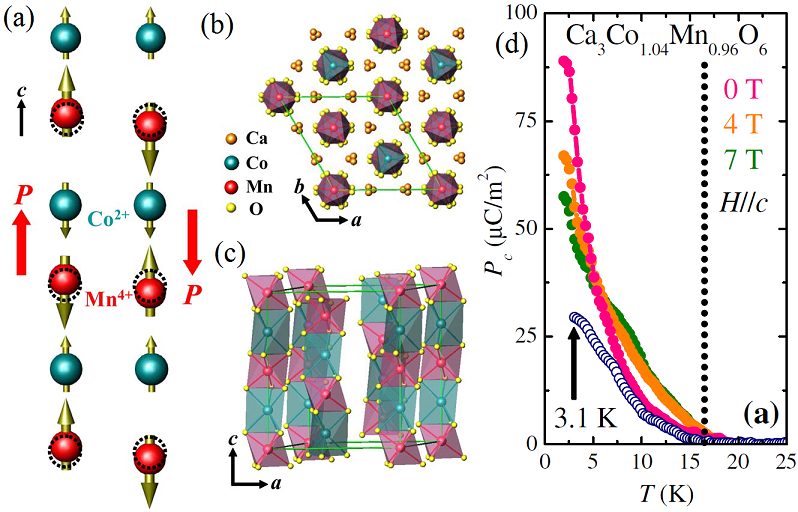}
\caption{(Colour online) Structures and multiferroic properties of Ca$_3$CoMnO$_6$. (a) Two Ising chains with the $\uparrow-\uparrow-\downarrow-\downarrow$ order along the $c$-axis. The ionic alignment takes the Mn-Co-Mn-Co order. The original ionic positions are shown as dashed circles, while the coherent displacements lead to a polarization $P$. (b-c) The projected views of crystal structure. (d) The pyroelectric polarization along the $c$-axis of Ca$_3$Co$_{1.04}$Mn$_{0.96}$O$_6$, which is suppressed by a magnetic field along the $c$-axis. Reprinted figure with permission from \href{http://dx.doi.org/10.1103/PhysRevLett.100.047601}{Y. J. Choi \textit{et al.}, Physical Review Letters, 100, p. 047601, 2008} \cite{Choi:Prl} Copyright \copyright (2008) by the American Physical Society.}
\label{CCMO}
\end{figure}

As an example, we discuss Ca$_3$CoMnO$_6$, as shown in Fig.~\ref{CCMO}(a-c), which consists of quasi-one-dimensional Co-Mn chains along the $c$-axis. Below $16.5$ K, the spins develop a particular $\uparrow\uparrow\downarrow\downarrow$ order \cite{Choi:Prl}. In consequence, the Co-Mn distance shrinks/elongates for the neighboring parallel/antiparallel spin pairs (Fig.~\ref{CCMO}(a)), respectively \cite{Wu:Prl,Zhang:Prb09.2}. Since the valences of Co$^{2+}$ and Mn$^{4+}$ are different, the ionic displacements along the Co-Mn chain are coherently aligned, inducing a net polarization along the $c$-axis (Fig.~\ref{CCMO}(a)), as first observed in Choi \textit{et al.}'s experiment \cite{Choi:Prl}.

This symmetric exchange striction can be understood in a simplified scenario. The exchange coefficient $J$ between the nearest-neighbor sites depends on the bond length. In the first order approximation, the Taylor expansion of $J(x)$ can be written as: $J(x)=J_0+\frac{\partial J}{\partial x}|_{x_0}\delta$ where $x$ is the bond length, $x_0$ is the original length, $J_0$ is the exchange at $x_0$, and $\delta=x-x_0$. By assuming a periodic ($\delta$, $-\delta$, $\delta$, $-\delta$) type displacement mode, the Hamiltonian for a given $\uparrow\uparrow\downarrow\downarrow$ spin chain is:
\begin{eqnarray}
\nonumber H&=&\sum_{<ij>}[J_{ij}(\delta_{ij}(\textbf{S}_{i}\cdot\textbf{S}_j)+\frac{\kappa}{2}(x_{ij}-x_0)^2]\\
&=&\sum_{n=4i}[4\frac{\partial J}{\partial x}|_{x_0}\delta+2\kappa\delta^2].
\label{exchangestrict}
\end{eqnarray}

A minimization of the energy leads to the equilibrium displacement $\delta$ of $-\frac{1}{\kappa}\frac{\partial J}{\partial x}|_{x_0}$. A ferroelectric polarization is generated if the charge at site $i$ is not identical to that at site $i+1$, just as the case here for Co$^{2+}$-Mn$^{4+}$ chain given Co$^{2+}$ at site $i$ and Mn$^{4+}$ at site $i+1$. This exchange striction mechanism is irrelevant with the weak spin-orbit coupling but determined by the spatial derivative of the exchange coupling, which can be sufficiently strong for a $3d$ electronic system. Indeed, the density functional calculation predicted a polarization up to $1.77$ $\mu$C/cm$^2$ along the chain (i.e. the $c$-axis) \cite{Zhang:Prb09.2}. Unfortunately, the measured value for single crystal Ca$_3$CoMnO$_6$ is only $\sim0.09$ $\mu$C/cm$^2$ (Fig.~\ref{CCMO}(d)). The order of magnitude difference between theoretical and experimental results may have something to do with the fragile stability of the $\uparrow\uparrow\downarrow\downarrow$ spin order in Ca$_3$CoMnO$_6$, which is competed from other orders \cite{Wu:Prl,Zhang:Prb09.2}. Interestingly, several subsequent experiments revealed that the $\uparrow\uparrow\downarrow\downarrow$ spin order does become unstable when Co and Mn are ideally stoichiometric $1:1$, associated with some structural ordering induced spin disordering. Instead, a slight nonstoichiometry of species benefits to the stability of this spin order and thus the enhancement of ferroelectric polarization \cite{Kiryukhin:Prl,Ding:Apl}.

\subsubsection{Two dimensional E-type antiferromagnets}
\begin{figure}
\centering
\includegraphics[width=\textwidth]{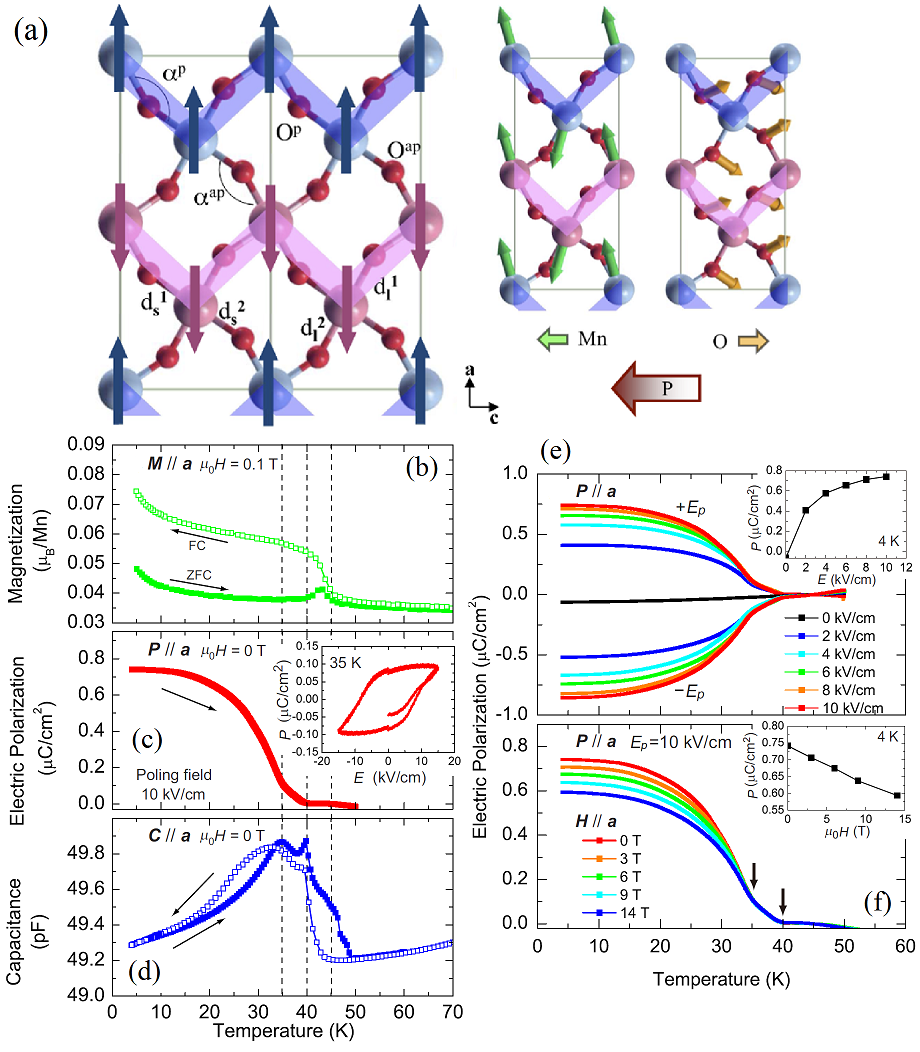}
\caption{(Colour online) E-type antiferromagnetism and multiferroicity. (a) The $ab$-plane structure (in the $Pbnm$ notation) and magnetic (E-type antiferromagnetic) structure of orthorhombic $R$MnO$_3$. The spins are denoted by arrows. The particular zigzag spin chains are highlighted by shaded areas. $\alpha^p$ and $\alpha^{ap}$ denote the Mn-O-Mn bond angles between the parallel and antiparallel spin pairs respectively, noting that they are different due to the exchange striction. (a) Reprinted figure with permission from \href{http://dx.doi.org/10.1103/PhysRevLett.99.227201}{S. Picozzi \textit{et al.}, Physical Review Letters, 99, p. 227201, 2007} \cite{Picozzi:Prl} Copyright \copyright (2007) by the American Physical Society. (b) The measured magnetization along the $a$-axis of orthorhombic YMnO$_3$ films (measuring field $0.1$ T). (c) The measured pyroelectric polarization along the $a$-axis direction. Inset: a ferroelectric loop measured at $35$ K. (d) The capacitance-temperature cycle measured along the $a$-axis direction. (e-f) The pyroelectric polarizations measured upon various electric poling fields ($E_p$). Inset: the pyroelectric polarization at $4$ K as a function of the poling field. (f) The pyroelectric polarizations under various magnetic fields. Inset: the pyroelectric polarization at $4$ K as a function of magnetic field. (b-f) Reprinted figure with permission from \href{http://dx.doi.org/10.1063/1.3555462}{M. Nakamura \textit{et al.}, Applied Physics Letters, 98, p. 082902, 2011} \cite{Nakamura:Apl} Copyright \copyright (2011) by the American Institute of Society.}
\label{E-AFM}
\end{figure}

Orthorhombic HoMnO$_3$ is another good example to illustrate the role of exchange striction in generating a polarization. HoMnO$_3$ prefers the hexagonal structure (a type-I multiferroic to be discussed later), but meta-stable orthorhombic structure can be obtained by proper synthesis routes, making it a type-II multiferroic. Below $42$ K, the Mn spins become ordered from a paramagnetic state to a sinusoidal antiferromagnetic state, similar to the case for isostructural TbMnO$_3$, then to the so-called E-type antiferromagnetic state at $26$ K \cite{Munoz:Ic}. This E-type antiferromagnetism is constructed by an alignment of the collinear zigzag chains in the $a-b$ plane, as shown in Fig.~\ref{E-AFM}(a). Along the pseudo-cubic $x$ and $y$ directions (the diagonal direction in the $a-b$ plane here), the E-type antiferromagnetic state shows the $\uparrow\uparrow\downarrow\downarrow$ order. Nevertheless, the consequence of the exchange striction here is not simply identical to that in Ca$_3$CoMnO$_6$, since the Mn cations here are isovalent. The Mn ionic displacements do not directly generate a polarization. Instead, a collective octahedral rotation associated with the displacements results in an alternative bending of the Mn-O-Mn bonds. Because the exchange $J$ depends on not only the bond length but also the bond angle, the exchange striction can modulate the Mn-O-Mn bond angles in coherence with the $\uparrow\uparrow\downarrow\downarrow$ order (see Fig.~\ref{E-AFM}(a)). Such a modulation induces the uniform displacements of oxygen ions, leading to a net polarization along the $a$-axis \cite{Sergienko:Prl}.

For orthorhombic HoMnO$_3$, the predicted polarization reaches $1.0-6.0$ $\mu$C/cm$^2$ \cite{Sergienko:Prl,Picozzi:Prl,Yamauchi:Prb}. Initially, the measured polarization in polycrystalline samples was small, e.g. $<0.01$ $\mu$C/cm$^2$ \cite{Lorenz:Prb}, probably due to the sample quality and big coercivity. With the advance of measurement techniques and improved sample quality, the measured values in orthorhombic HoMnO$_3$, YMnO$_3$, and analogous systems have been gradually enhanced to $0.8$ $\mu$C/cm$^2$ \cite{Ishiwata:Prb,Pomjakushin:Njp,Ishiwata:Jacs,Nakamura:Apl,Wadati:Prl}, as displayed in Fig.~\ref{E-AFM}(b-f), close to the lower end of the predicted values.

The discussions on the two model systems seem to suggest that the exchange striction mechanism based on the spin-lattice coupling only depends on the scalar production of spin pair ($\textbf{S}_{i}\cdot\textbf{S}_j$), leaving the polarization direction to be determined. This is conceptually different from the mechanisms based on the spin-orbit coupling. Here the direction of polarization is not related to the magnetic easy axis or easy plane, but determined by the crystal symmetry, e.g. the Mn-Co chain direction in Ca$_3$CoMnO$_6$ or the octahedral rotation mode in orthorhombic HoMnO$_3$. This argument obtained further support from recent studies on double-perovskites $R_2$CoMnO$_6$ and $R_2$NiMnO$_6$ ($R=Y$, or rare earths), in which the spins also develop the $\uparrow\uparrow\downarrow\downarrow$ order along the $c$-axis below certain temperature, similar to Ca$_3$CoMnO$_6$ \cite{Xin:Ra,Ma:Pccp,Zhou:Apl,Kumar:Prb}. However, the induced polarization is along the $b$-axis, different from the case of Ca$_3$CoMnO$_6$ where the polarization is aligned along the $c$-axis. As a result, the magnitude of polarization can be suppressed in accompanying with the suppression of $\uparrow\uparrow\downarrow\downarrow$ order, but its direction is hard to change by magnetic field (e.g. see Fig.~\ref{E-AFM}(f)), an interesting fact \cite{Ishiwata:Prb,Jo:Prb}.

Finally, one is reminded that the lattice distortion is not mandatory for generating ferroelectricity, at least from the theoretical point of view, although the title of this subsection is ``spin-lattice coupling". The distortion of electronic cloud may play a substitution role too, leading to a considerable electronic polarization. According to Picozzi \textit{et al.}'s density functional calculations, a pure electronic contribution in orthorhombic HoMnO$_3$ can reach up to $60\%$ of the total polarization, even larger than the ionic contribution \cite{Picozzi:Prl}. Such a big portion of polarization may be an advantage for ferroelectric applications in ultra-high frequency region where electronic dynamic response is not an issue at all. Sure enough, the electronic contribution may be an issue if it cancels out the ionic contribution.

Given the two frameworks of magnetoelectric coupling highlighted above, it is ready to discuss a number of multiferroics within the two frameworks, addressing the magnetoelectric phenomena. While the classification of multiferroics by Khomskii is followed, we pay attention to several representative materials by discussing the structure, magnetic order, and microscopic mechanisms for magnetoelectric properties in details, leaving brief introduction to other multiferroics in a major-minor sorting mode in due course.

\subsection{Magnetoelectric properties of some selected type-I multiferroics}
Again, we stress that BiFeO$_3$ is no doubt so far the most known type-I multiferroic material. Its prominent ferroelectricity above room temperature allows a manipulation of ferroelectric domains and electrically controllable magnetism to be possible, as extensively studied in the thin films and heterostructures. While the progress has been reviewed in details in literature, we will only give a brief summary in Sec. 4 of this article.

Besides BiFeO$_3$, there are several sub-families in the type-I multiferroics whose multiferroic phenomena are abundant deserving sustained attention. Most of them show a ferroelectric polarization induced by lattice geometry variants. In this sense, their polarizations are improper in spite of the irrelevance of spin-ordering. The magnetoelectric coupling is most likely mediated via the spin-lattice interactions.

\subsubsection{Hexagonal $R$MnO$_3$ \& $R$FeO$_3$}
\begin{figure}
\centering
\includegraphics[width=\textwidth]{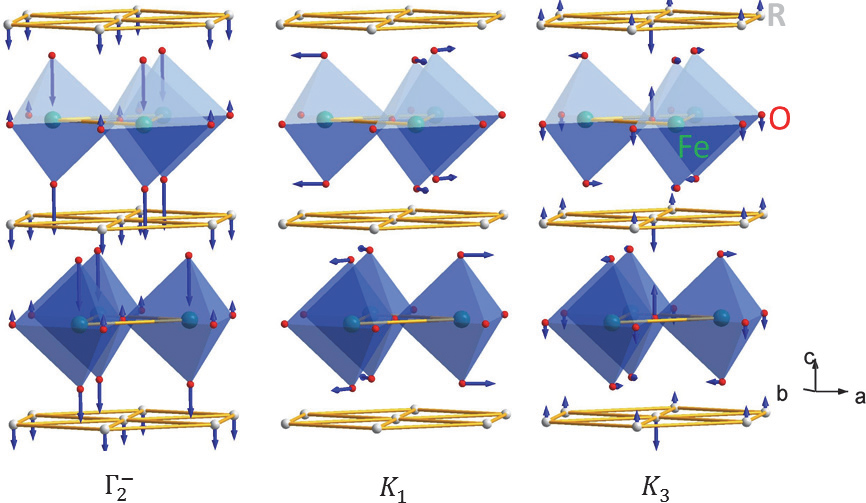}
\caption{(Colour online) A schematic of the three distortion modes ($\Gamma_2^-$, $K_1$, and $K_3$) related to the $P6_3/mmc$ - $P6_3cm$ structural transitions in e.g. hexagonal $R$MnO$_3$ and $R$FeO$_3$. Reprinted figure with permission from \href{http://dx.doi.org/10.1142/S0217984914300087}{X. S. Xu \textit{et al.}, Modern Physics Letters B, 28, p. 1430008, 2014} \cite{Xu:Mplb} Copyright \copyright (2014) by World Scientific Publishing Company.}
\label{hRMO}
\end{figure}

Hexagonal manganite $R$MnO$_3$ ($R$=Ho$^{3+}$, Y$^{3+}$ or other smaller rare earth ions) family is quite different from the perovskite counterpart in several aspects. First, for hexagonal $R$MnO$_3$, oxygen ions form the hexahedra (trigonal bipyramids) units instead of the octahedra in perovskite. The in-plane geometry of Mn sites is a triangle instead of a (distorted) square. The buckling of layered MnO$_5$ polyhedra drives the displacements of $R^{3+}$ ions, and the two-up-one-down profile of the $R^{3+}$ ionic movements results in a net polarization along the hexagonal $c$-axis \cite{Aken:Nm}, as sketched in Fig.~\ref{hRMO}.

Since the ferroelectric polarization originates from a structural geometry variation, hexagonal $R$MnO$_3$ family are sometimes also classified as geometric ferroelectrics \cite{Cheong:Nm}. Different from the proper ferroelectricity, a geometric ferroelectricity does not rely on the re-hybridization and covalency between ferroelectric-active ions and anions. Neither $R^{3+}$ nor Mn$^{3+}$ is ferroelectric-active. The improper ferroelectric polarization of hexagonal YMnO$_3$ is about $5-6$ $\mu$C/cm$^2$ \cite{Aken:Nm}, one order of magnitude smaller than that of typical proper ferroelectrics (e.g. PbTiO$_3$ and BiFeO$_3$). Even though, this ferroelectricity is much stronger than the magnetism driven ferroelectricity in the perovskite counterparts. The ferroelectric Curie temperatures are also high, e.g. $570-990$ K \cite{Fiebig:Nat}. It is noted that the space group of crystal structure transits from $P6_3/mmc$ (high temperature paraelectric state) to $P6_3cm$ (low temperature ferroelectric state), via yet the unit cell-tripling mode (trimerization) occurring at very high temperature into the $P6_3/mcm$ group. This trimerization at e.g. $\sim1350$ K for YMnO$_3$ \cite{Nenert:Jpcm}, does not immediately lead to a ferroelectric polarization which otherwise will not appear until a relatively low temperature, e.g. $880$ K for YMnO$_3$ \cite{Choi:Nm}. At this point, the ferroelectric transition takes place without further reduction in symmetry. This scenario includes more or less an ingredient of the collective ionic displacement in an extraordinary manner, while an alternative viewpoint claiming the one-step direct paraelectric-ferroelectric transition from the high-symmetric $P6_3/mmc$ to a polar $P6_3cm$ is available also \cite{Fennie:Prb}.

\begin{figure}
\centering
\includegraphics[width=0.75\textwidth]{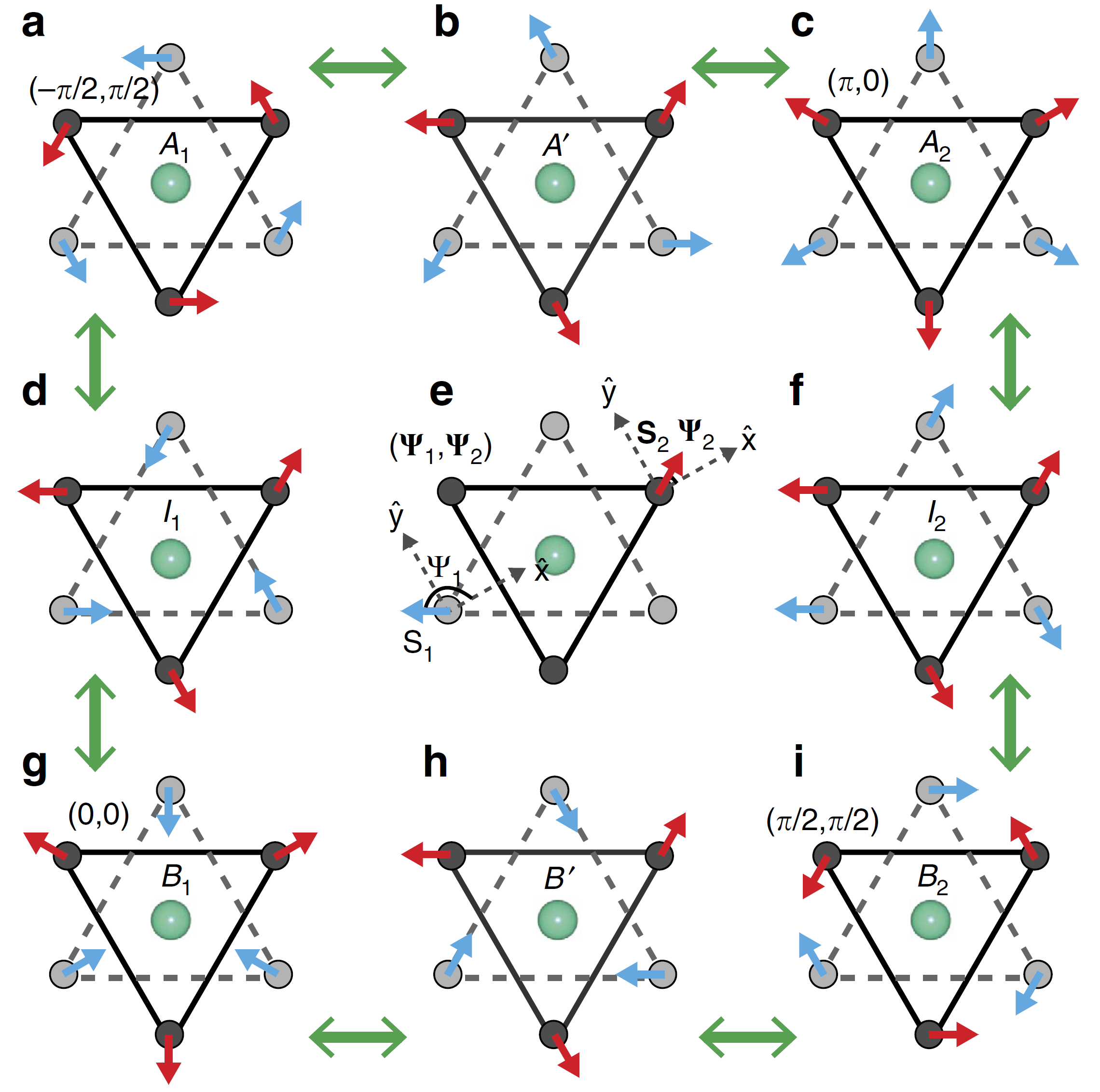}
\caption{(Colour online) Possible $120^\circ$ noncollinear spin configurations in hexagonal $R$MnO$_3$ and $R$FeO$_3$. Each unit cell contains two layers. The spin orientation (angles) and coordinate are defined in (e). The spin configurations are compatible with the crystal symmetry, including: $A_1$, $A_2$, $B_1$, and $B_2$. Only the $A_2$ and those configurations containing the $A_2$ component (i.e. $A'$ and $I_2$) allow a weak ferromagnetism due to the spin-canting, while the spin-canted moments between the two layers for other cases are cancelled. Reprinted by permission from Macmillian Publishers Ltd: \href{http://dx.doi.org/10.1038/ncomms3998}{H. Das \textit{et al.}, Nature Communications, 5, p. 2998, 2014} \cite{Das:Nc}. Copyright \copyright (2014).}
\label{sRMO}
\end{figure}

Besides the ferroelectricity, the magnetism of hexagonal $R$MnO$_3$ has been investigated from various aspects too. One source of magnetism is the Mn$^{3+}$ spins, which form a $120^\circ$ non-collinear patterns in triangular arrangement at low temperature (see Fig.~\ref{sRMO}), e.g. $75$ K for hexagonal HoMnO$_3$ \cite{Lottermoser:Nat}. Another source of magnetism may be related to the rare earth $R^{3+}$ $4f$ spins which usually don't order until a very low temperature, e.g. $4.6$ K for Ho$^{3+}$ in hexagonal HoMnO$_3$ \cite{Lottermoser:Nat}. Therefore, the magnetism and ferroelectricity in hexagonal $R$MnO$_3$ have different origins independent of each other. Nevertheless, a magnetoelectric response due to the spin-lattice coupling is available, as observed by the optical second harmonic generation (SHG) data \cite{Fiebig:Nat,Lottermoser:Nat}. A preliminary picture was proposed, where the correlation between the bond angles/bond lengths and the lattice distortion for ferroelectric polarization, is believed to modulate the inter-pair exchange coupling \cite{Aken:Nm}. This process is a reverse effect of the spin-lattice coupling driven ferroelectricity in orthorhombic HoMnO$_3$ \cite{Sergienko:Prl,Picozzi:Prl}. In fact, a giant magneto-elastic (i.e. spin-lattice) coupling was observed in hexagonal (Y$_{1-x}$Lu$_x$)MnO$_3$, characterized by its significant structural distortion at the antiferromagnetic transition \cite{Lee:Nat08}.

The unit cell-tripling mode (trimerization) is a specific structural characteristic of hexagonal $R$MnO$_3$ and its particular geometry should be one of the possible reasons for emergent real-space topological phenomena in these materials. In recent years, the topology of ferroelectric domains in hexagonal $R$MnO$_3$ has attracted a lot of research attention, which also extends the scope of multiferroic research. This topic will be discussed in Sec. 5.3.

\begin{figure}
\centering
\includegraphics[width=\textwidth]{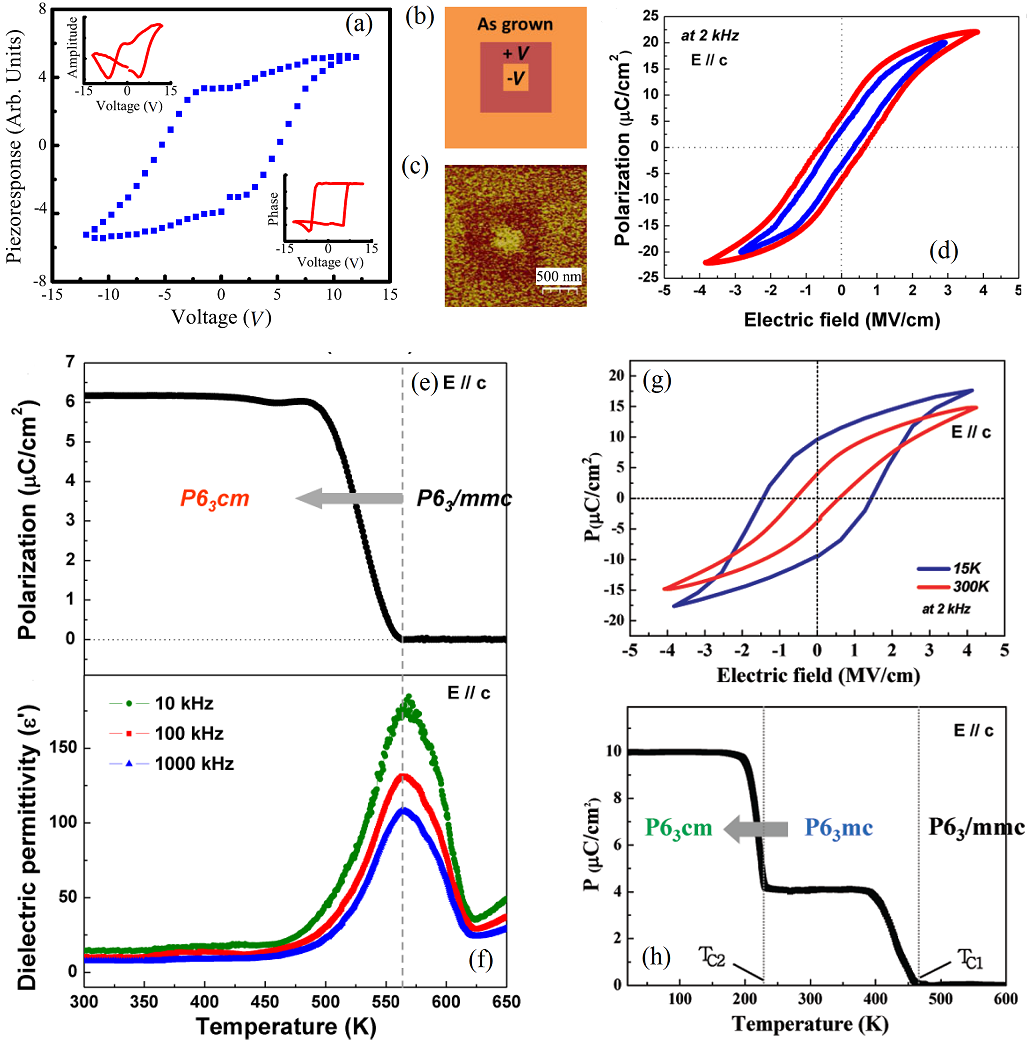}
\caption{(Colour online) Ferroelectric properties for several $R$FeO$_3$ thin films. (a-f) Hexagonal LuFeO$_3$. (a) A hysteresis loop measured by piezoresponse force microscopy. Insert: amplitude (upper) and phase (lower) contrasts. (b-c) A written ferroelectric domain pattern: (b) with a DC bias ($V=20V_{\rm DC}$) and (c) without this bias. (a-c) Reprinted figure with permission from \href{http://dx.doi.org/10.1103/PhysRevLett.110.237601}{W. B. Wang \textit{et al.}, Physical Review Letters, 110, p. 237601, 2013} \cite{Wang:Prl} Copyright \copyright (2013) by the American Physical Society. (d) A hysteresis loops measured at $300$ K. (e) The pyroelectric polarization showing the Curie temperature at $563$ K. (f) The dielectric permittivity data  measured at different frequencies show the corresponding peaks. (d-f) Reprinted figure with permission from \href{http://dx.doi.org/10.1021/cm300846j}{Y. K. Jeong \textit{et al.}, Chemistry of Materials, 24, pp. 2426-2428, 2012} \cite{Jeong:Cm} Copyright \copyright (2012) by the American Chemical Society. (g-h) Hexagonal YbFeO$_3$. (g) A hysteresis loop measured at $15$ K and $300$ K. (h) The pyroelectric polarization showing a characteristic two-step decay. (g-h) Reprinted figure with permission from \href{http://dx.doi.org/10.1021/ja210341b}{Y. K. Jeong \textit{et al.}, Journal of the American Chemical Society, 134, pp. 1450-1453, 2012} \cite{Jeong:Jacs} Copyright \copyright (2012) by the American Chemical Society. }
\label{PRMO}
\end{figure}

Besides hexagonal $R$MnO$_3$, some hexagonal ferrites $R$FeO$_3$ as isostructural partners are multiferroic \cite{Xu:Mplb} and share the similar physics. A distinct point is that an orthorhombic $R$FeO$_3$ is always more stable than a hexagonal $R$FeO$_3$, even when $R^{3+}$ is small, e.g. Lu$^{3+}$. Hexagonal $R$FeO$_3$ can be obtained in epitaxial thin films deposited on proper substrates \cite{Wang:Prl,Disseler:Prl}, or synthesized using special chemical methods \cite{Xu:Mplb}. The ferroelectricity of hexagonal $R$FeO$_3$ emerges above room temperature, with a moderate polarization, e.g. $560$ K and $\sim6.5$ $\mu$C/cm$^2$ at $300$ K for hexagonal LuFeO$_3$ \cite{Jeong:Cm}, as summarized in Fig.~\ref{PRMO}(a-f), noting another experimental finding of the structural transition from the non-polar ($P6_3/mmc$) one to polar ($P6_3cm$) one at a higher temperature ($1050$ K) \cite{Wang:Prl}. For hexagonal YbFeO$_3$, two ferroelectric transitions at $470$ K and $225$ K were observed, and the measured polarizations are about $\sim4$ $\mu$C/cm$^2$ and $\sim10$ $\mu$C/cm$^2$ respectively \cite{Jeong:Jacs}, as summarized in Fig.~\ref{PRMO}(g-h).

A recent neutron study revealed an antiferromagnetic ordering above room temperature ($\sim440$ K) in hexagonal LuFeO$_3$ films \cite{Wang:Prl}. Other experiments reported the onset of magnetic ordering between $115$ K$<T_{\rm N}<155$ K, depending on substrates \cite{Moyer:Aplm,Disseler:Prl}. The noncollinear spin patterns are believed although the reported N\'eel temperatures were authors-dependent. In parallel, a theoretical study on hexagonal $R$MnO$_3$ and $R$FeO$_3$ suggested the intrinsic magnetoelectric effect mediated by the spin-lattice coupling \cite{Das:Nc}. Due to the in-plane antiferromagnetic superexchange, the $120^\circ$ non-collinear spin configurations are expected and their absolute axes are fixed by the Dzyaloshinskii-Moriya interaction, as seen in Fig.~\ref{sRMO}. For some cases, e.g. all the hexagonal $R$FeO$_3$ members and some of the $R$MnO$_3$ series where the antiferromagnetic interlayer superexchange is non-negligible, a net magnetic moment along the $c$-axis arisen from a tiny spin canting along the $c$-axis, e.g. $0.02$ $\mu_{\rm B}$/Fe for hexagonal LuFeO$_3$, is obtained, due to the Dzyaloshinskii-Moriya interaction. This magnetoelectric coupling is a bulk effect \cite{Das:Nc}. In addition, a magnetoelectric effect associated with special domain structures in hexagonal $R$MnO$_3$ may be possible, to be discussed in Sec. 5.3.

\subsubsection{$A_3M_2$O$_7$}
\begin{figure}
\centering
\includegraphics[width=\textwidth]{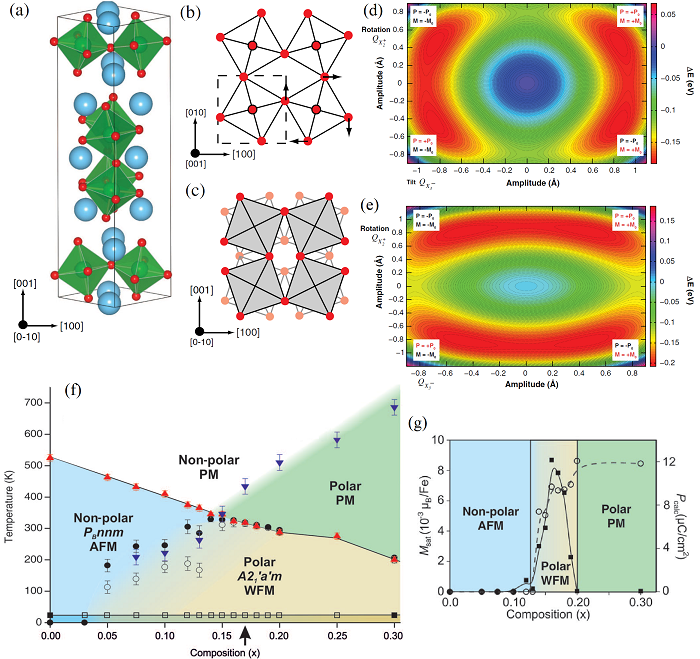}
\caption{(Colour online) Multiferroicity of $A_3M_2$O$_7$. (a) The crystal structure of $A_3M_2$O$_7$ (e.g. $A$=Ca, Sr etc; $M$=Mn, Fe, Ti etc) with the octahedra tilting and rotation. Blue: $A$, red: oxygen. (b) The $X^+_2$ rotation mode. (c) The $X^-_3$ tilt mode. (d-e) The calculated energy landscape (contour map) in the $X^+_2-X^-_3$ parameter space. Four structural domains ($\pm P$, $\pm M$) appear with the (equivalent) lowest energy. The possible paths for polarization switching (between $+P$ and $-P$) across the lowest barriers are marked (the red arcs). (d) A strain-free (bulk) case: the sign of $M$ is invariant upon the rotation. (e) The biaxial compressive case (film): the sign of $M$ can be reversible upon the tilting. (a-e) Reprinted figure with permission from \href{http://dx.doi.org/10.1103/PhysRevLett.106.107204}{N. A. Benedek \textit{et al.}, Physical Review Letters, 106, p. 107204, 2011} \cite{Benedek:Prl} Copyright \copyright (2011) by the American Physical Society. (f) The phase diagram of [$1-x$](Ca$_{0.6}$Sr$_{0.4}$)$_{1.15}$Tb$_{1.85}$Fe$_2$O$_7$-[$x$]Ca$_3$Ti$_2$O$_7$ which is multiferroic with a weak magnetization in the middle composition region. (g) The phase diagram at $300$ K as a function of composition. Solid squares: the measured saturated magnetic moment (per Fe ion); open circles: the calculated polarization. In the middle region, the material has a non-zero magnetization and a finite polarization. (f-g) From \href{http://dx.doi.org/10.1126/science.1262118}{M. J. Pitcher \textit{et al.}, Science, 347, pp. 420-424, 2015} \cite{Pitcher:Sci}. Reprinted with permission from the American Association for the Advancement of Science.}
\label{AMO}
\end{figure}

The geometric ferroelectricity is not limited in hexagonal manganites and ferrites, but also exists in some perovskite-derived systems. In perovskites, the rotation and tilting of oxide octahedra are ubiquitous, which usually suppress the ferroelectricity \cite{Benedek:Jpcc}. The rotation is a kind of antiferrodistortion (AFD), which bends the $M$-O-$M$ bonds and generates a local charge dipole for each bond. However, the bond-bending and charge-dipole between the nearest neighbor bonds must be along the opposite directions, leading to local polarization cancellation, as illustrated in Fig.~\ref{AMO}. It implies that a single antiferrodistortive mode is nonpolar.

In 2008, Bousquet \textit{et al.} predicted that a combination of two antiferrodistortive modes in PbTiO$_3$/SrTiO$_3$ superlattices may allow an improper ferroelectric polarization \cite{Bousquet:Nat}, and this prediction is a seminal progress, which stimulated Benedek and Fennie to re-visit this issue by taking Ruddlesden-Popper manganite Ca$_3$Mn$_2$O$_7$ as an object \cite{Benedek:Prl} (see Fig.~\ref{AMO}(a) for its structure). Here the two antiferrodistortive modes are respectively the $X_2^+$ rotation (path) and $X_3^-$ tilting (path), as illustrated in Fig.~\ref{AMO}(b-c). The cooperative $X_2^++X_3^-$ distortion establishes the $A2_1am$ space group, whose point group is polar. The Landau energy term accounting this effect can be expressed as: $aPQ_{X_2^+}Q_{X_3^-}$, where $a$ is the coefficient, $P$ is the polarization along the [010] axis, and $Q_{X_2^+}$ and $Q_{X_3^-}$ are the amplitudes of distortions. The polarization $P$ here is no longer the primary order parameter in the Landau energy, in contrast to the cases of proper ferroelectrics. Benedek and Fennie named this system as a ``hybrid improper ferroelectric", whose critical temperature is believed to be above room temperature, e.g. $500-600$ K in Ca$_3$Mn$_2$O$_7$, estimated from the magnitudes of $X_2^+$ and $X_3^-$ distortions \cite{Bendersky:Jssc}.

Density functional calculations predicted a polarization up to $5$ $\mu$C/cm$^2$ for Ca$_3$Mn$_2$O$_7$ and $20$ $\mu$C/cm$^2$ for Ca$_3$Ti$_2$O$_7$ (Ca$_3$Ti$_2$O$_7$ is ferroelectric without magnetism) \cite{Benedek:Prl}. For the former, a net magnetization ($\sim0.045$ $\mu_{\rm B}$/Mn) from the spin-canting along the [100] axis due to the Dzyaloshinskii-Moriya interaction was predicted, in spite of the G-type antiferromagnetic ground state below $\sim115$ K. Experimentally, this net magnetization seems to be only $\sim0.001$ $\mu_{\rm B}$/Mn, which does not depend on the direction of magnetic field mysteriously. Obviously, the $X_3^-$ mode distortion is responsible for this weak ferromagnetism \cite{Benedek:Prl}. Unfortunately, no sufficient experimental data on these properties are available at this moment.

It is interesting to note that a single reversal (not combined operation) of either the $Q_{X_2^+}$ path or the $Q_{X_3^-}$ path is sufficient to switch the polarization. However, only a reversal of the $Q_{X_3^-}$ path can switch the magnetization, suggesting that the $Q_{X_3^-}$ path is a key for a magnetoelectric manipulation. For Ca$_3$Mn$_2$O$_7$, an energy landscape of the distortions suggests that the polarization switching probably proceeds along the $Q_{X_2^+}$ path (Fig.~\ref{AMO}(d)), rendering an invariant magnetization. Even though, a magnetoelectric switch may be possible by applying a strain to the lattice, triggering the switching along the $Q_{X_3^-}$ path (Fig.~\ref{AMO}(e)). Recently, the geometric ferroelectricity and room temperature polarization switching were experimentally observed in (Ca,Sr)$_3$Ti$_2$O$_7$, a nonmagnetic isostructure of Ca$_3$Mn$_2$O$_7$ \cite{Oh:Nm}. Comprehensive experiments on (Ca$_y$Sr$_{1-y}$)$_{1.15}$Tb$_{1.85}$Fe$_2$O$_7$, also an isostructure of Ca$_3$Mn$_2$O$_7$, did reveal a weak magnetization above $300$ K and calculated polarization, as summarized in Fig.~\ref{AMO}(f-g) \cite{Pitcher:Sci}. Nevertheless, those predictions on polarization and magnetoelectric response in Ca$_3$Mn$_2$O$_7$ are still on the way for experimental checking \cite{Senn:Prl}, whose importance cannot be emphasized too much.

\subsubsection{Fluorides}
Transition metal fluorides are second to oxides to accommodate a number of (potentials) multiferroics reported so far. Scott and Blinc once published a brief review entitled ``Multiferroic magnetoelectric fluorides: why are there so many magnetic ferroelectrics?", in which many multiferroic fluorides beyond perovskite structures were introduced \cite{Scott:Jpcm}. For many non-perovskite non-oxides, the well-known $d^0$ condition for ferroelectricity may be avoided. Regardless of the huge number of fluorides, only ferroelectric fluoride sub-family Ba$M$F$_4$ ($M$=Zn, Mg, Mn, Ni, Co, Fe) will be briefly discussed here. Readers may find many others in Scott and Blinc's reviews \cite{Scott:Jpcm,Scott:Jpcm2}.

\begin{figure}
\centering
\includegraphics[width=0.7\textwidth]{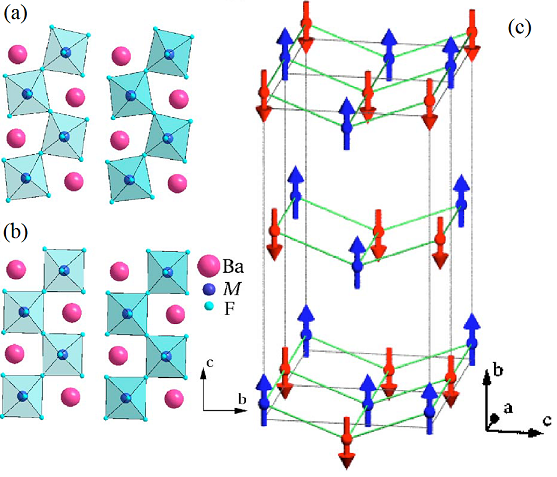}
\caption{(Colour online) (a) The crystal structure of Ba$M$F$_4$ viewed from the $a$ axis. (b) The corresponding symmetric prototype structure. (c) The magnetic structure ($M$=Mn, Fe, Ni). The magnetic moments are oriented along the $b$-axis, except the case of $M$=Co where the moments are along the $c$-axis. The magnetic ordering is uniform for all the $M$ species. Reprinted figure with permission from \href{http://dx.doi.org/10.1103/PhysRevB.74.024102}{C. Ederer \textit{et al.}, Physical Review B, 74, p. 024102, 2006} \cite{Ederer:Prb06} Copyright \copyright (2006) by the American Physical Society.}
\label{BaMF4}
\end{figure}

All the Ba$M$F$_4$ systems are magnetic except when $M$ is Zn or Mg. The common structure of B$M$F$_4$ is shown in Fig.~\ref{BaMF4}(a-b), consisting of quasi-layered $M$-F$_6$ octahedra separated by Ba layers. The ferroelectricity originates from the collective rotation of $M$-F$_6$ octahedra, occurring in very high temperatures ($>1000$ K), even higher than their melting points. The polarizations of Ba$M$F$_4$ are about $6.8-13.6$ $\mu$C/cm$^2$, as predicted by the density functional calculations \cite{Ederer:Prb06}. However, for most $M$ species except Zn, no polarization reversal has been observed, partially due to the large leakage and high coercive field. Alternatively, a pyroelectric measurement may be employed for evaluating the polarization in the future, as did for many magnetic ferroelectrics.

When $M$ is magnetic, Ba$M$F$_4$ favors the antiferromagnetic order (as sketched in Fig.~\ref{BaMF4}(c)) at low temperature. The N\'eel temperature ($T_{\rm N}$'s) for the three-dimensional antiferromagnetism is $26-29$ K when $M$ is Mn, $50$ K when $M$ is Ni, about $70$ K when $M$ is Co plus Ni \cite{Scott:Jpcm}. This order may be replaced by an in-plane two-dimensional magnetic order at a temperature higher than the N\'eel temperature, e.g. $T\sim2-3T_{\rm N}$. Recently, Zhou \textit{et al.} found the value of $T_{\rm N}$ for BaMnF$_4$ is $26$ K while the two-dimensional antiferromagnetism emerges at $65$ K, below which a magnetoelectric response was observed \cite{Zhou:Sr15}. Fox and Scott once predicted that the specific crystal structure of BaMnF$_4$ does allow a linear magnetoelectric coupling. The spontaneous polarization along the $a$-axis may induce a net magnetization along the $c$-axis \cite{Fox:Jpc}. However, recent measurements using different techniques gave divergences. The reported magnetic moment ranges from $0$ to $0.01$ $\mu_{\rm B}$/Mn \cite{Zhou:Sr15,Venturini:Aipcp,Yoshimura:Jmmm,Poole:Jpcm}, and further experiments in this area are needed to verify the results.

\subsubsection{Non-$d^0$ perovskites}
\begin{figure}
\centering
\includegraphics[width=\textwidth]{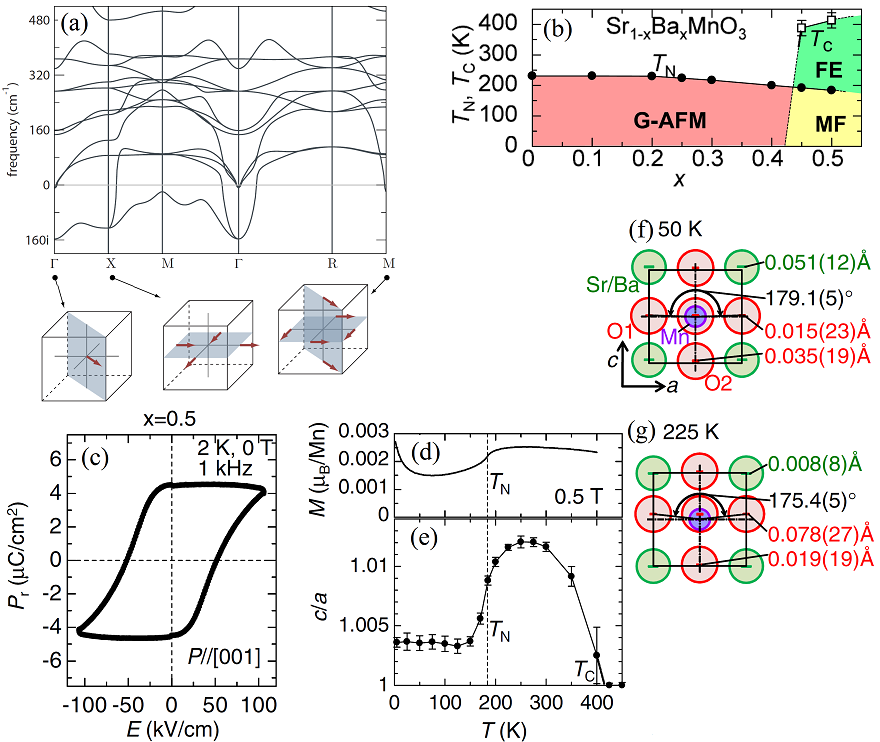}
\caption{(Colour online) Ferroelectricity and magnetic ionic displacements. (a) The phonon dispersions of G-type antiferromagnetic cubic BaMnO$_3$. The unstable phonon modes at $\gamma$, $X$, and $M$ are sketched in the cubic units with Mn at the cubic center. Reprinted figure with permission from \href{http://dx.doi.org/10.1103/PhysRevB.79.205119}{J. M. Rondinelli \textit{et al.}, Physical Review B, 79, p. 205119, 2009} \cite{Rondinelli:Prb09} Copyright \copyright (2009) by the American Physical Society. (b) The magnetic phase diagram of Sr$_{1-x}$Ba$_x$MnO$_3$. G-AFM: G-type antiferromagnetic; FE: ferroelectric; MF: multiferroic. (c) A ferroelectric hysteresis loop measured at $2$ K along the pseudocubic $c$-axis in a heavily twinned sample. (d-e) The temperature dependent magnetization under $0.5$ T field (d) and lattice tetragonality ($c/a$) (e). A schematic of the structure distortions and ionic displacements at $50$ K (f) and $225$ K (g) respectively. The data in (c-g) are for the $x=0.5$ sample. (b-g) Reprinted figure with permission from \href{http://dx.doi.org/10.1103/PhysRevLett.107.137601}{H. Sakai \textit{et al.}, Physical Review Letters, 107, p. 137601, 2011} \cite{Sakai:Prl} Copyright \copyright (2011) by the American Physical Society.}
\label{BaMnO3}
\end{figure}

A violation of the $d^0$ rule for ferroelectricity in perovskite oxides was first predicted for BaMnO$_3$ and strained CaMnO$_3$ films, thanks to the second-order Jahn-Teller distortion mechanism \cite{Rondinelli:Prb09,Bhattacharjee:Prl}. In general, when the O-$M$-O bond length is too long ($M$ is a magnetic ion like Mn$^{4+}$ or Fe$^{3+}$), the central $M$ ionic position becomes no longer stable and a spontaneous off-center displacement is preferred, which is evidenced by the imaging frequencies of phonon spectrum in the density functional calculation (see Fig.~\ref{BaMnO3}(a)). This tendency fits the mainstream argument for ferroelectricity in proper ferroelectric materials like titanates. To pursue such an unconventional proper ferroelectric source, two possible routes to reach a long enough O-$M$-O bond are: 1) using the strain coming from the substrates in the case of thin films; 2) using negative chemical pressure of big ion at the A-site. BaMnO$_3$ is certainly the best candidate for trying the second route, but unfortunately the perovskite structure is unstable. A partial substitution of Ba by Sr in BaMnO$_3$ benefits to the perovskite stability. Indeed, the Sr$_{1-x}$Ba$_x$MnO$_3$ at $x>0.4$ with perovskite structure was reported to be ferroelectric \cite{Sakai:Prl,Sakai:Prb}, as summarized in Fig.~\ref{BaMnO3}(b-g). The polarization of this set of heavily twinned samples reaches $4.5$ $\mu$C/cm$^2$, as shown in Fig.~\ref{BaMnO3}(c). Although the magnetism is not responsible for a ferroelectric polarization here, a variation of the polarization in response to the antiferromagnetic ordering was observed, and it was argued that the spin-lattice coupling, e.g. a soft-phonon mode coupled with the magnetic order, drives the magnetoelectric coupling. The exchange striction associated with the G-type antiferromagnetism suppresses the Mn-O-Mn bond bending and thus its polarization \cite{Rondinelli:Prb09,Sakai:Prl,Sakai:Prb,Giovannetti:Prl12}, as sketched in Fig.~\ref{BaMnO3}(f-g).

The significance of discovering ferroelectricity in Sr$_{1-x}$Ba$_x$MnO$_3$ when $x>0.4$ is no more than a violation of the $d^0$ rule for ferroelectricity in perovskite oxides, which was not realized until then. Similarly, a giant polarization observed in recently synthesized super-tetragonal BiFeO$_3$ thin films also has important contribution from displacements of the non-$d^0$ Fe$^{3+}$ ions from the central positions of the FeO$_6$ octahedra (to be further discussed in Sec. 4.1.1.) \cite{Zeches:Sci}. In addition, the density functional calculations predicted a similar polar instability in LaCrO$_3$ exhibits a similar polar instability \cite{Ederer:Prb11} and also in SrCrO$_3$ layer of the tensile-strained (SrCrO$_3$)$_1$/(SrTiO$_3$)$_1$ superlattice \cite{Zhou:Prl15}.

\subsubsection{Other typical type-I materials}
Apart from the above highlighted type-I multiferroics, some additional type-I multiferroics are mentioned here, such as Pb(Fe$_{\frac{1}{2}}$Nb$_{\frac{1}{2}}$)O$_3$ \cite{Yang:Prb04}, BiMnO$_3$ \cite{Kimura:Prb03}, CdCr$_2$S$_4$ \cite{Hemberger:Nat} which were discussed in details in literature. Some recently reported materials like 2H-BaMnO$_3$ \cite{Varignon:Prb}, vanadium doped La$_2$Ti$_2$O$_7$ \cite{Scarrozza:Prl}, and GeV$_4$S$_8$ \cite{Singh:Prl} also deserve concern. As an extraordinary case, an experiment based on magnetic and electric deflection measurements reported a multiferroicity in a pure metal rhodium cluster (Rh$_N$), a surprising result \cite{Ma:Prl14}. In addition, a recent calculation predicted that the vacancies on the surfaces of PbTiO$_3$ surfaces are multiferroic, needing an experimental verification \cite{Shimada:Prl}.

One also notes that the claimed multiferroicity of some materials has been questioned. LuFe$_2$O$_4$ was reported to be a special ferroelectric with charge-ordering driven electronic polarization \cite{Ikeda:Nat}. The conclusion was subsequently commented \cite{Groot:Prl,Niermann:Prl}, and no consensus has been reached yet to date. Another case is Pr$_{0.5}$Ca$_{0.5}$MnO$_3$, whose multiferroicity via the so-called site-bond-mixed charge-ordering as a novel origin of ferroelectricity was predicted long time ago \cite{Efremov:Nm,Giovannetti:Prl}. Although a few studies claimed indirect experimental evidences, a direct confirmation remains absent. Similar situation exists for charge-ordered Fe$_3$O$_4$ \cite{Alexe:Am}. It is seen that the multiferroic discipline nowadays seems to be a platform for freedom of speech. Quite a few conclusions already appeared in textbooks have been questioned and required re-visiting, while more theoretical and experimental efforts are certainly needed to clarify those extraordinary speculations on origins of ferroelectricity, such as the correlation between charge-ordering and ferroelectric polarization among others.

\subsection{Magnetoelectric properties of some selected type-II multiferroics}
In the past decade, even more exciting in the multiferroic discipline has been the discovery of a number of type-II multiferroics and relevant findings. The orthorhombic $R$MnO$_3$ family and Ca$_3$CoMnO$_6$ have been adopted as the model systems to illustrate these findings on magnetoelectric phenomena. We discuss some other type-II multiferroics in this subsection, focusing on the underlying microscopic physics.

\subsubsection{$R$Mn$_2$O$_5$}
\begin{figure}
\centering
\includegraphics[width=\textwidth]{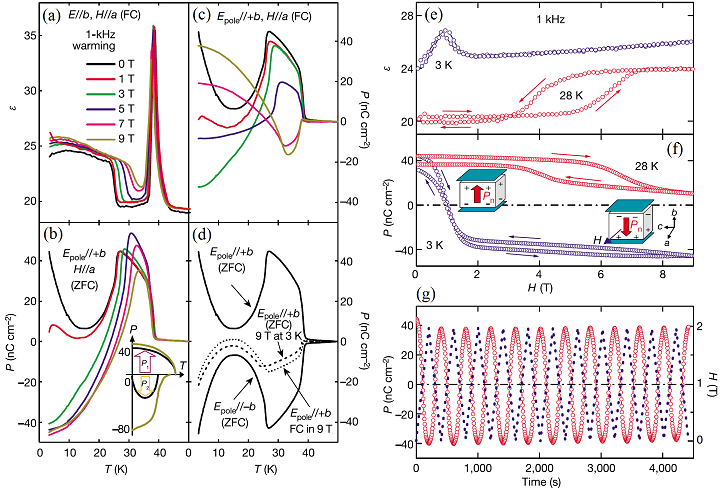}
\caption{(Colour online) Multiferroic behaviors of TbMn$_2$O$_5$. (a-d) The temperature and magnetic field (along the $a$-axis) dependences of dielectric and ferroelectric properties: (a) the dielectric constant along the $b$-axis measured in the warming run under various magnetic fields; (b) the ferroelectric polarizations along the $b$-axis measured under various magnetic fields (zero magnetic field cooling), inset: the total polarization is treated as a sum of a positive component ($P_1$) and a negative one ($P_2$); (c) the total polarization at zero magnetic field after the magnetic field cooling; (d) the total polarization measured under zero magnetic field after various cooling sequences, as indicated near the curves. Magnetic field imposes a permanent imprint in the polarization response, implying a magnetoelectric memory effect. (e-g) The polarization reversal triggered by a cycling magnetic field: (e) the dielectric constant as a function of magnetic field at $3$ K and $28$ K; (f) the total polarization as a function of magnetic field at $3$ K and $28$ K, calculated from the magnetoelectric current measured after the zero magnetic field cooling, insert cartoon: the polarization flip; (g) the polarization flip sequence at $3$ K triggered by linearly varying magnetic field cycles from $0$ to $2$ T, clearly displaying the highly reproducible polarization switching by magnetic field. Reprinted by permission from Macmillian Publishers Ltd: \href{http://dx.doi.org/10.1038/nature02572}{N. Hur \textit{et al.}, Nature, 429, pp. 392-395, 2004} \cite{Hur:Nat}. Copyright \copyright (2004).}
\label{RMn2O5}
\end{figure}

\begin{figure}
\centering
\includegraphics[width=0.88\textwidth]{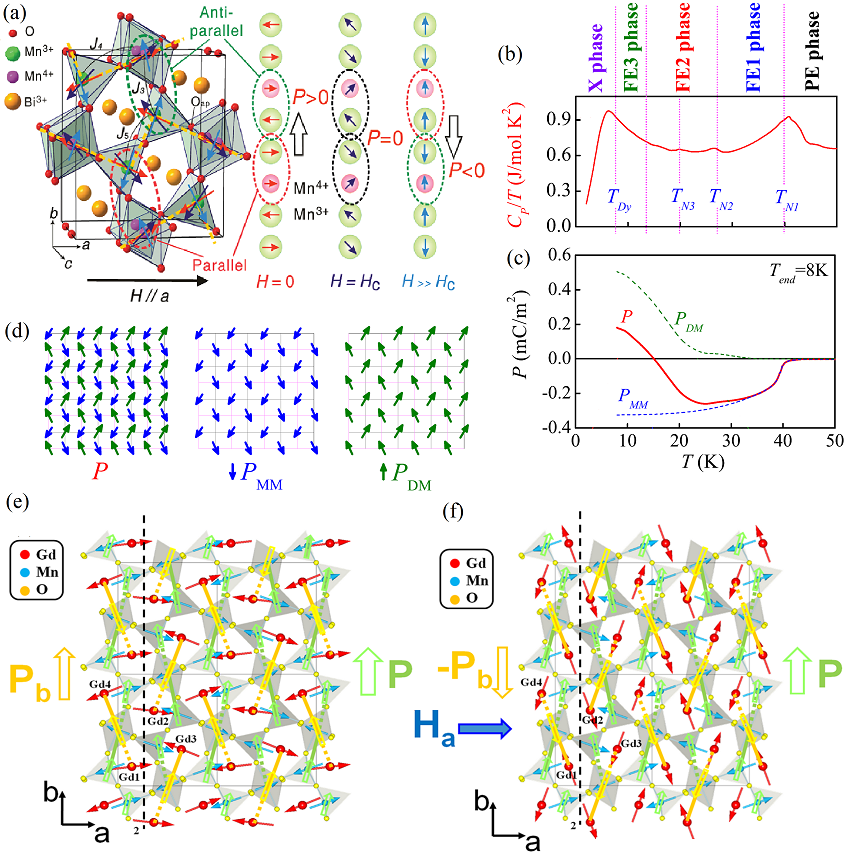}
\caption{(Colour online) The lattice and magnetic structures, and magnetoelectric response of $R$Mn$_2$O$_5$. (a) A schematic of $R$Mn$_2$O$_5$ structure (here $R$=Bi) and an evolution of the spin configuration upon increasing magnetic field. The polarization flips in accompanying with the rotation of magnetic moments. (a) Reprinted figure with permission from \href{http://dx.doi.org/10.1073/pnas.0907589106}{J. W. Kim \textit{et al.}, Proceedings of the National Academy of Sciences of the United States of America, 106, pp. 15573-15576, 2009} \cite{Kim:Pnas} Copyright \copyright (2009) by the National Academy of Sciences. (b) A schematic of the phase evolution of DyMn$_2$O$_5$ characterized by heat capacity as a function of $T$, suggesting the existence of several multiferroic states. (c) The two ferroelectric polarization components ($P_{\rm DM}$ and $P_{\rm MM}$) as evaluated from a ferrielectric model for DyMn$_2$O$_5$. The exchange striction between the Mn-Mn spin pairs gives rise to the $P_{\rm MM}$, while the exchange striction between the Dy-Mn spin pairs contributes to the $P_{\rm DM}$. The two roughly antiparallel components develop at different temperatures, leading to the unusual behavior of the pyroelectric polarization $P$ (from positive to negative with increasing $T$). (d) A schematic of the ferrielectric lattice, and corresponding $P_{\rm MM}$ lattice and $P_{\rm DM}$ lattice. (b-d) Reprinted by permission from Macmillian Publishers Ltd: \href{http://dx.doi.org/10.1038/srep03984}{Z. Y. Zhao \textit{et al.}, Scientific Reports, 4, p. 3984, 2014} \cite{Zhao:Sr}. Copyright \copyright (2014). (e) A schematic of magnetic structure of GdMn$_2$O$_5$ and the magnetostriction behavior at $5$ K. The strictions between the Gd-Mn and Mn-Mn pairs are shown by solid (attractive) and dotted (repulsive) lines, respectively. $P_b$ and $P$ denote the polarizations of the Gd and Mn magnetic sublattices, respectively. (f) Polarization $P_b$ can be reversed by a magnetic field along the $a$-axis. (e-f) Reprinted figure with permission from \href{http://dx.doi.org/10.1103/PhysRevLett.110.137203}{N. Lee \textit{et al.}, Physical Review Letters, 110, p. 137203, 2013} \cite{Lee:Prl} Copyright \copyright (2013) by the American Physical Society.}
\label{RMn2O52}
\end{figure}

The first family to be addressed here is $R$Mn$_2$O$_5$ ($R$=rare earth, Y, and Bi), another family of manganites. These manganites possesses more complex crystal structures than $R$MnO$_3$ family. The ferroelectricity was first reported in TbMn$_2$O$_5$ which has a pyroelectric polarization of about $0.04$ $\mu$C/cm$^2$ below $40$ K \cite{Hur:Nat} (summarized in Fig.~\ref{RMn2O5}). The magnetic phase transitions are also complicated, with one or more (two to five) magnetic phase transitions depending on $R$ \cite{Chapon:Prl,Cruz:Prb06,Chapon:Prl06,Kim:Pnas}. In general, upon temperature deceasing, $R$Mn$_2$O$_5$ transits first from the paramagnetic state to an incommensurate antiferromagnetic state (which does not appear in BiMn$_2$O$_5$ \cite{Kim:Pnas}), then to a commensurate antiferromagnetic state \cite{Cruz:Prb06}. In the commensurate state, the spins of the zigzag Mn$^{3+}$-Mn$^{4+}$ chains favor the $\uparrow\uparrow\downarrow\downarrow$-like configuration in the $a-b$ plane (see Fig.~\ref{RMn2O52}(a) for example), although they are not exactly collinear. For some others, e.g. DyMn$_2$O$_5$ and HoMn$_2$O$_5$, additional magnetic transitions, including a commensurate-to-incommensurate transition, the Mn spin-reorientation, and the $R^{3+}$ spin ordering, appear in sequence \cite{Cruz:Prb06} (see Fig.~\ref{RMn2O52}(b)). These magnetic transitions arise mainly from the competing multi-fold exchange interactions and lattice distortions, noting that Mn$^{3+}$, Mn$^{4+}$, and most $R^{3+}$, possess magnetic moments.

The most interesting magnetoelectric property of $R$Mn$_2$O$_5$ is the sensitive responses of ferroelectric polarization to the magnetic phase transitions. In particular, a pyroelectric current measurement detected a negative polarization at low temperatures or under a magnetic field when the poling electric field was positive \cite{Hur:Nat}, as shown in Fig.~\ref{RMn2O5}(b-c) and (f-g) for example. According to the studies on BiMn$_2$O$_5$ having the simplest magnetic order, the polarization reversal upon a magnetic field is due to the metamagnetic transition of Mn spins \cite{Kim:Pnas}, as sketched in Fig.~\ref{RMn2O52}(a). The situation is even more complicated in DyMn$_2$O$_5$ which involves the Mn$^{3+}$-Mn$^{4+}$ and Mn$^{3+}$/Mn$^{4+}$-Dy$^{3+}$ interactions as well as the ferrielectric nature \cite{Zhao:Sr}. All these characters imply more than one source of polarization, which compete with each other and render the observed positive-negative polarization switching. In spite of these observations, the underlying magnetoelectric physics of $R$Mn$_2$O$_5$ has not yet been well understood due to the complex crystal and magnetic structures. In general, the exchange striction between Mn$^{3+}$ and Mn$^{4+}$ is believed to contribute dominantly to the ferroelectric polarization \cite{Kim:Pnas}, while the exchange striction between $R^{3+}$-Mn$^{3+}$/Mn$^{4+}$ plays as the competitor \cite{Zhao:Sr}. A proposal of the ferrielectric scenario and corresponding sublattices are sketched in Fig.~\ref{RMn2O52}(c-d). For more details of the ferrielectric behavior in $R$Mn$_2$O$_5$ (e.g. DyMn$_2$O$_5$), readers can refer to a recent mini-review \cite{Liu:Jad}.

It should be mentioned that $R$Mn$_2$O$_5$ usually offers a giant tunability of ferroelectric polarization, and one example is GdMn$_2$O$_5$ where a tunability up to $0.5$ $\mu$C/cm$^2$ was observed \cite{Lee:Prl13}. The possible involved mechanism is sketched in Fig.~\ref{RMn2O52}(e-f). Moreover, an alternative evidence on ferroelectricity of $R$Mn$_2$O$_5$ was recently identified by X-ray and optical second harmonic generation techniques, claiming that a ``proper" ferroelectricity is already available at room temperature \cite{Baledent:Prl}, and the magnetism-induced polarization is merely an additional improper component to the proper ferroelectricity. The fascination of multiferroicity and relevant issues in $R$Mn$_2$O$_5$ family has been thus identified once more and it certainly deserves more theoretical and experimental works in the future.

\subsubsection{Orthoferrites \& orthochromites}
Besides orthorhombic and hexagonal manganites, orthorhombic $R$FeO$_3$ (orthoferrites) and $R$CrO$_3$ (orthochromites) are two additional families deserving exploration for promising multiferroicity. Some orthoferrites are indeed multiferroics. Tokunaga \textit{et al.} once reported gigantic magnetoelectric phenomena in DyFeO$_3$ below the antiferromagnetic ordering temperature of Dy$^{3+}$ moments ($3.5$ K) \cite{Tokunaga:Prl08}. A magnetic field along the $c$-axis can induce not only a weak ferromagnetic moment ($\geq0.5$ $\mu_{\rm B}$/formula unit), but also a ferroelectric polarization ($\geq0.2$ $\mu$C/cm$^2$), both of which are along the $c$-axis \cite{Tokunaga:Prl08}.

\begin{figure}
\centering
\includegraphics[width=0.99\textwidth]{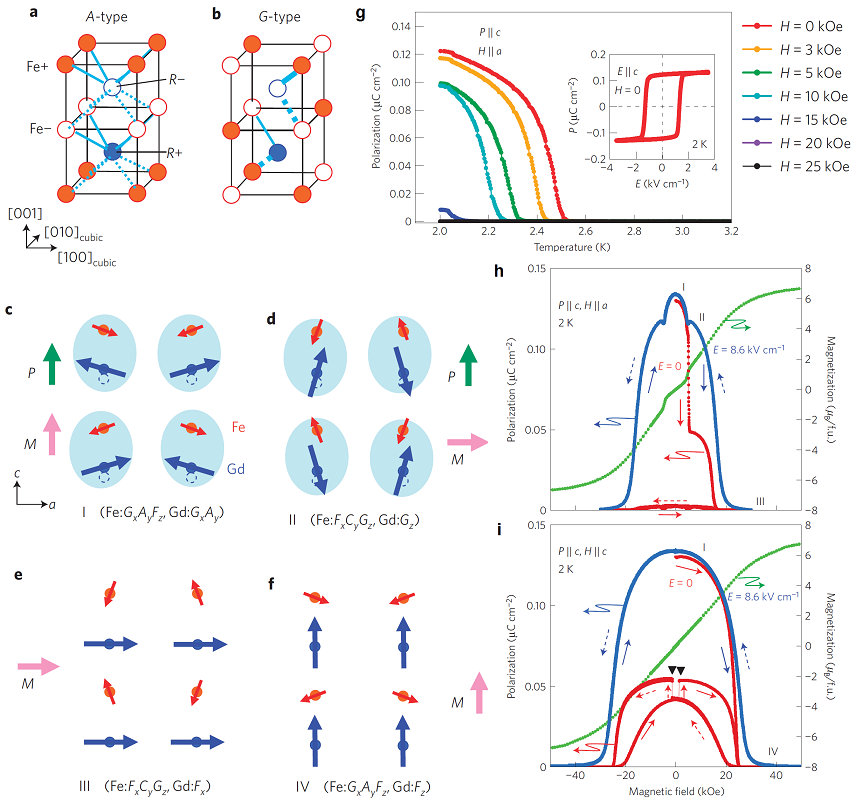}
\caption{(Colour online) The lattice and magnetic structures of GdFeO$_3$. (a-b) A sketch of the exchange strictions in orthorhombic $R$FeO$_3$ with the A-type (a) and G-type (b) antiferromagnetism. The solid (dotted) lines label the attractive (repulsive) forces. The spin orientations are distinguished by the solid/open circles and $+$/$-$ symbols. (c-f) The magnetic structures and charge dipoles of four states. The most strongly coupled Fe-Gd pairs are grouped by ellipsoids. The open circles represent the original positions without displacements. Here symbols $G$, $A$, $C$, and $F$ denote the corresponding antiferromagnetic and ferromagnetic components, respectively. (g) The measured pyroelectric polarizations along the $c$-axis under various magnetic fields applied along the $a$-axis. Inset: a standard ferroelectric hysteresis loop measured at $2$ K. (h-i) The magnetic field dependences of magnetization (green) and polarization along the $c$-axis at $2$ K, when magnetic field is along the $a$-axis (h) and $c$-axis (i). The phase regions I-IV are partitioned according to the magnetoelectric behaviors. Reprinted by permission from Macmillian Publishers Ltd: \href{http://dx.doi.org/10.1038/NMAT2469}{Y. Tokunaga, \textit{et al.}, Nature Materials, 8, pp. 558-562, 2009} \cite{Tokunaga:Nm}. Copyright \copyright (2009).}
\label{GdFeO3}
\end{figure}

\begin{figure}
\centering
\includegraphics[width=0.95\textwidth]{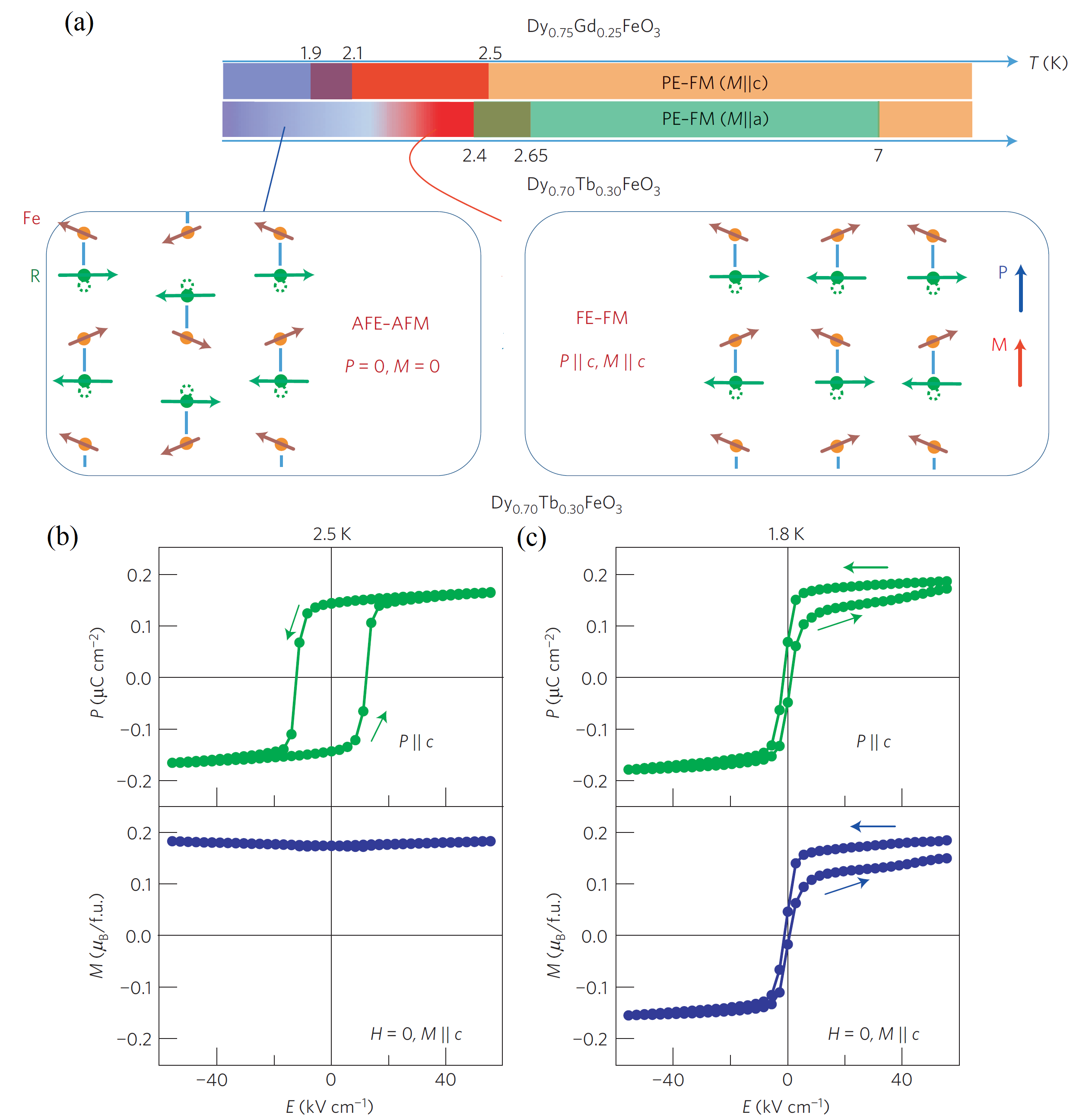}
\caption{(Colour online) A summary of results on the electric field control of canted magnetic moments in $R$FeO$_3$ ($R$=Dy$_{0.75}$Gd$_{0.25}$ and Dy$_{0.7}$Tb$_{0.3}$). (a) Upper: a magnetoelectric phase diagram of $R$FeO$_3$. The blue and red regions mark the antiferroelectric-antiferromagnetic (AFE-AFM) and ferroelectric-ferromagnetic (FE-FM) states, respectively. The paraelectric-ferromagnetic phases are marked by the orange and light-green colors, in which the net magnetic moments are along the $c$-axis and $a$-axis, respectively. Lower: a schematic of the magnetic configuration and $R$-ion displacements. Left: AFE-AFM; Right: FE-FM. Brown: Fe; Green: $R$. Dotted open circles: original positions without displacements. (b-c) The quasi-static ferroelectric and magnetoelectric hysteresis loops for Dy$_{0.7}$Tb$_{0.3}$FeO$_3$ measured at $2.5$ K and $1.8$ K, respectively. Reprinted by permission from Macmillian Publishers Ltd: \href{http://dx.doi.org/10.1038/NPHYS2405}{Y. Tokunaga, \textit{et al.}, Nature Physics, 8, pp. 838-844, 2012} \cite{Tokunaga:Np}. Copyright \copyright (2012).}
\label{DyGdFeO3}
\end{figure}

Later on, the same group observed a spontaneous polarization up to $0.12$ $\mu$C/cm$^2$ in GdFeO$_3$ below the antiferromagnetic ordering temperature of Gd$^{3+}$ ($2.5$ K), without an assistance of magnetic field \cite{Tokunaga:Nm}. The Fe$^{3+}$ moments form a G-type antiferromagnetic order with a tiny spin canting, which can be notated as the $G_xA_yF_z$ type (in Bertaut's notation). In other words, a net magnetization along the $c$-axis due to the $F_z$ component appears. The moments of Gd$^{3+}$ form the $G_xA_y$-type antiferromagnetic order. As sketched in Fig.~\ref{GdFeO3}(a-b), the $A_y$ component synchronization for both Gd$^{3+}$ and Fe$^{3+}$ allows a $\uparrow\uparrow\downarrow\downarrow$-like pattern along the $c$-axis, while the $G_x$ component synchronization results in a polar axis in the orthorhombic unit cell, noting that the exchange strictions in the cubic unit cell are fully cancelled. A magnetic field above $0.5$ T along the $a$-axis can re-orientate the Fe$^{3+}$/Gd$^{3+}$ spins to the $F_xC_yG_z$/$G_z$ patterns respectively, which leads to a sudden drop of polarization, as shown in Fig.~\ref{GdFeO3}(d-e) and (h). Further increasing magnetic field can fully suppress the polarization since the Gd$^{3+}$ moments are fully ferromagnetically aligned. If a magnetic field is applied along the $c$-axis (see Fig.~\ref{GdFeO3}(c), (f), and (i)), the ferroelectric polarization will decay monotonously in association with a gradual variation of magnetic structure of the Gd$^{3+}$ sublattice to the $F_z$. Another bright spot is that not only a magnetic-field-control of polarization but also an electric-field-control of magnetic moment were realized in $R$FeO$_3$ ($R$=Dy$_{0.7}$Tb$_{0.3}$ or Dy$_{0.75}$Gd$_{0.25}$) \cite{Tokunaga:Np}, as summarized in Fig.~\ref{DyGdFeO3}. The electro-control of magnetism is, in fact, becoming a highly concerned issue not only in multiferroics but also more as an emergent hot topic for next generation of spintronic devices.

It should be noted that the observed magnetoelectric effects in orthoferrites are intrinsically correlated with the ordering of rare-earth $4f$ spins. A disadvantage for this characteristic is the extremely low temperature at which this ordering occurs. One exception is SmFeO$_3$, and Lee \textit{et al.} reported a very high ferroelectric Curie temperature up to $670$ K, attributed to the canting of Fe$^{3+}$ spins \cite{Lee:Prl}. Similar high temperature multiferroicity has also been reported in LuFeO$_3$, in which Lu$^{3+}$ is nonmagnetic \cite{Chowdhury:Apl}. However, such a Fe$^{3+}$-induced ferroelectricity has been in question \cite{Johnson:Prl12,Lee:Prl12}. A recent neutron diffraction study has also provided a set of negative evidences against the high temperature ferroelectricity in SmFeO$_3$ \cite{Kuo:PRL}.

Some orthochromites were also reported to be multiferroics. For instance, Rajeswaran \textit{et al.} reported a switchable polarization (up to $0.2-0.8$ $\mu$C/cm$^2$) and a weak magnetization below the N\'eel temperature of Cr ($\sim130-200$ K) in polycrystalline $R$CrO$_3$ samples, where the rare-earth ion is magnetic \cite{Rajeswaran:Prb}. Several possible mechanisms were proposed, including the exchange striction between $R^{3+}$ and Cr$^{3+}$, similar to the case in GdFeO$_3$. Nevertheless, sufficient evidence with the ferroelectricity in these orthochromites is still needed and the difficulty originates from the large leakage due to the narrow band-gap of these materials. Readers can refer to a brief review by Meher \textit{et al.} for more details of multiferroic chromites and cobaltites \cite{Meher:Cm}.

\subsubsection{Quadruple perovskites}
As well known, Mn-based oxides host a considerable portion of the type-II multiferroics. However, most of them have a ferroelectric Curie temperature below $40$ K. In 2009, Dong \textit{et al.} predicted a spin-orthogonal stripe (SOS) phase in quarter ($1/4$) hole-doped perovskite manganites, which should be multiferroic due to the existence of a noncollinear spin order area (on the stripe-walls) and a collinear $\uparrow\uparrow\downarrow\downarrow$-like order area (inside the stripes) \cite{Dong:Prl}. The estimated characteristic temperatures for magnetic ordering and ferroelectricity may reach $\sim100$ K, about $3-4$ times of those of $R$MnO$_3$. The underlying mechanism to stabilize the noncollinear spin structure is not the traditional exchange frustration involving the next-nearest neighbor exchanges, as the case of TbMnO$_3$ \cite{Kimura:Prb}. Instead, the electronic self-organization is responsible for such spin-orthogonal stripes. A further study extended this spin-orthogonal stripe state to other doping concentrations, e.g. $1/2$, $1/3$, $1/5$, $1/6$, ..., $1/\infty$ \cite{Liang:Prb}.

\begin{figure}
\centering
\includegraphics[width=0.9\textwidth]{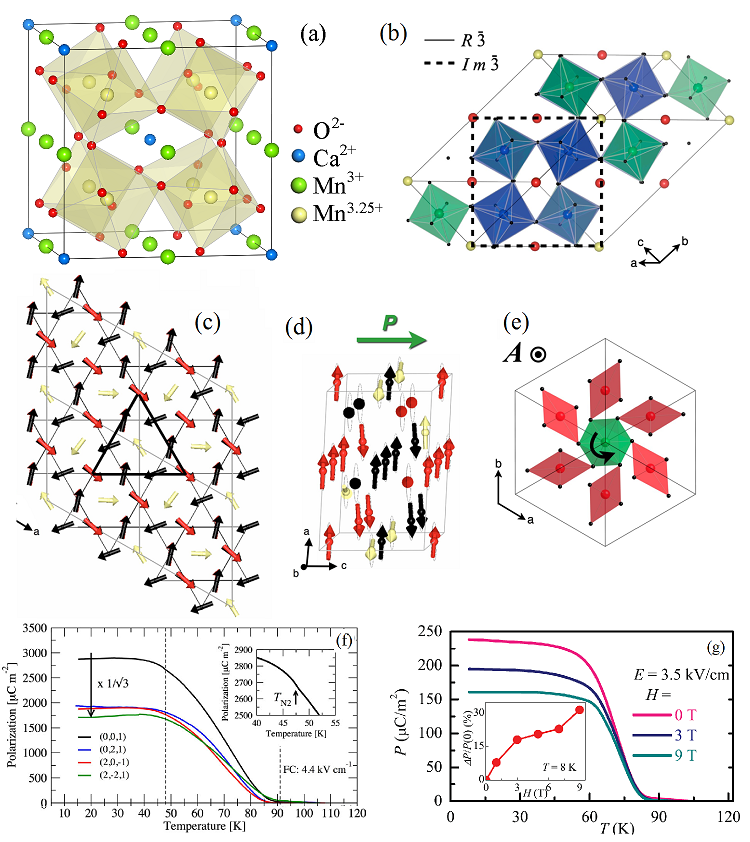}
\caption{(Colour online) The lattice and magnetic structures of CaMn$_7$O$_{12}$. (a) A cubic structure ($Im\bar{3}$) above $440$ K. The A(A')-site Ca$^{2+}$/Mn$^{3+}$ cations are ideally ordered. Due to the small sizes of Ca$^{2+}$/Mn$^{3+}$, the lattice is significantly shrunk with respect to a normal perovskite manganite. (b) The low temperature rhombohedral structure ($R\bar{3}$). The B-sites are occupied with Mn$^{3.25+}$ cations at high temperature but replaced by the 3 Mn$^{3+}$ + Mn$^{4+}$ charge-ordered network below $250$ K. (c-d) The noncollinear magnetic structure between $48$ K and $90$ K in the rhombohedral (c) $ab$ and (d) $ac$ planes. (e) A proposed ferroaxial vector model. (f) The measured pyroelectric polarizations along various axes. Insert: a tiny anomaly at $48$ K. (g) The measured pyroelectric polarizations of a polycrystalline sample under different magnetic fields. Insert: the magnetoelectric response at low temperature. (a) and (g) Reprinted figure with permission from \href{http://dx.doi.org/10.1103/PhysRevB.84.174413}{G. Q. Zhang \textit{et al.}, Physical Review B, 84, p. 174413, 2011} \cite{Zhang:Prb11} Copyright \copyright (2011) by the American Physical Society. (b-f) Reprinted figure with permission from \href{http://dx.doi.org/10.1103/PhysRevLett.108.067201}{R. D. Johnson \textit{et al.}, Physical Review Letters, 108, p. 067201, 2012} \cite{Johnson:Prl} Copyright \copyright (2012) by the American Physical Society.}
\label{CaMn7O12}
\end{figure}

The prerequisites for a possible SOS state are very rigor: 1) small lattice constants; 2) exact quarter ($1/4$) doping level; 3) zero or weak Jahn-Teller distortion; 4) sufficient band-gap (insulating); 5) zero or weak structure disorder. It is almost impossible to fulfill all these conditions in a conventional manganite. However, there exist a set of specific manganites, quadruple perovskite CaMn$_7$O$_{12}$, which can be also written as (CaMn$_3$)Mn$_4$O$_{12}$. The three Mn$^{3+}$ ions together with one Ca$^{2+}$ ion occupy the A-site in a fully ordered manner as shown in Fig.~\ref{CaMn7O12}(a-b). The average valence of the rest four Mn ions is $+3.75$, exactly at the quarter doping. The lattice constants are extremely small comparing with conventional ABO$_3$ manganites, while a quite weak Jahn-Teller distortion (the $Q_2$ mode) was reported. Surprisingly, all prerequisites seem to be satisfied!

Subsequently, Zhang \textit{et al.} synthesized CaMn$_7$O$_{12}$ polycrystalline samples and found a magnetism-related ferroelectric polarization below $90$ K \cite{Zhang:Prb11}. There exist two magnetic transitions at $90$ K and $48$ K respectively. Another experiment by Johnson \textit{et al.} confirmed the magnetic ferroelectricity emerging at $90$ K (Fig.~\ref{CaMn7O12}(f)) using single crystal samples and resolved the magnetic structure between $48$ K and $90$ K \cite{Johnson:Prl}, as sketched in Fig.~\ref{CaMn7O12}(c-d). The pyroelectric polarization reaches $0.287$ $\mu$C/cm$^2$, even larger than that of DyMnO$_3$ \cite{Kimura:Prb05}. Johnson \textit{et al.} argued a new physical mechanism for the ferroelectricity in CaMn$_7$O$_{12}$, in which the polarization couples to the ferroaxial component of the crystal structure \cite{Johnson:Prl} (see Fig.~\ref{CaMn7O12}(e)). This ferroaxial mechanism, expressed as $\textbf{P}\sim\sigma\textbf{A}$ where $\sigma$ is a measure of the macroscopic chirality of spin order and $\textbf{A}$ is the macroscopic axial vector of the crystal structure, was first proposed to understand the magnetism-induced ferroelectricity in Cu$_3$Nb$_2$O$_8$ \cite{Johnson:Prl11} and MnSb$_2$O$_6$ \cite{Johnson:Prl13}. More details regarding this ferroaxial concept can be found in a paper by Johnson and Radaelli \cite{Johnson:Armr}.

Alternatively, based on the first-principles calculations, Lu \textit{et al.} proposed that a combination of both the Dzyaloshinskii-Moriya interaction and the exchange striction is responsible to the giant improper ferroelectricity and its strong magnetoelectric response (e.g. see Fig.~\ref{CaMn7O12}(g)) \cite{Lu:Prl}. Very exceptionally, it is believed that the magnitude of polarization is determined mainly by the exchange striction but the orientation of polarization is controlled by the Dzyaloshinskii-Moriya interaction, explaining reasonably an unusual fact for this material that the polarization is large and the magnetoelectric response is strong. In addition, the particular orbital modulation was found to entangle with the magnetic helicity, which plays an important role in coupling the magnetism with ferroelectricity in CaMn$_7$O$_{12}$ \cite{Perks:Nc,Du:Prb14}.

CaMn$_7$O$_{12}$ is the first multiferroic member of the quadruple perovskite family but definitely not the last one. A recent calculation predicted its sister system SrMn$_7$O$_{12}$ to be multiferroic \cite{Liu:Epl}. Following experiment observed the pyroelectricity, but it was doubted due to the extrinsic contribution from high leakage \cite{Glazkova:Ic}. For both CaMn$_7$O$_{12}$ and SrMn$_7$O$_{12}$, the too small band gaps challenge the experimental verifications in relative high temperatures. Another example is LaMn$_3$Cr$_4$O$_{12}$, as reported by Wang \textit{et al.} \cite{Wang:Prl15}. Two unexpected ferroelectric transitions were surprisingly identified in this cubic structure which should otherwise be non-ferroelectric. The two transitions correspond to the B-site and A-site G-type antiferromagnetic orderings respectively. The symmetry analysis and density functional calculations suggested that the spin-orbit coupling together with the dual G-type antiferromagnetism can drive a weak polarization. Nevertheless, more investigations are definitely needed to understand the claimed ferroelectricity. It is also noted that the quadruple perovskite family has a huge number of members \cite{Vasilev:Ltp}, probably promising much beyond expectations.

\subsubsection{Cu-based oxides}
Cuprates are well known as high-$T_{\rm C}$ superconductors and strongly correlated electronic systems \cite{Dagotto:Rmp94}. Although most multiferroics are also strongly electron-correlated, not many multiferroic cuprates have been found. Several exceptions include the quantum magnets LiCu$_2$O$_2$ and LiCuVO$_4$ reported ten years ago \cite{Park:Prl,Naito:Jpsj}. Since the Cu$^{2+}$ spin moments are small, quantum fluctuations become inevitable and thus a multiferroicity if any may not appear until very low temperature, as shown in these cuprates. In spite of the conceptual difference between superconductivity and multiferroicity, search for high-$T_{\rm C}$ superconductors may help the multiferroic community to find additional multiferroics with better performance such as high ferroelectric Curie temperature $T_{\rm C}$.

\begin{figure}
\centering
\includegraphics[width=\textwidth]{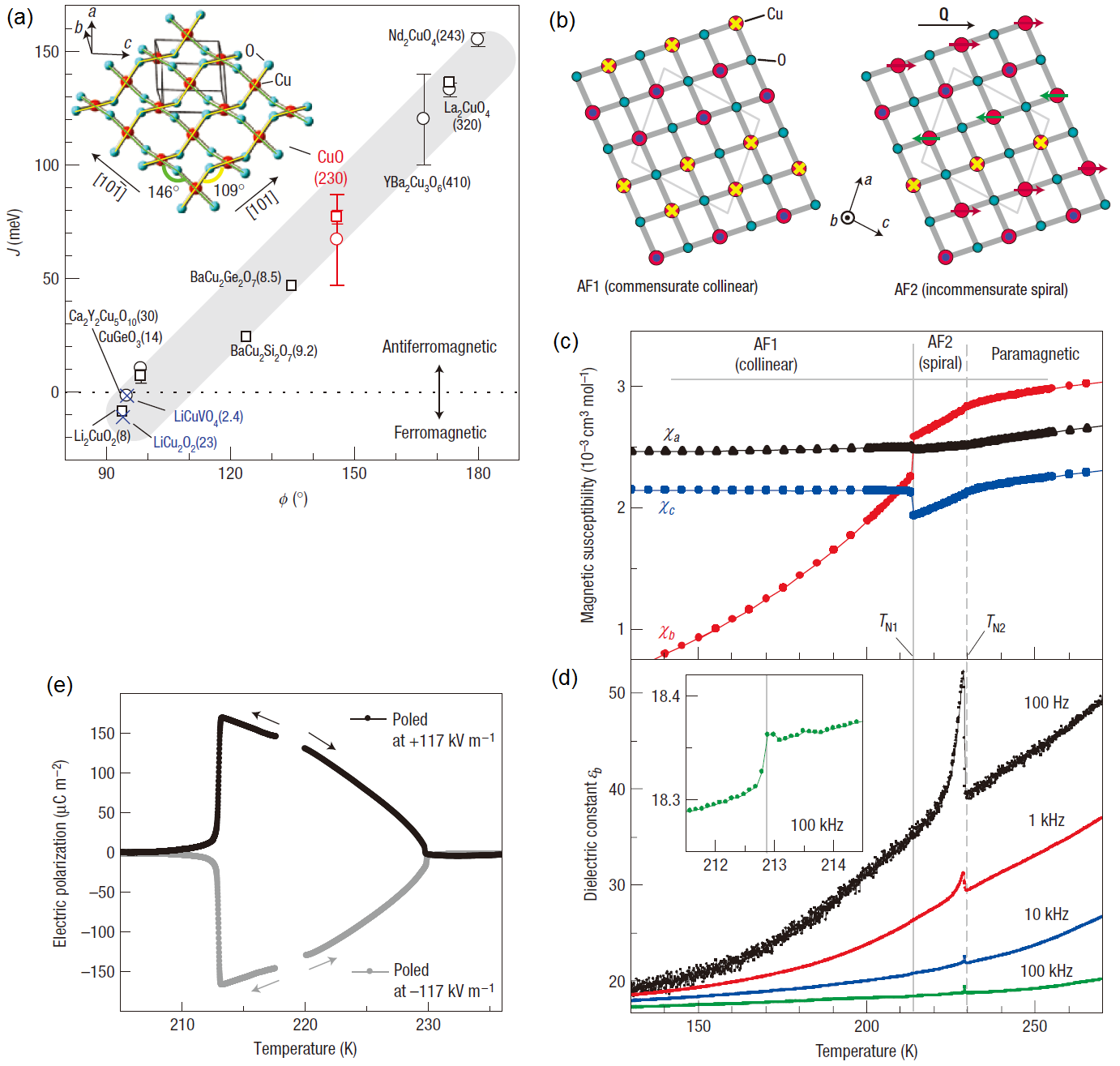}
\caption{(Colour online) Multiferroicity of CuO. (a) A magnetic phase diagram for a series of low-dimensional cuprates, where $J$ is the superexchange and $f$ the Cu-O-Cu bond angle. CuO  among various cuprates has a moderate $J$. The original $J-f$ plot for cuprates was presented in Refs.~\cite{Shizmizu:Jpsj,Mizuno:Prb}. LiCu$_2$O$_2$ and LiCuVO$_4$ are multiferroics with a spiral magnetic order. The numbers in the parentheses denote the magnetic ordering temperatures. Inset: a schematic of CuO crystal structure. (b) A sketch of the magnetic structures of CuO: commensurate collinear (AF1) and incommensurate spiral (AF2) phases. (c) The magnetic susceptibility as a function of temperature measured along the reciprocal lattice axes. (d) The dielectric constant measured along the $b^*$-axis at various frequencies. Inset: a magnified view at frequency of $100$ kHz around $T_{\rm N1}$. (e) The ferroelectric polarization along the $b^*$-axis as a function of temperature respectively under a positive (black) electric poling and a negative (grey) electric poling with the given electric fields. The poling started above $T_{\rm N2}$ till the sample cooling down to the lower boundary of the AF2 phase ($220$ K). Reprinted by permission from Macmillian Publishers Ltd: \href{http://dx.doi.org/10.1038/nmat2125}{T. Kimura, \textit{et al.}, Nature Materials, 7, pp. 291-294, 2008} \cite{Kimura:Nm}. Copyright \copyright (2008).}
\label{CuO}
\end{figure}

In 2008, Kimura \textit{et al.} reported the observations on CuO \cite{Kimura:Nm}, noting that CuO has very strong exchanges, according to earlier works (see Fig.~\ref{CuO}(a) for example). CuO has two antiferromagnetic transitions at $213$ K and $230$ K respectively, as sketched in Fig.~\ref{CuO}(b-c). It is paramagnetic above $230$ K and and then replaced by a collinear-antiferromagnetic order below $213$ K. Inside the narrow window between $213$ K and $230$ K is a noncollinear spiral spin order. Similar to other spiral magnets, CuO is a ferroelectric within this window with a weak polarization ($\sim0.016$ $\mu$C/cm$^2$), as evidenced by the dielectric and pyroelectric measurements (Fig.~\ref{CuO}(d-e)). The unusual feature here is the unexceptionally high temperature $T_{\rm C}$ for multiferroicity. The subsequent density functional calculations confirmed that such a high $T_{\rm C}$ does originate from the strong exchanges \cite{Giovannetti:Prl11,Jin:Prl}.

\subsubsection{Complex hexaferrites}
Empirically, the exchanges between Fe ions are usually strong and some high-temperature magnetic ferroelectrics from Fe-based oxides are expected. There have been some reported multiferroic hexaferrites, noting that hexaferrites are also a big family and many of them have been widely used in magneto-electronic industry \cite{Pullar:Pms}. In 2005, it was again Kimura \textit{et al.} who reported the magneto-control of ferroelectric polarization in a Y-type hexaferrite Ba$_{0.5}$Sr$_{1.5}$Zn$_2$Fe$_{12}$O$_{22}$ \cite{Kimura:Prl}. This material is a non-ferroelectric helimagnetic insulator in the zero-field ground state. A low magnetic field is sufficient to drive successive metamagnetic transitions. A concomitant ferroelectric polarization appears in some of the magnetic-field induced phases with a long-wavelength magnetic structure. The polarization can be rotated in $360^\circ$ by external magnetic field. This functionality opens up potential applications for not only room temperature magnetoelectric devices but also devices based on magnetically controlled electro-optical response.

\begin{figure}
\centering
\includegraphics[width=\textwidth]{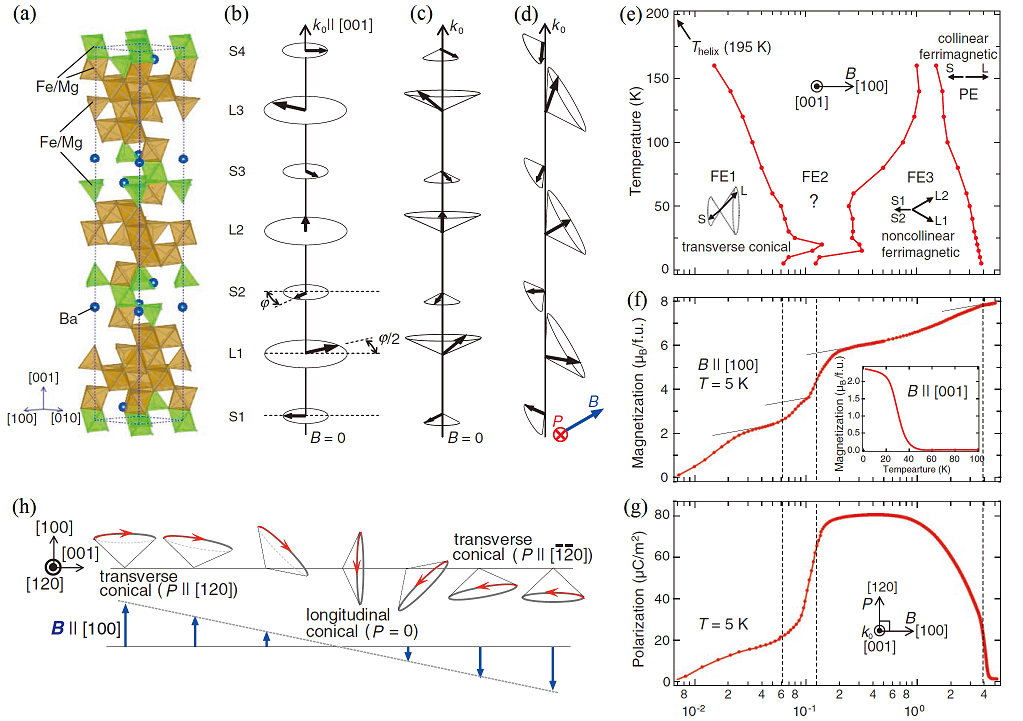}
\caption{(Colour online) Multiferroicity of Y-type hexaferrite Ba$_2$Mg$_2$Fe$_{12}$O$_{22}$. (a) A schematic of crystal structure. The magnetic structure consists of alternating stacks of L blocks (brown, large magnetic moment) and S blocks (green, small magnetic moment). (b-d) Illustrations of the helicoidal spin structures: (b) a proper screw structure between $50$ and $195$ K; (c) a longitudinal conical structure below $50$ K; (d) a slanted conical structure below $195$ K under a small magnetic field ($\sim30$ mT). (e) The magnetoelectric phase diagram. (f-g) The measured magnetization (f) and polarization (g) as a function of magnetic field at $5$ K respectively. (Inset) A temperature-dependent magnetization measured in a zero-field heating run after a $5$ T field cooling. A plausible spin configuration for each phase is shown with the phase boundaries in (f) and (g) marked the dashed lines. In (e-g), magnetic field is applied along the [001] axis. (h) A schematic of the evolutions of the spin cone and helicity with an oscillating magnetic field. From \href{http://dx.doi.org/10.1126/science.1154507}{S. Ishiwata \textit{et al.}, Science, 319, pp. 1643-1646, 2008} \cite{Ishiwata:Sci}. Reprinted with permission from the American Association for the Advancement of Science.}
\label{hexaferrite}
\end{figure}

In 2008, Ishiwata \textit{et al.} reported their observations on a low magnetic-field control of polarization vector in a Y-type hexaferrite Ba$_2$Mg$_2$Fe$_{12}$O$_{22}$ (Fig.~\ref{hexaferrite}(a)) \cite{Ishiwata:Sci}. As shown in Fig.~\ref{hexaferrite}(b-c), its spin structure is helimagnetic with the propagation vector $k_0$ parallel to the [001] axis. The magnetic field induced transverse conical spin structure (Fig.~\ref{hexaferrite}(d)) carries the $\textbf{P}$ vector directing perpendicular to the field and $k_0$, in accordance with the prediction of the KNB (Katsura-Nagaosa-Balatsky) spin current model. The magnetic phase diagram and associated magnetization/polarization are presented in Fig.~\ref{hexaferrite}(e-f). An oscillating or multidirectionally rotating magnetic field produces a cyclic displacement current via a flexible handling of the magnetic cone axis, as sketched in Fig.~\ref{hexaferrite}(h). Later on, the magnetic field induced transverse conical spin order was confirmed in neutron scattering studies \cite{Ishiwata:Prb10,Soda:Prl11}.

The advantage of multiferroicity of Ba$_2$Mg$_2$Fe$_{12}$O$_{22}$ is twofold. First, the magnetoelectric response is very sensitive. Different from the most type-II multiferroics which usually need a magnetic field up to several Tesla in order to reverse/rotate/suppress the ferroelectric polarization, the polarization of Ba$_2$Mg$_2$Fe$_{12}$O$_{22}$ exhibits sufficient response to a magnetic field as low as $0.03$ Tesla. Second, the ferroelectric Curie temperature is relatively high. It is noted that the proper screw spin structure can survive until $195$ K although it is non-ferroelectric. Probably, a non-ferroelectric screw order has a comparable energy as that of a ferroelectric conical order so that a low magnetic field is sufficient to favor one than the other, making the magneto-control of ferroelectric polarization a practically feasible event.

Furthermore, Chun \textit{et al.} found that the critical magnetic field for switching a polarization can be significantly reduced from $1.0$ T down to $1.0$ mT in a properly Al-substituted system, as shown in Ba$_{0.5}$Sr$_{1.5}$Zn$_2$(Fe$_{1-x}$Al$_x$)$_{12}$O$_{22}$ \cite{Chun:Prl,Noh:Prl}. A giant magnetoelectric susceptibility was observed in the sample with $x=0.08$. In this case, the more surprising is a magnetization reversal driven by an electric field without assistance of magnetic field, at a temperature up to $150$ K \cite{Chai:Nc}.

The above discussion on Ba-based hexaferrites is not an exceptional case and other types of hexaferrites may share the same physical picture. Kitagawa \textit{et al.} once reported the low-field magnetoelectric effect in Z-type hexaferrite Sr$_3$Co$_2$Fe$_{24}$O$_{41}$  at room temperature \cite{Kitagawa:Nm}. The magnetic transition temperatures are as high as $510$ K and $670$ K. At room temperature, this Sr-based hexaferrites shows a large magnetic moment of $6-8$ $\mu_{\rm B}$ per formula unit. A small magnetic field like $0.2$ Tesla can induce a polarization up to $0.002$ $\mu$C/cm$^2$ at room temperature. For more information about the progress on multiferroic hexaferrites, readers can refer to a recent review by Kimura \cite{Kimura:Arcp}.

\subsubsection{Organic molecules \& polymers}
\begin{figure}
\centering
\includegraphics[width=0.8\textwidth]{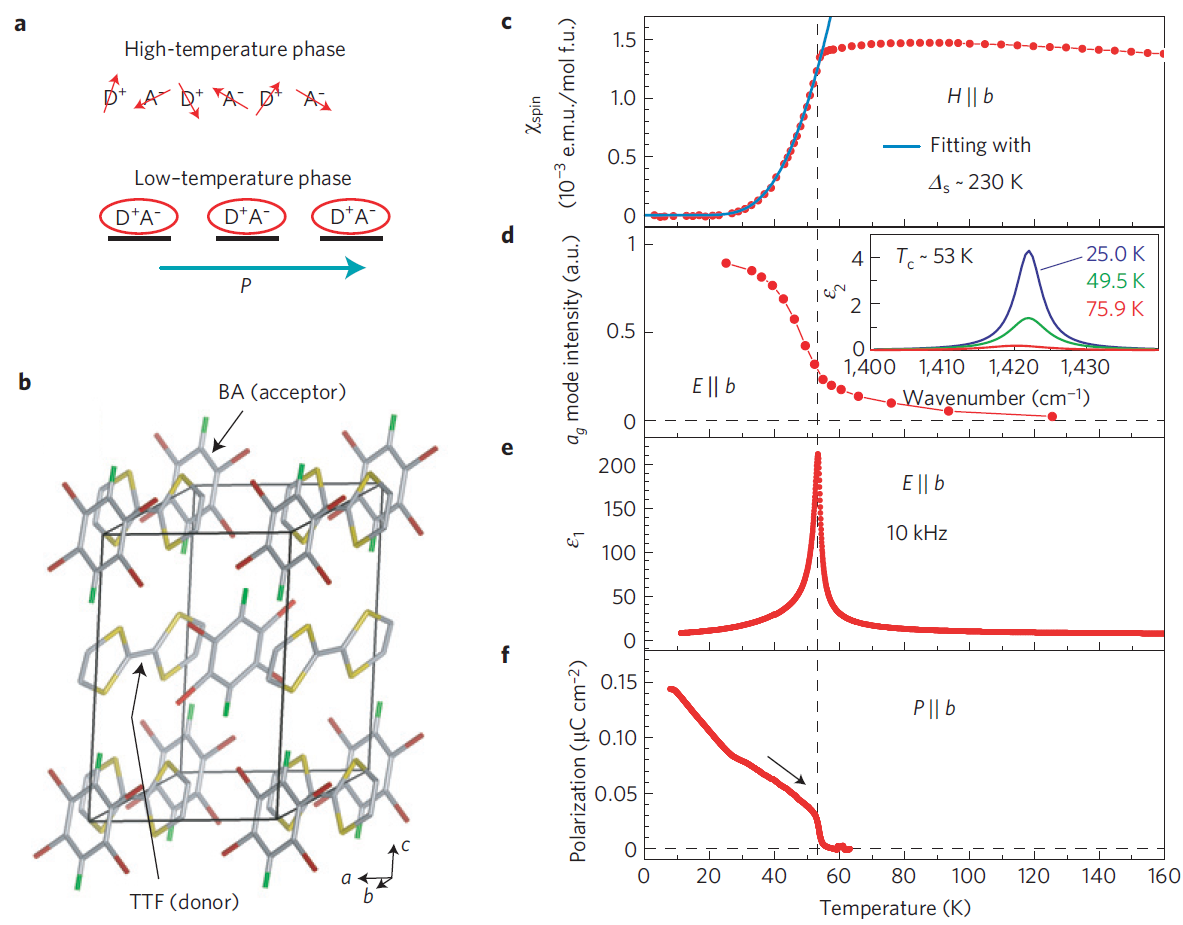}
\caption{(Colour online) Ferroelectric spin-Peierls transition in TTF-BA. (a) A schematic of ionic donor ($D^+$) and acceptor ($A^-$) in the high- and low-temperature phases of TTF-BA. The spins are denoted by arrows. The ellipsoids with underlines represent the dimers. (b) The crystal structure of TTF-BA. (c-f) The measured temperature dependences of relevant physical properties: (c) the spin susceptibility; (d) the normalized spectral weight of the infrared $a_g$ mode as a fingerprint of local $D^+A^-$ dimerization, insert: temperature dependence of the $a_g$ mode; (e) the dielectric constant; (f) the ferroelectric polarization along the $b$ axis. Reprinted by permission from Macmillian Publishers Ltd: \href{http://dx.doi.org/10.1038/NPHYS1503}{F. Kagawa, \textit{et al.}, Nature Physics, 6, pp. 169-172, 2010} \cite{Kagawa:Np}. Copyright \copyright (2010).}
\label{TTF}
\end{figure}

So far all of the multiferroics we have discussed are inorganic matters, but some organic compounds, in spite of not many, were reported to be multiferroic. The organic charge-transfer salt TTF-BA (tetrathiafulvalene-p-bromanil) was investigated in 2010 by Kagawa \textit{et al.} who identified a set of magnetic ferroelectric behaviors in such a distinct structure \cite{Kagawa:Np}. As schematically shown in Fig.~\ref{TTF}(a-b), TTF-BA contains the ionic TTF donor ($D^+$) and BA acceptor ($A^-$) molecules, forming the $D^+A^-D^+A^-$ mixed stacks along the $a$- and $b$-axes, respectively. The dimerization occurs in both the stacks below $53$ K. The TTF and BA molecules in TTF--BA are almost ionic in the whole temperature region (the ionicity is as high as $0.95$). In this sense, the $D^+A^-$ stack can be regarded as a one-dimensional Heisenberg chain with spin $-1/2$, similar to inorganic Ca$_3$CoMnO$_6$. Due to the spin-Peierls transition, i.e. the dimerization between $D^+A^-$, ferroelectricity emerges below $53$ K. The anomalies of spin susceptibility and infrared reflectivity coincide with the sharp peak of dielectric constant and emergency of spontaneous polarization along the $b$-axis, as presented in Fig.~\ref{TTF}(c-f).

Besides TTF-BA, there are a variety of organic multiferroics (molecules \& polymers) verified or predicted \cite{Ren:Am,Wu:Jacs,Qin:An,Qin:An15}, but further experimental studies on those predicted materials are needed.

\subsubsection{Other typical type-II multiferroics}
In addition to the typical type-II multiferroics which are categorized into different subgroups (sub-families), there are indeed more which have been experimentally/theoretically investigated. A short list of them are MnWO$_4$ \cite{Arkenbout:Prb,Yu:Prb13}, CoCr$_2$O$_4$ \cite{Yamasaki:Prl06}, CuFeO$_2$ \cite{Kimura:Prb06}, Ni$_3$V$_2$O$_8$ \cite{Lawes:Prl}, Cu$_3$Nb$_2$O$_8$ \cite{Johnson:Prl11}, $R_2$CoMnO$_6$ \cite{Ma:Pccp,Xin:Ra}, $R_2$NiMnO$_6$ \cite{Zhou:Apl,Kumar:Prb10}, Ba$_2$CoGe$_2$O$_7$ \cite{Murakawa:Prl10,Soda:Prl,Nakajima:Prl}, Ba$_3$NiNb$_2$O$_9$ \cite{Hwang:Prl}, RbFe(MoO$_4$)$_2$ \cite{Hearmon:Prl,Mitamura:Prl}, DyVO$_3$ \cite{Zhang:Prb14}, CdV$_2$O$_4$ \cite{Mun:Prl},  Ni$_3$TeO$_6$ \cite{Hudl:Prb,Oh:Nc}, and so on. More candidates were predicted by the density functional calculations, e.g. $R$NiO$_3$ \cite{Giovannetti:Prl09}. Some sulfides and selenides as type-II multiferroics, e.g. FeCr$_2$S$_4$ \cite{Bertinshaw:Sr,Lin:Sr}, CoCr$_2$S$_4$ \cite{Dey:Prb}, and Cu$_2$MnSnS$_4$/Cu$_2$MnSnSe$_4$ (only predicted but not yet experimentally confirmed), were once discussed too \cite{Fukushima:Prb}. These materials usually show complicated magnetic structures but the multiferroic mechanisms are not very different from the established frameworks.

One should be reminded that a pursuit of more multiferroic compounds with desirable performance will continue in the future. In fact, the performances of so far available single phase multiferroic materials are still unsatisfied for practical applications, although one or two are indeed superior in terms of one or more sides of the multiferroic functionality polyhedron. Opportunities are ubiquitous but unforseen to some extent. In the next subsection we discuss an additional aspect of the relevant progress before we visit other issues.

\subsection{Magnetoelectric coupling III: multiple contributions}
As a general framework of magnetoelectricity, the spin-orbit coupling and spin-lattice coupling are the two major ingredients with different characteristics, as described one by one in previous subsections. Our conventional understanding is that neither the spin-orbit coupling nor the spin-lattice coupling alone can give fully ideal magnetoelectricity. If they can be incorporated into one material, a large polarization from the spin-lattice coupling may become magnetically sensitive due to the spin-orbit coupling. Namely, a hybrid mechanism combining the advantages from the spin-orbit coupling and spin-lattice coupling would benefit to the enhancement of magnetoelectric performance. In fact, such a hybrid mechanism widely exists in many multiferroics.

We may take the perovskite $R$MnO$_3$ family as a model. In the previous subsections, TbMnO$_3$ and DyMnO$_3$ are classified into the zoo of spin-orbit coupling, while o-HoMnO$_3$ and o-YMnO$_3$ belong to the zoo of spin-lattice coupling. However, later experimental and theoretical investigations suggested that such a classification is not a rigorous treatment and these materials in fact accommodate multiple contributions to the magnetoelectricity, i.e. a hybrid mechanism is responsible for the magnetoelectricity. For example, for those manganites with spiral spin orders, which were treated as the prototypes driven by the spin-orbit coupling, the exchange striction may also make non-negligible contribution. Therefore, these materials are also the representatives for illustrating this hybrid mechanism. For more details, we revisit DyMnO$_3$ even though it as a prototype for evidencing the spin-orbit coupling induced ferroelectricity has been briefly reviewed before.

\begin{figure}
\centering
\includegraphics[width=\textwidth]{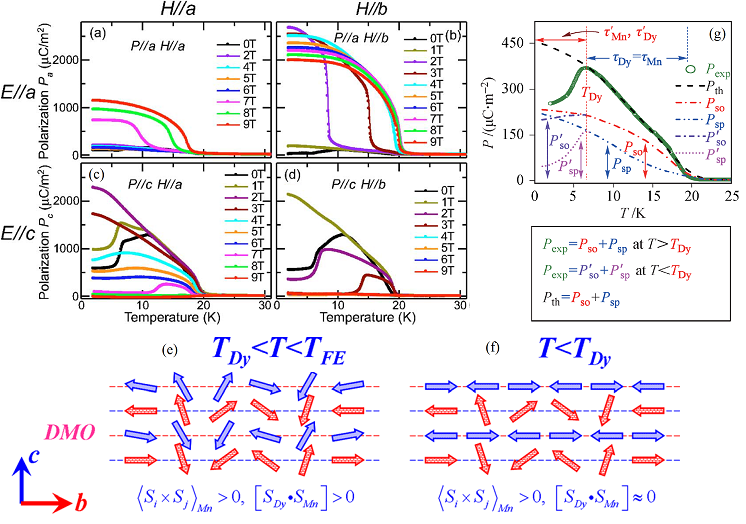}
\caption{(Colour online) Ferroelectricity of DyMnO$_3$. (a-d) the pyroelectric polarizations ($P$) of DyMnO$_3$ along the $a$- and $c$-axes as a function of temperature at various magnetic fields ($H$). At zero magnetic field, the polarization along the $c$-axis is suppressed significantly at low temperature, as shown in (c-d). (a-d) Reprinted figure with permission from \href{http://dx.doi.org/10.1103/PhysRevB.71.224425}{T. Kimura \textit{et al.}, Physical Review B, 71, p. 224425, 2005} \cite{Kimura:Prb05} Copyright \copyright (2005) by the American Physical Society. (e-f) The Dy$^{3+}$ and Mn$^{3+}$ spin configurations in DyMnO$_3$: (e) in the middle temperature region ($T_{\rm Dy}<T<T_{\rm FE}$); (f) in the low temperature region ($T<T_{\rm Dy}$). (e-f) Reprinted figure with permission from \href{http://dx.doi.org/10.1063/1.3636399}{N. Zhang \textit{et al.}, Applied Physics Letters, 99, p. 102509, 2011} \cite{Zhang:Apl11.2} Copyright \copyright (2011) by the American Institute of Physics. (g) A sketch of the two polarization components in  DyMnO$_3$: $P_{\rm so}$ ($P_{\rm so}'$) is the contribution from the spiral order of Mn spins, $P_{\rm sp}$ ($P_{\rm sp}'$) is the contribution from the spin-lattice coupling for the Dy-Mn spin pairs. Here $P_{\rm exp}$ is the measured polarization, $P_{\rm th}$ is the evaluated polarization assuming no independent Dy spin order at low temperature. (g) Reprinted figure with permission from \href{http://dx.doi.org/10.1007/s11467-011-0225-9}{N. Zhang \textit{et al.}, Frontiers of Physics, 7, pp. 408-417, 2012} \cite{Zhang:Fp} Copyright \copyright (2012) by the Higher Education Press and Springer-Verlag Berlin Heidelberg.}
\label{DyMnO3}
\end{figure}

DyMnO$_3$ is very similar to TbMnO$_3$, with a spiral spin order of Mn lying in the $b-c$ plane below $18$ K. In spite of the low critical temperature, the polarization of DyMnO$_3$ is larger than that of other multiferroics with the spiral magnetism, reaching $0.2$ $\mu$C/cm$^2$ \cite{Kimura:Prb05}. Another strange character of DyMnO$_3$ is the sudden drop of polarization below $\sim8$ K (see Fig.~\ref{DyMnO3}(a-d) for details). Subsequent experimental measurements revealed that the Dy$^{3+}$ moments modulate synchronously with the Mn$^{3+}$ moments above $7-8$ K, below which the independent Dy$^{3+}$ moment ordering appears \cite{Schierle:Prl,Feyerherm:Jpco}, as sketched in Fig.~\ref{DyMnO3}(c-f). The strong exchange striction between Dy$^{3+}$ and Mn$^{3+}$ spins, i.e. the strong Dy$^{3+}$-Mn$^{3+}$ coupling, contributes a considerable portion to the total polarization along the $c$-axis within the window from $18$ K to $8$ K, as sketched in Fig.~\ref{DyMnO3}(g) \cite{Zhang:Apl11}. The sudden drop of polarization below $8$ K is resulted from the decoupled Dy$^{3+}$-Mn$^{3+}$ moment synchronous modulation, which disables the exchange striction mechanism for polarization. A suppression of the independent Dy$^{3+}$ spin ordering at low temperature by a magnetic field or other stimuli can restore the dropped polarization so long as the Dy$^{3+}$-Mn$^{3+}$ moment synchronous modulation can be restored \cite{Zhang:Apl11.2}.

Such the $R^{3+}$-Mn$^{3+}$ coupling is also strong in orthorhombic HoMnO$_3$, a similar hybrid mechanism would make sense. However, for some cases the E-type antiferromagnetism is not sufficiently stable so as to be replaced by an incommensurate sinusoidal antiferromagnetism persisting down to low temperature \cite{Lee:Prb11R}. The underlying reason for this unusual behavior, in comparison with DyMnO$_3$, remains unclear. One possibility is the sensitive dependence of the magnetic ground state on sample quality. Anyway, such an incommensurate Mn$^{3+}$ sinusoidal antiferromagnetic order, which should have no spin-orbit coupling induced ferroelectricity, still allows a polarization along the $c$-axis again due to the exchange striction between Ho$^{3+}$ and Mn$^{3+}$. The prerequisite for this mechanism is the synchronous modulation of Ho$^{3+}$ moments and Mn$^{3+}$ moments.

The $b-c$ plane spiral of $R$MnO$_3$ can be rotated onto the $a-b$ plane spiral by magnetic field along the $c$-axis. This rotation makes the exchange striction between neighboring $R^{3+}$ and Mn$^{3+}$ spins fail to generate a polarization along the $a$-axis. Even though, Mochizuki \textit{et al.} suggested that the exchange striction between neighboring Mn spins is still active because the spin spiral is distorted, namely the angles between the nearest-neighbor Mn spins are modulated \cite{Mochizuki:Prb11}. A similar conclusion, i.e. the spin-spiral inhomogeneity, was also obtained by the Hartree-Fock treatment of the Hubbard model \cite{Solovyev:Prb11}. Such an exchange striction induces a polarization along the $a$-axis, applicable to orthorhombic HoMnO$_3$. It is thus understandable why the polarization from the $a-b$ plane spiral is larger than that from the $b-c$ plane, since for the latter case the local polarizations arisen from the exchange striction are canceled. On the other hand, Mochizuki \textit{et al.} also predicted that a noncollinear spin spiral as a distortion to the E-type antiferromagnetic order, has a contribution to the total polarization via the spin-orbit coupling \cite{Mochizuki:Prl2}.

Furthermore, experimental evidences with the coexistence of spiral component and E-type antiferromagnetic component in some $R$MnO$_3$ are available \cite{Lu:Apa,Ishiwata:Prb}, and this coexistence was first predicted theoretically \cite{Dong:Prb08.2}. Most likely, the whole $R$MnO$_3$ family is an ideal platform for illustrating such multiple contributions to the magnetism-driven ferroelectricity.

Of course, the hybrid mechanism is not limited to the $R$MnO$_3$ family, and would be highly possible in those multiferroics with complicated crystal structure and magnetism. One case is the $R$Mn$_2$O$_5$ family if $R^{3+}$  ion is magnetic, where both the Mn$^{3+}$-Mn$^{4+}$ and the $R^{3+}$-Mn$^{3+}$/Mn$^{4+}$ couplings contribute to the total polarization \cite{Zhao:Sr}. For CaMn$_7$O$_{12}$, the density functional calculations predicted that the large polarization is mainly due to the exchange striction, while the direction of polarization is determined by the spin-orbit coupling, also rendering a hybrid character, once discussed earlier \cite{Lu:Prl}.

As a summary, although a classification of multiferroic materials and underlying magnetoelectric mechanisms can be made, in many cases such a classification may not be rigorous and exceptions which contain multiple contributions to magnetoelectricity can be found.

\subsection{Thermodynamic formulations \& phase diagrams}
In the earlier sections, the Landau phenomenological theory of magnetoelectric coupling and the details of microscopic mechanisms for multiferroicity in a broad spectrum of multiferroic materials have been introduced. Physically, a fundamental question is how to understand and obtain these special magnetic orders. A complete formulation of every exchange term/coupling term contributing to the multiferroicity would be more appreciated. In this subsection, the thermodynamics of typical magnetic orders involved in the type-II multiferroics will be reviewed, while discussion on the ground state formulation at zero temperature is also given for some cases. Consequently, a set of multiferroic phase diagrams will be presented.

\begin{figure}
\centering
\includegraphics[width=0.5\textwidth]{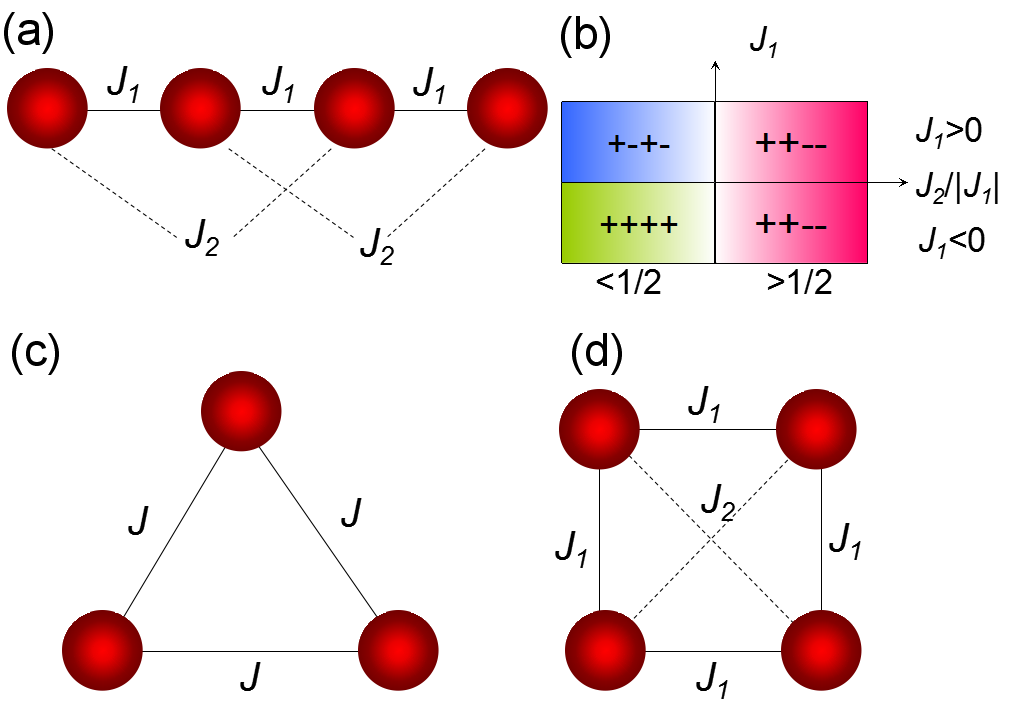}
\caption{(Colour online) A schematic of exchange frustrations in a spin chain. (a) A one-dimensional chain with the nearest-neighbor and next-nearest-neighbor exchanges ($J_1$ and $J_2$, respectively). (b) A simplified ground state phase diagram for an infinite spin chain as (a), if all the spins are Ising-type. The phase boundaries are at $J_2/|J_1|=1/2$ and $J_1=0$. (c) The geometric frustration in a regular triangle lattice if $J$ is antiferromagnetic. (d) The exchange frustration in a square lattice with competing $J_1$ and $J_2$.}
\label{J}
\end{figure}

In general, for a magnetic system, a simple Ising/Heisenberg model Hamiltonian can be written as:
\begin{equation}
H=J_1\sum_{<ij>}\textbf{S}_i\cdot\textbf{S}_j+J_2\sum_{[ik]}\textbf{S}_i\cdot\textbf{S}_k,
\label{j1j2}
\end{equation}
where $J_1$ and $J_2$ denote the exchanges between the nearest-neighbor and next-nearest-neighbor spin pairs, respectively; subscripts $i$/$j$/$k$ are the site indices. For a (quasi-) one-dimensional classical spin chain (Fig.~\ref{J}(a)), only the nearest-neighbor exchange is considered (i.e. $J_2=0$) and the ground state is fully ferromagnetic if $J_1<0$ or antiferromagnetic ($\uparrow\downarrow\uparrow\downarrow$) if $J_1>0$. If the next-nearest-neighbor exchange $J_2$ is nonzero, the ground state can be different. For an Ising-type spin chain, the ground state becomes $\uparrow\uparrow\downarrow\downarrow$ if $J_2/|J_1|>0.5$ regardless of the sign of $J_1$ (see Fig.~\ref{J}(b) for the phase diagram). For a Heisenberg-type spin chain with a ferromagnetic $J_1$ (negative) and an antiferromagnetic $J_2$ (positive), the ground state becomes a spiral order with the wavelength as $2\pi/\arccos(-J_1/4J_2)$ if $J_2>-J_1/4$.

Actually, one can find several discussions on one-dimensional chains in the previous subsections, which are some simple presentations of spin frustration. Spin frustration can also occur in a two-dimensional (three-dimensional) lattice. For example, in a two-dimensional triangular lattice, if all the nearest-neighbor exchanges are identical and positive (see Fig.~\ref{J}(c)), the three spins in each triangle cell will be frustrated, giving the $\uparrow\downarrow\uparrow$ or $\uparrow\downarrow\downarrow$ pattern for Ising-type spins, or the $120^\circ$ type noncollinear pattern for Heisenberg-type spins. This type of frustration is not due to the competing next-nearest-neighbor interactions but due to the structural geometry \cite{Ramirez:Arms}. For a square lattice, the next-nearest-neighbor exchange may bring spin frustration into the lattice (see Fig.~\ref{J}(d)). In addition, if there are multiple exchanges, for example the double-exchange and superexchange in manganites, spin frustration is possible even only the nearest-neighbor interactions are considered \cite{Dagotto:Prp}.

\begin{figure}
\centering
\includegraphics[width=\textwidth]{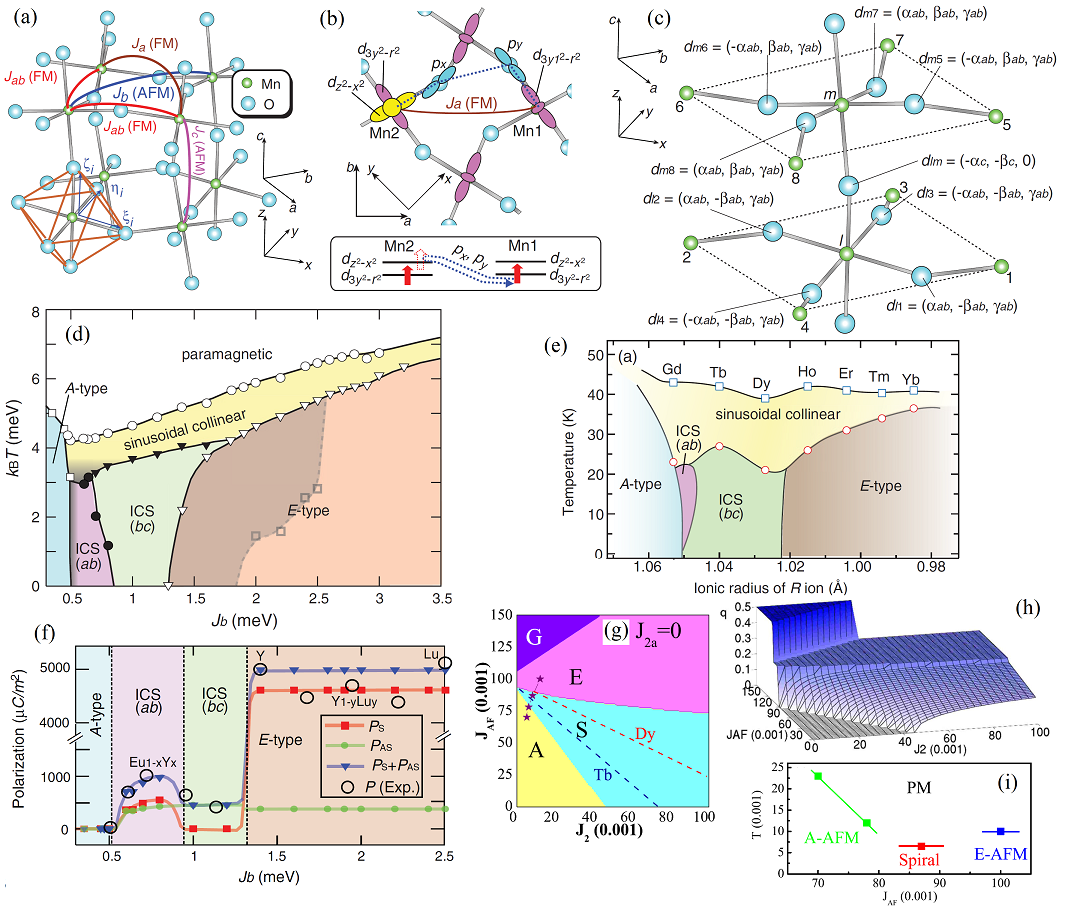}
\caption{(Colour online) Multiferroic phase diagrams of $R$MnO$_3$. (a) A schematic of MnO$_6$ octahedra in orthorhombic $R$MnO$_3$, with marked spin exchange paths. Here FM (AFM) denotes the (anti-)ferromagnetic exchange interaction. (b) The proposed main paths for the next-neighbor ferromagnetic exchange $J_a$ along the $a$-axis. (c) The Dzyaloshinskii-Moriya vectors $d_{ij}$ of Mn$_i$-O-Mn$_j$ bonds. (d) A predicted phase diagram of $R$MnO$_3$ using the Mochizuki-Furukawa classical model. Here ICS denotes the incommensurate spiral phase. In the shadow area, an incommensurate spin state coexists with the E-type antiferromagnetic state. (e) The measured phase diagram of $R$MnO$_3$ in comparison with the predicted diagram in (d). (f) The zero-temperature polarization as a function of $J_b$ (in proportional to decreasing $R^{3+}$ size). $P_S$: contribution from the symmetric exchange striction ($\textbf{S}\cdot\textbf{S}$); $P_{AS}$: contribution from the asymmetric Dzyaloshinskii-Moriya interaction ($\textbf{S}\times\textbf{S}$). $P(Exp.)$: experimental data on Eu$_{1-x}$Y$_x$MnO$_3$ and Y$_{1-y}$Lu$_y$MnO$_3$, in good agreement with $P_S+P_{AS}$. (a-d) Reprinted figure with permission from \href{http://dx.doi.org/10.1103/PhysRevB.84.144409}{M. Mochizuki \textit{et al.}, Physical Review B, 84, p. 144409, 2011} \cite{Mochizuki:Prb11} Copyright \copyright (2011) by the American Physical Society. (g) The ground state phase diagram of $R$MnO$_3$ as a function of the nearest-neighbor superexchange ($J_{\rm AF}$) and the next-nearest-neighbor superexchange ($J_2$) along the $b$-axis, calculated using the two-orbital double-exchange model. A/E/G: A/E/G-type antiferromagnetism; $S$: spiral spin order. (h) The corresponding wave vector along the pseudocubic direction, i.e. $x$ or $y$ as indicated in (c). (i) A simplified finite-temperature phase diagram of the two-orbital double-exchange model, obtained using Monte Carlo simulation. (g-i) Reprinted figure with permission from \href{http://dx.doi.org/10.1103/PhysRevB.78.155121}{S. Dong \textit{et al.}, Physical Review B, 78, p. 155121, 2008} \cite{Dong:Prb08.2} Copyright \copyright (2008) by the American Physical Society.}
\label{RMnO3}
\end{figure}

We take perovskite $R$MnO$_3$ family as model systems to describe various stages of research in formulating a complete thermodynamic approach to the multiferroicity. First, we show how frustration drives abundant magnetic orders. Usually, $R$MnO$_3$ accommodates the GdFeO$_3$-type lattice distortion which becomes more serious with decreasing A-site ionic size, and thus the Mn-O-Mn bonds become bent more, as sketched in Fig.~\ref{RMnO3}(a). This bond-bending suppresses the exchanges between the nearest-neighbor Mn sites, but enhances the exchanges between the next-nearest-neighbor Mn sites along the $b$-axis. In sequence, the ground state of $R$MnO$_3$ changes from the A-type antiferromagnetism as in LaMnO$_3$ to a spiral-spin order as in TbMnO$_3$, and finally to the E-type antiferromagnetism as in HoMnO$_3$ \cite{Goto:Prl}.

Second, in subsequence to these works in earlier years \cite{Kimura:Prb,Hotta:Prl,Sergienko:Prb}, a comprehensive and semi-quantum grasp of various aspects of the phase diagram started from 2008. Based on the two-orbital double-exchange model, Dong \textit{et al.} reproduced the sequential phase transitions \cite{Dong:Prb08.2,Dong:Epjb}. It was revealed that a weak next-nearest-neighbor superexchange along the $b$-axis is required to obtain a realistic spiral wavelength for the spiral spin order (see Fig.~\ref{RMnO3}(g-i)), and the Jahn-Teller distortion is also crucial. Later on, Kumar \textit{et al.} argued that even a weak $J_2$ was not necessary if a finite Hund coupling (between the $t_{\rm 2g}$ spins and $e_{\rm g}$ spin), instead of an infinite coupling assumed usually in the double-exchange physics of manganites, is considered \cite{Kumar:Prl10}. This later argument seems not yet well confirmed since the proposed phase diagram does not obtain verification, e.g. using an unbiased Monte Carlo method.

Third, a comprehensive Heisenberg spin model (Mochizuki-Furukawa model) accounting of every interaction/coupling identified in orthorhombic $R$MnO$_3$ family was proposed in 2009 \cite{Mochizuki:Prb,Mochizuki:Jpsj}. This model considers not only the exchanges (both the nearest-neighbor and next-nearest-neighbor ones), but also the single-ion anisotropy, Dzyaloshinskii-Moriya interaction, and cubic anisotropy, as shown in Fig.~\ref{RMnO3}(b-c). The measured phase diagrams (Fig.~\ref{RMnO3}(e)) on multiferroic manganites can be well reproduced with a set of properly selected parameters (see Fig.~\ref{RMnO3}(d)). The key point to access the spiral spin order is still the next-nearest-neighbor spin exchanges enhanced by the orthorhombic lattice distortion, while the spiral plane ($a-b$ versus $b-c$) is exquisitely controlled by tuning the competition between the single-ion anisotropy and Dzyaloshinskii-Moriya interaction. Using this model, Mochizuki \textit{et al.} studied a series of multiferroicity-related properties of $R$MnO$_3$, such as polarization reorientation \cite{Mochizuki:Prb11}, electromagnons \cite{Mochizuki:Prl}, magnetostrictions \cite{Mochizuki:Prl2}, picosecond optical switching of spin chirality \cite{Mochizuki:Prl3}, and magnetic switching of ferroelectricity \cite{Mochizuki:Prl4} etc. This model was also applied to simulate magnetic-field induced rotation of spiral plane (and the corresponding ferroelectric polarization) in $R$MnO$_3$, and cycloidal spin structures of $R$MnO$_3$ thin films on various magnetic substrates \cite{Qin:Apl,Qin:Apl2}. All these simulations explained reasonably the experimental observations on $R$MnO$_3$ on one hand and gave some valuable predictions on the other hand. For example, a recent experiment on TbMnO$_3$ under high pressure ($>5$ GPa) revealed a considerably enhanced improper ferroelectric polarization ($1.0-1.8$ $\mu$C/cm$^2$) due to the entrance of the E-type antiferromagnetic phase which would be absent in ambient pressure \cite{Aoyama:Nc,Aoyama:Prb}. This finding can be well understood based on these spin models, because the lattice compression could enhance the GdFeO$_3$-type distortion, shifting the ground state from the spiral spin order to the E-type antiferromagnetism \cite{Dong:Prb08.2,Mochizuki:Prb}. In addition, a strain to orthorhombic LuMnO$_3$ may also tune the the GdFeO$_3$-type distortion, and thus affect its magnetism and ferroelectricity \cite{Windsor:Prl}, which can be understood according to these models. Furthermore, both the double-exchange model and Mochizuki-Furukawa classical spin model predicted the phase coexisting tendency around the spiral-E-type phase boundary, which has been experimentally evidenced \cite{Lu:Apa,Ishiwata:Prb,Ishiwata:Jacs}.

Different from the above discussions, the phase diagram of $R$MnO$_3$ was also constructed using phenomenological Landau theory \cite{Harris:Prb,Ribeiro:Prb} in an alternative presentation to e.g. the Mochizuki-Furukawa model. Besides the most studied $R$MnO$_3$, spin models on other multiferroic systems, e.g. $R$Mn$_2$O$_5$, have been proposed \cite{Cao:Prl}. The phase diagram and spin excitation phenomena were discussed. Xiang developed an efficient four-state scheme to extract the exchanges and Dzyaloshinskii-Moriya interactions, as well as their derivatives with respect to atomic displacements on the basis of density functional calculations, so that these spin models can be used for quantitative predictions on various multiferroics \cite{Xiang:Prb}.

\subsection{Magnetoelectric excitations \& dynamics}
Besides the static and quasi-static magnetoelectric coupling phenomena in multiferroics, magnetoelectric excitations/dynamics are also the core ingredients of multiferroic physics. The magnetoelectric dynamics deals with particular multiferroic quantities $X$ behaving as the nontrivial functions of time ($t$), i.e., $dX/dt\neq0$. These quantities cover electromagnons, domain dynamics, and other multiferroic excitons in various frequency ranges of excitations. We mainly review recent progress in understanding the electromagnons and dynamic evolutions of multiferroic domain structures.

\subsubsection{Electromagnons: spin-lattice coupling vs spin-orbit coupling}
Following the symmetry consensus discussed in Sec. 2.2, any energy term for magnetoelectric coupling must be invariant under the time-reversal and space-inversion symmetry operations. In this sense, a simple time-dependent (dynamic) combination of $\textbf{M}$ and $\textbf{P}$, $\textbf{M}\cdot d\textbf{P}/dt$ or its variants, is allowed in the energy expression. It is noted that the order of this term is lower than all those time-independent (static or quasi-static) magnetoelectric coupling terms discussed earlier, implying a possible strongly-coupled magnetoelectric dynamic effect associated with this energy term.

To obtain a nonzero $d\textbf{P}/dt$, an AC electric field can be employed to excite the dynamic polarization. Due to this allowed dynamic magnetoelectric coupling, a spatial modulation of spin moments in a multiferroic, i.e., magnon, is present. This type of excitations is named electromagnons. For realistic multiferroics, such an AC electric field is usually the electric component of electromagnetic wave (light) in the Terahertz or far-infrared range.

\begin{figure}
\centering
\includegraphics[width=\textwidth]{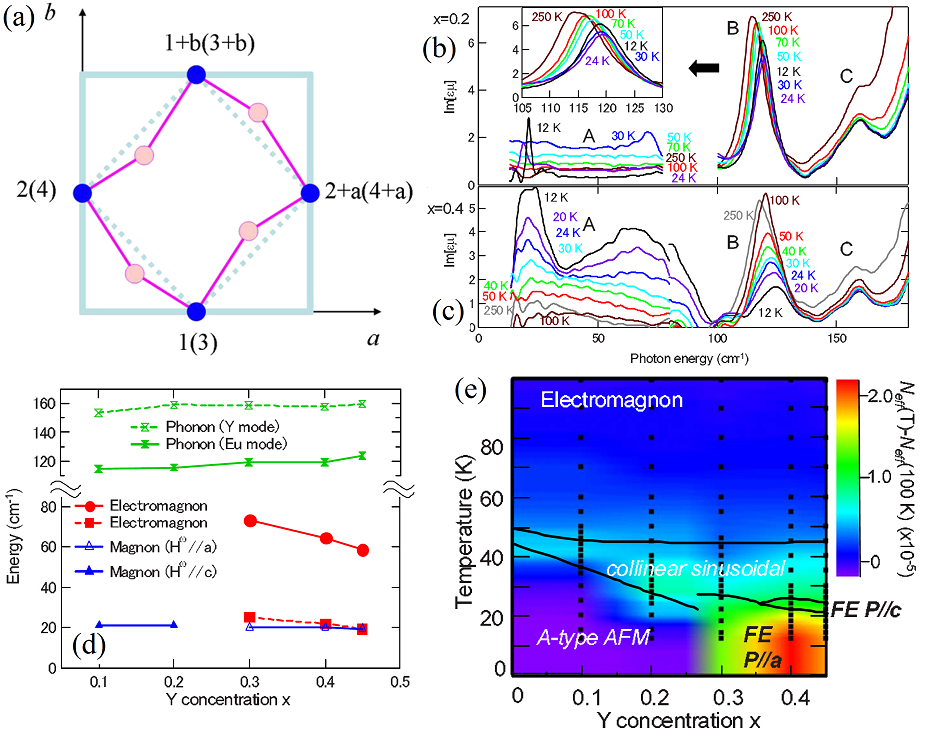}
\caption{(Colour online) Electromagnon excitations of $R$MnO$_3$ with the spiral spin order. (a) A schematic of the magnetostriction-induced electromagnon. The $a-b$ plane of the $Pbnm$ unit cell, blue: Mn, pink: oxygen. The oxygen displacements from the midpoints of the Mn-Mn bonds can be expressed as: $\delta r_{1+b,2+a}=-\delta r_{1,2}=(\delta x, \delta y, \delta z)$, $\delta r_{1,2+a}=-\delta r_{1+b,2}=(-\delta x, \delta y, \delta z)$. In the next layer (with $z=\pm c/2$), the Mn index is given in the parentheses and the displacements of oxygen ions have opposite $\delta z$. The AC electric component along the $a$-axis distorts the Mn-O-Mn ionic bonds. For example, along the $b$-axis, one Mn-O-Mn bond becomes more bent while the next bond becomes straighter. Such a distortion will excite magnons. (a) Reprinted figure with permission from \href{http://dx.doi.org/10.1103/PhysRevLett.102.047203}{R. Vald\'es Aguilar \textit{et al.}, Physical Review Letters, 102, p. 047203, 2009} \cite{Aguilar:Prl} Copyright \copyright (2009) by the American Physical Society. (b-c) The temperature dependence of the terahertz absorption spectra for Eu$_{1-x}$Y$_x$MnO$_3$: (c) $x=0.2$ (A-type antiferromagnetic); (d) $x=0.4$ (spiral). The AC electric component $E^\omega$ is along the $a$-axis. The electromagnon and two phonon modes are denoted as A, B, and C, respectively. As expected, the electromagnon is negligible in the $x=0.2$ case. (d) The substitution-dependent energies of phonon, magnon, and electromagnon in Eu$_{1-x}$Y$_x$MnO$_3$. (e) A contour mapping of the electromagnon (spectral weight) on the phase diagram. The dots indicate the measured data and the color contour was obtained by a linear interpolation from these data points. (b-e) Reprinted figure with permission from \href{http://dx.doi.org/10.1103/PhysRevB.79.214431}{Y. Takahashi \textit{et al.}, Physical Review B, 79, p. 214431, 2009} \cite{Takahashi:Prb} Copyright \copyright (2009) by the American Physical Society.}
\label{Electromagnon}
\end{figure}

As early as 2006, Pimenov \textit{et al.} reported an observation of the Terahertz-magnetodielectric effect and electromagnon spectra in GdMnO$_3$ and TbMnO$_3$ \cite{Pimenov:Np}. The imaginary part of the terahertz-dielectric function shows a broad relaxation-like excitation with characteristic frequency $n_0=23\pm3$ cm$^{-1}$ in GdMnO$_3$ and $n_0=20\pm3$ cm$^{-1}$ in TbMnO$_3$ when the electric component is along the $a$-axis. Such an excitation is magnetism-related since magnetic field can effectively suppress this excitation. The magnitude of excitations increases with decreasing temperature and saturates in the low-temperature magnetic phase. However, no significant variations were observed in TbMnO$_3$ passing across the magnetic/ferroelectric phase boundary at $28$ K. Instead, when the electric component is rotated to the $b$-axis, this excitation can be significantly suppressed. For example, the temperature dependence of the terahertz absorption spectra for analogous Eu$_{1-x}$Y$_x$MnO$_3$ is summarized in Fig.~\ref{Electromagnon}(b-e). All these characters suggest that electromagnon is independent of the existence of static ferroelectric polarization, although it is associated with the coupling between magnetism and lattice degrees of freedom. Surely, in some cases, a coupling of magnon modes to electric polarization is possible, as evidenced experimentally \cite{Senff:Prl}.

Soon, in 2007, Katsura, Nagaosa, and Balatsky (KNB) proposed a dynamic theory based on their seminal model on spiral-driven-ferroelectricity, in order to explain the electromagnons observed in orthorhombic $R$MnO$_3$ \cite{Katsura:Prl07}. As known, for TbMnO$_3$, the $b-c$ plane spin spiral propagating along the $b$-axis generates a polarization along the $c$-axis. Such a spiral and associated polarization can be rotated to the $a-b$ plane and $a$-axis, respectively. It is natural to expect that an electric field along the $a$-axis can excite this spiral-plane rotation, rendering the electromagnons observed experimentally. A calculation of the eigenmodes of magnetic cycloid predicts an approximately correct spectrum for electromagnons. In this theory, term $d\textbf{P}/dt$ describes a flip of the spiral-spin-driven polarization. However, the calculated dielectric contribution from such electromagnons is much smaller than the measured value. Another serious contradiction is that the observed excitation spectrum does not always follow the prediction. Taking Eu$_{0.75}$Y$_{0.25}$MnO$_3$ as an example, its spin spiral is in the $a-b$ plane and thus the induced polarization is along the $a$-axis. The expected electric component should be along the $c$-axis to excite the flip of polarization. Unfortunately, this prediction fails and the active electric component still aligning along the $a$-axis was observed, as shown in TbMnO$_3$ \cite{Aguilar:Prb}. Later on, a two-magnon excitation was proposed to understand the experimental observations \cite{Takahashi:Prl}.

In 2009, Vald\'es Aguilar \textit{et al.} measured the far infrared absorption of TbMnO$_3$ \cite{Aguilar:Prl}. Their experiment clearly indicated that the electric component was along the $a$-axis, no matter whether the spiral plane was in the $b-c$ plane or the $a-b$ plane. More importantly, they developed an alternative model based on the Heisenberg exchange to explain the electromagnons. Different from the KNB theory based on the spin-orbit coupling, their model is based on the spin-lattice coupling, although the static ferroelectric polarization is from the spin-orbit coupling. As shown in Fig.~\ref{Electromagnon}(a), the GdFeO$_3$-type distortion drives the oxygen anions in the $a-b$ plane to deviate from the middle point of the Mn-Mn bonds in a zigzag manner. An electric field applied along the $a$-axis synchronously shift all the oxygen anions to one direction. In the other words, here the dynamic polarization component involved in the $d\textbf{P}/dt$ term is from the electric-field driven electric displacements in such an ionic crystal, instead of its spontaneous ferroelectric polarization. Therefore, here TbMnO$_3$ (or Eu$_{0.75}$Y$_{0.25}$MnO$_3$) was used as a magnetic dielectric material instead of a magnetic ferroelectric. In details, the AC electric field distorts the Mn-O-Mn bond angles, giving rise to modulated Heisenberg exchanges, which distort the noncollinear spin configuration and excite magnons. Considering that this mechanism cannot distort the collinear spin pairs, the electric component along the $b$-axis or $c$-axis cannot activate single-magnon excitations. In short, the much stronger Heisenberg exchange term, instead of the spin-orbit coupling term, dominate the electromagnons (corresponding to the zone-edge magnon mode) in TbMnO$_3$, while the KNB theory on the spiral-plane flop is basically a minor effect and only responsible for the low energy electromagnons.

Furthermore, the electromagnon mode in TbMnO$_3$ can be used to estimate the stiffness of spin-lattice coupling. According to the excitation energy, the exchange striction intensity in orthorhombic $R$MnO$_3$ can be evaluated \cite{Aguilar:Prl}. The as-generated ferroelectric polarization in those E-type antiferromagnetic manganites (e.g. orthorhombic HoMnO$_3$ and YMnO$_3$), as estimated, should be $\sim1$ $\mu$C/cm$^2$, agreeing with recently measured data \cite{Nakamura:Apl}.

\begin{figure}
\centering
\includegraphics[width=\textwidth]{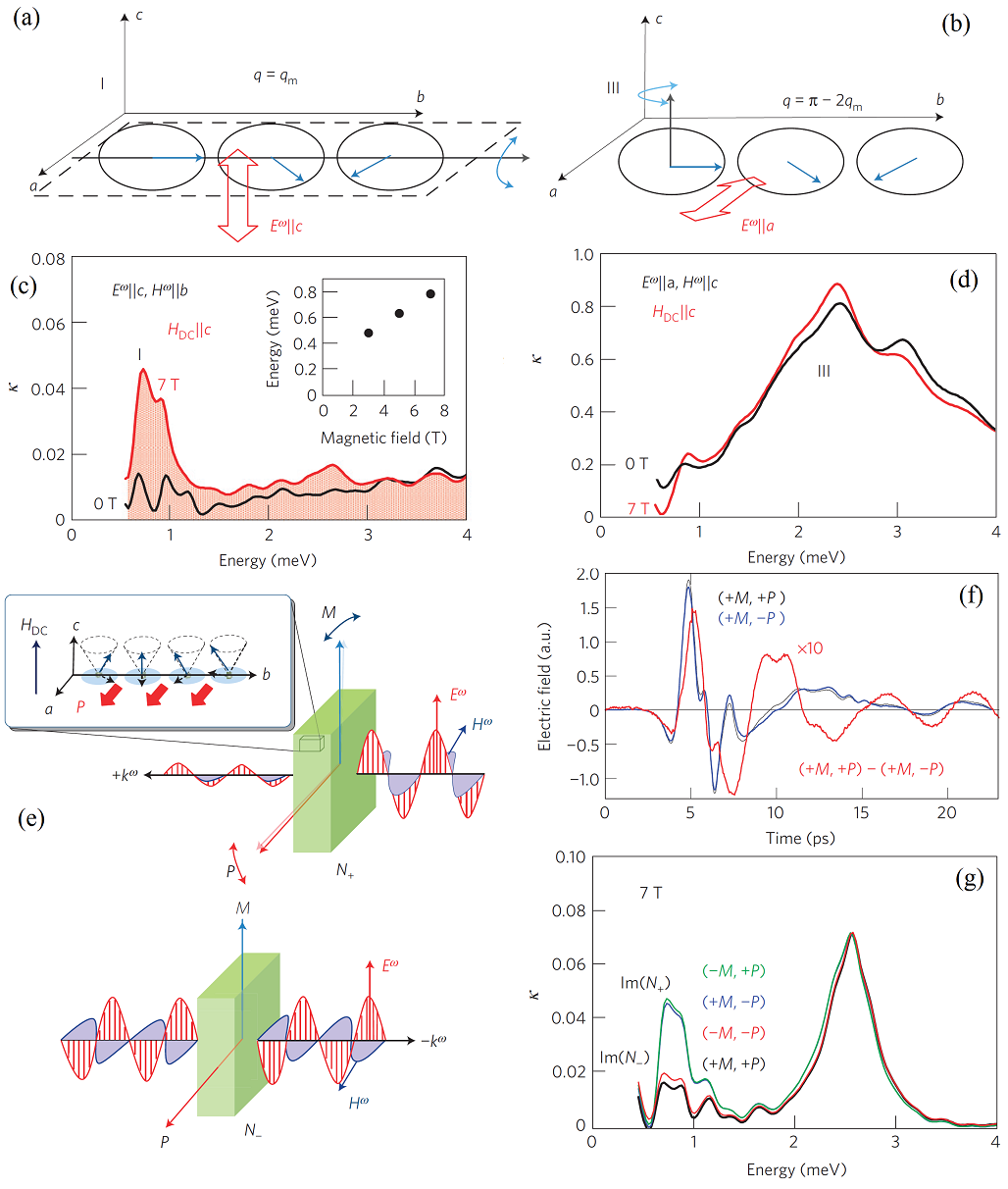}
\caption{(Colour online) The dynamics of electromagnons in multiferroic manganites. (a-b) A schematic of two electromagnon modes in the $ab$-plane cycloidal spin state of Eu$_{0.55}$Y$_{0.45}$MnO$_3$. $q$: wave numbers of magnons. (a) The cross-coupled electromagnon: the AC electric field along the $c$-axis induces a rotational vibration of the helical spin plane around the $b$ axis. (b) The magnetostriction-induced electromagnon:  the AC electric field along the $a$-axis distorts the Mn-O-Mn bonds and induces a rotational vibration of the spins around the $c$ axis. (c-d) The corresponding imaginary part $\kappa$ of the refractive index at $4$ K. A static magnetic field ($\textbf{H}_{\rm DC}$) is applied along the $c$-axis. (c) ($\textbf{E}^{\omega}||c$, $\textbf{H}^\omega||a$), corresponding to the cross-coupled electromagnon. (d) ($\textbf{E}^{\omega}||a$, $\textbf{H}^\omega||c$), corresponding to the magnetostriction-induced electromagnon. The cross-coupled electromagnon (peaked $<1.0$ meV) is much weaker than the magnetostriction-induced electromagnon (peaked $2.4$ meV). The peak at $2.7$ meV in (c) is due to the conventional antiferromagnetic resonance driven by the $\textbf{H}^\omega||a$ component of light (not discussed here). (e) The experimental configuration for directional dichroism measurement. The polarization ($\textbf{P}$) and magnetization ($\textbf{M}$) under a static magnetic field (see insert) are shown, plus the light components ($\textbf{E}^{\omega}$, $\textbf{H}^\omega$, and wave vector $k^\omega$). In the $+k^\omega$ case, ($\textbf{E}^{\omega}$, $\textbf{H}^\omega$) can drive the vibration of both $\textbf{P}$ and $\textbf{M}$ (upper panel), whereas in the $-k^\omega$ case light suppresses the vibrations of both $\textbf{P}$ and $\textbf{M}$ (lower panel). (f) The time-domain waveform of the transmitted terahertz pulses in different configurations (black and blue) and their difference (red). (g) The imaginary part of refractive index for ($\textbf{E}^{\omega}||c$, $\textbf{H}^\omega||a$) under a magnetic field ($\textbf{H}_{\rm DC}$) along the $c$ axis at $4$ K. Reprinted by permission from Macmillian Publishers Ltd: \href{http://dx.doi.org/10.1038/NPHYS2161}{Y. Takahashi, \textit{et al.}, Nature Physics, 8, pp. 121-125, 2012} \cite{Takahashi:Np}. Copyright \copyright (2012).}
\label{Electromagnon2}
\end{figure}

However, the spin-lattice coupling mediated electromagnons are magnetic-dipole inactive, indicating the absence of cross-coupling between magnetism and ferroelectricity. The subsequent studies on electromagnons revealed more than one type of excitations in the spiral order states of $R$MnO$_3$ \cite{Lee:Prb09}. A later on experiment on Eu$_{0.55}$Y$_{0.45}$MnO$_3$ by Takahashi \textit{et al.} unraveled the cross-coupled electromagnon driven by the spin-orbit coupling \cite{Takahashi:Np}, as proposed by Katsura, Balatsky, and Nagaosa \cite{Katsura:Prl07}. Here, the absorbed photons for $E_\omega||c$ are to excite the spiral-plane flop (see Fig.~\ref{Electromagnon2}(a)), whose peak position is at $\sim0.4-0.8$ meV (Fig.~\ref{Electromagnon2}(c)), depending on DC magnetic field. For a comparison, the magnetostriction-induced electromagon (Fig.~\ref{Electromagnon2}(b)) (selective rule $E_\omega||a$) for identical material is at $2.4$ meV and its amplitude is much stronger (Fig.~\ref{Electromagnon2}(d)).

One of the most intriguing aspects of this type of electromagnons is its dynamic magnetoelectric effect. The incident light of wave vector $\textbf{k}$ can couple with magnetization $\textbf{M}$ (under magnetic field) and polarization $\textbf{P}$ (Fig.~\ref{Electromagnon2}(e)). The coupling term $\textbf{k}\cdot(\textbf{P}\times\textbf{M})$ biases the refractive index, making the directional non-reciprocity for light transmission (Fig.~\ref{Electromagnon2}(f-g)). Similar directional dichroism was also observed in Ba$_2$CoGe$_2$O$_7$ \cite{Kezsmarki:Prl,Bordacs:Np}. This issue seems to be a topic deserving specific attention in the future, noting that a detailed description on electromagnons in orthorhombic $R$MnO$_3$ before 2011 can be found in the review article by Shuvaev, Mukhin, and Pimenov \cite{Shuvaev:Jpcm}.

In dynamics, the excitation of electromagnons can be simulated using the Landau-Lifshitz-Gilbert equation for a classical spin model. By considering a whole set of interactions/couplings, a dynamic Mochizuki-Furukawa-Nagaosa spin model as an extension to the Mochizuki-Furukawa model was proposed for multiferroic manganites, which can not only successfully reproduce the phase diagram of $R$MnO$_3$, e.g., the spin spiral and its spin plane flop, but also simulate the electromagnon excitations \cite{Mochizuki:Prl}. In addition to the main electromagnons' modes, higher harmonics of the spiral spin configuration were found to impose crucial influence on the magnon dispersion and electromagnon spectrum. This finding explains the puzzling low-energy peak in the optical spectrum. Furthermore, this spin model also predicted that picosecond optical pulses with terahertz frequency can switch the spin chirality (clockwise/counterclockwise $ab$/$bc$-plane spirals) by intensely exciting the electromagnons \cite{Mochizuki:Prl3}. Besides, neutron scattering was once used to probe the electromagnon excitation too and two types of electromagnons, i.e., the high energy one due to the exchange striction and the low energy one due to the cycloid rotation were found in DyMnO$_3$ \cite{Finger:Prb}, in consistent with the optical absorption measurements.

Besides orthorhombic TbMnO$_3$ and Eu$_{1-x}$Y$_x$MnO$_3$, there have been a number of experimental studies on electromagnons in other multiferroics, e.g. DyMnO$_3$ \cite{Finger:Prb,Shuvaev:Prl}, E-type antiferromagnetic $R$MnO$_3$ \cite{Takahashi:Prb10}, CuO \cite{Jones:Nc}, $R$Mn$_2$O$_5$ \cite{Sushkov:Prl,Kim:Prl,Sushkov:Prb}, CaMn$_7$O$_{12}$ \cite{Kadlec:Prb}, CuFe$_{1-x}$Ga$_x$O$_2$ \cite{Seki:Prl}, Ba$_2$Mg$_2$Fe$_{12}$O$_{22}$ \cite{Kida:Prb}, BiFeO$_3$ \cite{Cazayous:Prl}, Ba$_2$CoGe$_2$O$_7$ \cite{Kezsmarki:Prl,Bordacs:Np}, and so on. The dynamic simulations using the Landau-Lifshitz-Gilbert equation based on classical spin models are also available for some of these materials, e.g. $R$Mn$_2$O$_5$ \cite{Cao:Jpcm}, CuO \cite{Cao:Prl15}, and BiFeO$_3$ \cite{Fishman:Prb,Chen:Prl15}.

\subsubsection{Dynamics of multiferroic domains in o-$R$MnO$_3$}
\begin{figure}
\centering
\includegraphics[width=\textwidth]{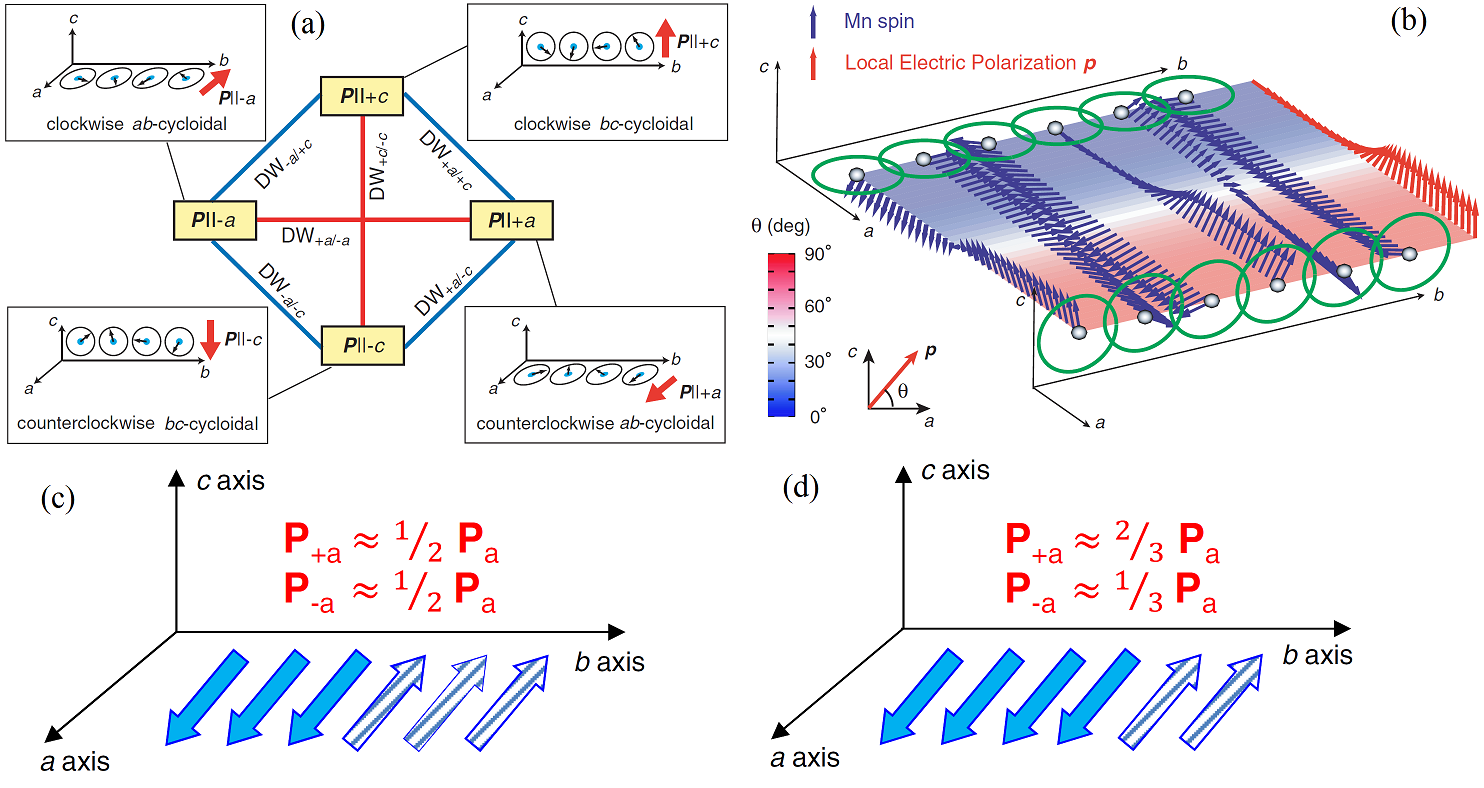}
\caption{(Colour online) The multiferroic domains. (a) A schematic of multiferroic domain walls in orthorhombic $R$MnO$_3$ with the spiral spin orders ($ab$-plane cycloidal \textit{vs} $bc$-plane cycloidal). The red lines denote the walls (DW$\pm a$ and DW$\pm c$) across which the neighboring spin helicities and polarizations are opposite. The blue lines denote the walls (DW$\pm a/\pm c$) across which the neighboring spin helicity and polarization are perpendicular. (b) The calculated domain wall structure between the $bc$-plane cycloidal and $ab$-plane cycloidal domains. Blue arrows: Mn spins; red arrows: local electric polarizations. (a-b) Reprinted figure with permission from \href{http://dx.doi.org/10.1103/PhysRevLett.102.057604}{F. Kagawa \textit{et al.}, Physical Review Letters, 102, p. 057604, 2009} \cite{Kagawa:Prl} Copyright \copyright (2009) by the American Physical Society. (c-d) Two ferroelectric domains: (c) random helicity and equivalent positive/negative polar domains; (d) the poling effect is partially memorized, with more positive polar domains than negative domains, after an electric poling and then spiral plane rotations (to the $bc$-plane and then back to the $ab$-plane). (c-d) Reprinted figure with permission from \href{http://dx.doi.org/10.1103/PhysRevLett.107.257601}{I. Fina \textit{et al.}, Physical Review Letters, 107, p. 257601, 2011} \cite{Fina:Prl} Copyright \copyright (2011) by the American Physical Society.}
\label{domain}
\end{figure}

In accompanying with electromagnon excitations, the other aspects of dynamic magnetoelectric coupling include multiferroic domain dynamics driven by external stimulations. Taking the most studied orthorhombic manganites as examples, the $ab$-plane spiral and $bc$-plane spiral, corresponding to the ferroelectric polarizations along the $a$-axis and $c$-axis, coexist in the spiral spin order region respectively. Three types of domain walls can be defined as: DW$\pm a$, DW$\pm c$, DW$\pm a/\pm c$ (Fig.~\ref{domain}(a)). The first two denote the domain walls between the co-plane spirals with opposite helicity (also opposite $\textbf{P}$'s), while the third one is the domain wall between the $ab$-plane spiral and $bc$-plane spiral (with perpendicular $\textbf{P}$, as sketched in Fig.~\ref{domain}(b)).

We discuss one case, and start from a virgin multiferroic state without electric poling. The helicity is randomly chosen, with equivalent weights of the positive and negative helicities as shown in Fig.~\ref{domain}(c). The electric poling can change the weights of clockwise and anticlockwise spirals, as experimentally observed in Gd$_{0.7}$Tb$_{0.3}$MnO$_3$ \cite{Yamasaki:Prl}. Fina \textit{et al.} found the memory effects during the spiral flop in a strained orthorhombic YMnO$_3$ \cite{Fina:Prl} thin film. Although bulk orthorhombic YMnO$_3$ favors more likely the E-type antiferromagnetic order, the YMnO$_3$ film on Nb:SrTiO$_3$ substrate favors the $bc$-plane spiral as the ground state. A magnetic field may drive the spiral to the $ab$-plane, generating the DW$\pm a/\pm c$. Without an electric poling, the clockwise and anticlockwise $ab$-plane cycloidal domains under a magnetic field are similar in volume, implying the random choice of domain switching. The nontrivial dynamic process happens after an electric poling. The electric poling in the $ab$-plane cycloidal state is memorized after the spiral plane is flopped to the $bc$-plane and then back to the $ab$-plane by tuning the magnetic field \cite{Fina:Prl}. In the other words, the helicity polarized state in the $ab$-plane can be partially sustained even the spiral plane is changed, as sketched in Fig.~\ref{domain}(d). This memory effect is attributed to the dynamic effect of domain wall DW$\pm a/\pm c$.

The dynamics of multiferroic domain walls can be elucidated using the frequency-dependent dielectric susceptibility. In the $10^3-10^9$ Hz range, all domain walls in DyMnO$_3$ exhibit the relaxation-like motion \cite{Kagawa:Prl,Kagawa:Prb}. This relaxation gives rise to a giant magnetocapacitance effect \cite{Kagawa:Prl,Schrettle:Prl}. With respect to the dynamics of conventional ferroelectric domain walls, the motion of multiferroic domain walls are much faster (with a high relaxation rate $\sim 10^7$ s$^{-1}$), reflecting its magnetic origin \cite{Kagawa:Prl}.

More details on the dynamic magnetoelectric response in the type-II multiferroics, not limited to electromagnons but also the dynamics of multiferroic domains, can be found in a topical review by Tokura and Kida \cite{Tokura:Ptrsa}.

\section{Magnetoelectric coupling in thin films and heterostructures: from science to devices}
In parallel to extensive research on synthesizing single phase multiferroics and exploring relevant magnetoelectric phenomena, artificial design and manufacturing of novel magnetoelectric structures from already existent transition metal compounds represent the other track receiving similarly intensive attention. In recent years, interfaces between functional oxides have drawn lot of research attention not only for their fascinating physics which is simply absent in the component materials, but also for promising perspectives of applications which can be designed more or less as one wishes \cite{Dagotto:Sci07,Hammerl:Sci,Hwang:Nm}. While correlated electronic materials themselves show many emergent physical properties beyond traditional functional materials such as semiconductors, introduction of interfaces by well controlled manners, as done in thin films and heterostructures, provides an immense space for novel functionalities and relevant physics \cite{Takagi:Sci,Mannhart:Sci}. This is a truth for multiferroics too. In this section, the state-of-the-art of interfacial physics of magnetoelectric thin films and heterostructures in the past decade will be reviewed.

Certainly, for thin films and heterostructures, additional ingredients besides a combination and coupling of various ferroic orders from the corresponding components are crucial for magnetoelectricity, including strain effect (piezo-effect) and interfacial carrier modulation. In these low-dimensional structures, domain walls and other emergent interfacial features may also give their prominence to the bulk effects in terms of magnetoelectric performance. These issues will be discussed separately but they coexist in many cases.

\subsection{Strain-mediated magnetoelectricity}
Although the pure strain-mediated magnetoelectric coupling is not the emphasis of the present review, some interesting topics involving the quantum-level mechanisms are introduced here.

\subsubsection{Strain generated non-$d^0$ ferroelectric polarization}
The violation of the $d^0$ rule for ferroelectricity in perovskites was first predicted in bulk BaMnO$_3$, as mentioned in Sec. 3.3.4. In addition, the non-$d^0$ ferroelectricity can be obtained in compressive BiFeO$_3$ films. The pseudo-cubic lattice constants of original rhombohedral structure are $a=b=c=3.96$ \AA{} \cite{Wang:Sci}. As the mostly used substrate, SrTiO$_3$ is cubic with lattice constant of $3.905$ \AA{}, which has a good proximity with rhombohedral BiFeO$_3$. However, for LaAlO$_3$ (001) and YAlO$_3$ (110) substrates whose pseudo-cubic lattice constants are $3.79$ \AA{} and $3.69$ \AA{} respectively \cite{Schlom:Armr}, the in-plane lattice of deposited BiFeO$_3$ films is strongly compressed while the out-of-plane lattice is elongated, giving a tetragonal-like phase. The super-tetragonality is characterized by the big $c$/$a$ ratio reaching up to $1.27$ \cite{Zeches:Sci}.

\begin{figure}
\centering
\includegraphics[width=\textwidth]{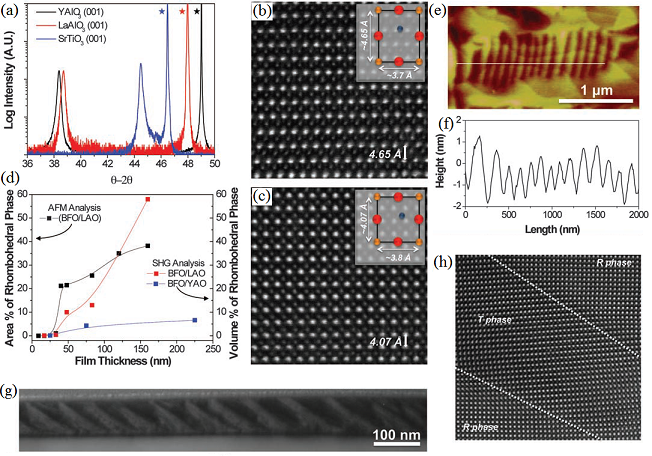}
\caption{(Colour online) The super-tetragonal BiFeO$_3$. (a) The X-ray diffraction patterns of the pseudo-cubic $002$-peak of BiFeO$_3$ thin films on three substrates: SrTiO$_3$(001), LaAlO$_3$(001), and YAlO$_3$(110). Each substrate peak is marked with a star. The long $c$-axis tetragonal phase appears in the films deposited on LaAlO$_3$(001) and YAlO$_3$(110) substrates. (b-c) The transmission electron microscopy images of the tetragonal phase and rhombohedral phase, respectively. Insets: a schematic of the unit cell. (d) The thickness-dependent structural parameters. Left axis: the area fraction of rhombohedral phase measured by atomic force microscopy. Right axis: the volume fraction of rhombohedral phase evaluated from the second harmonic generation signals. (e) A high-resolution atomic force microscopy image of a mixed phase region. The atomic height profile along the white line is shown in (f), which shows a $2-3$ nm variation in height from the tetragonal (bright) phase to the rhombohedral (dark) phase. (g) A low-resolution cross-sectional transmission electron microscopy image of a mixed phase region in a BiFeO$_3$ film deposited on LaAlO$_3$(001). Light: tetragonal phase; dark: rhombohedral phase. (h) A high-resolution transmission electron microscopy image of the boundaries between the two phases in (g). From \href{http://dx.doi.org/10.1126/science.1177046}{R. J. Zeches \textit{et al.}, Science, 326, pp. 977-980, 2009} \cite{Zeches:Sci}. Reprinted with permission from the American Association for the Advancement of Science.}
\label{TBFO}
\end{figure}

Such a giant tetragonality can significantly enhance the ferroelectric polarization along the $c$-axis. In spite of the $d^5$ nature, the Fe ion can have an off-center displacement like the $d^0$ Ti in ferroelectric BaTiO$_3$, due to the extremely elongated Fe-O bonds along the $c$-axis. In fact, as early as 2004, Yun \textit{et al.} found a giant polarization beyond $150$ $\mu$C/cm$^2$ in BiFeO$_3$ thin films grown on Pt/TiO$_2$/SiO$_2$/Si substrates, which might be due to this giant tetragonality \cite{Yun:Jjap}. Later, Ederer and Spaldin performed the density functional calculations on the strain effects in various ferroelectrics, and found that a large polarization up to $\sim150$ $\mu$C/cm$^2$ for tetragonal BiFeO$_3$ is possible \cite{Ederer:Prl}. In 2006, Ricinschi \textit{et al.} confirmed the density functional calculation results, and reported the evidence for a tetragonal phase in their BiFeO$_3$ films grown on SrTiO$_3$, although the stability of a tetragonal phase on this substrate may be questioned \cite{Ricinschi:Jpcm}. In 2009, using the high-resolution transmission electron microscopy and other imaging techniques, Zeches \textit{et al.} observed the tetragonal-like phase in the films deposited on LaAlO$_3$ and YAlO$_3$ substrates (see Fig.~\ref{TBFO}(a-c)) \cite{Zeches:Sci}. They also found the coexistence of tetragonal-like phase and rhombohedral phase (see Fig.~\ref{TBFO}(d-h)) upon different deposition conditions, e.g. thickness of film and underlying substrate. The two phases can be converted reversibly by electric field, rendering a strong piezoelectric effect which was claimed to be a substitute for lead-based materials in the future piezoelectric applications \cite{Zhang:Nn}.

\subsubsection{Strain driven ferromagnetic \& ferroelectric EuTiO$_3$}
Besides BiFeO$_3$, it is known that some perovskite titantes ($A$TiO$_3$, where $A$ is a divalent cation) are proper ferroelectrics, e.g. PbTiO$_3$ and BaTiO$_3$. In the other words, Ti$^{4+}$ is a ferroelectric-active ion. It was thus speculated that $A$TiO$_3$ can be multiferroic if cation $A^{2+}$ is magnetic. EuTiO$_3$ is just the case, where the Eu's valence is $+2$ instead of normal $+3$ for most rare earth ions. Based on the density functional calculations, Fennie and Rabe predicted that the paraelectric antiferromagnetic EuTiO$_3$ would become ferromagnetic and ferroelectric, given a compressive strain exceeding $1.2\%$ \cite{Fennie:Prl}. However, due to some practical reasons (e.g. no proper substrate), this prediction has yet to be well confirmed.

\begin{figure}
\centering
\includegraphics[width=\textwidth]{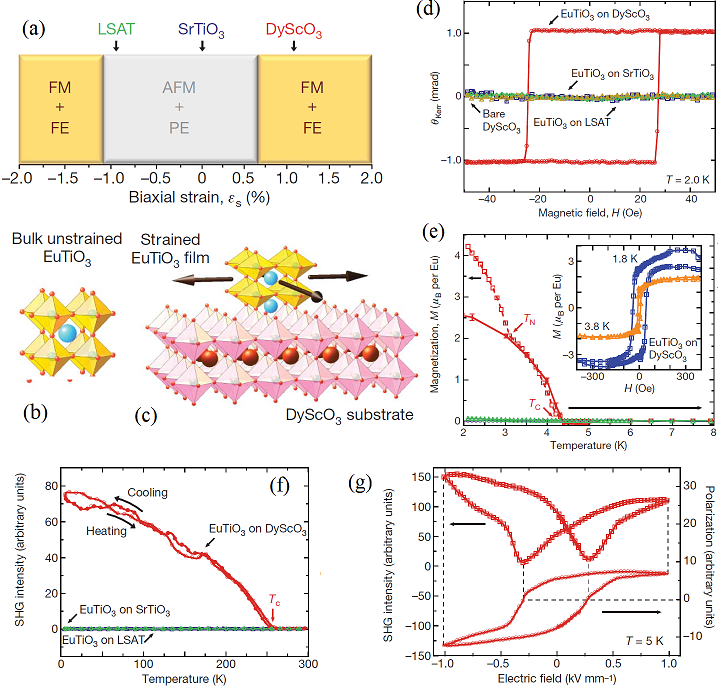}
\caption{(Colour online) Multiferroic behaviors of EuTiO$_3$ under biaxial strain. (a) A calculated phase diagram. PE: paraelectric; FE: ferroelectric; AFM: antiferromagnetic; FM: ferromagnetic. (b) The structure of EuTiO$_3$. (c) A schematic of an epitaxially tensile EuTiO$_3$ thin-film on DyScO$_3$ substrate. (d) The measured magnetic hysteresis loops of EuTiO$_3$ thin films on various substrates: non-ferromagnetic EuTiO$_3$ on SrTiO$_3$ (blue) and EuTiO$_3$ on LSAT ((LaAlO$_3$)$_{0.29}$-(SrAl$_{1/2}$Ta$_{1/2}$O$_3$)$_0.71$) (green), and ferromagnetic EuTiO$_3$ on DyScO$_3$ (red). The data for bare DyScO$_3$ substrate (gold) are inserted for reference. The $\theta_{\rm Kerr}$ angle ($\theta_{Kerr}$) is proportional to the in-plane magnetization. (e) The temperature dependence of magnetization. Inset, the isothermal magnetization hysteresis loops at $1.8$ K and $3.8$ K. (f) The temperature dependent second harmonic generation (SHG) intensity, indicating that only the EuTiO$_3$ on DyScO$_3$ (red) is polar below $250$ K and others are nonpolar. (g) The hysteresis loop (top) and corresponding polarization loop (bottom) for a EuTiO$_3$ film on DyScO$_3$ substrate at $5$ K. Reprinted by permission from Macmillian Publishers Ltd: \href{http://dx.doi.org/10.1038/nature09331}{J. H. Lee, Nature, 466, pp. 954-958, 2010} \cite{Lee:Nat}. Copyright \copyright (2010).}
\label{EuTiO3}
\end{figure}

Later, their density functional calculations predicted that a weak tensile strain ($>0.75\%$) could also induce the ferromagnetic $+$ ferroelectric ground state, as summarized in Fig.~\ref{EuTiO3}(a-c). The ferromagnetism comes from the Eu$^{2+}$ $4f$ spins while the ferroelectricity is due to the Ti$^{4+}$ ions \cite{Lee:Nat}. In this sense, this strained EuTiO$_3$ is a type-I multiferroic. Consequently, Lee \textit{et al.} fabricated such a tensile thin film on DyScO$_3$ substrate which can provide a biaxial tensile strain of $1.1\%$ \cite{Lee:Nat}. The temperature-dependent second harmonic generation measurement found that its ferroelectric Curie temperature is as high as $250$ K (see Fig.~\ref{EuTiO3}(f-g)), while the magneto-optic Kerr angle measurement found that the ferromagnetism is below $4.24$ K (see Fig.~\ref{EuTiO3}(d-e)) \cite{Lee:Nat}. A subsequent experiment suggested an even more complex scenario: an inhomogeneous state with the coexistence of a ferromagnetic state and a possibly paramagnetic state in the low magnetic field range, implying a degeneracy of the magnetic states \cite{Geng:Prb}.

\subsubsection{Piezostrain tuning of magnetism}
The strain effect described above is static, which paves a basis for an electro-control of multiferroicity using the ``dynamic" strain. In principle, all ferroelectrics are piezoelectric materials, thus allowing an opportunity for an electro-tunable strain using a piezoelectric substrate. A pioneer work along this line was the synthesis of BaTiO$_3$-CoFe$_2$O$_4$ ordered nanostructures \cite{Zheng:Sci}. The magnetization of CoFe$_2$O$_4$ shows an anomaly at the ferroelectric Curie temperature although no electro-control of magnetism was mentioned. Later, many experimental attempts have been reported and some of them reported in earlier years were reviewed \cite{Nan:Jap,Ma:Am,Wang:Mt}. Here some selected works in recent years are briefly highlighted to cover the latest progress.

($1-x$)[PbMg$_{1/3}$Nb$_{2/3}$O$_3$]-$x$[PbTiO$_3$] (PMN-PT) is a widely used piezoelectric material with prominent piezoelectric coefficient. With increasing $x$, the structure evolves from a rhombohedral phase to a tetragonal, generating a morphotropic phase boundary (MPB) with the best piezoelectricity around $x=33\%$ \cite{Noheda:Prb}. PMN-PT has been often adopted as substrates in attempting piezostrain tuning of magnetism in heterostructures. Other piezoelectric materials used as substrates or components in heterostructures include PbTiO$_3$, BaTiO$_3$, and Pb(Zr$_{1-x}$Ti$_x$)O$_3$.

For the piezoelectric-ferromagnetic heterostructures, piezostrain can tune the orientation of magnetization since magnetic anisotropy is usually associated with lattice distortion. As a representative approach, a complete reversal of magnetization may be realized via a two-step scheme. As the first step, a perpendicular ($90^\circ$) rotation of magnetization was observed respectively in, e.g. Ni/PMN-PT \cite{Buzzi:Prl}, Co$_{40}$Fe$_{40}$B$_{20}$/PMN-PT \cite{Li:Am,Zhang:Sr}, which can be understood as a result of the in-plane piezostrain distortion. A full flip ($180^\circ$) of magnetization was realized as a successive deterministic perpendicular rotation, as demonstrated in Ni/BaTiO$_3$ \cite{Ghidini:Nc} and Co/PMN-PT \cite{Yang:Am}. This effect is a process-dependent dynamic sequence of the magnetic moments \cite{Hu:Nl}, since the static energy cannot distinguish the $180^\circ$ flipped state and the initial non-flipped state unless an assistance of magnetic field) is given \cite{Wang:Sr}. The hysteresis and coercivity of magnetism and pizeoelectricity are crucial and a careful tuning to realize this electro-control-magnetism functionality is needed.

Other types of piezostrain tuning of magnetism can be assigned too, e.g. utilizing some magnetic phase transitions. For example, in PMN-PT/manganite heterostructures, the strain can tune the phase competition of manganites, by which the magnetism and transport behavior of a manganite layer can be significantly modulated \cite{Thiele:Prb,Zheng:Prb,Yang:Apl12,Zheng:Prb14,Zheng:Apl}. Besides manganites, the charge-ordering transition (i.e. the Verwey transition) of Fe$_3$O$_4$ can be tuned by the piezostrain of PMN-PT substrate \cite{Liu:Sr}. An electric field induced shifting of the ferromagnetic resonance in Fe$_3$O$_4$/Pb(Zr$_{1-x}$Ti$_x$)O$_3$ and Fe$_3$O$_4$/PMN-PT heterostructures was reported too, rendering a giant microwave tunability of magnetism \cite{Liu:Afm}. The piezostrain tuning of magnetism in other magnetic oxides, e.g. CoFe$_2$O$_4$, NiFe$_2$O$_4$, and BiFeO$_3$, using PMN-PT substrates for most cases, was discussed by various approaches \cite{Yang:Apl,Park:Prb,Park:Apl}. A high-density magnetoresistive random access memory operating at ultra-low voltage and room temperature was also designed based on the piezostrain effect \cite{Hu:Nc}.

It is noted that all of the above examples deal with the piezostrain tuning of magnetism and proper ferroelectricity. The strain effect on improper ferroelectricity was also reported. For example, the ferroelectricity and magnetic order of orthorhombic $R$MnO$_3$ can be re-shaped using strain \cite{Windsor:Prl,Fina:Prl,Li:Sr}.

\subsection{Carrier-mediated interfacial magnetoelectricity}
For an interface or a surface, the space-inversion symmetry is naturally broken even the underlying material is nonpolar. If any interfacial/surface magnetism is involved to break the time-reversal symmetry, it will be quite promising to access the magnetoelectric coupling due to the simultaneously broken time-reversal and space-inversion symmetries. Such an interfacial/surface magnetism can be realized via carrier-mediation by several simplest approaches. We outline four of them here.

\subsubsection{Spin-dependent screening, bonding, \& oxidization}
The simplest model system in this category is the surface of ferromagnetic metals, which breaks both the time-reversal and space-inversion symmetries. The magnetoelectric effect originates from the spin-dependent carrier screening of an electric field which leads to notable variations of the surface magnetization and surface magnetocrystalline anisotropy \cite{Duan:Prl}. This magnetoelectric effect can be substantially magnified at the interface between a ferromagnetic film and a high-$\kappa$ dielectric material, e.g. SrRuO$_3$/SrTiO$_3$ \cite{Rondinelli:Nn}.

The carrier-mediated magnetoelectric effect can be further enhanced in ferromagnet/ferroelectric heterostructures. A ferroelectric polarization near an interface is equivalent to an amount of surface charges. The charge density can be estimated from the value of polarization, e.g. $10$ $\mu$C/cm$^2$ corresponds to $\sim0.1$ electron per unit cell if the pseudo-cubic lattice constant is $\sim4$ \AA{} \cite{Dong:Prb11}. This effect can be categorized as the ferroelectric field effect (Fig.~\ref{FET}(a-b)), which was predicted to appear in SrRuO$_3$/BaTiO$_3$/SrRuO$_3$ heterostructure \cite{Niranjan:Apl}. As a result, the magnetizations of SrRuO$_3$ at the two end interfaces become significantly different due to the ferroelectricity in BaTiO$_3$. The obtained change of magnetic moment caused by a polarization reversal is $0.31$ $\mu_{\rm B}$ per Ru per interface. Meanwhile, Cai \textit{et al.} proposed a strategy to realize the room-temperature magnetoelectric effect in a tri-component ferromagnet/ferroelectric/normal-metal superlattice \cite{Cai:Prb}. Due to the broken inversion symmetry between the ferromagnet/ferroelectric and normal-metal/ferroelectric interfaces, an additional magnetization caused by spin-dependent screening accumulates at the ferromagnet/ferroelectric interface, which is not canceled by the depletion at the normal-metal/ferroelectric interface. A large global magnetization over the whole superlattice can be induced. Based on this model, Lee \textit{et al.} predicted a robust magnetoelectric coupling in Fe/BaTiO$_3$/Pt and Fe/PbTiO$_3$/Pt superlattices \cite{Lee:Prb10}.

The microscopic mechanism for such a spin-dependent screening behavior can be complicated. As early as 2006, Duan \textit{et al.} studied the Fe/BaTiO$_3$ interface, in which the hybridization between the Ti's $3d$ and Fe's $3d$ states near the interface should be responsible for the large magnetoelectric effect \cite{Duan:Prl06}. This strong hybridization leads to the formation of bonding states which result in a net magnetic moment on the Ti site. The induced moment is sensitive to the interfacial bonding strength and can be controlled by switching the ferroelectric polarization of BaTiO$_3$: for a polarization pointing toward the interface, the Fe-Ti bond becomes shorter, which enhances the bonding coupling and thus pushes the minority-spin bonding states down to a lower energy. These bonding states are more populated than the others, resulting in a large magnetic moment on the Ti site. The opposite alignment of the polarization, pointing away from the interface, results in a small magnetic moment on the Ti site. It was calculated that the difference in moment for the two opposite alignments reaches $0.22$ $\mu_{\rm B}$. The first-principles calculations revealed that this ``interfacial bonding-driven" magnetoelectric coupling, as a quite general effect, occurs in many other heterostructures, including Fe$_3$O$_4$/BaTiO$_3$ \cite{Niranjan:Prb}, Co$_2$MnSi/BaTiO$_3$ \cite{Yamauchi:Apl}, Fe/PbTiO$_3$ \cite{Fechner:Prb}, and Co/PbZr$_x$Ti$_{1-x}$O$_3$ \cite{Borisov:Prb}.

Following these predictions, a large magnetoelectric effect at the Fe/BaTiO$_3$ interface was indeed observed recently \cite{Radaelli:Nc}. Using the X-ray magnetic circular dichroism in combination with the high-resolution electron microscopy and first-principles calculations, an ultrathin oxidized iron layer near the interface was observed, whose magnetization could be reversibly switched on and off at room temperature by reversing the polarization of the underlying BaTiO$_3$ layer. An interfacially oxidized Fe monolayer is the key to obtain a large magnetoelectric effect. The similar interfacial oxidation controlled by electric field was also achieved in Co/GdO$_x$ bilayer via the voltage-driven O$^{2-}$ migration, and consequently a modulation of magnetism was confirmed \cite{Bauer:Nm}.

Phenomenologically, the magnetoelectric energy term accounting of the screening physics can be expressed as $\textbf{M}^2d\textbf{P}/d\textbf{r}$ or its variants. The discontinuity of polarization at a surface/interface defines a particular coefficient $d\textbf{P}/d\textbf{r}$, which can be giant. This energy term is invariant under the space-inversion operation, allowing a coupling with the even (not odd) power orders of magnetization. This coupling scenario makes a change of magnetization near the surface/interface possible if a polarization reversal is triggered by electric stimuli or others.

\subsubsection{Electrical tuning of magnetocrystalline anisotropy}
As shown in Sec. 4.1.3, a strong piezostrain effect can tune the magnetocrystalline anisotropy. In fact, there is one more mechanism for electrically tuning the magnetocrystalline anisotropy, i.e. via modulation of interfacial carrier density. Since the anisotropy determines the preferential orientation of magnetization, this mechanism is an alternative approach to the electrical control of magnetism.

A manipulation of the magnetocrystalline anisotropy by applying an electric field has recently been studied and realized in a number of thin films and heterostructures with ferromagnetic-metallic interfaces (surfaces) \cite{Tsujikawa:Prl,Nakamura:Prl,Niranjan:Apl10,He:Apl,Weisheit:Sci,Maruyama:Nn,Nozaki:Apl} as well as at the interfaces (surfaces) of dilute magnetic semiconductors \cite{Ohta:Apl,Chiba:Nat}. Alternatively, the interfacial magnetocrystalline anisotropy can be more efficiently controlled by a ferroelectric polarization from an adjacent ferroelectric film, as demonstrated by a set of \textit{ab initio} studies and experiments \cite{Duan:Apl,Sahoo:Prb,Fechner:Prl,Lukashev:Jpcm}. For example, Lukashev \textit{et al.} studied BaTiO$_3$/Fe$_4$ multilayers and found a big change (about $30\%$) of the perpendicular interfacial magnetic anisotropy energy in responding to a reversal of ferroelectric polarization of BaTiO$_3$ \cite{Lukashev:Jpcm}. In addition, the magnetic anisotropy of manganites can be tunable by a ferroelectric polarization, as characterized by the anisotropic magnetoresistance of manganite/ferroelectric heterostructures \cite{Preziosi:Prb}. The underlying physics can be understood by analyzing the charge density and spin-orbital-resolved density of states. A redistribution of electrons over the $3d$ orbitals at the interface driven by polarization reversal is responsible for the variation of magnetocrystalline anisotropy. As stated already, the magnetocrystalline anisotropy is a result of the spin-orbit coupling between occupied valence bands and unoccupied conducting bands. A modulation of orbital occupation disturbs the spin-orbit coupling and thus the magnetocrystalline anisotropy \cite{Nakamura:Prl}.

In addition, other than traditional transition metal oxide ferroelectrics, organic ferroelectrics have been adopted to tune the magnetocrystalline anisotropy in relevant organic ferroelectrc/magnetic heterostructures. Several systems include poly (vinylidene fluoride) PVDF/Co \cite{Lukashev:An}, PVDF/Fe \cite{Wang:Jap}, and $70\%$ vinylidene fluoride with $30\%$ trifluoroethylene copolymer P(VDF-TrEE)/Co/Pd \cite{Mardana:Nl}. In these heterostructures, the significant variations of magnetocrystalline anisotropy energy, driven by ferroelectric polarization reversal, were observed, some of them even reached up to $50\%$. These observations shed a new light on possibilities to utilize the magnetoelectric coupling mechanism at organic ferroelectric/ferromagnet interfaces for flexible and cost-competitive devices.

\subsubsection{Ferroelectric control of magnetic phases}
\begin{figure}
\centering
\includegraphics[width=0.75\textwidth]{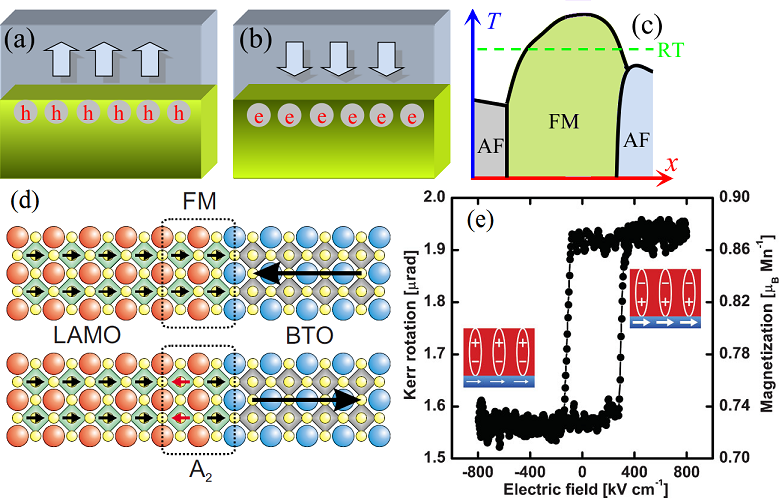}
\caption{(Colour online) Electro-control of interfacial magnetism. (a-b) A schematic of ferroelectric field effect. The arrows denote ferroelectric polarization. The holes (denoted by symbol h) or electrons (denoted by symbol e) are attracted to the interfaces upon the polarization direction. (c) A schematic phase diagram of La$_{1-x}$Sr$_x$MnO$_3$. $x$: carrier density (dopant concentration); $T$: temperature; $RT$: room temperature. (a-c) Reprinted figure with permission from \href{http://dx.doi.org/10.1103/PhysRevB.88.140404}{S. Dong \textit{et al.}, Physical Review B, 88, p. 140404(R), 2013} \cite{Dong:Prb13.2}. Copyright \copyright (2013) by the American Physical Society. (d) An electrically induced magnetic reconstruction at the La$_{1-x}A_x$MnO$_3$/BaTiO$_3$ interface, as proposed by the density functional calculations. Long arrows: ferroelectric polarization. Small arrows: magnetic moment. The Mn spin order changes from the ferromagnetic (FM) state to the A-type antiferromagnetic state ($A_2$) in the first two interfacial layers if the polarization points away from the interface. (d) Reprinted figure with permission from \href{http://dx.doi.org/10.1103/PhysRevB.80.174406}{J. D. Burton \textit{et al.}, Physical Review B, 80, p. 174406, 2009} \cite{Burton:Prb}. Copyright \copyright (2009) by the American Physical Society. (e) The measured magnetoelectric hysteresis of PbZr$_{0.2}$Ti$_{0.8}$O$_3$/La$_{0.8}$Sr$_{0.2}$MnO$_3$ measured at $100$ K. Insets show the magnetic and charge states in La$_{0.8}$Sr$_{0.2}$MnO$_3$ and PbZr$_{0.2}$Ti$_{0.8}$O$_3$ layers, respectively. (e) Reprinted figure with permission from \href{http://dx.doi.org/10.1002/adma.200900278}{H. J. A. Molegraaf \textit{et al.}, Advanced Materials, 21, pp. 3470-3474, 2009} \cite{Molegraaf:Am}. Copyright \copyright (2009) by the WILEY-VCH Verlag GmbH \& Co. KGaA.}
\label{FET}
\end{figure}

In addition to the above described continuous tuning of magnetic properties, more dramatic magnetoelectric effect can be realized in heterostructures with a ferroelectric component and a magnetic correlated electronic component such as manganite La$_{1-x}A_x$MnO$_3$ ($A$=Ca, Sr, or Ba) (Fig.~\ref{FET}(c)). Correlated electronic materials show rich phase diagrams depending on the doped carrier density, consisting of plethoric competing phases with different resistivities, structures, and magnetic orders \cite{Dagotto:Sci}. In particular, if the doped carrier density is close to any phase boundary, it would be possible to realize the switching between two phases (e.g. ferromagnetic \textit{vs} antiferromagnetic) upon an external electric field, and therefore a gigantic magnetoelectric effect is expected. Although the driving force is also based on the charge screening, the underlying mechanism goes beyond the simple addition of spin-polarized carriers as in the previous two types of mechanisms, but more relies on the doping modification of the magnetic ground state.

Indeed, dramatic magnetoelectric coupling was predicted in La$_{0.5}A_{0.5}$MnO$_3$/BaTiO$_3$ heterostructures by the first-principles calculations \cite{Burton:Prb}. The polarization direction of BaTiO$_3$ was used to electrostatically modulate the hole-carrier density in La$_{0.5}A_{0.5}$MnO$_3$ and thus the magnetism of La$_{0.5}A_{0.5}$MnO$_3$ between the ferromagnetic state and antiferromagnetic state at the interface, as shown in Fig.~\ref{FET}(d). If the polarization points away from the interface, An apparent upward shift of local density of states happens, implying that the electron population at the interface is reduced, corresponding to the hole accumulation state. The opposite situation occurs if the polarization points toward the interface, leading to a downward shift of local density of states and the hole depletion state. Hence, the interfacial layer favors either the antiferromagnetic state (hole accumulation) or ferromagnetic state (hole depletion), depending on the polarization orientation. A large variation of magnetic moment corresponds to a large magnetoelectric effect.

Experimentally, this particular magnetoelectric coupling was confirmed in  La$_{0.8}$Sr$_{0.2}$MnO$_3$/PbZr$_{0.2}$Ti$_{0.8}$O$_3$ heterostructures \cite{Molegraaf:Am,Vaz:Prl,Vaz:Apl}. The variation of magnetic hysteresis loop was identified by the magneto-optical Kerr measurement, in response to the polarization switching of PbZr$_{0.2}$Ti$_{0.8}$O$_3$, as shown in Fig.~\ref{FET}(e). A bigger coercivity and a smaller saturated magnetization were indeed observed for the accumulation state as compared to the depletion one. It is seen that the magnetic reconstruction only occurs within a few atomic layers near the interface, while the rest La$_{1-x}$Sr$_x$MnO$_3$ layers sustain a robust ferromagnetic order against the polarization switching \cite{Vaz:Jap,Yi:Prl,Lu:Apl12,Jiang:Apl}, in consistent with the theoretical prediction.

In parallel to the first-principles calculations, a microscopic model based on the two-orbital double-exchange was once proposed in order to understand the ferroelectric screening effect in ferroelectric/manganite heterostructures \cite{Dong:Prb11}. The model simulation confirmed that the charge accumulation/depletion near the interface can drive the interfacial phase transitions, giving rise to robust magnetoelectric responses and bipolar resistive switching, in qualitative agreement with the density functional calculations. However, The interfacial candidate phases predicted by the density functional calculations and model simulations are different: an uncompensated magnetic interfacial layer was predicted in the first principles calculations while a compensated one was suggested in the model simulation. A recent experiment seemed to support the compensated interfacial magnetism \cite{Ma:Apl}.

Even without ferroelectric layers, the carrier-driven ferromagnetic switching can be achieved solely by electric field in some charge transferred heterostructures, e.g. CaRuO$_3$/CaMnO$_3$ \cite{Grutter:Prl}.

\begin{figure}
\centering
\includegraphics[width=\textwidth]{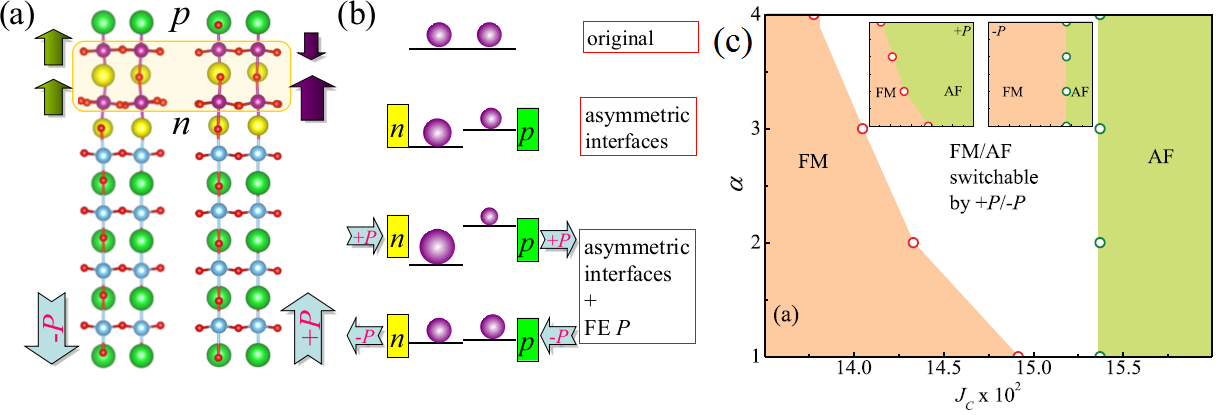}
\caption{(Colour online) Ferroelectric control of manganite magnetism.(a) A bilayered manganite-ferroelectric junction structure. Purple: Mn; yellow:  $R_{1-x}A_x$, green: Ba; red: O; cyan:Ti. The asymmetric $n$-$p$-type interfaces are adopted. $P$: ferroelectric polarization. The magnetization of each Mn layer is indicated by arrow. (b) The proposed mechanism of electro-control of $e_{\rm g}$ density (spheres) and local potential (bars), both of which are modulated by the asymmetric interfaces (bricks) and ferroelectric polarization (arrows). (c) An approximate phase diagram of the bilayer manganite. Insets: the phase diagrams under the $+P$ (left) and $-P$ (right) states respectively. A comparison of the two phase diagrams (insert) clearly illustrates the magnetic switching (different colors) by ferroelectric polarization: a FM/AF switchable white region. Reprinted figure with permission from \href{http://dx.doi.org/10.1103/PhysRevB.88.140404}{S. Dong \textit{et al.}, Physical Review B, 88, p. 140404(R), 2013} \cite{Dong:Prb13.2}. Copyright \copyright (2013) by the American Physical Society.}
\label{FETb}
\end{figure}

Based on this model, Dong \textit{et al.} proposed a scheme to pursue a full control of magnetism by reversing the ferroelectric polarization in a bilayer manganite-ferroelectric superlattice \cite{Dong:Prb13.2}. The design is sketched in Fig.~\ref{FETb}(a). Here, two asymmetric polar interfaces: one $n$-type and one $p$-type, are adopted. The $n$-type interface will attract more electrons to its nearest-neighbour Mn layer, while the $p$-type interface will repel electrons away from the interface. Therefore, even in absence of external ferroelectricity, the asymmetric interfaces already impose a modulation of the electronic density distribution and electrostatic potential within the bilayer manganite. If a ferroelectric polarization points to the $n$-type interface (the $+P$ case), the electrostatic potential difference between the two MnO$_2$ layers is split, further enhancing the charge disproportion. In contrast, the electrostatic potential from the polar interfaces will be partially or fully compensated by a ferroelectric polarization pointing to the $p$-type interface (the $-P$ case), suppressing the electronic disproportionality, as shown in Fig.~\ref{FETb}(b). Due to the biggest interface/volume ratio (up to $100\%$) for the bilayer structure, the magnetic and electronic states of every manganite layer can be fully controlled by the polarization. A remarkable variation in total magnetization (up to $\sim90\%$) tuned by the polarization was predicted, as shown in Fig.~\ref{FETb}(c).

It should be noted that the piezostrain effect and the field effect may exist simultaneously in many heterostructures if a ferroelectric layer instead of a pure piezoelectric layer is used \cite{Alberca:Prb}. The field effect usually plays the dominant role for the cases with ultra-thin magnetic films (e.g. a few atomic layers), while the piezostrain effect takes the place for relatively thick magnetic films.

\subsubsection{Ferroelectric-magnetic tunneling junctions}
The magnetoelectric effect prominent in La$_{0.8}$Sr$_{0.2}$MnO$_3$/PbZr$_{0.2}$Ti$_{0.8}$O$_3$ heterostructures originates from the charge-driven interfacial magnetic transitions. Actually, an apparent variation of the in-plane conductivity in the magnetic layer for a magnetic/ferroelectric heterostructure, upon a polarization switching in the ferroelectric layer, is also possible and thus attractive, as demonstrated in La$_{0.8}$Sr$_{0.2}$MnO$_3$/PbZr$_{0.2}$Ti$_{0.8}$O$_3$ heterostructures \cite{Vaz:Prl,Vaz:Apl,Jiang:Apl}. The two polarization states enable a big variation of resistivity in La$_{0.8}$Sr$_{0.2}$MnO$_3$ layer via the interfacial mediation mechanism already discussed. In the other words, the prominent variation in the transport behavior is driven by the ferroelectric layer, since a non-ferroelectric capping layer (LaAlO$_3$ or SrTiO$_3$) makes no difference in the transport behavior of the neighboring La$_{0.8}$Sr$_{0.2}$MnO$_3$ layer from that of a bare La$_{0.8}$Sr$_{0.2}$MnO$_3$ layer.

\begin{figure}
\centering
\includegraphics[width=\textwidth]{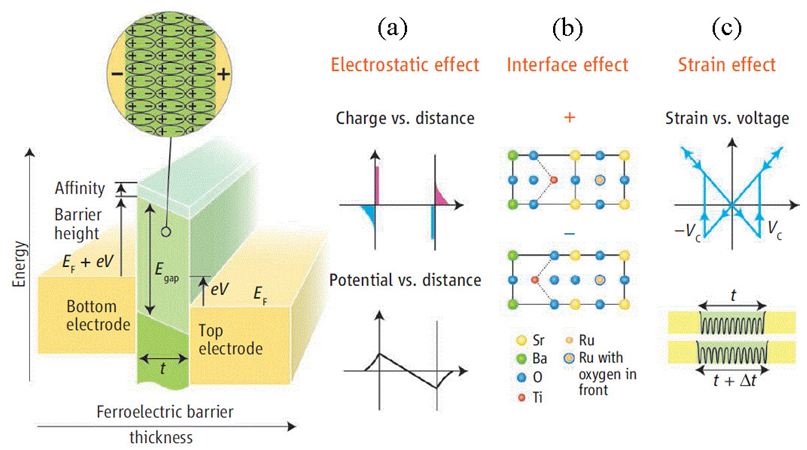}
\caption{(Colour online) A schematic diagram of a ferroelectric tunnel junction, which consists of two electrodes separated by a nanometer-thick ferroelectric barrier layer. $E_{\rm gap}$ is the energy gap. $E_{\rm F}$ is the Fermi energy, $V$ is the applied voltage, $V_{\rm c}$ is the coercive voltage, $t$ is the barrier thickness, and $\Delta_t$ is the thickness variation driven by an applied field. (a)-(c) The proposed mechanisms modulating the tunneling behavior in a ferroelectric tunnel junction: (a) the electrostatic potential across a junction, (b) the interfacial bonding and (c) strain effect associated with the piezoelectric response. From \href{http://dx.doi.org/10.1126/science.1126230}{E. Y. Tsymbal \textit{et al.}, Science, 313, pp. 181-183, 2006} \cite{Tsymbal:Sci}. Reprinted with permission from The American Association for the Advancement of Science.}
\label{FETJ}
\end{figure}

In addition to the observed in-plane conductance modulation, a tunneling electroresistance effect should not be a surprise in ferroelectric junctions. In short, three critical elements for determining the interface transmission function and thus tunneling electroresistance need to be concerned \cite{Tsymbal:Sci}, as summarized in Fig.~\ref{FETJ}: (a) the electrostatic potential across the junction, (b) the interface bonding strength, and (c) the strain associated with piezoelectric effect. The electrostatic effect is resulted from the incomplete screening of polarization charges at the interface. This leads to an electrostatic potential that are superimposed onto the contact potential in the tunnel junction, creating an asymmetric potential profile across the two end electrodes. The interfacial effect comes from the ferroelectric displacements at the boundary between the ferroelectric layer and two electrodes \cite{Fong:Prb}. The atomic displacements alter the orbital hybridizations at the interface and the transmission across it \cite{Duan:Prl,Wortmann:Prb}. The strain effect originates from the piezoelectric effect of all ferroelectrics. A piezoelectric distortion along the junction axis deforms the transport characteristics of the barrier such as barrier thickness and attenuation constant.

Experimentally, a very large tunneling electroresistance effect was indeed observed in BaTiO$_3$ \cite{Gruverman:Nl,Garcia:Nat,Chanthbouala:Nn} and PbZr$_{1-x}$Ti$_x$O$_3$ \cite{Maksymovych:Sci,Pantel:Apl} ferroelectric thin films sandwiched with different end electrodes, using the scanning probe techniques. Garcia revealed a giant electroresistance effect in BaTiO$_3$/La$_{0.67}$Sr$_{0.33}$MnO$_3$ ferroelectric tunnel junctions \cite{Garcia:Nat}. They found that the tunnel electroresistance ratio scales exponentially with the ferroelectric film thickness, reaching $\sim10,000\%$ and $\sim75,000\%$ at a thickness of $2$ nm and $3$ nm, respectively. These experimental findings unambiguously prove that the ferroelectric polarization can control the tunneling electroresistance effect, supporting earlier theoretical predictions \cite{Zhuravlev:Prl,Kohlstedt:Prb}.

\begin{figure}
\centering
\includegraphics[width=0.8\textwidth]{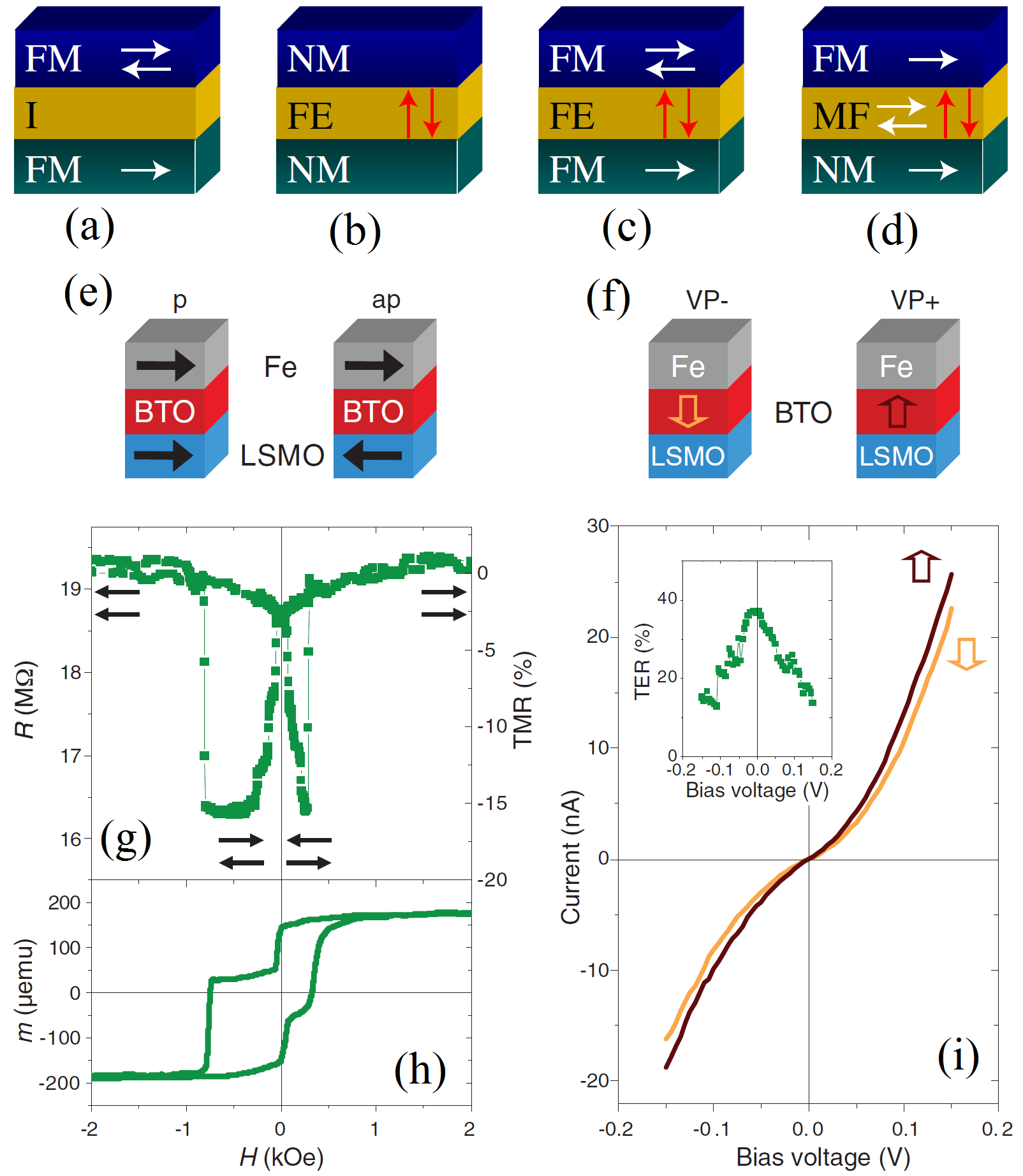}
\caption{(Colour online) Schematic configurations of several tunnel junctions: (a) a magnetic tunnel junction, (b) a ferroelectric tunnel junction, (c-d) two multiferroic tunnel junctions. FE: ferroelectric layer; FM: ferromagnetic layer; I: insulator; NM: nonmagnetic layer; MF: multiferroic layer. The ferroic moments are denoted by arrows. (a-d) Reprinted figure with permission from \href{http://dx.doi.org/10.1098/rsta.2010.0344}{J. P. Velev \textit{et al.}, Philosophical Transactions of the Royal Society A, 369, pp. 3069-3097, 2011} \cite{Velev:Ptrsa}. Copyright \copyright (2011) by The Royal Society. (e-i) The measured magnetoresistive and electroresistive behaviors of a Fe/BaTiO$_3$/La$_{0.67}$Sr$_{0.33}$MnO$_3$ multiferroic tunnel junction. (e-f) The schematic of magnetic and ferroelectric profiles. The resistance (g) and magnetic moment (h) as a function of magnetic field. (i) The current-voltage characteristics of the junction measured at $4$ K after the upward and downward electrical poling of the ferroelectric BaTiO$_3$ barrier respectively. (e-i) From \href{http://dx.doi.org/10.1126/science.1184028}{V. Garcia \textit{et al.}, Science, 327, pp. 1106-1110, 2010} \cite{Garcia:Sci}. Reprinted with permission from The American Association for the Advancement of Science.}
\label{MFTunnel}
\end{figure}

A ferroelectric junction becomes a multiferroic junction if the metallic electrodes are magnetic, which has the capability to control both the charge and spin tunneling via the ferromagnetic and ferroelectric components in the junction (Fig.~\ref{MFTunnel}(a-d)). It is noted that the resistance of such a multiferroic tunnel junction can be significantly modulated when the polarization in the ferroelectric barrier is reversed and/or when the magnetization of electrodes is switched from the parallel alignment to the antiparallel alignment, possibly rendering a four-state resistance device where the resistance state is controlled both by electric field and by magnetic field \cite{Zhuravlev:Apl}, as sketched in Fig.~\ref{MFTunnel}(e-f).

The four-state resistance behavior in the SrRuO$_3$/BaTiO$_3$/SrRuO$_3$ multiferroic tunnel junctions with the asymmetric interfaces was predicted using the first-principles theory \cite{Velev:Nl}. The tunneling magnetoresistance effect has the same origin as that of ordinary magnetic tunnel junctions. When the two electrodes are in the parallel magnetic configuration, both spin channels (the majority-spin and minority-spin) contribute to the conductance, while the conductance is seriously suppressed for the antiparallel magnetic configuration, similar to the mechanism for a tunneling magnetoresistance device. The tunneling electroresistance effect originates from the asymmetric interfaces with asymmetric potential profile upon the ferroelectric polarization switching. Subsequently, the four-state prototype devices were fabricated and the four states were observed experimentally \cite{Garcia:Sci,Hambe:Afm,Yin:Fp,Pantel:Nm}. Figure~\ref{MFTunnel}(g-i) summaries the magnetoresistive and electroresistive properties of the Fe/BaTiO$_3$/La$_{0.67}$Sr$_{0.33}$MnO$_3$ multiferroic tunnel junctions \cite{Garcia:Sci}. The tunneling magnetoresistance effect ($\sim-17\%$) was achieved by tuning the magnetization direction of Fe and La$_{2/3}$Sr$_{1/3}$MnO$_3$. A set of short voltage pulses of $\pm1$ V triggered the tunnel resistance ($\sim30\%$)  in a reversible way, linked to the variation of the barrier height.

However, these multiferroic tunnel junctions could only work at low temperatures and the observed magnetoresistance ratio was much smaller than the predicted value \cite{Velev:Nl}. Recently, Yin \textit{et al.} fabricated La$_{0.7}$Ca$_{0.3}$MnO$_3$/(Ba,Sr)TiO$_3$ or La$_{0.7}$Sr$_{0.3}$MnO$_3$/(Ba,Sr)TiO$_3$ multiferroic tunnel junctions \cite{Yin:Fp,Yin:Jap}, and their junctions showed the four resistance states and a larger magnetoresistance effect (up to $\sim300\%$). In addition, the operating temperature of their junctions can be as high as room temperature since (Ba,Sr)TiO$_3$ can sustain the ferroelectricity up to room temperature.

It is interesting to note that inserting a nanometer-thick La$_{0.5}$Ca$_{0.5}$MnO$_3$ interlayer in-between the two components in the La$_{0.7}$Sr$_{0.3}$MnO$_3$/BaTiO$_3$ ferroelectric tunnel junction enhanced the tunnel electroresistance ratio up to $\sim10,000\%$ \cite{Yin:Nm}. Here, La$_{0.5}$Ca$_{0.5}$MnO$_3$ acts as a switchable magnetic layer supplemented to the ferroelectric barrier. In fact, Burton \textit{et al.} demonstrated that the magnetoelectric interaction between BaTiO$_3$ and La$_{1-x}$Sr$_x$MnO$_3$ magnetic electrode can generate a giant tunneling electroresistance effect due to the magnetic phase transitions of La$_{1-x}$Sr$_x$MnO$_3$ \cite{Burton:Prl}. This idea was recently realized experimentally in La$_{1-x}$Sr$_x$MnO$_3$/PbZr$_{1-x}$Ti$_x$O$_3$ heterostructures \cite{Jiang:Nl}. The ferroelectricity induced modulation at the interface ultimately results in an enhanced electroresistance effect. Another type of multiferroic tunnel junctions employ single phase multiferroics as the tunnel barriers. Gajek \textit{et al.} first fabricated the La$_{2/3}$Sr$_{1/3}$MnO$_3$/La$_{0.1}$Bi$_{0.9}$MnO$_3$/Au multiferroic tunnel junctions in which the tunneling resistance is controlled by both electric and magnetic fields \cite{Gajek:Nm}. In these junctions, the tunneling magnetoresistance arises from the spin filtering effect, and the tunneling electroresistance is related to the variation of barrier potential upon the polarization reversal.

More details regarding the ferroelectric and multiferroic tunneling junctions can be found in Ref.~\cite{Tsymbal:Mrs} by Tsymbal \textit{et al.}.

\subsection{Electrically controllable exchange bias}
The exchange bias effect is a shift of magnetic hysteresis loop away from the center of symmetry, which has been used in a variety of magnetic storage and sensor devices \cite{Nogues:Prp,Kiwi:Mmm,Nogues:Mmm}. This effect can be understood theoretically as a spin pinning at the ferromagnetic/antiferromagnetic interface. A control of the exchange bias utilizing magnetoelectric coupling would provide a new path to control magnetism.

The first demonstration of the electric field controlled exchange bias was reported in [Co/Pt]/Cr$_2$O$_3$ (111) heterostructures \cite{Borisov:Prl,He:Nm}. Application of an electric field to Cr$_2$O$_3$ generates a net magnetization whose direction depends on the direction of electric field. Subsequently, experimental studies of the exchange bias control in a number of multiferroic materials, such as YMnO$_3$ \cite{Dho:Apl05,Marti:Apl,Laukhin:Prl} and BiFeO$_3$ \cite{Dho:Am}, were reported. In particular, Chu \textit{et al.} and B\'ea \textit{et al.} studied the FeCo/BiFeO$_3$ and CoFeB/BiFeO$_3$ heterostructures and found an apparent change of the exchange bias by tuning the electric field at room temperature \cite{Chu:Nm,Bea:Prl}. A close link of the exchange bias with particular ferroelectric domain walls in BiFeO$_3$ was suggested \cite{Martin:Nl}.

\begin{figure}
\centering
\includegraphics[width=\textwidth]{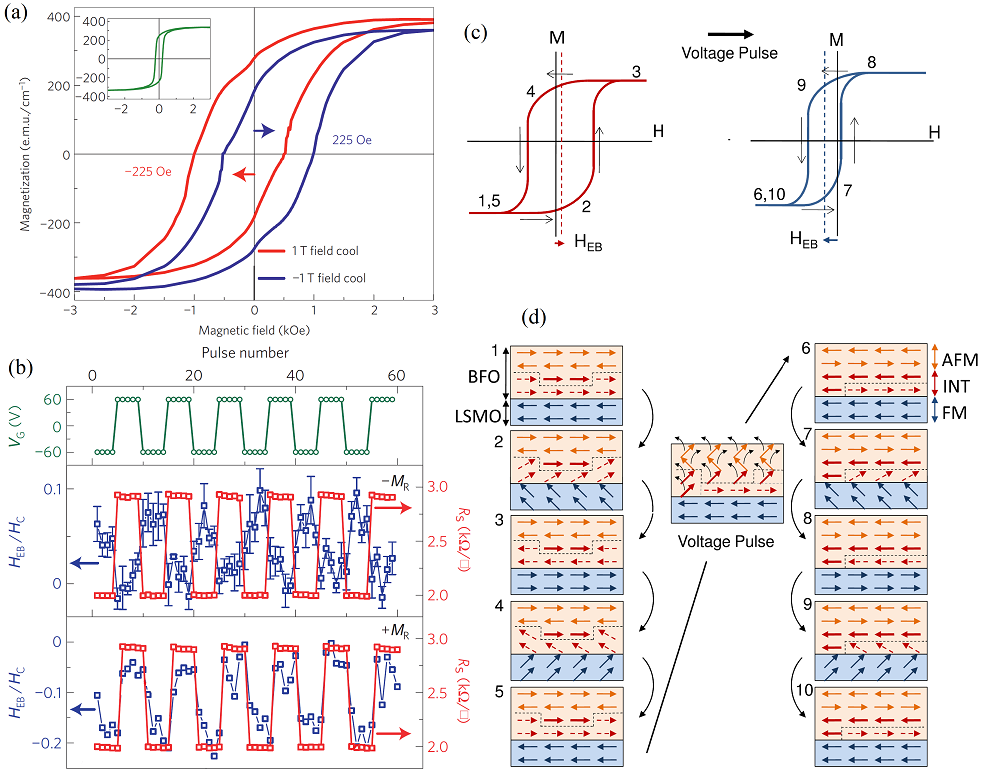}
\caption{(Colour online) Electro-control of exchange bias. (a) The magnetic hysteresis loops of BiFeO$_3$/La$_{0.7}$Sr$_{0.3}$MnO$_3$ heterostructures measured at $7$ K, upon the +/- $1$ T field cooling from $350$ K respectively. The inset is a loop for a BiFeO$_3$/SrTiO$_3$/La$_{0.7}$Sr$_{0.3}$MnO$_3$ structure, showing no exchange bias after the field cooling. (b) The electric field control of exchange bias. Top panel: gate-voltage-pulse sequence used for measurements. Middle and bottom panels: normalized exchange bias and peak resistance in response to the gate-pulse sequence shown in the top panel. Reprinted by permission from Macmillian Publishers Ltd: \href{http://dx.doi.org/10.1038/NMAT2803}{S. M. Wu \textit{et al.}, Nature Materials, 9, p. 756-761, 2010} \cite{Wu:Nm}. Copyright \copyright (2010). (c) The magnetic hysteresis loops of La$_{0.7}$Sr$_{0.3}$MnO$_3$ before and after the ferroelectric polarization reversal of BiFeO$_3$. The measurement sequence is denoted by numbers and arrows as magnetic field is swept. (d) A depiction of the interfacial spin configurations corresponding respectively to the numbered states in (c). Reprinted figure with permission from \href{http://dx.doi.org/10.1103/PhysRevLett.110.067202}{S. M. Wu \textit{et al.}, Physical Review Letters, 110, p. 067202, 2013} \cite{Wu:Prl13}. Copyright \copyright (2013) by the American Physical Society.}
\label{EB}
\end{figure}

\begin{figure}
\centering
\includegraphics[width=\textwidth]{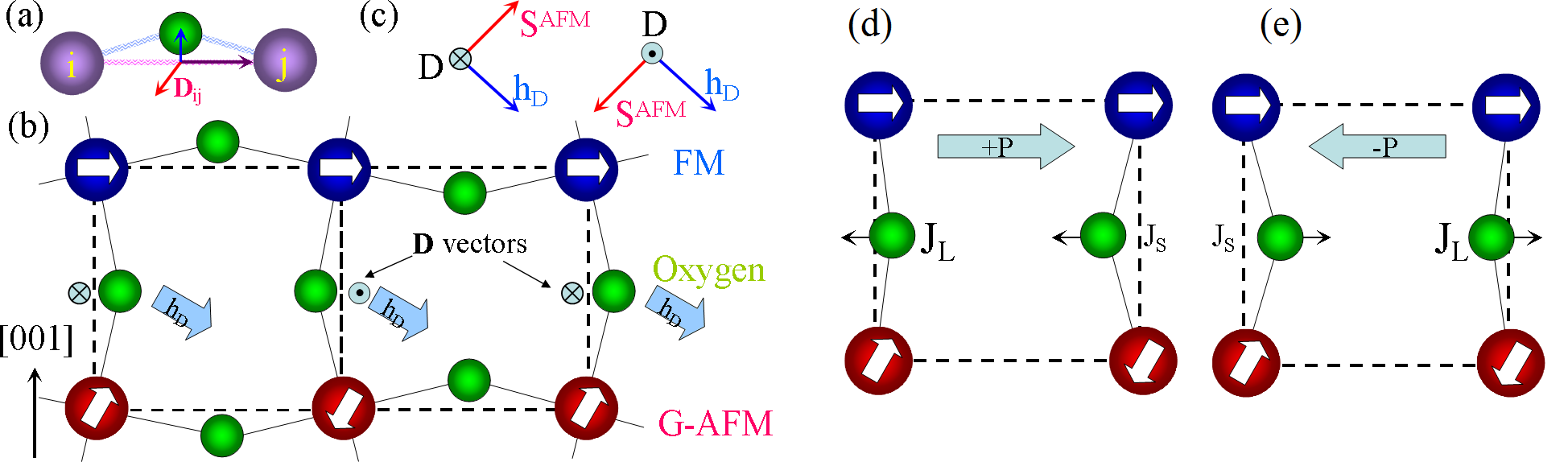}
\caption{(Colour online) (a) The relationship between the $M_i$-O-$M_j$ bond, oxygen displacement, and corresponding $\vec{D}_{ij}$ vector in perovskites. (b) A schematic of the interface between the ferromagnetic and G-type antiferromagnetic perovskites, including the oxygen octahedral tilting. (c) A homogeneous $\vec{h}_{D}$ should be perpendicular to $\vec{S}^{\rm AFM}$ and $\vec{D}$. (d-e) A schematic of the ferroelectric polarization driven asymmetric bond angles. This asymmetry allows a modulation of the superexchange coupling at the interface. Reprinted figure with permission from \href{http://dx.doi.org/10.1103/PhysRevLett.103.127201}{S. Dong \textit{et al.}, Physical Review Letters, 103, p. 127201, 2009} \cite{Dong:Prl2}. Copyright \copyright (2009) by the American Physical Society.}
\label{EBDM}
\end{figure}

The electrically controlled exchange bias was also found in BiFeO$_3$/La$_{0.7}$Sr$_{0.3}$MnO$_3$ heterostructures \cite{Wu:Nm10,Yu:Prl}. Figure~\ref{EB}(a) shows the typical magnetic hysteresis loops of BiFeO$_3$/La$_{0.7}$Sr$_{0.3}$MnO$_3$ heterostructures, exhibiting the corresponding shift after the field cooling. This shift is due to the antiferromagnetic-ferromagnetic coupling, since it is absent if a thin SrTiO$_3$ layer is inserted in-between BiFeO$_3$ and La$_{0.7}$Sr$_{0.3}$MnO$_3$, as shown in insert of Fig.~\ref{EB}(a). More exciting is an observation of the two distinct exchange-bias states which can be operated reversibly by switching the ferroelectric polarization of BiFeO$_3$, as shown in Fig.~\ref{EB}(b). It is clear that the magnitude of exchange bias is modulated by electric field switching between a high value and a low one (Fig.~\ref{EB}(c)), corresponding to a modulation of channel resistance and magnetic coercivity \cite{Wu:Prl13}. The possible evolution of interfacial spin structure in the exchange bias process is sketched in Fig.~\ref{EB}(d).

A recent work demonstrated a deterministic switching of weak ferromagnetism for BiFeO$_3$ at room temperature using an electric field \cite{Heron:Nat}. Due to the spin canting driven by the Dzyaloshinskii-Moriya interaction, a weak magnetization can be obtained in G-type antiferromagnetic BiFeO$_3$ thin films. Although a symmetry consideration asserts a forbidden direct $180^\circ$ switching of the Dzyaloshinskii-Moriya vector by ferroelectric polarization \cite{Ederer:Prb}, a two-step switching process of the polarization can lead to the switching of Dzyaloshinskii-Moriya vector and thus a deterministic reversal of the net magnetization \cite{Heron:Nat}. In Co$_{0.9}$Fe$_{0.1}$/BiFeO$_3$ heterostructures, this magnetoelectric switching effect can be further amplified by the magnetic exchange between ferromagnet and BiFeO$_3$. Namely, the ferromagnetic domains in Co$_{0.9}$Fe$_{0.1}$ can be tuned via this mechanism. However, no exchange bias was observed in this work, implying that the intrinsically canted moments of BiFeO$_3$ may not be the origin of exchange bias.

To understand the electrically controllable exchange bias, Dong \textit{et al.} proposed two (related) mechanisms respectively based on the Dzyaloshinskii-Moriya interaction and on the ferroelectric polarization \cite{Dong:Prl2}. The latter mechanism should be applicable only in (multiferroic) heterostructures with a large ferroelectric polarization, while the former one can be more general. Without losing a generality, the spin-spin interaction in perovskites can be described by a simplified Hamiltonian (Eq.~\ref{jd}) including both the superexchange and Dzyaloshinskii-Moriya interaction. This problem was once discussed in Sec. 3.1.1. For a perovskite, the Dzyaloshinskii-Moriya vector ($\textbf{D}_{ij}$) is perpendicular to the $M_i$-O-$M_j$ bond \cite{Sergienko:Prb,Moriya:Pr}, as shown in Fig.~\ref{EBDM}(a). Because of the collective tilting/rotation of oxygen octahedra, the nearest-neighbor oxygens in the same direction will move away from the midpoint to the opposite directions, namely, the nearest-neighbor displacements are staggered, as shown in Fig.~\ref{EBDM}(b). A combination of the two staggered components $\textbf{D}_{ij}$ and $\textbf{S}_{\rm AFM}$ will give rise to a homogeneous Dzyaloshinskii-Moriya effect at the interface (Fig.~\ref{EBDM}(b-c)), which can be described by an effective Hamiltonian:
\begin{equation}
H^{\rm interface}_{\rm DM} = \sum_{<ij>}\textbf{D}_{ij}\cdot (\textbf{S}_{i}^{\rm FM}\times\textbf{S}_{j}^{\rm AFM})=-\textbf{h}_{D}\cdot\sum_{i}\textbf{S}_{i}^{\rm FM},
\label{exd}
\end{equation}
where $\textbf{h}_{D}$ can be regarded as a biased magnetic field which may be fixed by a field-cooling process and frozen at low temperatures during the hysteresis loop measurement.

Furthermore, if one component of the heterostructure has a ferroelectric polarization which corresponds to a uniform displacement between the cations and anions, the bond angles at the interface become no longer symmetric. Since the magnitude of a normal super-exchange coupling depends on the bond angle, the modulated bond angles certainly induce a pair of staggered interfacial superexchange couplings, which are respectively denoted as $J_{\rm L}$ and $J_{\rm S}$, as shown in Fig.~\ref{EBDM}(d-e). Once again, the staggered superexchange couplings at the interface will also induce a homogeneous biased field $\textbf{h}_J$ in the presence of G-type antiferromagnetic spin order, which is described by:
\begin{equation}
H^{\rm interface}_{\rm DM} = \sum_{<ij>}J_{ij}(\textbf{S}_{i}^{\rm FM}\cdot\textbf{S}_{j}^{\rm AFM})=-\textbf{h}_{J}\cdot\sum_{i}\textbf{S}_{i}^{\rm FM}.
\label{exj}
\end{equation}

This model on exchange bias relies on the interactions between lattice distortion and magnetism rather than uncompensated antiferromagnetic moments anymore. Next, using the first-principles calculations, Dong \textit{et al.} chose SrRuO$_3$/SrMnO$_3$ heterostructures to verify the two mechanisms \cite{Dong:Prb11.2}. In parallel, other models in terms of the domain concept to understand this electrically controllable exchange bias were proposed \cite{Livesey:Prb,Belashchenko:Prl}, and the carrier-mediated screening besides the interfacial bonding coupling may also play an important role in tuning the exchange bias since the exchanges across the interface depend on the interfacial electronic structure.

More information on the exchange bias in BiFeO$_3$-based heterostructures can be found in a recent review by Yu, Chu, and Ramesh \cite{Yu:Ptrsa}.

\subsection{Hybrid improper ferroelectricity in superlattices}
As discussed in Sec. 3.3.1-3.3.2, improper ferroelectricity can emerge as a result of a combination of non-polar distortion modes in hexagonal $R$MnO$_3$, hexagonal $R$FeO$_3$, Ca$_3$Mn$_2$O$_7$, and Ca$_3$Ti$_2$O$_7$. The core physics for this improper ferroelectricity as a means of enlightening can be extended to oxide heterostructures or superlattices.

\begin{figure}
\centering
\includegraphics[width=\textwidth]{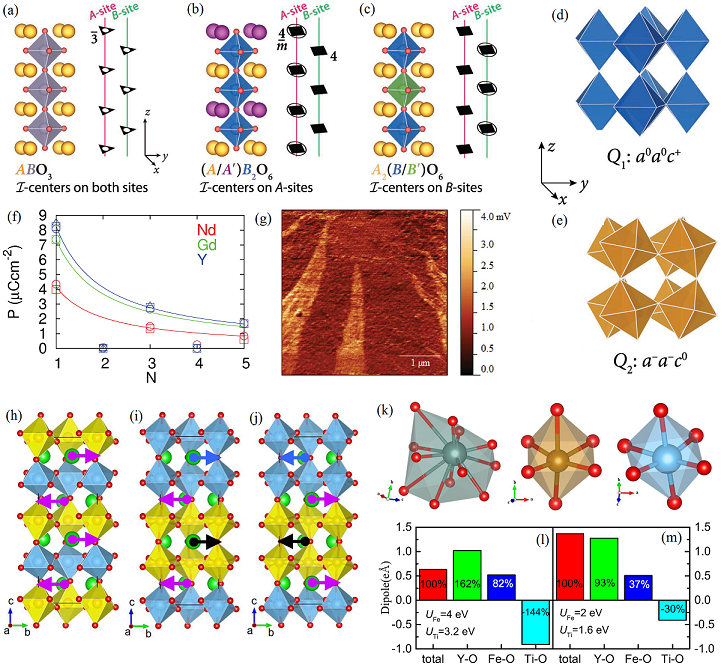}
\caption{(Colour online) Hybrid improper ferroelectricity. (a-c) A schematic of the chemical criterion for the collective rotation induced ferroelectricity in perovskite superlattices. (a) The inversion centers are on both the A and B sites for bulk $AB$O$_3$ perovskites. (b) For ($AB$O$_3$)$_1$/($A'B$O$_3$)$_1$ superlattices, the inversion centers are only on the A site. (c) For ($AB$O$_3$)$_1$/($AB'$O$_3$)$_1$ superlattices, the inversion centers are only on the B site. Only the ($AB$O$_3$)$_1$/($A'B$O$_3$)$_1$ superlattices host the hybrid improper ferroelectricity since the octahedra rotations on the B sites preserve the inversion symmetry but those on the A-sites may not. (d-e) The two primary rotational modes $Q_1$ and $Q_2$. A combination of the two rotation modes produces the common tilt pattern ($a^-a^-c^+$) in orthorhombic perovskites. (a-e) Reprinted figure with permission from \href{http://dx.doi.org/10.1002/adma.201104674}{J. Rondinelli \textit{et al.}, Advanced Materials, 24, pp. 1961-1968, 2012} \cite{Rondinelli:Am12}. Copyright \copyright (2012) by the Wiley-Vch Verlag GmbH \& Co. KGaA. (f) The calculated polarization ($P$) as a function of the number of perovskite unit cells ($N$) in (LaFeO$_3$)$_N$/($R$FeO$_3$)$_N$ ($R$=Nd, Gd, Y) superlattices. The curves represent the fittings by function $P_1/N$, where $P_1$ is the calculated polarization for the $N=1$ superlattices. Here only the odd $N$ cases can give rise to a finite polarization. (g) The piezoelectric force microscopy image of (LaFeO$_3$)$_5$/(YFeO$_3$)$_5$ film. Bright areas indicate the polar domains while the dark areas are the non-polar regions. (f-g) Reprinted figure with permission from \href{http://dx.doi.org/10.1039/c3sc53248h}{J. Alaria \textit{et al.}, Chemical Science, 5, pp. 1599-1610, 2014} \cite{Alaria:Cs}. Copyright \copyright (2014) by the Royal Society of Chemistry. (h-j) The schematic ($AB$O$_3$)$_N$/($AB'$O$_3$)$_N$ superlattices. The arrows denote the displacements of the A site ions. (h) $N=1$, where the displacements between two neighboring layers are compensated. (i-j) $N=2$, where the distortions cannot be fully compensated, giving rise to a net polarization. Along the stacking direction, the B-site ions form the $B-B-B'-B'$ pattern, in analog to the $\uparrow\uparrow\downarrow\downarrow$-type antiferromagnetic order. (i) The case with a positive polarization; (j) the case with a negative polarization. (k) The schematic Y-O-icosahedron (left), Fe-O-octahedra (middle), and Ti-O-octahedra (right) in (YFeO$_3$)$_2$/(YTiO$_3$)$_2$ superlattices. (l-m) The point-charge model estimated individual contributions to the total polarization, given different $U_{\rm eff}$ values in the density functional calculations, where the Fe$^{2+}$-Ti$^{4+}$ pair is adopted. (l) The insulating case; (m) the metallic case, where the total polarization can be surprisingly enhanced since the (negative) proper ferroelectric distortions of Ti-O-octahedra are suppressed by the metallicity, while the improper ferroelectric distortions from the Y-O-icosahedron robustly persist independent of the metallicity. (h-m) Reprinted figure with permission from \href{http://dx.doi.org/10.1103/PhysRevB.91.195145}{H. M. Zhang \textit{et al.}, Physical Review B, 91, p. 195145, 2015} \cite{Zhang:Prb15}. Copyright \copyright (2015) by the American Physical Society.}
\label{AABBO}
\end{figure}

Indeed, this concept was once ``created" in artificial superlattices. As mentioned, the first work along this line was done on pure ferroelectric PbTiO$_3$/SrTiO$_3$ superlattices \cite{Bousquet:Nat}, leading to the prediction of the hybrid improper ferroelectricity in Ca$_3$Mn$_2$O$_7$. In 2012, Rondinelli and Fennie proposed a general design rule for improper ferroelectricity in $AB$O$_3$/$A'B'$O$_3$ superlattices \cite{Rondinelli:Am12}. The prerequisites include: 1) chemical criterion; 2) energetic criterion. According to the chemical criterion, the ($AB$O$_3$)$_1$/($A'B$O$_3$)$_1$ can break the inversion symmetry but the ($AB$O$_3$)$_1$/($AB'$O$_3$)$_1$ cannot, since the A and B sites have different local symmetries (see Fig.~\ref{AABBO}(a-c)). The energetic criterion defines two cooperative primary modes of octahedral rotation: $Q_1$ ($=a^0a^0c^+$) and $Q_2$ ($=a^-a^-c^0$) in the Glazer notation (see Fig.~\ref{AABBO}(d-e)). More details regarding this octahedral rotation driven ferroelectricity in superlattices, can be found in a topical review by Benedek, Mulder, and Fennie \cite{Benedek:Jssc}.

This design rule stems from a pure chemical and structural discussion and does not exclude magnetism. It is thus straightforward to design multiferroic superlattices with improper ferroelectric polarizations so long as magnetic perovskites with desired distortions are used. For example, Bristowe \textit{et al.} predicted a ferromagnetic-ferroelectric superlattice ($R$TiO$_3$)$_1$/($A$TiO$_3$)$_1$ \cite{Bristowe:Nc}. The charge ordering of Ti$^{3+}$-Ti$^{4+}$ is responsible for the insulating property and ferromagnetism.

In 2014, Alaria \textit{et al.} performed a systematic investigation on (YFeO$_3$)$_n$/(LaFeO$_3$)$_n$ superlattices where $n$ is the number of unit cells \cite{Alaria:Cs}. Their density functional calculations predicted that the improper ferroelectricity is indeed odd-even interleaved, namely the polarization is nonzero only in the odd $n$ superlattices (see Fig.~\ref{AABBO}(f)). Their experimental data based on the second harmonic generation and piezoelectric force microscopy confirmed the existence of polarization in the $n=5$ superlattice (Fig.~\ref{AABBO}(g)).

Later on, Zhang \textit{et al.} studied (YFeO$_3$)$_n$/(YTiO$_3$)$_n$ superlattices \cite{Zhang:Prb15}. Although this ($AB$O$_3$)$_1$/($AB'$O$_3$)$_1$ type superlattice does not show ferroelectricity as expected, the ($AB$O$_3$)$_2$/($AB'$O$_3$)$_2$ can be ferroelectric, which is just a complementary case to the ($AB$O$_3$)$_n$/($A'B$O$_3$)$_n$ superlattices (see Fig.~\ref{AABBO}(h-j)). This $\mathbb{Z}_2$ behavior is due to the geometric origin of hybrid improper ferroelectricity. Furthermore, in (YFeO$_3$)$_n$/(YTiO$_3$)$_n$, a complete electron transfer from Ti$^{3+}$ to Fe$^{3+}$ implies that the ferroelectricity in the $n=2$ superlattice can also contain the (negative) contribution from Ti$^{4+}$ which is a proper ferroelectric active ion (Fig.~\ref{AABBO}(k)). In Zhang \textit{et al.}'s calculations, the metallicity can suppress this proper ferroelectric component while the improper ferroelectric part remains less affected, rendering enhanced total ferroelectric distortion (see Fig.~\ref{AABBO}(l-m)). This so-called ferroelectric metal, unusual in bulk crystals, may be obtained with a careful design of the charge-transferred perovskite superlattices \cite{Shi:Nm,Puggioni:Nc}.

In addition, charge transfer in pure magnetic superlattices without any ferroelectric-active element, e.g. NdMnO$_3$/SrMnO$_3$/LaMnO$_3$, can contribute an unconventional polarization. The asymmetric charge transfer across the NdMnO$_3$/SrMnO$_3$ and SrMnO$_3$/LaMnO$_3$ interfaces benefits to the ionic displacements, giving rise to electric dipoles. These dipoles can be reversed although the two opposite states are not fully symmetric according to the density functional calculations and symmetry analysis. Since the charge transfer is naturally magnetic-ordering dependent, the electric dipoles in this superlattice are correlated with magnetism, even though this magnetoelectric coupling is not directly magnetism-driven \cite{Rogdakis:Nc}.

\subsection{Magnetoelectricity at domain walls}
We highlight an issue related to domain walls in the end of this section. It is known that ferroic materials usually accommodate domain structure, giving rise to domain walls which may be treated as ``spontaneous" interfaces besides ``artificial" interfaces generated in heterostructures. Although a ferroic order certainly breaks a particular symmetry, domain structure usually emerges as the residual fingerprints of symmetry. In multiferroics, ferroelectric domains are often clamped with magnetic domains, rendering one more manifestation of magnetoelectricity. However, the significance of domain walls may not be prominent in bulk multiferroics, whereas it can be outshot in the utmost sense in heterostructures and thin films due to the dimensionality confinement and easy controllability of domain walls.

In fact, as early as 2002, using the second harmonic generation, Fiebig \textit{et al.} observed the coupled antiferromagnetic and ferroelectric domains in hexagonal YMnO$_3$ platelets \cite{Fiebig:Nat}. This phenomenon allows a possibility to control the magnetic phase by electric field in hexagonal HoMnO$_3$ \cite{Lottermoser:Nat}. The bridge linking the ferroelectric domains and antiferromagnetic domains is the magneto-elastic (spin-lattice) coupling \cite{Lee:Nat08}.

The similar coupling effects were observed also in BiFeO$_3$ \cite{Zhao:Nm}. It is known that the easy-plane of Fe magnetic moment is perpendicular to the ferroelectric polarization, as mediated by the Dzyaloshinskii-Moriya interaction \cite{Ederer:Prb}. For BiFeO$_3$, the eight polarization directions, i.e. the eight $<111>$ axes in the pseudocubic framework allow the $71^\circ$ and $109^\circ$ rotations of ferroelectric polarization. If the magnetic easy plane of a local magnetic domain keeps perpendicular to the polarization of the linked ferroelectric domain, a net magnetization is inevitably generated at the ferroelectric domain walls, implying the domain wall magnetoelectricity. Furthermore, this net magnetization can be coupled with the neighboring ferromagnetic layer in heterostructures, which can magnify the domain wall magnetoelectricity.

In particular, Chu \textit{et al.} demonstrated that the ferromagnetic domain structure of Co$_{0.9}$Fe$_{0.1}$ can be controlled by an electrical field applied to the underlying BiFeO$_3$ layer \cite{Chu:Nm}. The one-to-one correspondence between the ferroelectric domains and antiferromagnetic domains in BiFeO$_3$ makes a control of magnetism via domain engineering possible. Because the ferroelectric domain pattern can be controlled during the growth process, a well-ordered $2$-variant striped domain structure with solely the $71^\circ$ or $109^\circ$ domain walls in BiFeO$_3$ films deposited on (110) DyScO$_3$ substrates can be obtained \cite{Chu:Nl}.

On the other hand, it was demonstrated that different ferroelectric domain patterns in BiFeO$_3$ can be obtained by controlling the growth rate, leading to the stripe-like domains (at low growth rates) vs. the mosaic-like domains (at high growth rates). The former favors the $71^\circ$ domain walls while dominant $109^\circ$ domain walls are observed in the latter case. Using this technique, Martin \textit{et al.} found a direct correlation between the domain structure and exchange bias of Co$_{0.9}$Fe$_{0.1}$/BiFeO$_3$ heterostructures \cite{Martin:Nl}, i.e. the magnitude of exchange bias is linearly proportional to the length of $109^\circ$ domain walls.

In fact, the domain walls in BiFeO$_3$ offer a set of novel phenomena. For example, the domain walls can be more conductive than the domains \cite{Seidel:Nm,Farokhipoor:Prl,Chiu:Am}. Interestingly, more insulating domain walls in hexagonal YMnO$_3$ \cite{Choi:Nm} and strained perovskite SrMnO$_3$ \cite{Becher:Nn} were reported. The enhanced electrical conductivity in the ferroelectric vortex cores (the cross points of multiple domain walls) of BiFeO$_3$ films was also observed \cite{Balke:Np}. The underlying mechanism may be a ferroelectricity-related tuning of the bandgap at domain walls \cite{Lubk:Prb}, or defects trapped at the domain walls \cite{Farokhipoor:Prl}. The domain walls in  BiFeO$_3$ films can be even a key player for anomalous photovoltaic effect with high open circuit voltages \cite{Yang:Nn,Bhatnagar:Nc}. A more detailed discussion on domain structures in hexagonal $R$MnO$_3$ will be given in Sec. 5.3 below, concentrating on the nontrivial topology.

Theoretically, Daraktchiev, Catalan, and Scott once developed a Landau theory of the domain wall magnetoelectricity \cite{Daraktchiev:Prb}. In addition to the conventional free energy terms like $P^2$, $P^4$, $M^2$, $M^4$, and $P^2M^2$, a term e.g. $(\nabla P)^2$ and $(\nabla M)^2$ is added to account for the contribution from the domain walls. They predicted an enhancement in the local magnetization near a ferroelectric domain wall. Of course, their phenomenological theory is somewhat over-simplified, e.g. using the ferromagnetic order parameter instead of the antiferromagnetic order parameter. An application of this prediction to, e.g. the antiferromagnetic BiFeO$_3$, may not be assured. Even though, a neutron scattering experiment on hexagonal HoMnO$_3$ indeed found that electric field can influence the uncompensated magnetization at the antiferromagnetic domain walls \cite{Ueland:Prl}.

At the same time, a theoretical attempt on the domain wall magnetoelectricity in PbTiO$_3$/LaAlO$_3$ heterostructures was made \cite{Zhou:Sr}, where the extra carriers (from the polar-discontinuous interfaces) confined in ferroelectric PbTiO$_3$ layer can stabilize the head-to-head domain walls (or tail-to-tail walls depending on the carrier type). Consequently, electrons will stay around the head-to-head domain walls, contributing a spontaneous magnetic moment. In this case, the magnetoelectric coupling energy should be written as $(\nabla\cdot\textbf{P})\textbf{M}^2$, similar to the aforementioned field-induced magnetoelectric effect. Furthermore, phase-field method has been proven to be powerful to simulate the ferroelectric phase transitions and related domain structures \cite{Chen:Jacs}, supplementary to many experimental works. By considering more ingredients associated with the magnetoelectric couplings, the domain wall magnetoelectricity should be an interesting issue to be investigated using the phase-field method \cite{Liu:Ns,Liu:Jmps}.

\section{Magnetoelectric topology: monopole \& vortex}
Recent years, a pleasing progress aside the main streams of multiferroic discipline, has been a profit of the intimate interaction of multiferroicity and magnetoelectricity with other highly concerned disciplines such as topological physics. In fact, attention to topology in condensed matter physics has been intensively stimulated in recent years, not only due to the predictions and observations of topological insulators, topological superconductors, and Weyl semimetals etc. A key advantage of the topological materials is that some of the particular physical properties are protected by the intrinsic topology. Since topology as a mathematical and geometrical concept can be characterized by a set of discrete topological numbers instead of continuous order parameters, the affiliated physical properties are robust against weak disturbances \cite{Hasan:Rmp,Qi:Rmp,Wehling:Ap}.

In fact, topological electronic states in some specific condensed matters were first discovered to be responsible for quantum Hall effect. The flat Landau levels make a two-dimensional electronic gas to be ``insulating" since a gap is opened between the Landau levels. However, the quantized conductance persists due to the edge transport, which is robust against impurity scattering. This metallic edge state originates from the nontrivial topology of Landau bands, which can be characterized by the first Chern numbers and are originated from the Berry's phase \cite{Xiao:Rmp}. The topological insulator state is a new manifestation of topological bands, in which the spin-orbit coupling plays the similar role as the magnetic vector potential in the quantum hall effect. In addition, the (quantum) anomalous Hall effect is also a topological consequence \cite{Nagaosa:Rmp}.

In fact, the concept of topology is not strange in condensed matter physics. It is either not limited to nontrivial topological band structures. Real space topology phenomena are widely observable in condensed matters. For examples, topological ``particles"/``defects" in magnetoelectric materials have been well demonstrated, all of which can be visualized in real space instead of momentum space. Of course, the momentum space topology can also result in topological magnetoelectric surface states, which has been recently evidenced  in topological insulators.

In this section, we concentrate on those conceptual and observed topological phenomena associated with multiferroicity and magnetoelectric coupling, and in some cases those topological structures in proper ferroelectrics will also be briefly discussed. Nevertheless, a comprehensive overview may be still too early since this frontier field is still under rapid developing.

\subsection{Monopole}
\subsubsection{Magnetoelectric response of monopole}
\begin{figure}
\centering
\includegraphics[width=\textwidth]{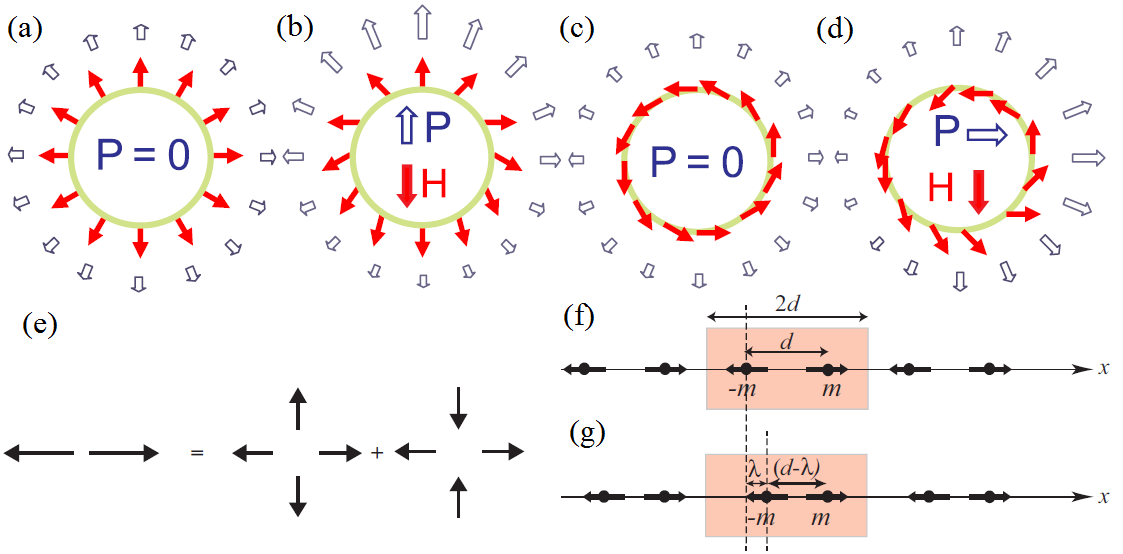}
\caption{(Colour online) Magnetic vortex and magnetoelectricity. (a) A magnetic vortex (monopole). Thin solid arrows: spins; thick open arrows: local electric dipoles. (b) A vortex distorted by a magnetic field ($\textbf{H}$). A net polarization $\textbf{P}$ in parallel or antiparallel to magnetic field $\textbf{H}$ is induced by distorting the spin texture and associated local dipole texture. (c) A magnetic vortex with a toroidal moment. (d) A net polarization perpendicular to field $\textbf{H}$ is induced for a toroidal vortex. In summary, although an ideal monopole or an ideal toroidal spin texture is non-ferroelectric, they have a linear magnetoelectric response. (a-d) Reprinted figure with permission from \href{http://dx.doi.org/10.1103/PhysRevLett.102.157203}{K. T. Delaney \textit{et al.}, Physical Review Letters, 102, p. 157203, 2009} \cite{Delaney:Prl}. Copyright \copyright (2009) by the American Physical Society. (e) A collinear antiferromagnetic spin pair can be decomposed into a monopolar contribution and a quadrupolar contribution. (f-g) A monopolization calculation for two different one-dimensional antiferromagnetic chains. The unit cell is indicated by the shaded areas. (f) A non-monopolar state, which is space-inversion symmetric with respect to each moment site. (g) A monopolar state in which the space-inversion symmetry with respect to each moment site is broken. (e-g) Reprinted figure with permission from \href{http://dx.doi.org/10.1103/PhysRevB.88.094429}{N. A. Spaldin \textit{et al.}, Physical Review B, 88, p. 094429, 2013} \cite{Spaldin:Prb}. Copyright \copyright (2013) by the American Physical Society.}
\label{monopole}
\end{figure}

According to the Maxwell equations, a magnetic monopole as a counterpart of charge is expected, both of which are the topological ``defects" of electromagnetic field. However, real magnetic monopole as an elementary particle has not yet been confirmed so far. In condensed matter physics, the concept of monopole can be used to explain some particular spin patterns. As shown in Fig.~\ref{monopole}, a monopole stands for a magnetic texture with divergent orientations of spins, while an anti-monopole denotes the texture with convergent spins orientations.

In 2009, Delaney, Mostovoy, and Spaldin demonstrated an unusually large linear magnetoelectric response of magnetic vortices coupled to a distorted polar lattice via superexchange \cite{Delaney:Prl}. A magnetic vortex can be a monopole-type or toroidal-type. An ideal vortex with uniform angles between the nearest neighbor magnetic moments does not imply a finite electric polarization due to the inter-cancellation. However, the magnetic texture of a vortex can be distorted by an external magnetic field, and thus a net electric polarization is generated due to the nonuniform angles between the nearest neighbor magnetic moments. For a monopole-type vortex, such a local polarization is parallel to magnetic field, implying a diagonal magnetoelectric tensor. In contrast, the induced electric polarization is perpendicular to the magnetic field for the toroidal-type vortex, implying an anti-symmetric magnetoelectric tensor, as found in spin glass system Ni$_x$Mn$_{1-x}$TiO$_3$ \cite{Yamaguchi:Prl}.

The reversal process is also available for these vortices. In a mechanism similar to the electromagnons in TbMnO$_3$, an external electric field can distort the ionic bonds and thus modulate the inter-spin exchanges. Consequently, the spin configuration may be distorted, yielding a large linear magnetoelectric response. Theoretically, such a magnetoelectric response for a monopole-type magnetic texture can be formalized by accounting the macroscopic monopolization, which is not limited to noncollinear spin texture. The order parameter $A$ for a monopolization can couple with polarization and magnetization ($A\textbf{P}\cdot\textbf{M}$) \cite{Spaldin:Prb}. Similarly, the toroidal-like magnetoelectric energy can be expressed as $\textbf{T}\cdot(\textbf{P}\times\textbf{M})$ where $\textbf{T}$ is a coefficient breaking both the time-reversal and space-inversion symmetries.

\subsubsection{Monopole excitation in spin ices}
\begin{figure}
\centering
\includegraphics[width=0.85\textwidth]{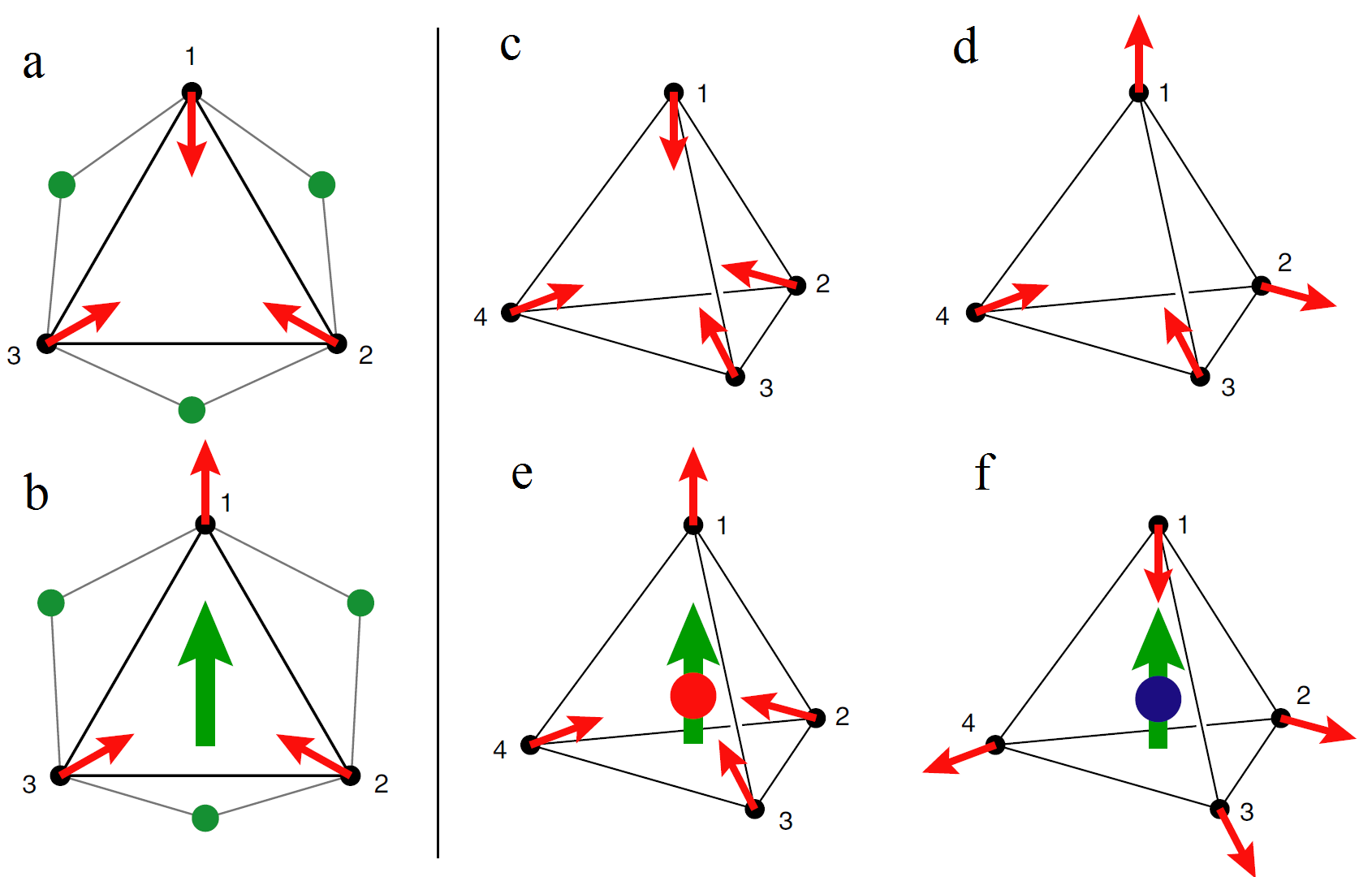}
\caption{(Colour online) The local electric dipoles (green arrows) induced by magnetostriction mechanism in triangle/tetrahedron lattices. Black circles: cations with spins (red arrows); green circles: anions. (a) The $3$-in spin configuration with symmetric oxygen locations and equivalent bonds. (b) The $2$-in-$1$-out spin configuration with asymmetric oxygen locations. An electric dipole is induced. (c) The $4$-in spin configuration. (d) The $2$-in-$2$-out spin ice configuration. (e) The $3$-in-$1$-out configuration (monopole). (f) The $1$-in-$3$-out configuration (antimonopole). The local dipoles appear in cases (e, f) where dipoles align along the same direction, but no dipole is present in cases (c, d). Reprinted by permission from Macmillian Publishers Ltd: \href{http://dx.doi.org/10.1038/ncomms1904}{D. I. Khomskii, Nature Communications, 3, pp. 904, 2012} \cite{Khomskii:Nc}. Copyright \copyright (2012).}
\label{spinice}
\end{figure}

An ideal candidate for demonstrating magnetic monopoles is the pyrochlorite family with spin ice texture, e.g. Dy$_2$Ti$_2$O$_7$ and Ho$_2$Ti$_2$O$_7$ \cite{Gardner:Rmp}. As shown in Fig.~\ref{spinice}, the corner-shared $4f$ magnetic moments form the $2$-in-$2$-out configuration for each tetrahedron at low temperature, which is called the ice-rule \cite{Bramwell:Sci}. This particular spin texture is a consequence of the strong single-axis anisotropy and frustrated exchanges. However, the spin configuration is not unique under this ice rule, leaving a macroscopic degeneracy, namely there are many possible configurations fulfilling the $2$-in-$2$-out rule. A nonzero entropy is thus expected in these spin ice pyrochlorites even at the lowest temperature.

The $2$-in-$2$-out spin texture can be thermally excited to the $3$-in-$1$-out plus $1$-in-$3$-out pair, and this pair may be spatially separated with nearly no energy cost. The higher level excitation is the $4$-in and $4$-out pair. All these excitations own divergent (or convergent) orientations of magnetic moments, which are accessible using the monopole (or anti-monopole) language \cite{Castelnovo:Nat,Jaubert:Np,Morris:Sci}.

In 2012, Khomskii proposed that a monopole may carry a magnetic charge and thus a finite electric dipole, just like an electron with an electric charge and a magnetic dipole \cite{Khomskii:Nc}. The underlying mechanism is similar to the theory proposed by Delaney, Mostovoy, and Spaldin \cite{Delaney:Prl}. The $4$-in/$4$-out/$2$-in-$2$-out configurations are highly symmetric and do not generate a net dipole, but the $3$-in-$1$-out and $1$-in-$3$-out configurations can do via the exchange striction. The magnitude of charge transfer into/out a tetrahedron (i.e. the local dipole) can be expressed as \cite{Khomskii:Nc}:
\begin{equation}
n\sim\textbf{S}_1\cdot(\textbf{S}_2+\textbf{S}_3+\textbf{S}_4)-(\textbf{S}_2\cdot\textbf{S}_3+\textbf{S}_2\cdot\textbf{S}_4+\textbf{S}_3\cdot\textbf{S}_4),
\label{emonopole}
\end{equation}
which aligns along the $1$-in (or $1$-out) direction, as shown in Fig.~\ref{spinice}. According to this equation, the $180^\circ$ flops of all the four spins will not change the dipole, although the local net magnetization is flopped ($3$-in-$1$-out $\leftrightarrow$ $1$-in-$3$-out). In this sense, although the electric dipole is magnetism-driven, it may be robust against external magnetic field along the [111] direction. At a finite temperature (not too high) or under a magnetic field, the $3$-in-$1$-out/$1$-in-$3$-out monopoles/anti-monopoles always exist, carrying local electric dipoles. These dipoles can be polarized by an electric field, and a magnetic field along various orientations except the [111]-axis surely can modulate these dipoles, i.e. magnetoelectricity.

Experimental evidences of the magnetoelectricity in spin ice systems include the dielectric anomalies in Dy$_2$Ti$_2$O$_7$ \cite{Saito:Prb} and pyroelectricity in Ho$_2$Ti$_2$O$_7$ and Dy$_2$Ti$_2$O$_7$ \cite{Dong:Apl10,Lin:Njp}. The measured pyroelectric polarizations emerge at several systems but further confirmation is needed. Conversely, the multiferroicity offers a promising mechanism to control topological magnetism in spin ice pyrochlorites \cite{Jaubert:Prb,Lin:Njp}. A major issue is the electric dipole ordering in these spin ice systems, which has not yet obtained sufficient evidences.

\subsection{Skyrmion}
\begin{figure}
\centering
\includegraphics[width=0.9\textwidth]{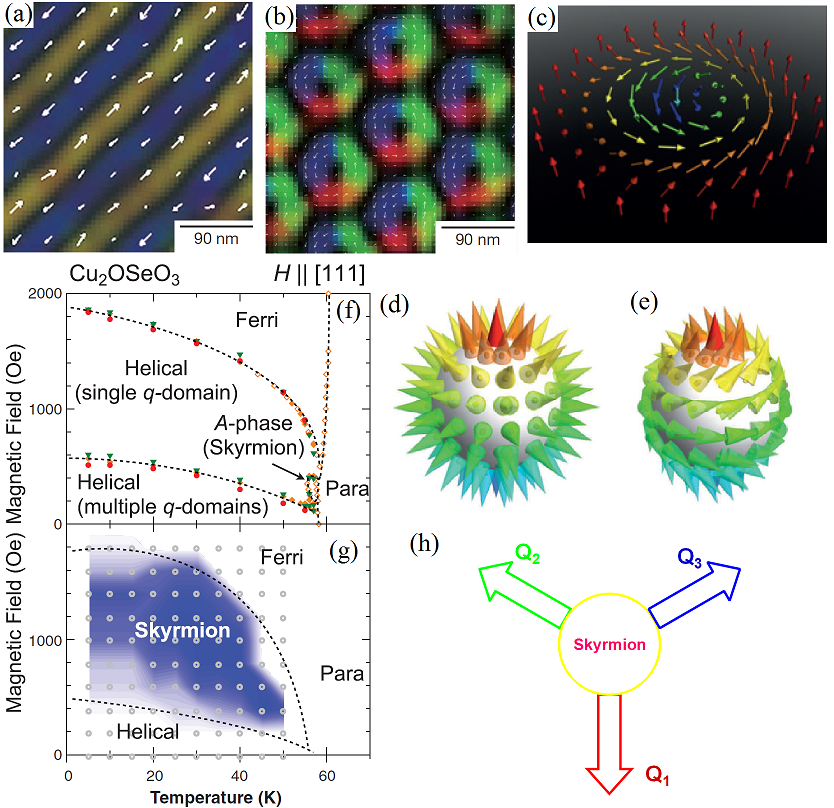}
\caption{(Colour online) Skyrmion and magnetoelectricity. (a-b) The observed spin texture in a helical magnet (Fe$_{0.5}$Co$_{0.5}$Si). (a) The helical structure under zero magnetic field; (b) The skyrmions under a weak magnetic field. The colour map and arrows denote the local magnetization directions. (c) A schematic of the spin texture in a skyrmion. (a-c) Reprinted by permission from Macmillian Publishers Ltd: \href{http://dx.doi.org/10.1038/nature09124}{X. Z. Yu, \textit{et al.}, Nature, 465, pp. 901-904, 2010} \cite{Yu:Nat}. Copyright \copyright (2010). (d-e) The spin texture of a skyrmion shown in (c) can be mapped to a hairy sphere (d) or a vortex (e). They are topologically equivalent. (d-e) Reprinted by permission from Macmillian Publishers Ltd: \href{http://dx.doi.org/10.1038/nphys2081}{C. Pfleiderer, Nature Physics, 7, pp. 673-674, 2011} \cite{Pfleiderer:Np}. Copyright \copyright (2011). (f-g) The magnetic phase diagram of Cu$_2$OSeO$_3$ in (f) bulk form and (g) thin-film form. Similar to many helical magnets, the helical spin structure in Cu$_2$OSeO$_3$ transits to the skyrmion phase upon a weak magnetic field. However, different from most other metallic helical magnets, Cu$_2$OSeO$_3$ is an insulator which can have a microscopic ferroelectric polarization. (f-g) From \href{http://dx.doi.org/10.1126/science.1214143}{S. Seki \textit{et al.}, Science, 336, pp. 198-201,  2012} \cite{Seki:Sci}. Reprinted with permission from The American Association for the Advancement of Science. (h) A two-dimensional skyrmion spin texture which can be constructed using three spirals with $120^\circ$ angles between two wavevectors.}
\label{skyrmion}
\end{figure}

Skyrmion was first proposed as an elementary particle. In 1980s, Bogdanov and Yablonskii started from a theoretical magnetic model and proposed a picture of magnetic skyrmion \cite{Bogdanov:Jetp}. Skyrmion is a special twist of magnetic moments with a non-coplanar pattern (Fig.~\ref{skyrmion}(a-c)), where the directions of all the spins in a skyrmion can be mapped onto a sphere surface (Fig.~\ref{skyrmion}(d-e)). An intuition view of a two-dimensional skyrmion is that the outer spins point up (down) and the most inner spins point down (up), while an outer-inner transition is via the continuous rotation of spins in the three-dimensional space, as sketched in Fig.~\ref{skyrmion}(c). The topological number to characterize a skyrmion in a two-dimensional space can be defined by:
\begin{equation}
N_{\rm sk}=\frac{1}{4\pi}\int\int\textbf{M}\cdot(\frac{\partial\textbf{M}}{\partial x}\times\frac{\partial \textbf{M}}{\partial y})d^2\textbf{r}
\label{sk1}
\end{equation}
where vector $\textbf{M}$ denotes the magnetic orientation, $\textbf{r}(x,y)$ is the space vector. The number $N_{\rm sk}$ counts how many skyrmions (or anti-skyrmions) in a given area. One skyrmion corresponds to a topological number $+1$, while one anti-skyrmion gives $-1$. Skyrmions in solids can be observed using neutron scattering \cite{Muhlbauer:Sci}, Lorentz transmission electron microscopy \cite{Yu:Nat}, and spin-resolved scanning tunneling microscopy \cite{Heinze:Np,Romming:Sci} etc. More detailed description on skyrmions can be found in recent reviews by Nagaosa \textit{et al.} \cite{Nagaosa:Ptrsa,Nagaosa:Nn}.

There is a common ingredient of physics between skyrmion and some type-II multiferroics: noncollinear spin textures with chirality. In fact, for many real skyrmion materials, the skyrmion order evolves from a spiral spin order upon a moderate magnetic field. Furthermore, the noncoplanar spin texture of a skyrmion can be decomposed into three spirals (Fig.~\ref{skyrmion}(h)) \cite{Muhlbauer:Sci,Nagaosa:Nn}:
\begin{equation}
\textbf{M}\approx\textbf{M}_0+\sum_{i=1}^3\textbf{M}_{\textbf{Q}_i}(\textbf{r}),
\label{sk2}
\end{equation}
where $\textbf{M}_0$ is the uniform background. The spiral components can be written as:
\begin{equation}
\textbf{M}_{\textbf{Q}_i}(\textbf{r})=\textbf{M}_{\textbf{Q}_i}^0[\cos(\textbf{Q}_i\cdot\textbf{r}+\phi_i)+\sin(\textbf{Q}_i\cdot\textbf{r}+\phi_i)],
\label{sk3}
\end{equation}
where $\sum_{i=1}^3\textbf{Q}_i=0$.

In addition, some multiferroic oxides with frustrated exchanges, e.g. Cu$_2$OSeO$_3$, do favor a skyrmion phase in a finite temperature region (see Fig.~\ref{skyrmion}(f-g) for its phase diagrams) \cite{Adams:Prl,Seki:Sci}. Nevertheless, different from most skyrmion materials which are metallic, Cu$_2$OSeO$_3$ is an insulator and thus can have a spontaneous ferroelectric polarization, not driven by the long wavelength noncollinear spin order but the single-spin-site contribution due to the spin-orbit coupling \cite{Yang:Prl12}.

\begin{figure}
\centering
\includegraphics[width=\textwidth]{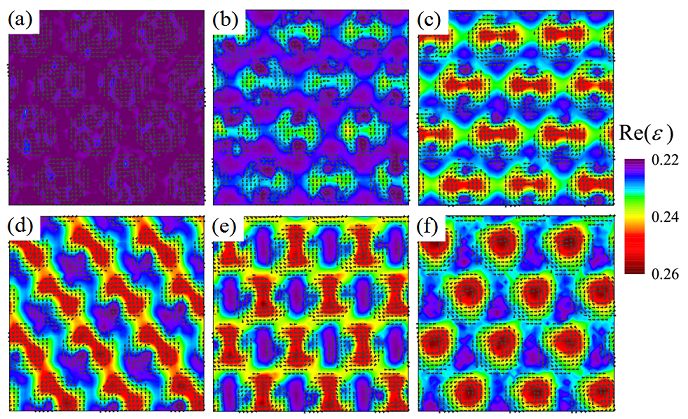}
\caption{(Colour online) Simulated snapshot dielectric permittivity (real part $R(\omega)$) patterns of a two-dimensional skyrmion lattice (in the $x-y$ plane). The AC electric field is along (a-c) the $y$-axis, (d) the $x-y$-diagonal-axis, (e) the $x$-axis, and (f) the $z$-axis. The AC frequency is respectively (a) $30$, (b) $2$, and (c-f) $0.2$, in unit of $\tau^{-1}$ which is determined by magnetic coefficients. Reprinted by permission from Macmillian Publishers Ltd: \href{http://dx.doi.org/10.1038/srep08318}{P. Chu \textit{et al.}, Scientific Reports, 5, p. 8318,  2015} \cite{Chu:Sr}. Copyright \copyright (2015).}
\label{skyrmion2}
\end{figure}

The size of a skyrmion quasi-particle can vary from several nanometers to hundreds nanometers. For most cases, the size depends on the ratio between the Dzyaloshinskii-Moriya interaction and the normal exchanges. The atomic size skyrmion was identified in Fe monolayer deposited on (111) Ir, which is driven by a mechanism beyond the conventional physics based on the weak Dzyaloshinskii-Moriya interaction \cite{Heinze:Np}. Because of such a small skyrmion size, the topological skyrmion number may not be exactly an integer since the spatial modulation of spin moments cannot be quasi-continuous but discrete. Even though, the strong noncollinear (in fact noncoplanar) spin texture in this case will be an very interesting playground to host magnetoelectric phenomena. For example, the noncollinear spin pairs between the nearest-neighbor sites will generate a local electric dipole according to the spin current model. Based on this idea, Chu \textit{et al.} performed a simulation to show that the skyrmion lattice can be visualized using a microwave-frequency dielectric detection technique (see Fig.~\ref{skyrmion2}), which may provide an alternative method to monitor skyrmions in a noncollinear multiferroic \cite{Chu:Sr}. In fact, experimentally, Okamura \textit{et al.} observed the microwave magnetoelectric effect via the skyrmion resonance modes with oscillating magnetic field at a frequency of $1-2$ GHz in Cu$_2$OSeO$_3$ \cite{Okamura:Nc}. The nonreciprocal directional dichroism was also revealed at the resonant mode, i.e. oppositely propagating microwaves exhibit different absorptions, similar to the electromagnon excitations driven by the spin-orbit coupling in Eu$_{0.55}$Y$_{0.45}$MnO$_3$. Furthermore, White \textit{et al.} showed that electric field could control the skyrmion lattice rotations in Cu$_2$OSeO$_3$ due to an electric-field-induced skyrmion distortion \cite{White:Prl}. Thus, the intrinsic magnetoelectric coupling in a skyrmion lattice can provide an alternative manipulation concept rather than the traditional spin-transfer torques.

\subsection{Ferroelectric domain vortex in hexagonal $R$MnO$_3$}
\begin{figure}
\centering
\includegraphics[width=\textwidth]{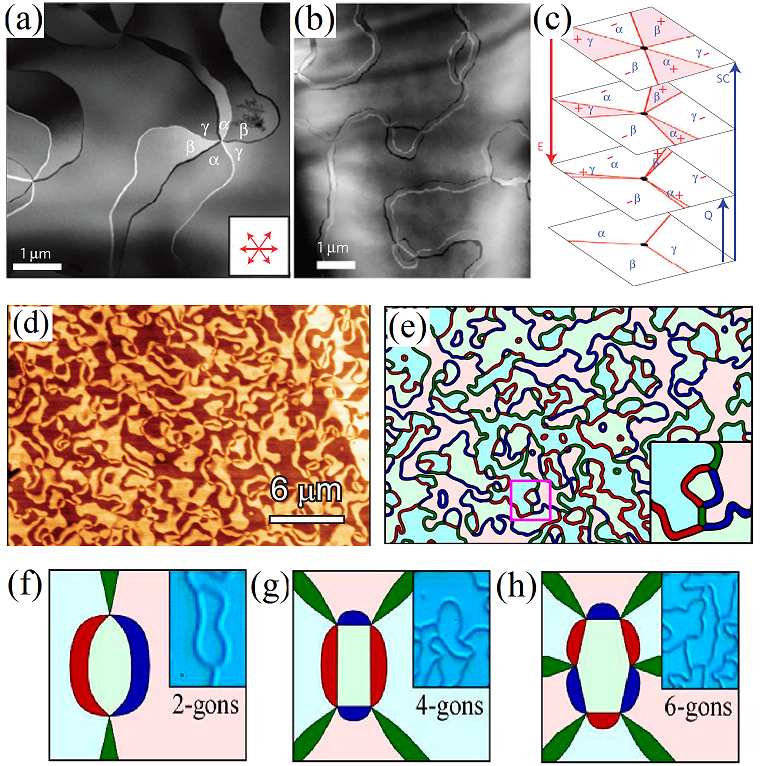}
\caption{(Colour online) Vortex domain structure in hexagonal $R$MnO$_3$. (a) A dark-field transmission electron microscopic image ($R$=Y). The six structural antiphase domains emerge from one central point (so-called vortex core), ordered in the $\alpha^-\beta^-\gamma^-\alpha^-\beta^-\gamma^-$ sequence. (b) A transmission electron microscopic image of an almost fully electric-poled region. The three domains of negative ferroelectric polarizations become seriously shrunk into narrow stripes but cannot be completely melted away due to the robustness of the domain walls, demonstrating that the vortex core is topologically protected. (c) A schematic of the evolution of a cloverleaf ferroelectric domain driven by a downward electric field $E$. Arrow $E$ denotes the electric field, arrow $Q$ refers to a quenching and arrow $SC$ refers to a slow cooling, both from high temperature (above the ferroelectric Curie temperature). (a-c) Reprinted by permission from Macmillian Publishers Ltd: \href{http://dx.doi.org/10.1038/NMAT3786}{T. Choi, \textit{et al.}, Nature Materials, 9, pp. 253-258, 2010} \cite{Choi:Nm}. Copyright \copyright (2010). (d) A piezoelectric force microscope image of ErMnO$_3$. (e-h) The representation of domain structure using the famous four-color theorem. The $\alpha$, $\beta$, and $\gamma$ regions are colored using three colors: red, blue, and green. The two directions of polarization up (face) and polarization down (edge) are distinguished by normal colors and light colors. (d) Reprinted figure with permission from \href{http://dx.doi.org/10.1103/PhysRevLett.108.167603}{S. C. Chae \textit{et al.}, Physical Review Letters, 108, p. 167603, 2012} \cite{Chae:Prl}. Copyright \copyright (2012) by the American Physical Society. (e) All adjacent edges or faces have different colors. Inset: an enlarged view of the pink rectangle area. (f-h) The schematic two-, four-, and six-gons with one, two, and three vortex-antivortex pairs, respectively. Insets: the corresponding real examples imaged using an optical microscope. (e-h) Reprinted figure with permission from \href{http://dx.doi.org/10.1073/pnas.1011380107}{S. C. Chae \textit{et al.}, Proceedings of the National Academy of Sciences of the United States of America, 107, pp. 21366-21370, 2010} \cite{Chae:Pnas} Copyright \copyright (2010) by the National Academy of Sciences.}
\label{votex}
\end{figure}

Besides skyrmion as a real space topological particle (defect), other topological defects are possible in ferroelectrics. A recent hot topic is ferroelectric domain vortex as first observed and then well investigated in hexagonal $R$MnO$_3$ ($R$=Y, Ho, Lu, ......). In 2010, Choi \textit{et al.} reported an interlocked ferroelectric and structural antiphase domain structure in multiferroic YMnO$_3$ \cite{Choi:Nm}. As discussed in Sec. 3.3.1, the structural trimerization (so-called $K_3$ mode distortion) of Mn triangles gives rise to three types of ferroelectric/antiphase domains: $\alpha$, $\beta$, $\gamma$, which break the $60^\circ$ rotation symmetry of hexagonal lattice. The ferroelectric polarization, which is also a result of trimerization, comes from the $\uparrow-\uparrow-\downarrow$ shifts of the $R$ spin trimers and aligns along the $c$-axis ($+$ or $-$). This configuration breaks an additional space inversion symmetry. Thus, the ferroelectric state belongs to the the $\mathbb{Z}_3\times\mathbb{Z}_2$ group, resulting in the cloverleaf pattern of six domains emerging from one point, as shown in Fig.~\ref{votex}.

\begin{figure}
\centering
\includegraphics[width=\textwidth]{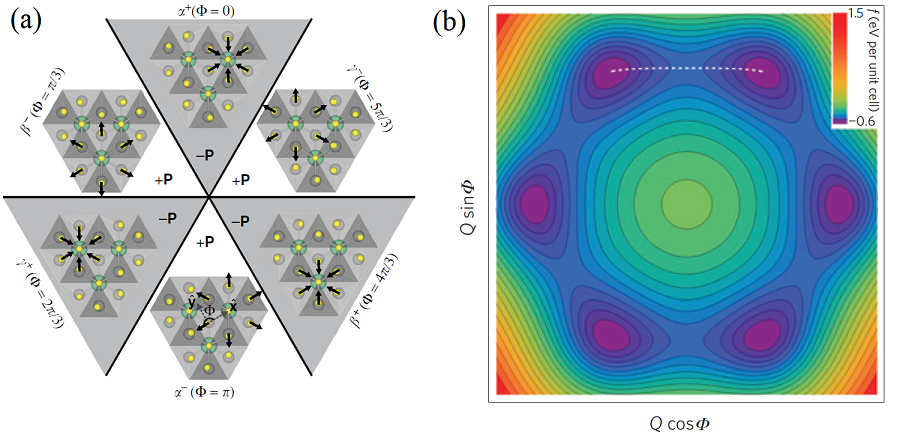}
\caption{(Colour online) Structural trimerization \& energy landscape of topological domain structure. (a) A schematic of a structural trimer domain. The six domains are distinguished by a phase $\Phi$ of the $K_3$ mode distortion. An anticlockwise rotation around the vortex core gives rise to a $\pi/3$ phase step across each domain wall. $\textbf{P}$: local polarization along the $c$-axis (perpendicular to the plane). Arrows indicate the direction of partial apical oxygen motion of the trimer distortions. Reprinted by permission from Macmillian Publishers Ltd: \href{http://dx.doi.org/10.1038/ncomms3998}{H. Das \textit{et al.}, Nature Communications, 5, p. 2998, 2014} \cite{Das:Nc}. Copyright \copyright (2014). (b) A contour plot of the free energy for uniformly trimerized states as a function of trimerization (amplitude $Q$ and phase $\Phi$). The six lowest energy positions ($\Phi=0$, $\pm\pi/3$, $\pm2\pi/3$, $\pi$) correspond to the six domains. The lowest-energy structural domain wall is marked using the white dashed line, connecting two neighbouring energy minima with a phase shift $\Delta\Phi\pm\pi/3$. Due to the Mexican hat shape-like potential well, any domain wall beyond $\Delta\Phi\pm\pi/3$ is much higher in energy and thus unstable against the spontaneous decomposition into multi-domain walls. Reprinted by permission from Macmillian Publishers Ltd: \href{http://dx.doi.org/10.1038/NMAT3786}{S. Artyukhin, \textit{et al.}, Nature Materials, 13, pp. 42-49, 2014} \cite{Artyukhin:Nm}. Copyright \copyright (2014).}
\label{votex2}
\end{figure}

\begin{figure}
\centering
\includegraphics[width=\textwidth]{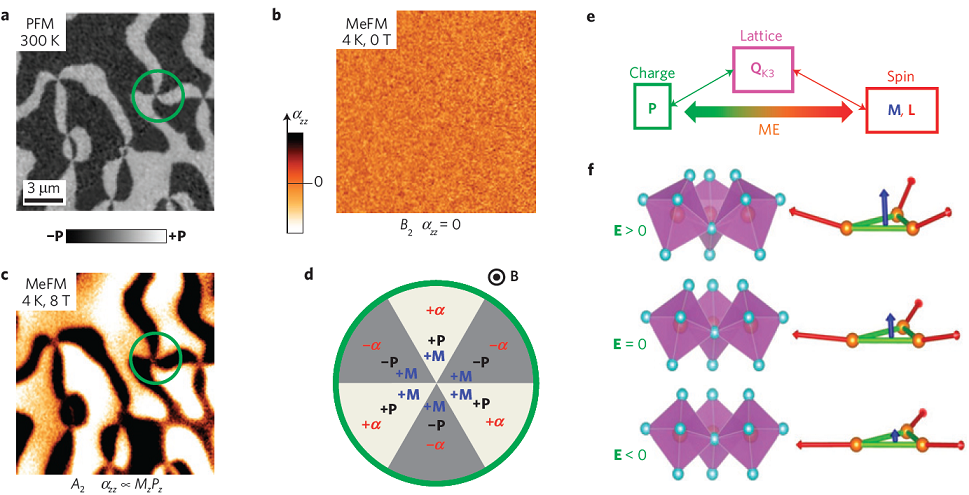}
\caption{(Colour online) Magnetoelectric domain structure of a hexagonal ErMnO$_3$ single crystal. (a) A room-temperature piezo-response force microscopy image. The white and dark colors denote the up and down ferroelectric polarizations, respectively. (b-c) The low-temperature ($4$ K) magnetoelectric force microscopy images: (b) at zero magnetic field and (c) at a magnetic field of $8$ T. The images in (a-c) were taken at the same location on the (001) surface. (d) A cartoon listing the magnetoelectric coefficient ($\alpha$) of the $A_2$ phase in different ferroelectric domains. It is clear that the sign of $\alpha$ changes with polarization $P_z$ although the canted magnetic moment $M$ is positive in all domains. (e) A cartoon showing the effective magnetoelectric coupling through structural trimerization. (f) Cartoons showing a variation of the buckling of MnO$_5$ polyhedra (trimer mode $Q_{K_3}$) induced by electric fields via polarization component $P_z$, resulting in a variation of the canted moment $M_z$ of Mn spins. Reprinted by permission from Macmillian Publishers Ltd: \href{http://dx.doi.org/10.1038/nmat3813}{Y. Geng \textit{et al.}, Nature Materials, 13, pp. 163-167, 2014} \cite{Geng:Nm}. Copyright \copyright (2014).}
\label{ErMnO3}
\end{figure}

For a sample grown below the ferroelectric Curie temperature, the trimerization induced ferroelectric domains can be in stripe-like pattern \cite{Chae:Prl}. In contrast, if the sample is grown above the ferroelectric Curie temperature, the vortex patterns due to the structural trimerization appear above the ferroelectric Curie temperature. When temperature passes across the ferroelectric Curie point, these vortex patterns interlock the ferroelectric domains, leading to a frozen domain structure during the quenching process. Around the center point of six domains, the domain circling sequence restricted by very high energy barriers is always $\alpha^+-\beta^--\gamma^+-\alpha^--\beta^+-\gamma^-$ or $\alpha^+-\gamma^--\beta^+-\alpha^--\gamma^+-\beta^-$, corresponding to a vortex or an anti-vortex, respectively. The cloverleaf domains, or namely the vortex structure, is robust against thermal fluctuations over high temperature.

Theoretically, the complex topological ferroelectric domain structures can be understood in the framework of Landau theory with the help of density functional calculations \cite{Artyukhin:Nm}. The clamping between the polarization domains and structural antiphase domains can be understood by a nonlinear coupling between structural trimerization (characterized by the amplitude $Q_{K_3}$ and phase $\Phi$) and polarization along the $z$-axis ($P_z$):
\begin{equation}
f_{trimer}\sim-Q_{K_3}^3P_z\cos(3\Phi)
\label{tri}
\end{equation}
where $f_{trimer}$ denotes the free energy term. Given a distortion mode, a spontaneous ferroelectric polarization is generated. To lower the energy, the preferred $\Phi$ values are: $0$, $\pm\pi/3$, $\pm2\pi/3$, and $\pi$. In other words, six structural antiphase domains will be formed if a trimerization appears ($Q_{K_3}>0$), giving six $\Phi$ values, as sketched in Fig.~\ref{votex2}(a). For $\Phi=0$ and $\pm\pi/3$, the polarization should align upward, and it would align downward for $\Phi=\pm2\pi/3$ and $\pi$.

Considering the other energy terms and proper coefficients extracted from the density functional calculations, the energy landscape in the ($Q_{K_3}$, $\Phi$) plane is indeed a Mexican hat shape like with six minima (six preferred $\Phi$ values), as sketched in Fig.~\ref{votex2}(b). The direct paths connecting any two nearest-neighbor minima give the lowest energy barriers comparing with other paths connecting the non-nearest-neighbor minima. Therefore, parameter $\Phi$ across any domain wall must be stepped by $\pm\pi/3$, while all domain walls with other steps ($\pm2\pi/3$ and $\pi$) are much higher in energy and will decay to multiple domain walls with a $\pm\pi/3$ step. In this sense, once a cloverleaf vortex emerges, its six domain walls cannot merge to change the sequence of domains. Of course, these domain walls can terminate at the surface or an anti-vortex, leading to the vortex-antivortex pairs connected by domain wall strings \cite{Choi:Nm}.

Experimentally, an electric field polarizes a ferroelectric polarization via domain wall shifting (expanding the aligned domains and shrinking the opposite domains) \cite{Choi:Nm}. Nevertheless, a vortex/antivortex core cannot be annihilated individually until a dielectric breakdown, as shown in Fig.~\ref{votex}(b-c) \cite{Choi:Nm,Han:Am}. Even though, a shear strain can impose a Magnus-type force which pulls vortices and antivortices in opposite directions and unfolds them into a topological stripe domain state \cite{Wang:Prl14}, analogous to the current-driven dynamics of vortices in superconductors and superfluids. In addition, the annihilation between a vortex and an anti-vortex is possible \cite{Chae:Pnas}, analogous to the annihilation between an elementary particle/antiparticle.

In hexagonal ErMnO$_3$ where Er$^{3+}$ is magnetic, magnetic signals from these domain walls were observed at low temperature \cite{Geng:Nl}. Although these signals directly originate from the Er$^{3+}$ spins, they depend on Mn antiferromagnetic structure, implying that the Mn antiferromagnetic domains are also clamped to the structural-antiphase/ferroelectric domains, a manifestation of the domain wall magnetoelectricity.

Besides this domain wall magnetoelectricity, the bulk magnetoelectric effect in hexagonal $R$MnO$_3$/$R$FeO$_3$ mediated by the Dzyaloshinskii-Moriya interaction may be possible, if the Mn/Fe magnetic order take the so-called $A_2$ pattern (see Fig.~\ref{sRMO}) \cite{Das:Nc}. However, for many hexagonal $R$MnO$_3$ materials, e.g. ErMnO$_3$, the magnetic ground state is the $B_2$ pattern (see Fig.~\ref{sRMO}), which forbids any bulk magnetoelectricity. Even though, a magnetic field can drive the transition of magnetic state to the $A_2$ pattern. In this case, the magnetoelectric domains (not the magnetic domains but defined by the magnetoelectric coefficient) can be visualized, as demonstrated in ErMnO$_3$, where the magnetoelectric domains are clamped to the structural-antiphase domains mentioned above (see Fig.~\ref{ErMnO3} for more details) \cite{Geng:Nm}.

Finally, it is noted that ferroelectric vortices have also been observed and simulated in BiFeO$_3$ films and heterostructures besides $R$MnO$_3$ \cite{Balke:Np,Nelson:Nl}.

\subsection{Magnetoelectricity of topological surface state}
In condensed matter physics, a phase of matter can be defined by its response to some stimuli. The electromagnetic Lagrangian of a topological insulator, which has a form analogous to the theory of axion electrodynamics, contains a scalar product of electric field and magnetic field, $\textbf{E}\cdot\textbf{B}$ \cite{Wilczek:Prl,Moore:Nat}. This linear magnetoelectric term can generate a number of novel physical phenomena, e.g. monopole-like behavior. Regarding the magnetoelectric effect, an electric field can induce a magnetic dipole and vice versa. This magnetoelectric effect purely originates from the orbital motion of electrons, which can be highly sensitive and reproducible without fatigue.

Such an intrinsic magnetoelectric effect can be intuitively visualized using a theoretical scenario. Although the $\mathbb{Z}_2$ topological insulators are in principle nonmagnetic, the surface state hides the intrinsic noncollinear spin texture in moment space, e.g. the spin rotates around the surface Dirac cone. This coupling is glued by the spin-orbit coupling and no net magnetic moment is available by integrating the whole surface state. Nevertheless, an electric field will shift the Dirac surface state to one momentum direction a little bit, resulting in a change of the Fermi surface. Since this surface state is spin-momentum-bound, a net magnetization will be generated, implying an intrinsic linear magnetoelectric response \cite{Zhu:Prl,JWang:Prl15}.

In addition, Qi \textit{et al.} predicted a mirror magnetic monopole which can be generated in a topological insulator by a charge near the surface \cite{Qi:Sci}. Experimentally, the magnetic order and magnetoresistance of a magnetic ion doped topological insulator (Cr$_{0.15}$(Bi$_{0.1}$Sb$_{0.9}$)$_{1.85}$Te$_3$) can be tuned using electrical method \cite{Zhang:Nc}.

\section{Summary \& Perspective}
We summarized the great progress on multiferroicity and magnetoelectricity in the past decade by presenting not only a series of new physical concepts and experimental findings but also discovery of a large number of multiferroic materials. This progress clearly shows that multiferroicity and magnetoelectricity is one of the hot topics in condensed matter physics. Even though the progress by the multiferroic research community is profound, it seems that more questions and challenges are continuously emerging, which on the other hand act as the driving force for future research. Therefore, a perspective rather than a conclusive summary may be more appropriate.

\subsection{Relook on multiferroicity \& magnetoelectricity}
In this article, almost each plane of the polyhedron of magnetoelectricity and multiferroicity have been reviewed. Now it is time to give a summary on this divergent discipline, based on the fundamental physics.

First, we give an overlook on multiferroic materials. On one hand, since in most cases the origin of magnetic moments is solely from spins, multiferroics are basically classified according to the different sources for electric dipoles, which has been widely adopted \cite{Khomskii:Phy,Cheong:Nm}. On the other hand, it is widely recognized that the multiple degrees of freedom (spin, lattice, orbital, and charge) are all active in these multiferroic systems. Thus, one may categorize all the multiferroics according to the driving force for ferroelectricity, as shown in Table~\ref{table1}, where only a few representative materials are listed as examples in each category. It should be noted that this classification is not rigorous since the ferroelectricity in real materials may have multiple contributions, as introduced in Sec. 3.5. It should be noted that there are still debates on the ferroelectricity driven by charge-ordering. It is curious that no multiferroics whose ferroelectricity is driven by orbital degree of freedom have been demonstrated. In some multiferroics, the orbital degree of freedom does play a crucial role but not the leading force to break the space inversion symmetry. This ``deficiency" certainly deserves further investigations, and those systems of strong orbital ordering should be concerned.

\begin{sidewaystable}
\caption{A brief summary of multiferroics classified by the main driving forces involved. ``ME link" denote the medium linking magnetism and electricity. In each category, only four typical materials are listed. There are still some debates on the charge ordering driven ferroelectricity.}
\centering
\begin{tabular*}{0.72\textwidth}{|*{6}{c|}}
\hline
\multirow{2}{*}{\textbf{Origin}} & \multirow{2}{*}{\textbf{Implementation}} & \multirow{2}{*}{\textbf{Category}} & \multirow{2}{*}{\textbf{Materials}} & \multirow{2}{*}{\textbf{ME link}} & \multirow{2}{*}{\textbf{Ferroelectricity}}\\
    &  &  &  & & \\
\hline
\multirow{8}{*}{Lattice} & \multirow{4}{*}{polar distortion} & \multirow{4}{*}{proper} & BiFeO$_3$ &  &\\
  &  &  & strained EuTiO$_3$  & spin-lattice & mostly high $T_{\rm C}$\\
  &  &  & Ba$_{1-x}$Sr$_x$MnO$_3$ & spin-orbit & moderate-large $P$\\
    &  &  & $M$TiO$_3$  & & \\
\cline{2-6}
& & \multirow{4}{*}{geometric} & $R_3M_2$O$_7$ & &\\
& non-polar &  & $h$-$RM$O$_3$ & spin-lattice  & high $T_{\rm C}$\\
& distortions &  & Ba$M$F$_4$ & spin-orbit & moderate $P$\\
 &  &  & $AB$O$_3$/$A'B'$O$_3$ & &\\
\hline
\multirow{12}{*}{Spin} & \multirow{4}{*}{$\uparrow\uparrow\downarrow\downarrow$-like} & \multirow{12}{*}{magnetic} & $o$-HoMnO$_3$ & &\\
 &  &  & Ca$_3$CoMnO$_6$ & spin-lattice & mostly low $T_{\rm C}$\\
  &  &  & $R$Mn$_2$O$_5$ & spin-orbit & small-moderate $P$ \\
    &  &  & GdFeO$_3$ & & \\
\cline{2-2} \cline{4-6}
  & \multirow{4}{*}{cycloid/conical} &  & TbMnO$_3$ & \multirow{8}{*}{spin-orbit} & \\
  &  &  & hexaferrites &  & \\
  &  &  & CuO &  & \\
  &  &  & CoCr$_2$O$_4$ & & mostly low $T_{\rm C}$ \\
\cline{2-2} \cline{4-4}
 & \multirow{4}{*}{misc} & & LaMn$_3$Cr$_4$O$_{12}$ &  & small $P$\\
 &  &  & CuFeO$_2$ & & \\
 &  &  & Ba$_2$CoGe$_2$O$_7$ & & \\
  &  &  & Ba$_3$NiNb$_2$O$_9$  & & \\
\hline
\multirow{4}{*}{Charge} & \multirow{4}{*}{ordering} & \multirow{4}{*}{electronic} & LuFe$_2$O$_4$ & \multirow{4}{*}{spin-charge} &\\
 &  &  & Fe$_3$O$_4$ & & moderate $T_{\rm C}$\\
 &  &  & Pa$_{1-x}$Ca$_x$MnO$_3$ & & moderate $P$\\
 &  &  & LiFe$_2$F$_6$  & & \\
\hline
Orbital & unknown & unknown & unknown & unknown & unknown\\
\hline
\end{tabular*}
\label{table1}
\end{sidewaystable}

Second, besides the single phase multiferroics, magnetoelectricity may be available in non-multiferroic systems. The involved physics can be even more complicated. In accordance with the style of Table~\ref{table1}, we present a rough classification according to the main driving forces, as shown in Table~\ref{table2}. It is seen that the manifestation of magnetoelectricity can be quite divergent due to the multiple physical couplings among various degrees of freedom. The orbital, e.g. the ``circular''-like movement of electrons, can contribute to novel magnetoelectric effects, as demonstrated by the surface state of topological insulators.

\begin{sidewaystable}
\caption{A brief summary of magnetoelectric effects which are classified by the main driving forces involved. ``ME link" denote the medium linking magnetism and electricity. In each category, only a few of representative materials are listed. The manifestation denotes the related magnetoelectric phenomena. The orbital here denotes a general ``circular''-like movement of electrons but not limited to the atomic orbitals.}
\centering
\small
\begin{tabular*}{0.8\textwidth}{|*{5}{c|}}
\hline
\multirow{2}{*}{\textbf{Driving forces}} & \multirow{2}{*}{\textbf{Implementation}} & \multirow{2}{*}{\textbf{Materials}} & \multirow{2}{*}{\textbf{ME link}} & \multirow{2}{*}{\textbf{Manifestation}}\\
    &  &  &  &\\
\hline
\multirow{4}{*}{Lattice} & strain via &  composites, & spin-lattice & pizeoelectric + \\
  & interface & heterostructures & spin-orbit & magnetostrictive \\
\cline{2-3} \cline{5-5}
  & structural & multiferroics,  & lattice-field & spin canting, \\
  & distortion & magnetic dielectrics & spin-field & electromagnon \\
\hline
\multirow{6}{*}{Spin} & & multiferroics  & spin-orbit & polarization rotation\\
  & axis/plane & with noncollinear  & spin-field & /suppression by magnetic\\
  &  & spin orders &  &  field, electronmagnon\\
\cline{2-5}
  & & multiferroics with & spin-lattice & polarization suppression\\
  & ordering & $\uparrow\uparrow\downarrow\downarrow$-like & spin-field &  by magnetic \\
    &  & orders & lattice-field &  field, electromagnon  \\
\hline
\multirow{5}{*}{Charge} & ordering & multiferroics & charge-lattice & \\
  &  & with charge ordering, & spin-charge & \\
\cline{2-5}
  &  & heterostructures, & field effect & electric tuning \\
  & carrier & domain wall, & spin-orbit & of magnetization/magnetic \\
  &  & polar surface & lattice-field & axis/magnetic transition \\
\hline
\multirow{4}{*}{Orbital} & & & spin-orbit & charge/electric  \\
 & topology of & topological-band & spin-field & field tuning of \\
 & bands & materials & orbit-field &  magnetic dipole/ \\
 &  &  & & magnetic order \\
\hline
\end{tabular*}
\label{table2}
\end{sidewaystable}

\begin{figure}
\centering
\includegraphics[width=0.75\textwidth]{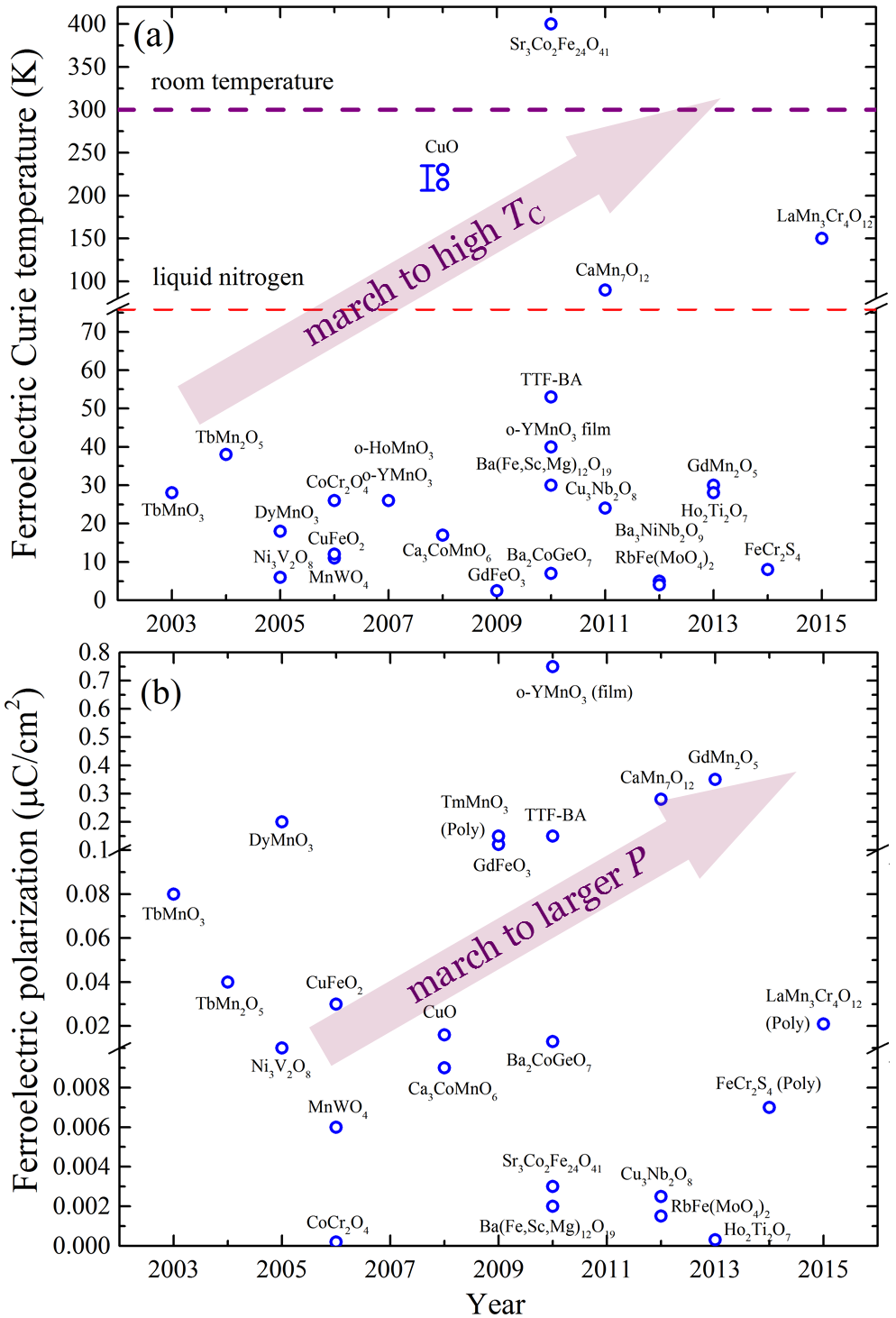}
\caption{(Colour online) A brief summary of ferroelectricity [(a) Curie temperature and (b) polarization] of partial typical type-II multiferroics. Here only the experimental confirmed materials are shown. The values of polarization are mostly measured by pyroelectric method for single crystals. For some polycrystalline samples, the abbreviation ``Poly" is indicated.}
\label{TcP}
\end{figure}

As a supplement to Tables~\ref{table1} \&~\ref{table2}, we also present a chronological portrait of the type-II multiferroics in terms of the ferroelectric Curie temperature and polarization magnitude, as shown in Fig.~\ref{TcP}, where representative materials are listed for reference. Certainly, we are in a good position to rejoice over the gradual progress of multiferroic research over the years, substantial efforts along the imaginary upright arrows are needed, to be presented below.

\subsection{Novel multiferroics on the way}
In fact, the pursuit of new multiferroics is still on the way. For applications, the desired properties of a ``good" multiferroic material include: 1) high serving temperatures; 2) considerable magnetization and polarization; 3) strong coupling and cross-control between magnetization and polarization. The available materials up to date do not satisfy all these requirements. Nevertheless, there have been indeed some incipient and fanciful efforts devoted to searching for new multiferroics with better performance or new physics, three of which will be highlighted as examples.

\subsubsection{BaFe$_2$Se$_3$: crossover with superconductor family}
Iron-based pnictides and chalcogenides have become new hot-spots in the condensed matter community since 2008, because of their superconducting properties \cite{Johnston:Ap,Dagotto:Rmp}. However, no any in the two families of materials has been found to be multiferroic until recently when Dong \textit{et al.} predicted that iron-selenide BaFe$_2$Se$_3$ is a magnetic ferrielectric with attractive multiferroic performance \cite{Dong:PRL14}.

\begin{figure}
\centering
\includegraphics[width=\textwidth]{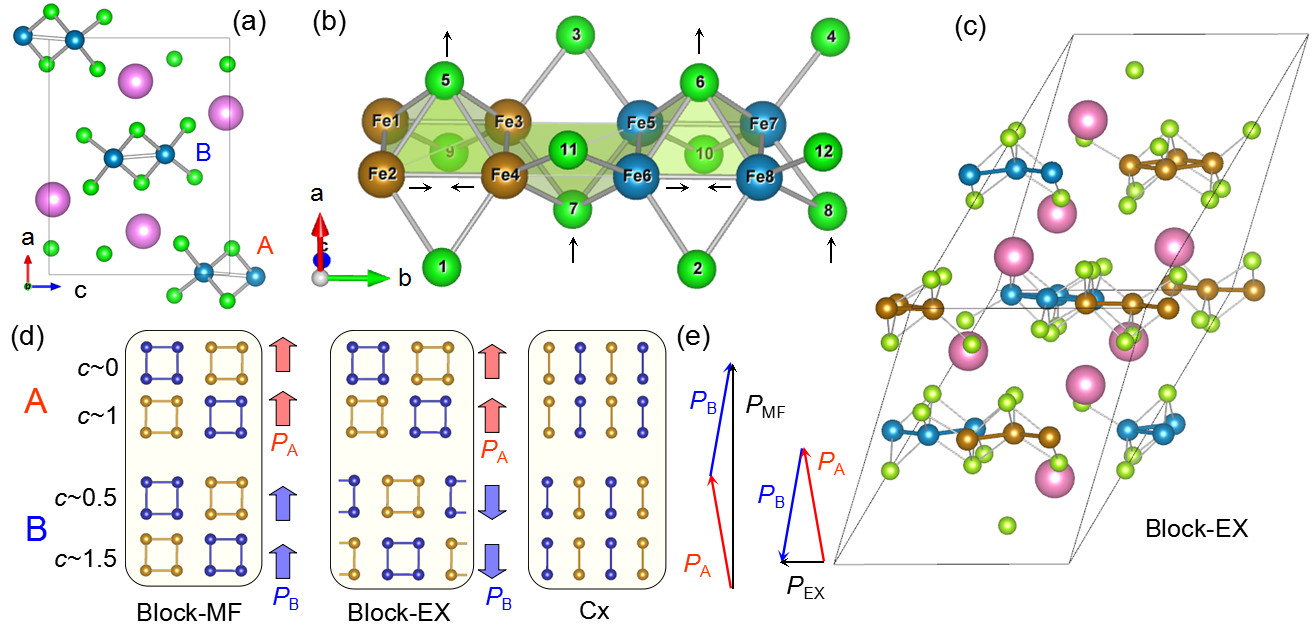}
\caption{(Colour online) The crystal and magnetic structures of BaFe$_2$Se$_3$. (a) A side view along the $b$-axis. Blue: Fe; green: Se; pink: Ba. (b) A Fe-Se ladder along the $b$ axis and its magnetic order. The partial ionic displacements driven by the exchange striction are marked by black arrows. (c) A unit cell considering the AFM order. (d) The spin structures. Left: Block-MF; middle: Block-EX; right: Cx. The side arrows denote the local dipole of each ladder. In (b-d), the Fe spins ($\uparrow$ and $\downarrow$) are distinguished by colors. (e) A vector addition of the dipoles of ladders A and B. Reprinted figure with permission from \href{http://dx.doi.org/10.1103/PhysRevLett.113.187204}{S. Dong \textit{et al.}, Physical Review Letters, 111, p. 187204, 2014} \cite{Dong:PRL14} Copyright \copyright (2014) by the American Physical Society.}
\label{BaFeSe}
\end{figure}

The crystal structure of BaFe$_2$Se$_3$ is shown in Fig.~\ref{BaFeSe}(a-c), which has two iron ladders (labeled as A and B) in each unit. The long-range block-type antiferromagnetic order (Fig.~\ref{BaFeSe}(b) and (d)) is established below $256$ K, confirmed by both neutron studies and first-principles calculations \cite{Caron:Prb,Medvedev:Jl}.

The block antiferromagnetic order is a variant of $\uparrow\uparrow\downarrow\downarrow$ spin chain. Therefore it induces exchange striction. Indeed, as revealed by neutron studies, the iron displacements are prominent: the nearest-neighbor distance between Fe($\uparrow$)-Fe($\uparrow$) [or Fe($\downarrow$)-Fe($\downarrow$)] at $200$ K becomes $2.593$ \AA{}, shorter than the Fe($\uparrow$)-Fe($\downarrow$) distance of $2.840$ \AA{} \cite{Caron:Prb}. As shown in Fig.~\ref{BaFeSe}(b), the Se(5) and Se(7) heights do not need to be antisymmetric with respect to the Fe ladder plane anymore. The same mechanism works for the edge Se, e.g. Se(1) and Se(11). As a consequence, the Se atomic positions break the space inversion symmetry, generating a local electric dipole aligned perpendicular to the iron ladders plane \cite{Dong:PRL14}.

According to neutron studies, the block antiferromagnetic pattern shows a $\pi/2$-phase shift between the nearest-neighbor A-B ladders but a $\pi$-phase shift between the nearest-neighbor A-A ladders (and the nearest-neighbor B-B ladders), as in the Block-EX shown in Fig.~\ref{BaFeSe}(d). The $\pi/2$-phase shift between A-B ladders will induce (nearly) opposite dipoles, as sketched in Fig.~\ref{BaFeSe}(d-e). A full cancellation does not occur due to a small canting angle between the ladders A and B planes (Fig.~\ref{BaFeSe}(a)), leading to a residual ferrielectric polarization ($\textbf{P}_{\rm EX}$) pointing roughly along the $c$-axis (Fig.~\ref{BaFeSe}(e)).

If the $\pi/2$-phase shift between the ladders A and B is eliminated, the magnetic structure prefers the Block-MF state. Then, the magnetism-induced electric dipoles of all ladders will coherently produce a combined polarization $\textbf{P}_{\rm MF}$ pointing along the $a$-axis (Fig.~\ref{BaFeSe}(e)).

In short, this Block-EX state may induce a ferrielectricity, with a net $\textbf{P}_{\rm EX}$ pointing along the $c$-axis, while the Block-MF state induces ferroelectricity with an order of magnitude larger polarization $\textbf{P}_{\rm MF}$ along the $a$-axis. As confirmed by the density functional calculations, both the Block-EX and Block-MF are multiferroic. $\textbf{P}_{\rm MF}$ is large and along the $a$-axis ($\sim2-3$ $\mu$C/cm$^2$). The net polarization of the Block-EX is along the $c$-axis and its magnitude is about $0.19$ $\mu$C/cm$^2$ (obtained from the pure GGA calculation), which is one order of magnitude smaller than $\textbf{P}_{\rm MF}$, as expected \cite{Dong:PRL14}.

\subsubsection{High temperature ferrimagnetic ferroelectrics by design}
Benefitted from density functional calculations which are becoming powerful, new to be synthesized materials with promising properties, can be ``designed" by computers. A recent example is on multiferroic double-perovskites.

In 2011, Le\v{z}ai\'{c} and Spaldin predicted that $3d$-$5d$ double-perovskite Bi$_2$NiReO$_6$ and Bi$_2$MnReO$_6$ are room-temperature ferrimagnetic ferroelectrics (strain-driven needed) \cite{Lezaic:Prb}. The design principle is intuitively based on two ingredients: 1) Bi$^{3+}$ is a highly polarizable ion for proper ferroelectricity; 2) the difference between $3d$ and $5d$ magnetic moments can give rise to a considerable magnetization even if they are antiparallel.

Along this way, Wang \textit{et al.} predicted a ferrimagnetic-ferroelectric phase of Zn$_2$FeOsO$_6$ with strong magnetoelectric coupling up to room temperature \cite{PSWang:Prl}. The genetic algorithm \cite{Lu:Cms} generated a polar ground state (space group $R3$). Using the coefficients extracted from the density functional calculations, Monte Carlo simulation and molecular dynamics simulations suggested the ferroelectric and ferrimagnetic transitions above room temperature. Due to the strong spin-orbit coupling of $5d$ orbitals, a strong correlation between polarization and magnetic easy axis is expected, as shown in BiFeO$_3$. In summary, this is a BiFeO$_3$-like material with multiferroic properties probably superior to BiFeO$_3$ \cite{PSWang:Prl}.

The similar prediction based on the density functional calculations was also done for $R_2$NiMnO$_6$/La$_2$NiMnO$_6$ superlattices \cite{Zhao:Nc}. The near room temperature ferromagnetism due to the strong Ni-Mn exchanges was predicted, and a hybrid improper ferroelectric polarization driven by the geometric mechanism as reviewed before is believed.

\subsubsection{Ferromagnetic \& ferroelectric titanates}
The perovskite titanate family is attractive because they can be either ferroelectric (e.g. BaTiO$_3$) or ferromagnetic (e.g. YTiO$_3$ \cite{Mochizuki:Njp}), depending on the valence of Ti. The $d^0$ state is likely to be ferroelectric while the $d^1$ state can be ferromagnetic or antiferromagnetic.

By properly designing a [001]-oriented ($A^{2+}$TiO$_3$)$_1$/($R^{3+}$TiO$_3$)$_1$ superlattice, a ferromagnetic $+$ ferroelectric titanate was predicted \cite{Bristowe:Nc}. The ferroelectricity, as reviewed in Sec. 3.3.2 and Sec. 4.3, originates from the improper structural-ferroelectric trilinear coupling, as first proposed for SrTiO$_3$/PbTiO$_3$ superlattice \cite{Bousquet:Nat} and further applied to $A_3M_2$O$_7$ system \cite{Benedek:Prl} and $AB$O3/$A'B'$O3 superlattices \cite{Rondinelli:Am12}. The entangled charge ordering (Ti$^{3+}$+Ti$^{+4}$) and orbital ordering are crucial to determine the ferromagnetism and insulating behavior. Thus, this system turns to be ferromagnetic and ferroelectric \cite{Bristowe:Nc}.

\subsection{New mechanisms needed}
So far there have been a number of multiferroic materials whose ferroelectricity should be magnetism-driven. However, the magnetic origins of ferroelectricity in these multiferroics seem to be beyond currently known mechanisms. Examples include those two dimensional triangular lattice antiferromagnets with $120^\circ$ noncollinear spin order (e.g. CuCrO$_2$ and related materials \cite{Seki:Prl08,Kimura:Prb08}, RbFe(MoO$_4$)$_2$ \cite{Kenzelmann:Prl07}, Ba$_3$NiNb$_2$O$_9$ and related materials \cite{Hwang:Prl,Lee:Prb14.2}), and recently discovered LaMn$_3$Cr$_4$O$_{12}$ with the collinear spin order \cite{Wang:Prl15}. Although the multiferroicity of these materials can be covered with the phenomenological theory framework, the decent microscopic mechanisms remain to be accessible. The unified model proposed by Xiang \textit{et al.} \cite{Xiang:Prl11,Xiang:Prb13} can be a hint but not an end to reveal the microscopic mechanisms in the quantum level.

Besides, a number of issues regarding the multiferroicity of $R$Mn$_2$O$_5$ remain unsolved, although they were discovered ten years ago. The latest experimental findings, including the large improper ferroelectric polarization \cite{Lee:Prl13} and room-temperature ferroelectricity which is magnetism-irrelevant \cite{Baledent:Prl}, makes this family of multiferroics more attractive but still mysterious.

In addition, our understanding of the domain wall vortex in hexagonal $R$MnO$_3$ and magnetoelectricity at domain walls is far from comprehensive. A realistic microscopic theory is not yet available in spite of several simplified phenomenological models reported.

\subsection{New territories to explore}
In general, the magnetoelectricity is not necessarily related to the multiferrocity, and vice versa. Some magnets, like Cr$_2$O$_3$, show linear magnetoelectric response but they do not own spontaneous ferroelectric polarizations (thus not multiferroics). This linear magnetoelectric response also occurs in magnetic monopole systems. Thus, magnetoelectricity can be a concept extendable to broader territories and cover more materials. For example, skyrmions and topological-band materials also exhibit the magnetoelectric effects.

Furthermore, even for traditional multiferroics, the magnetoelectric physics is being updated, going far beyond pure magnetism plus ferroelectricity. More degrees of freedom are found to participate in the magnetoelectric phenomena. For example, the orbital physics, which was once emphasized in colossal magnetoresistive manganites \cite{Tokura:Sci,Hotta:Rpp}, are found to play important role in recently discovered multiferroics, e.g. CaMn$_7$O$_{12}$ \cite{Perks:Nc,Du:Prb} and ($A^{2+}$TiO$_3$)$_1$/($R^{3+}$TiO$_3$)$_1$ superlattice \cite{Bristowe:Nc}. Even for the simplest Ising-chain type multiferroic Ca$_3$CoMnO$_6$, the spin-crossover, namely the transition between high-spin state and low-spin state of Co$^{2+}$, involves a subtle balance of the orbital degrees of freedom \cite{Fina:Prl}. For the E-type antiferromagnetic manganites, the orbital order can twist with ferroelectricity and magnetism \cite{Barone:Prl}. Besides the charge and spin degrees of freedom which contribute to magnetoelectricity via their correlation with lattice, it is yet an interesting issue whether the orbital degree of freedom can make a deterministic/direct contribution to ferroelectricity/magnetoelectricity generation. In addition, other important concepts in condensed matter physics and thermodynamics, such as quantum phase transitions, quantum fluctuations, and critical slowing-down etc, are finding their figures in this exciting discipline \cite{Niermann:Prl15,Kim:Nc}.

Besides, the significance of some non-ferroelectric lattice distortion modes, as glue to bridge magnetism and ferroelectric polarization in a number of multiferroics including hexagonal $R$MnO$_3$, $R$FeO$_3$, $A_3M_2$O$_7$, and $AB$O$_3$/$A'B'$O$_3$ superlattices, is under investigations.

Traditionally, metallicity contradicts ferroelectricity. However, the latest studies have predicted and found some exotic metallic ferroelectric materials, e.g. LiOsO$_3$ \cite{Shi:Nm}. The underlying physics for this metallic ferroelectricity \cite{Puggioni:Nc} does not exclude magnetism, rendering metallic multiferroics to be a topic for exploration, as discussed recently in magnetic superlattices \cite{Zhang:Prb15}. In fact, according to the first-principles calculation, LiOsO$_3$ locates at the edge between metallic paramagnetism and Mott-insulating antiferromagnetism. Tiny modifications of LiOsO$_3$ can render it to be a Mott multiferroic \cite{Puggioni:Prl}.

One more new area for multiferroic materials is green energy harvesting. Some multiferroics, like BiFeO$_3$, are good ferroelectrics and proper semiconducting absorbers for solar energy, which may be used as ferroelectric photovoltaics and photocatalyst \cite{Choi:Sci,Yang:Nn,Alexe:Nc,Guo:Nc,Nechache:Nph,Gao:Am}. A recent theoretical calculation predicted hexagonal manganites to be promising photovoltaics \cite{Huang:Prb}. The photoferroelectricity, although not directly related to magnetism in the current stage, crosses with multiferroicity in physics and broadens the perspective of applications.

Another relevant subtopic is the physics of defects and domain walls in multiferroics. In general, defects may create impurity levels in the forbidden band, tune the local structures, and pin the domain walls in ferroic materials. The emergent physical properties, including magnetoelectricity here, may be altered around the domain walls. The state-of-the-art experimental techniques can now detect the magnetoelectric domain in typical type-II multiferroics whose ferroelectric signals are faintly weak, helpful for extending the domain wall nanoelectronics \cite{Matsubara:Sci}. The defect enrichment at ferroelectric domain walls has been fully confirmed \cite{Becher:Nn,Farokhipoor:Prl}, while new phases with different chemical components at the structural domain walls can be created by defect engineering for TbMnO$_3$ films deposited on SrTiO$_3$ \cite{Farokhipoor:Nat}.

Recently, Shang \textit{et al.} proposed that the two-terminal model device made of magnetoelectric material is one of the fourth fundamental circuit elements, which completes the relational graph together with resistor, capacitor, and inductor \cite{Shang:Cpb}. Previously, a memristor was proposed as the fourth element \cite{Chua:Ieee}, and later achieved in experiments \cite{Strukov:Nat}. However, the emergence of magnetoelectric materials may change this scenario. A magnetoelectric system can establish a direct correspondence between charge and magnetic flux, while a memristor cannot. Such a conceptual revision, not only re-emphasizes the fundamental physics of magnetoelectricity, but also initializes exploration of magnetoelectrics as circuit devices.

In short, a number of new issues and challenges have been brought into the multiferroic community. The magnetoelectricity is becoming a highly inter-discipline territory in which new concepts, new phenomena, and new physics are emerging.

\subsection{New applications}
Naturally, one of the ultimate aims of research on multiferroicity and magnetoelectricity is the application of multiferroic materials. With no doubt, researches on multiferroics are moving towards applications-oriented targets. While some magnetoelectric composite materials have already found extensive territories in which many prototype devices and integrated systems based on the strong magnetoelectric coupling effects are under development, e.g. magnetoelectric transducers, sensors, actuators, and microwave active/passive components etc \cite{Wang:Mt}. A number of epitaxial ferromagnetic plus ferroelectric heterostructures, multilayers, and superlattices have been prepared and their application potentials as multi-state field effect transistors, magnetic sensors, and tunneling junctions among many others, are under investigations \cite{Tsymbal:Mrs}.

Definitely not beyond our grasp, BiFeO$_3$ among various single phase multiferroics is on the verge of practical applications for its prominent ferroelectricity and environmentally friendly property. It is also a ferroelectric semiconductor and ferroelectric photovoltaic/photocatalytic candidate. Besides, the complex but well-controllable domain structures in BiFeO$_3$ thin films present a number of application potentials not only in domain electronics. Hexagonal manganites/ferrites ($R$MnO$_3$/$R$FeO$_3$) may play similar roles considering their high ferroelectric critical temperatures, moderate polarizations, and narrow band gaps.

For other single phase multiferroics and heterostructures, major interests are currently onto fundamental aspects more than application sides. While the low ferroic-ordering temperatures and weak electric/magnetic signals are disadvantageous in terms of immediate application targeting, the established framework of multiferroic physics and plenty of magnetoelectric phenomena do provide an innovative integration of knowledge on designing/fabricating high performance materials for elevated temperature magnetoelectric devices. It is indubitability that the continuous advantage of this lifeful discipline is looking forward to breakthroughs in near future.

\section*{Acknowledgment \& Funding}
We thank E. Dagotto, Y.H. Chu, X.G. Li, H.J. Xiang, Y.S. Chai, P. Yu among many others for helpful discussions. This work was partially supported by the National Natural Science Foundation of China (Grant Nos. 51322206, 11234005, 11274060, 51431006, 51332006), the National 973 Projects of China (Grant No. 2015CB654602), the Jiangsu Key Laboratory for Advanced Metallic Materials (Grant No. BM2007204). S.W.C. is supported in part by the Department of Energy under Grant No. DE-FG02-07ER46382, and also by the Visiting Distinguished Professorship of Nanjing University sponsored by the State Administration of Foreign Experts Affairs of China. The work performed at the University of Houston is funded the Department of Energy, Basic Energy Science, under Grant Nos. DE-FG02-13ER46917 / DE-SC0010831.

\bibliographystyle{tADP}
\bibliography{../../ref}

\end{document}